\documentstyle[12pt,newcomh,epsfig]{ansgar}

\newsavebox{\blobdz}
\newsavebox{\blobza}
\newsavebox{\Vt}
\newsavebox{\Vr}
\newsavebox{\St}
\newsavebox{\Sr}
\newsavebox{\Fr}
\newsavebox{\Fl}
\newsavebox{\Ft}
\newsavebox{\Vtr}
\newsavebox{\Vbr}
\newsavebox{\Str}
\newsavebox{\Sbr}
\newsavebox{\Ftr}
\newsavebox{\Fbr}
\newsavebox{\Vtbr}
\newsavebox{\Ftbr}
\newsavebox{\Stbr}
\newsavebox{\Gr}
\newsavebox{\Gtbr}
\newsavebox{\Lt}
\newsavebox{\Lr}
\newsavebox{\Ltr}
\newsavebox{\Lbr}
\newsavebox{\Lb}
\newsavebox{\Lbl}
\newsavebox{\Ltl}
\newsavebox{\Vp}
 
\begin{document}
                   
\pagestyle{empty}

\title{Techniques for the calculation of electroweak \\[1ex]
radiative corrections at the one-loop level \\[1ex]
and results for $W$-physics at LEP200} 
\subtitle{}
 
\author{{\sc A.~Denner}
\\[1ex]
{\it Physikalisches Institut,
Universit\"at W\"urzburg} \\[-0.1ex]
{\it 8700 W\"urzburg, Germany}}
 
\date{\today}
 
\nopagebreak
 
\abstract{
We review the techniques necessary for the calculation of virtual 
electroweak and soft photonic corrections at the one-loop level. 
In particular we describe renormalization, calculation of one-loop 
integrals and evaluation of one-loop Feynman amplitudes. We summarize 
many explicit results of general relevance. We give 
the Feynman rules and 
the explicit form of the counter terms of the electroweak standard model,
we list analytical expressions for scalar one-loop integrals and reduction 
of tensor integrals,  we present the decomposition of the invariant matrix 
element 
for processes with two external fermions and we give the analytic form of soft 
photonic corrections. These techniques are applied to physical processes 
with external $W$-bosons. We present 
the full set of analytical formulae and 
the corresponding numerical results for the decay width of the 
$W$-boson and the top quark. We discuss the cross section for the 
production of $W$-bosons in $e^{+}e^{-}$-annihilaton including all 
$O(\alpha )$ radiative corrections and finite width effects. 
Improved Born approximations for these processes are given.}
 
\maketitle

\tableofcontents
 
\clearpage
 
\pagestyle{plain}
 
\setcounter{page}{1}
 
\pagestyle{plain}
 
\setcounter{page}{1}
 
\chapter{Introduction}
 
All known experimental facts about the electroweak interaction are in agreement
with the Glashow-Salam-Weinberg (GSW) model \cite{Gl61,We67,Sa68,Gl70}.
Therefore, this theory is called the standard model (SM) of electroweak physics.
Despite its extraordinary experimental success it is by no means tested
in its full scope. Many more experimental and
theoretical efforts are needed for its further confirmation.
 
An important step in this direction is provided by the $e^{+}e^{-}$
colliders SLC and LEP100 which started a new era of precision
experiments. The first important results from these experiments were the
determination of the number of light neutrinos and the
precise measurement of the mass of the neutral weak gauge boson, the
$Z$-boson \cite{Dy90}. Furthermore the total and partial widths of the
$Z$-boson and various on-resonance asymmetries have been determined and will
be measured with increasing accuracy. These experiments will
uniquely allow to study in great detail all the properties of the
$Z$-boson and its couplings to fermions.
 
There are, however, ingredients of the electroweak SM, which are
not directly accessible at SLC and LEP100. The most important one is
probably the gauge boson self-interaction which is crucial for the
nonabelian structure of the GSW model. It will be
directly tested for the first time at LEP200, the upgraded version of LEP. There
the center of mass energy will be high enough to produce pairs of
charged weak gauge bosons, the $W$-bosons, such that one can study the
reaction $e^{+}e^{-}\to W^{+}W^{-}$ in great detail. It will allow
the investigation of the nonabelian three-gauge boson interactions
$\gamma W^{+}W
^{-}$ and $ZW^{+}W^{-}$ at the classical level of the theory. Moreover,
all the properties of the $W$-boson, like its mass and its total and
partial widths can be measured directly there. The statistics
will not be as good as on the $Z$-peak. One expects of the order of $10^{4}$
$W$-pairs and thus an accuracy at the percent level. The
examination of several independent methods indicates that an error
of about 0.1\% for the $W$-mass determination can be reached \cite{Da87}.
 
Theoretical predictions should have an accuracy comparable to or
even better than the experimental errors. If the experimental
precision is of the order of one percent the classical level of the theory is
no longer sufficient. One is forced to take into account quantum
corrections: the radiative corrections. In the case of the electroweak
SM these can reach several percent. For the high precision
experiments at LEP100 even the first order corrections are
inadequate, one has to take into account leading higher order
corrections, too.
 
Radiative corrections are not only compelling for the precise comparison
between the theoretical predictions and the experimental results, but
offer the possibility to get informations about sectors of the theory
that are not directly observable. While the direct investigation of
certain objects may not be possible because the available energy is too
small to produce them they may affect the radiative corrections
noticeably.
 
In the electroweak SM there are at least two such objects.
The top quark, the still undiscovered constituent of the third fermion
generation,
and the Higgs boson, the physical remnant of the Higgs-Kibble
mechanism of spontaneous symmetry breaking. Both particles seem to be too
massive to be produced directly in the existing colliders. However, the
high precision experiments performed so far together with the precise knowledge
of the radiative corrections of the electroweak SM already allow
to derive limits on the mass of the top quark within the SM
\cite{Ve77,Dy90}.
Since the sensitivity of radiative corrections to the
mass of the Higgs boson is weaker, the restrictions on this
parameter are at present only marginal \cite{El90}. The situation may
improve with increasing experimental accuracy. While direct
determinations of physical parameters
are in general to a large extent model independent, the
information extracted from radiative corrections depends
on the entire structure of the underlying theory.
 
Finally there is a third important issue concerning radiative
corrections. It is likely that the electroweak SM, despite
its experimental success, is only an effective theory, the low-energy
approximation of a more general structure. This would
manifest itself typically in small deviations from the
SM predictions. Furthermore most of the presently discussed new physics is
connected with scales bigger than the experimentally accessible
energies. Therefore new phenomena will show up predominantly via indirect
effects rather than via direct production of new particles.
In order to disentangle these
small effects one has to know once again the predictions of the SM
accurately and thus needs radiative corrections.
 
The actual evaluation of the radiative corrections is a
tedious and time consuming task.
It requires extensive calculations involving many different techniques, like
renormalization, evaluation of loop integrals, Dirac algebra
calculations, phase space integrations and so on. Fortunately the whole
procedure can be organized into different independent steps.
Furthermore many steps can be facilitated with
the help of computer algebra \cite{Ye90,Il90,Ku90,Me91}.
 
For the interesting processes at LEP100 radiative corrections have been
calculated by many authors \cite{LEP89}. Their structure is
relatively simple since
the masses of the external fermions can be neglected. Calculations for
gauge boson production processes at LEP200 are already more
complicated because the masses of the external gauge bosons are nonnegligible.
Such calculations have been performed by several groups and we will give
the most important results in the second part of this review.
The whole complexity of one-loop corrections will show up when considering
reactions where all external particles are massive like e.g.\ gauge boson
scattering processes which may be investigated at the LHC or SSC. The
calculation of radiative corrections to these processes has just started.
 
In the first part of this review we collect the relevant formulae and
techniques necessary for the calculation of electroweak one-loop
radiative corrections. Although we discuss everything in the context of
the SM the presented material is --  apart from the explicit form of the
renormalization
constants -- applicable to extended models as well.
In the second part these methods are applied to physical processes
with external $W$-bosons. This part not only gives
examples for the calculation of one-loop electroweak corrections,
but also provides a survey on the status of radiative corrections for
the production and the decay of $W$-pairs in $e^{+}e^{-}$ annihilation.
The corresponding experiments will be carried through in a few years at
LEP200.
 
The general techniques described in this paper are restricted to the virtual
part of the electroweak corrections and soft photon bremsstrahlung.
We do not consider
the methods appropriate for hard photon bremsstrahlung. This can be
efficiently treated using spinor techniques \cite{Be91} and Monte Carlo
simulations \cite{Kl89}. Furthermore we do not touch the methods
developed for calculating higher order QCD corrections.
 
This paper is organized as follows:
 
In chapter 2 we specify the Lagrangian of the electroweak SM. Chapter 3
outlines the on-shell
renormalization for the physical sector of the electroweak SM
and provides explicit expressions for the counter terms. All
relevant formulae for the calculation of one-loop Feynman integrals are
collected in chapter 4. In chapter 5 we introduce the standard matrix
elements, a concept which allows to represent the results for one-loop
diagrams in a systematic and simple way. In chapter 6 we show how
everything is put together in the actual calculation of one-loop
amplitudes and provide first simple examples. The relevant formulae for
the calculation of the soft photon corrections are summarized in
chapter 7. Chapter 8 serves to define our input parameters and the way
of resumming higher order corrections.
 
The remaining chapters are devoted to applications. In chapter 9 we give
results for the width of the $W$-boson, in chapter 10 for the width of
the top quark.
Finally the radiative corrections to the production of $W$-pairs in
$e^{+}e^{-}$ annihilation are discussed in chapter 11.
 
The appendices contain the Feynman rules of the electroweak SM,
the explicit expressions for the self energies of the physical
particles and the vertex
functions as well as the bremsstrahlung integrals relevant for the
$W$-boson and top quark decay width.

\chapter{The Glashow-Salam-Weinberg Model}
\label{chaSM}
 
The Glashow-Salam-Weinberg (GSW) model  of the electroweak interaction
has been proposed by Glashow \cite{Gl61}, Weinberg \cite{We67},
and Salam \cite{Sa68} for
leptons and extended to the hadronic degrees of freedom by Glashow,
Iliopoulos and Maiani \cite{Gl70}.
It is the presently most comprehensive formulation of a theory of
the unified electroweak interaction: theoretically consistent and in
agreement with all experimentally known phenomena of electroweak origin. For
energies that are small compared to the electroweak scale it reproduces quantum
electrodynamics and the Fermi
model, which already accomplished a good description of the
electromagnetic and weak interactions
at low energies. It is minimal in the sense that it contains the
smallest number of degrees of freedom necessary to describe the known
experimental facts.
 
The electroweak standard model (SM) is a nonabelian gauge theory based on
the non-simple group $SU(2)_{W} \times U(1)_{Y}$.
From experiment we know that three out of the four
associated gauge bosons have to be massive. This is implemented via the
Higgs-Kibble mechanism \cite{Hi64}.
By introducing a scalar field with nonvanishing
vacuum expectation value the $SU(2)_{W} \times U(1)_{Y}$ gauge symmetry is
spontaneously broken in such a way that invariance under the electromagnetic
subgroup $U(1)_{em}$ is preserved. The SM is chiral since right- and left-handed
fermions transform according to different representations of the gauge group.
Consequently fermion masses are forbidden in the symmetric theory.
They are generated through spontaneous symmetry breaking from the
Yukawa couplings. Diagonalization of the fermion mass matrices introduces the
quark mixing matrix in the quark sector. This can give rise to CP-violation.
Fermions appear in generations. The model does not fix their number, but from
experiment we know that there are exactly three with light neutrinos
\cite{Dy90}.
 
The SM is a consistent quantum field theory.
It is renormalizable, as was proven by 't Hooft \cite{tH71}, and free
of anomalies. Therefore it allows to calculate unique quantum corrections.
Given a finite set of input parameters measurable quantities can be
predicted order by order in perturbation theory.
 
The classical Lagrangian ${\cal L}_{C}$ of the SM is composed of a
Yang-Mills, a Higgs and a fermion part
\beq
{\cal L}_{C} = {\cal L}_{YM} + {\cal L}_{H} + {\cal L}_{F} \; .
\label{LC}
\eeq
Each of them is separately gauge invariant. They are specified as
follows:
 
\section{The Yang-Mills-part}
 
The gauge fields are four vector fields transforming according to the
adjoint representation of the gauge group $SU(2)_{W} \times U(1)_{Y}$.
The isotriplet $W_{\mu }^{a}$, $a=1,2,3$ is associated with the
generators $I_{W}^{a}$ of the weak isospin group $SU(2)_{W}$, the isosinglet
$B_{\mu }$ with the weak hypercharge $Y_{W}$ of the group $U(1)_{Y}$.
The pure gauge field Lagrangian reads
\begin{equation}
{\cal L}_{YM} =  -\frac{1}{4} \left( \partial_{\mu
}W^{a}_{\nu } - \partial_{\nu }W^{a}_{\mu } + g_{2} \varepsilon ^{abc}
W^{b}_{\mu }W^{c}_{\nu } \right)^{2}
- \frac{1}{4} \left( \partial_{\mu }B_{\nu } - \partial_{\nu }
B_{\mu } \right)^{2} ,
\end{equation}
where $\varepsilon ^{abc}$ are the totally antisymmetric structure constants
of $SU(2)$. Since the gauge group is non-simple there are two gauge coupling
constants, the $SU(2)_{W}$ gauge coupling $g_{2}$ and the $U(1)_{Y}$ gauge
coupling $g_{1}$. The covariant derivative is given by
\begin{equation} \label{covder}
D_{\mu } = \partial_{\mu } -i g_{2} I_{W}^{a} W^{a}_{\mu }
+ i g_{1} \frac{Y_{W}}{2} B_{\mu }.
\end{equation}
 
The electric charge operator $Q$ is composed of the weak isospin generator
$I_{W}^{3}$ and the weak hypercharge according to the Gell-Mann Nishijima
relation
\beq
Q=I_{W}^{3}+\frac{Y_{W}}{2}.
\label{GMN}
\eeq
 
\section{The Higgs part}
 
The minimal Higgs sector consists of a single complex scalar $SU(2)_{W}$
doublet field with hypercharge $Y_{W} = 1$
\begin{equation}
\Phi(x)  = \left( \begin{array}{c} \phi^{+}(x) \\ \phi^{0}(x)
\end{array} \right) .
\end{equation}
It is coupled to the gauge fields with the covariant derivative
(\ref{covder}) and has a self coupling resulting in the Lagrangian
\begin{equation}
{\cal L}_{H} = \left( D_{\mu }\Phi\right)  ^{\dagger} \left( D^{\mu }\Phi
\right) -V(\Phi).
\end{equation}
The Higgs potential
\beq
V(\Phi) = \frac{\lambda }{4} \left(\Phi^{\dagger}\Phi\right)^{2}
- \mu ^{2} \Phi^{\dagger}\Phi
\eeq
is constructed in such a way that it gives rise to spontaneous symmetry
breaking. This means that the parameters $\lambda $ and $\mu $ are chosen
such that the potential $V(\Phi)$ takes its minimum for a
nonvanishing Higgs field, i.e.\ the vacuum expectation value $\langle
\Phi \rangle$ of the Higgs field is nonzero.
 
\section{ Fermionic Part}
 
The left-handed fermions of each lepton ($L$) and quark ($Q$)
generation are grouped
into $SU(2)_{W}$ doublets (we suppress the colour index)
\beq
{L'}_{j}^{L}=\omega _{-}  {L'}_{j} =
\left( \barr{l} {\nu'}_{j}^{L} \\ {l'}_{j}^{L} \earr \right) , \quad
{Q'}_{j}^{L}=\omega _{-}  {Q'}_{j} =
\left( \barr{l} {u'}_{j}^{L} \\ {d'}_{j}^{L} \earr \right) ,
\eeq
the right-handed fermions into singlets
\beq
{l'}_{j}^{R}=\omega _{+}  {l'}_{j}, \quad
{u'}_{j}^{R}=\omega _{+}  {u'}_{j}, \quad
{d'}_{j}^{R}=\omega _{+}  {d'}_{j},
\eeq
where $\omega _{\pm}=\frac{1\pm\gamma _{5}}{2}$ is the projector on
right- and left-handed fields, respectively,
$j$ is the generation index and $\nu$, $l$, $u$ and $d$ stand for
neutrinos, charged leptons, up-type quarks and down-type quarks,
respectively.
The weak hypercharge of the right- and left-handed multiplets is chosen
such that the known electromagnetic charges of the fermions are
reproduced by the Gell-Mann-Nishijima relation (\ref{GMN}).
There are no right-handed neutrinos. These could be easily added, but
they would induce nonvanishing neutrino masses, which have not been observed
experimentally so far.
 
The fermionic part of the Lagrangian reads
\begin{equation}
\begin{array}{lll}
{\cal L}_{F} &=&
\displaystyle \sum_{i}\,\left(
 \overline{L'}_{i}^{L} i \gamma^{\mu } D_{\mu } {L'}_{i}^{L}
+\overline{Q'}_{i}^{L} i \gamma^{\mu } D_{\mu } {Q'}_{i}^{L} \right)\\[1em]
&&\displaystyle \mbox{} + \sum_{i}\,\left(
 \overline{l'}_{i}^{R} i \gamma^{\mu } D_{\mu } {l'}_{i}^{R}
+\overline{u'}_{i}^{R} i \gamma^{\mu } D_{\mu } {u'}_{i}^{R}
+\overline{d'}_{i}^{R} i \gamma^{\mu } D_{\mu } {d'}_{i}^{R}\right) \\[1em]
&&\displaystyle -\sum_{ij} \left(
 \overline{L'}_{i}^{L}G^{l}_{ij}{l'}_{j}^{R}\Phi
+\overline{Q'}_{i}^{L}G^{u}_{ij}{u'}_{j}^{R}\tilde{\Phi}
+\overline{Q'}_{i}^{L}G^{d}_{ij}{d'}_{j}^{R}\Phi
+ h.c. \right) .
\end{array}
\end{equation}
Note that in the covariant derivative $D_{\mu }$ acting on right-handed
fermions the term involving $g_{2}$ is absent, since they are
$SU(2)_{W}$ singlets.
The primed fermion fields are by definition eigenstates of the
electroweak gauge interaction, i.e.\ the covariant  derivatives are
diagonal in this basis with respect to the generation indices.
$G_{ij}^{l}$, $G_{ij}^{u}$ and $G_{ij}^{d}$ are the
Yukawa coupling matrices,
$\tilde{\Phi } = \left(\phi^{0*}, -\phi^{-} \right)^{T}$ is the charge
conjugated Higgs field and $\phi ^{-}=\left(\phi^{+}\right)^{*}$.
The $SU(2)_{W}\times U(1)_{Y}$ symmetry forbids explicit mass terms for
the fermions. The masses of the fermions are generated through the
Yukawa couplings
via spontaneous symmetry breaking.
 
\section{Physical fields and parameters}
\label{phyfap}
 
The theory is constructed such that the
classical ground state of the scalar field satisfies
\begin{equation}
|\langle\Phi\rangle|^{2} = \frac{2\mu ^{2}}{\lambda } = \frac{v^{2}}{2}
\ne 0 \; .
\end{equation}
In perturbation theory one has to expand around the ground state.
Its phase is chosen such that the electromagnetic gauge invariance $U(1)_{em}$
is preserved and the Higgs field is written as
\begin{equation}
\Phi (x) = \left( \begin{array}{c}
\phi ^{+}(x) \\ \frac{1}{\sqrt{2}}\bigl(v + H(x) +i \chi (x) \bigr)
\end{array} \right)  ,
\label{higdec}
\end{equation}
where the components $\phi ^{+}$, $H$ and $\chi$ have zero vacuum expectation
values.
$\phi ^{+},\phi ^{-}$ and $\chi $ are unphysical degrees of freedom
and can be eliminated by a suitable gauge transformation. The gauge in
which they are absent is called unitary.  The field $H$ is the physical
Higgs field with mass
\begin{equation}
M_{H} = \sqrt{2} \mu.
\label{Hmass}
\end{equation}
Inserting (\ref{higdec}) into ${\cal L}_{C}$ the vacuum expectation
value $v$ introduces couplings with mass dimension and mass terms for
the gauge bosons and fermions.
 
The physical gauge boson and fermion fields are obtained by diagonalizing
the corresponding mass matrices
\begin{equation}
\begin{array}{cll}
W_{\mu }^{\pm} & = & \disp \frac{1}{\sqrt{2}}
\left( W_{\mu }^{1} \mp i W_{\mu }^{2} \right) , \\[1em]
\left(\barr{l} Z_{\mu } \\ A_{\mu } \earr \right) & = &
\left(\barr{rr} c_{W} & s_{W} \\ -s_{W} & c_{W}  \earr \right)
\left(\barr{l} W_{\mu }^{3} \\ B_{\mu } \earr \right) , \\[2em]
f^{L}_{i} & = & U_{ik}^{f,L}{f'}_{k}^{L} , \\[1em]
f^{R}_{i} & = & U_{ik}^{f,R}{f'}_{k}^{R} ,
\end{array}
\end{equation}
where \beq
c_{W}=\cos\theta_{W}=\frac{g_{2}}{\sqrt{g_{2}^{2}+g_{1}^{2}}},
\quad s_{W}=\sin\theta_{W} ,
\eeq
with the weak mixing angle $\theta _{W}$
and $f$ stands for $\nu $, $l$, $u$ or $d$.
The resulting masses are
\begin{equation}
\begin{array}{llll}
M_{W} &\disp = \frac{1}{2}g_{2}v,  &
\quad M_{Z} &\disp = \frac{1}{2}\sqrt{g_{1}^{2}+g_{2}^{2}} \; v, \\[1em]
M_{\gamma } & = 0 ,&
\quad m_{f,i} & =  U_{ik}^{f,L} G_{km}^{f} U_{mi}^{f,R\dagger}
\disp \frac{v}{\sqrt{2}} \, .
\end{array}
\label{SMmass}
\end{equation}
The neutrinos remain massless since the absence of the right-handed
neutrinos forbids the Yukawa couplings which would generate their
masses. With (\ref{SMmass}) we find for the weak mixing angle
\beq
c_{W} = \frac{M_{W}}{M_{Z}}.
\eeq
 
Identifying the coupling of the photon field $A_{\mu}$ to the electron
with the electrical charge $e=\sqrt{4\pi \alpha }$ yields
\begin{equation}
e = \frac{g_{1}g_{2}}{\sqrt{g_{1}^{2}+g_{2}^{2}}} \; ,
\label{eSM}
\end{equation}
or
\beq
g_{1}=\frac{e}{c_{W}}, \qquad g_{2}=\frac{e}{s_{W}}.
\eeq
The diagonalization of the fermion mass matrices introduces a
matrix into the quark-W-boson couplings, the unitary quark mixing
matrix
\beq
V_{ij}=U^{u,L}_{ik} U^{d,L\dagger}_{kj}  .
\label{QMM0}
\eeq
There is no corresponding matrix in the lepton sector. Since there is no
neutrino mass matrix, $U^{\nu,L}$ is completely arbitrary and can be
chosen such that it cancels $U^{l,L}$ in the lepton-W-boson couplings.
The same would also be true for the quark sector if all up-type or
down-type quarks would be degenerate in masses. For
degenerate masses one can choose
$U^{L} = U^{R\dagger}$ arbitrary without destroying the diagonality of
the corresponding mass matrix and thus eliminate $V_{ij}$.
 
The above relations
(\ref{Hmass}, \ref{SMmass}, \ref{eSM}, \ref{QMM0}) allow to replace
the original set of \mbox{parameters}
\beq \label{sympar}
g_{1},\;g_{2},\;\lambda, \;\mu ^{2}, \; G^{l},\; G^{u},\; G^{d}
\eeq
by the parameters
\beq \label{phypar}
e,\;M_{W},\;M_{Z},\;M_{H},\;m_{f,i},\;V_{ij}
\eeq
which have a direct physical meaning.
Thus we can express the Lagrangian (\ref{LC}) in terms of physical
parameters and fields.
 
Inserting (\ref{higdec}) into ${\cal L}_{C}$ generates a term linear in
the Higgs field $H$ which we denote by $t H(x)$ with
\beq
t = v(\mu^{2}-\frac{\lambda }{4}v^{2}).
\label{deft}
\eeq
The tadpole $t$ vanishes at lowest order due to the choice of $v$.
We use $t$ instead of
$v$ in the following. Choosing $v$ as the correct vacuum expectation
value of the Higgs field $\Phi$ is equivalent to the vanishing of $t$.
 
\section{Quantization}
 
Quantization of ${\cal L}_{C}$ and higher order calculations require the
specification of a gauge. We choose a renormalizable 't Hooft gauge with the
following linear gauge fixings
\begin{equation}
\begin{array}{ll}
F^{\pm} &\displaystyle = (\xi _{1}^{W})^{-\frac{1}{2}} \partial^{\mu }
W^{\pm}_{\mu } \mp i M_{W} (\xi _{2}^{W})^{\frac{1}{2}} \phi ^{\pm} , \\[1em]
F^{Z}&\displaystyle = (\xi _{1}^{Z})^{-\frac{1}{2}} \partial ^{\mu }Z_{\mu }
- M_{Z} (\xi _{2}^{Z})^{\frac{1}{2}} \chi , \\[1em]
F^{\gamma } &\displaystyle = (\xi _{1}^{\gamma })^{-\frac{1}{2}}
\partial^{\mu } A_{\mu },
\end{array}
\end{equation}
leading to the following gauge fixing Lagrangian
\begin{equation}
{\cal L}_{fix} = -\frac{1}{2} \left[ (F^{\gamma })^{2} + (F^{Z})^{2} + 2
F^{+} F^{-} \right] .
\end{equation}
${\cal L}_{fix}$ involves the unphysical components of the gauge fields.
In order to compensate their effects one introduces Faddeev
Popov ghosts $u^{\alpha}(x)$, $\bar{u}^{\alpha}(x)$
($\alpha =\pm,\gamma ,Z$) with the Lagrangian
\begin{equation}
{\cal L}_{FP} = \bar{u}^{\alpha }(x) \frac{\delta F^{\alpha }}{\delta \theta
^{\beta }(x)} u^{\beta }(x).
\end{equation}
$\disp\frac{\delta F^{\alpha }}{\delta \theta^{\beta }(x)}$ is the variation of
the gauge fixing operators $F^{\alpha }$ under infinitesimal gauge
transformations characterized by $\theta ^{\beta}(x)$.
 
The 't Hooft Feynman gauge $\xi ^{\alpha } =1$ is particularly simple. At
lowest order the poles of the ghost fields, unphysical Higgs fields and
longitudinal gauge fields coincide with the poles of the corresponding
transverse gauge fields.  Furthermore no gauge-field-Higgs mixing occurs.
 
With ${\cal L}_{fix}$ and ${\cal L}_{FP}$ the complete renormalizable Lagrangian
for the electroweak SM reads
\begin{equation}
{\cal L}_{GSW} = {\cal L}_{C} + {\cal L}_{fix} + {\cal L}_{FP} \; .
\end{equation}
 
The corresponding Feynman rules are given in App.~A.

\chapter {Renormalization}
\label{charen}
 
The Lagrangian (\ref{LC}) of the minimal $SU(2)_{W}\times U(1)_{Y}$
model involves a certain number of free parameters (\ref{phypar}) which
have to be determined experimentally.
These are chosen such that they have an
intuitive physical meaning at tree level (physical masses, couplings),
i.e.\ they are directly related to experimental quantities.
This direct relation
is destroyed through higher order corrections. Moreover the parameters
of the original Lagrangian, the so-called bare parameters, differ from
the corresponding physical
quantities by UV-divergent contributions. However, in renormalizable
theories these divergencies cancel in relations between physical
quantities, thus allowing meaningful predictions. The renormalizability
of nonabelian gauge theories with spontaneous symmetry breaking and thus
of the SM was proven by 't Hooft \cite{tH71}.
 
One possibility to evaluate predictions of a renormalizable model is the
following:
\bit
\item Calculate physical quantities in terms of the bare parameters.
\item Use as many of the resulting relations as bare parameters exist to
express these in terms of physical observables.
\item Insert the resulting expressions into the remaining relations.
\eit
Thus one arrives at predictions for physical observables in terms of
other physical quantities, which have to be determined from experiment.
In these predictions all UV-divergencies cancel in any order of
perturbation theory. 
The predictions obtained from different input parameters differ in finite
orders of perturbation theory by higher order contributions.
This treatment of renormalization has been pioneered by
Passarino, Veltman and Consoli \cite{Pa79}
and is the basis of the so-called
'star' scheme of Kennedy and Lynn \cite{Ke89}.
 
We use the counterterm approach. Here the UV-divergent bare
parameters are expressed by finite renormalized parameters and
divergent renormalization constants (counterterms). In addition the
bare fields may be replaced by renormalized fields. The counterterms
are fixed through renormalization conditions.
These can be chosen arbitrarily, but determine the relation between
renormalized and physical parameters. Further evaluation proceeds like
described above. The results depend in finite orders of perturbation theory not
only on the choice of the input parameters but also on the choice of
the renormalized parameters. Clearly the physical results are
unambiguous up to the orders which have been taken into account
completely. The renormalization
procedure can be summarized as follows:
\bit
\item Choose a set of independent parameters (e.g.\ (\ref{phypar})
in the SM).
\item Separate the bare parameters (and fields) into renormalized
parameters (fields) and renormalization constants (see Sect.\
\ref{secrct}).
\item Choose renormalization conditions to fix the counterterms (see
Sect.\ \ref{secrcd}).
\item Express physical quantities in terms of the renormalized
parameters.
\item Choose input data in order to fix the values of the renormalized
parameters.
\item Evaluate predictions for physical quantities as functions of the input
data.
\eit
The first three items in this list specify a renormalization scheme.
 
Putting the counterterms equal to zero, the renormalized parameters
equal the bare parameters and we recover the first approach.
 
However, we can choose the counterterms such that the finite renormalized
parameters are equal to physical parameters in all orders of
perturbation theory. This is the so-called on-shell renormalization
scheme. In the SM one uses
the masses of the physical particles $M_{W},M_{Z},M_{H},m_{f}$, the
charge of the electron $e$ and the quark mixing matrix $V_{ij}$
as renormalized parameters. This scheme was proposed by Ross and
Taylor \cite{Ro79} and is widely used in the electroweak theory.
The advantage of the on-shell scheme is that all parameters have a clear
physical meaning and can be measured directly in suitable experiments\footnote
{This is not the case for the quark masses, due to the
presence of the strong interaction. In practice these are replaced by
suitable experimental input parameters (see Sect.~\ref{secpar}).}.
Furthermore the Thomson cross section from which $e$ is obtained is exact
to all orders of perturbation theory. However, not all of the
particle masses are known experimentally with good accuracy. Therefore
other schemes may sometimes be advantageous.
 
Renormalization of the parameters
is sufficient to obtain finite S-matrix elements, but it leaves
Green functions divergent. This is due to the fact that radiative
corrections change the normalization of the fields by an infinite amount.
In order to get finite propagators and vertex functions the
fields have to be renormalized, too.
Furthermore radiative corrections provide nondiagonal corrections to the
mass matrices so that the bare fields are no longer mass eigenstates.
In order to rediagonalize the mass matrices one has to introduce matrix
valued field renormalization constants. These allow to define
the renormalized fields in such a way that they are the correct
physical mass eigenstates in all orders of perturbation theory.
If one does not renormalize the fields in this way,
one needs a nontrivial wave function renormalization for the external
particles. This is required
in going from Green functions to S-matrix elements in order to
obtain a properly normalized S-matrix.
 
The results for physical S-matrix elements are independent of the
specific choice of field renormalization. There exist many different
treatments in the literature \cite{Ba80,Fl81,Sa81,Ao82,Bo86}.
Calculations without field renormalization were performed by \cite{Si80}.
 
\section {Renormalization constants and counterterms}
\label{secrct}
 
In the following we specify the on-shell renormalization scheme for the
electroweak SM quantitatively. As independent parameters we choose the
physical parameters specified in (\ref{phypar}). The
renormalized quantities and renormalization constants are defined as
follows (we denote bare quantities by an index $0$)
\beqar
\nn e_{0}\quad  & = &  Z_{e} e \;=\; (1+\delta Z_{e})e ,    \\[1ex]
\nn M_{W,0}^{2} & = &  M_{W}^{2} + \delta M_{W}^{2} ,    \\[1ex]
    M_{Z,0}^{2} & = &  M_{Z}^{2} + \delta M_{Z}^{2} ,    \\[1ex]
\nn M_{H,0}^{2} & = &  M_{H}^{2} + \delta M_{H}^{2} ,    \\[1ex]
\nn m_{f,i,0}\;  & = &  m_{f,i}   + \delta m_{f,i}   ,    \\[1ex]
\nn V_{ij,0}\;   & = &  (U_{1}VU_{2}^{\dagger})_{ij}
\;=\;V_{ij} + \delta V_{ij}.
\eeqar
$U_{1}$ and $U_{2}$ are unitary matrices since $V_{ij,0}$ and
$V_{ij}$ are both unitary.
 
Radiative corrections affect the Higgs potential in such a way that
its minimum is shifted. In order to correct for this
shift one introduces a counterterm to the vacuum expectation
value of the Higgs field, which is determined such that the renormalized $v$
is given by the actual minimum of the effective Higgs potential.
Since we have replaced $v$ by $t$ (\ref{deft}) we must introduce a
counterterm $\delta t$. This is fixed such that it cancels all tadpole
diagrams, i.e. that the effective potential contains no term linear in
the Higgs field $H$.
 
The counterterms defined above are sufficient to render all S-matrix
elements finite. In order to
have finite Green functions we must renormalize the fields, too.
As explained above we need field renormalization
matrices in order to be able to define renormalized fields which
are mass eigenstates
\beq
\barr{cll}
W_{0}^{\pm}  & = & Z_{W}^{1/2} W^{\pm}=
(1+\frac{1}{2}\delta Z_{W}) W^{\pm} , \\[1em]
\left(\barr{l} Z_{0} \\ A_{0} \earr \right)  & = &
\left(\barr{ll} Z_{ZZ}^{1/2} & Z_{ZA}^{1/2}  \\[1ex]
                Z_{AZ}^{1/2} & Z_{AA}^{1/2}  \earr \right)
\left(\barr{l} Z \\ A \earr \right)   =
\left(\barr{cc} 1 + \frac{1}{2}\delta Z_{ZZ} & \frac{1}{2}\delta Z_{ZA}
\\ [1ex]
               \frac{1}{2}\delta Z_{AZ}  & 1 + \frac{1}{2}\delta Z_{AA}
\earr \right)
\left(\barr{l} Z \\[1ex] A \earr \right)  , \\[2em]
H_{0} & = & Z_{H}^{1/2} H = 
(1+\frac{1}{2}\delta Z_{H}) H, \\[1em]
f_{i,0}^{L} & = & Z^{1/2,f,L}_{ij} f_{j}^{L} 
=(\delta _{ij}+\frac{1}{2}\delta Z_{ij}^{f,L})  f_{j}^{L} , \\[1em]
f_{i,0}^{R} & = & Z^{1/2,f,R}_{ij} f_{j}^{R} 
=(\delta _{ij}+\frac{1}{2}\delta Z_{ij}^{f,R})  f_{j}^{R}.
\earr
\eeq
 
We do not discuss the renormalization constants of the unphysical ghost
and Higgs fields. They do not affect Green functions of physical
particles and are not relevant for the calculation of physical one-loop
amplitudes. Furthermore the renormalization of the unphysical sector
decouples from the one of the physical sector.  It is governed by the
Slavnov-Taylor identities. A discussion of this subject
can be found e.g.\ in \cite{Ao82,Bo86}.
 
In writing $Z=1+\delta Z$ for the multiplicative renormalization constants
(matrices) we can split the bare Lagrangian ${\cal L}_{0}$ into the
basic Lagrangian ${\cal L}$ and the counterterm Lagrangian $\delta {\cal L}$
\beq
{\cal L}_{0} = {\cal L} + \delta {\cal L} .
\eeq
${\cal L}$ has the same form as ${\cal L}_{0}$ but depends on
renormalized parameters and fields instead of unrenormalized ones.
$\delta {\cal L}$ yields the counterterms. The corresponding Feynman rules
are listed in App.~A. They give rise to counterterm diagrams
which have to be added to the loop graphs. 
Since we are only interested in one-loop corrections, we neglect terms
of order $(\delta Z)^{2}$ everywhere.
 
\section {Renormalization conditions}
\label{secrcd}
 
The renormalization constants introduced in the previous section are
fixed by imposing renormalization conditions. These decompose into
two sets. The conditions which define the renormalized parameters and the
ones which define the  renormalized fields. While the choice of the
first affects physical predictions to finite orders of perturbation
theory, the second are only relevant for Green functions and drop out
when calculating S-matrix elements. Nevertheless their use is very
convenient in the on-shell scheme.
They not only allow to eliminate the explicit wave function
renormalization of the external particles, but also
simplify the explicit form of the renormalization conditions for the
physical parameters considerably.
 
In the on-shell scheme all renormalization conditions are formulated for
on mass shell external fields. The field renormalization constants, the
mass renormalization constant and the renormalization constant
of the quark mixing matrix are fixed using the
one-particle irreducible two-point functions. For the charge
renormalization we need one three-point function. For this we choose the
$ee\gamma $-vertex function. In the following renormalized quantities
are denoted by the same symbols as the corresponding unrenormalized 
quantities, but with the superscript $\hat{}$.
 
\savebox{\blobza}(28,28)[lb]{
\put(14,14){\circle{28}}
\put(4,4){\line(1,1){19.5}}
\put(14,0){\line(1,1){14}}
\put(0,14){\line(1,1){14}}
\put(8,1){\line(1,1){18.5}}
\put(1,8){\line(1,1){18.5}}}
 
\savebox{\blobdz}(32,32)[lb]{
\put(16,16){\circle{32}}
\put(5,5){\line(1,1){22}}
\put(16,0){\line(1,1){16}}
\put(0,16){\line(1,1){16}}
\put(9.5,1.5){\line(1,1){21}}
\put(1.5,9.5){\line(1,1){21}}}
 
\savebox{\Vr}(48,0)[bl]
{\multiput(3,0)(12,0){4}{\oval(6,4)[t]}
\multiput(9,0)(12,0){4}{\oval(6,4)[b]} }
\savebox{\Vtr}(32,24)[bl]
{\multiput(4,0)(8,6){4}{\oval(8,6)[tl]}
\multiput(4,6)(8,6){4}{\oval(8,6)[br]} }
\savebox{\Vbr}(32,24)[bl]
{\multiput(4,24)(8,-6){4}{\oval(8,6)[bl]}
\multiput(4,18)(8,-6){4}{\oval(8,6)[tr]}}
\savebox{\Vtbr}(32,48)[bl]
{\put(00,24){\usebox{\Vtr}}
\put(00,00){\usebox{\Vbr}}}
 
\savebox{\Sr}(48,0)[bl]
{ \multiput(0,0)(12.5,0){4}{\line(4,0){10}} }
\savebox{\Str}(32,24)[bl]
{ \multiput(-2,-1.5)(12,9){3}{\line(4,3){10}} }
\savebox{\Sbr}(32,24)[bl]
{\multiput(-2,25.5)(12,-9){3}{\line(4,-3){10}} }
\savebox{\Stbr}(32,48)[bl]
{\put(00,24){\usebox{\Str}}
\put(00,00){\usebox{\Sbr}}}
 
\savebox{\Fr}(48,0)[bl]
{ \put(0,0){\vector(1,0){26}} \put(24,0){\line(1,0){24}} }
\savebox{\Ftr}(32,24)[bl]
{ \put(0,0){\vector(4,3){18}} \put(16,12){\line(4,3){16}} }
\savebox{\Fbr}(32,24)[bl]
{ \put(32,0){\vector(-4,3){19}} \put(16,12){\line(-4,3){16}} }
\savebox{\Ftbr}(32,48)[bl]
{\put(00,24){\usebox{\Ftr}}
\put(00,00){\usebox{\Fbr}}}
 
As discussed above the first renormalization condition involves the
tadpole $T$, the Higgs field one-point amputated renormalized Green
function
\beq
\barr{lll}
\hat{T} &=& \quad \rule[-16pt]{0pt}{26pt}
\begin{picture}(80,32)(0,13)
\put(5,20){$H$}
\put(0,16){\usebox{\Sr}}
\put(48,00){\usebox{\blobdz}}
\end{picture} \;,
\earr
\eeq
and simply states
\beq
 \hat{T} = T + \delta t = 0.
\label{RCT}
\eeq
As a consequence of this condition no tadpoles need to be considered in
actual calculations.
 
Next we need the renormalized one-particle irreducible two-point
functions. These are defined as follows (we are using the 't
Hooft-Feynman gauge)
\beqar
\nn &&\rule[-16pt]{0pt}{26pt}
\begin{picture}(128,32)(0,13)
\put(5,22){$W_{\mu }$}
\put(5,1){$k$}
\put(0,16){\usebox{\Vr}}
\put(48,00){\usebox{\blobdz}}
\put(80,16){\usebox{\Vr}}
\put(115,22){$W_{\nu }$}
\end{picture} \quad
=\hat{\Gamma }^{W}_{\mu \nu }(k)\\[1.2em]
\nn &&\qquad=-ig_{\mu \nu }(k^{2}-M^{2}_{W})
-i\left(g_{\mu \nu} -\disp\frac{k_{\mu} k_{\nu }}{k^{2}}\right)
\hat{\Sigma}^{W}_{T}(k^{2})
-i\disp\frac{k_{\mu} k_{\nu }}{k^{2}}\hat{\Sigma}^{W}_{L}(k^{2}),  \\[1em]
\nn &&\begin{picture}(128,32)(0,13)
\put(5,22){$a,{\mu }$}
\put(5,2){$k$}
\put(0,16){\usebox{\Vr}}
\put(48,00){\usebox{\blobdz}}
\put(80,16){\usebox{\Vr}}
\put(115,21){$b,{\nu }$}
\end{picture} \quad
=\hat{\Gamma }^{ab}_{\mu \nu }(k)\\[1.2em]
\nn &&\qquad=-ig_{\mu \nu }(k^{2}-M^{2}_{a})\delta_{ab}
-i\left(g_{\mu \nu} -\disp\frac{k_{\mu} k_{\nu }}{k^{2}}\right)
\hat{\Sigma}^{ab}_{T}(k^{2})
-i\disp\frac{k_{\mu} k_{\nu }}{k^{2}}\hat{\Sigma}^{ab}_{L}(k^{2}),  \\[1em]
&&\quad \mbox{where} \quad
a,b = A,Z, \quad M_{A}^{2}=0,\\[1em]
\nn &&\rule[-16pt]{0pt}{26pt}
\begin{picture}(128,32)(0,13)
\put(5,20){$H$}
\put(5,3){$k$}
\put(0,16){\usebox{\Sr}}
\put(48,00){\usebox{\blobdz}}
\put(80,16){\usebox{\Sr}}
\put(115,20){$H$}
\end{picture}  \quad
=\hat{\Gamma}^{H}(k)=i(k^{2}-M^{2}_{H}) + i\hat{\Sigma
}^{H}(k^{2}),\\[1.2em]
\nn &&\rule[-16pt]{0pt}{26pt}
\begin{picture}(128,32)(0,13)
\put(5,22){$f_{j}$}
\put(5,4){$p$}
\put(0,16){\usebox{\Fr}}
\put(48,00){\usebox{\blobdz}}
\put(80,16){\usebox{\Fr}}
\put(115,22){$f_{i}$}
\end{picture} \quad
=\hat{\Gamma }_{ij}^{f}(p)
\\[1em]
\nn &&\qquad =i\delta_{ij}(\ps-m_{i})
+i\left[\ps \omega_{-} \hat{\Sigma}_{ij}^{f,L}(p^{2})
       +\ps \omega_{+} \hat{\Sigma}_{ij}^{f,R}(p^{2})
       +(m_{f,i}\omega_{-}+m_{f,j}\omega_{+})\hat{\Sigma}_{ij}^{f,S}(p^{2})
\right].
\eeqar
The corresponding propagators are obtained as the inverse of these
two-point functions. Note that we have to invert matrices for the neutral
gauge bosons and for the fermions.
 
The renormalized mass parameters of the physical particles are
fixed by the requirement that they are equal to the physical masses,
i.e.\ to the real parts of the poles of the corresponding
propagators which are equivalent to the zeros of the one-particle
irreducible two-point functions. In case of mass matrices these
conditions have to
be fulfilled by the corresponding eigenvalues resulting in complicated
expressions. These can be considerably simplified by requiring
simultaneously the on-shell conditions for the
field renormalization matrices. These state that the
renormalized one-particle irreducible two-point functions are diagonal if
the external lines are on their mass shell.
This determines the
nondiagonal elements of the field renormalization matrices. The
diagonal elements are fixed such that the renormalized fields are
properly normalized, i.e.\ that the residues of the renormalized
propagators are equal to one. This choice of field renormalization
implies that the renormalization conditions for the mass parameters
(in all orders of perturbation theory)
involve only the corresponding diagonal self energies. Thus we arrive at
the following
renormalization conditions for the two-point functions for on-shell
external physical fields
\beq
\barr{ll}
\disp\left. \widetilde{\Re}\,\hat{\Gamma }^{W }_{\mu \nu } (k)
\varepsilon ^{\nu }(k)
\right\vert_{k^{2}= M_{W}^{2}} = 0 , \\[1.2em]
\disp\left. \Re\,\hat{\Gamma }^{ZZ}_{\mu \nu } (k) \varepsilon ^{\nu }(k)
\right\vert_{k^{2}= M_{Z}^{2}}= 0 , &\disp
\left. \Re\,\hat{\Gamma }^{AZ}_{\mu \nu } (k) \varepsilon ^{\nu }(k)
\right\vert_{k^{2}= M_{Z}^{2}}= 0 , \\[1.2em]
\disp\left. \hat{\Gamma }^{AZ}_{\mu \nu } (k) \varepsilon ^{\nu }(k)
\right\vert_{k^{2}= 0}= 0 , &\disp
\left. \hat{\Gamma }^{AA}_{\mu \nu } (k) \varepsilon ^{\nu }(k)
\right\vert_{k^{2}= 0}= 0 , \\[1.2em]
\disp\lim_{k^{2}\to M_{W}^{2}} \disp\frac{1}{k^{2}-M_{W}^{2}}
\widetilde{\Re}\,\hat{\Gamma }^{W }_{\mu \nu } (k) \varepsilon ^{\nu }(k)
= -i\varepsilon_{\mu}(k) , \\[1.2em]
\disp\lim_{k^{2}\to M_{Z}^{2}} \disp\frac{1}{k^{2}-M_{Z}^{2}}
\Re\, \hat{\Gamma }^{ZZ}_{\mu \nu } (k) \varepsilon ^{\nu }(k) =
-i\varepsilon_{\mu}(k) ,
\quad &\disp
\disp\lim_{k^{2}\to 0} \disp\frac{1}{k^{2}}
\Re\, \hat{\Gamma }^{AA}_{\mu \nu } (k) \varepsilon ^{\nu }(k) =
-i\varepsilon_{\mu}(k) , \\[1.2em]
\disp\left. \Re\,\hat{\Gamma }^{H } (k)
\right\vert_{k^{2}= M_{H}^{2}} = 0 , &\disp
\lim_{k^{2}\to M_{H}^{2}} \frac{1}{k^{2}-M_{H}^{2}}
\Re\; \hat{\Gamma }^{H } (k)    = i ,\\[1.2em]
\disp\left. \widetilde{\Re}\,\hat{\Gamma }_{ij}^{f}(p) u_{j}(p)
\right\vert_{p^{2}= m_{f,j}^{2}} = 0 ,  &\disp
\left. \widetilde{\Re}\,\bar{u}_{i}(p') \hat{\Gamma }_{ij}^{f}({p'})
\right\vert_{p'^{2}= m_{f,i}^{2}} = 0 , \\[1.2em]
\disp\lim_{p^{2} \to m_{f,i}^{2}} \frac{\ps+m_{f,i}}{p^{2}-m_{f,i}^{2}}
\widetilde{\Re}\,\hat{\Gamma }^{f}_{ii} (p) u_{i}(p) = iu_i(p) , &\disp
\disp\lim_{p'^{2} \to m_{f,i}^{2}} \bar{u}_{i}(p') \widetilde{\Re}\,
\hat{\Gamma }^{f}_{ii} ({p'})\frac{\ps'+m_{f,i}}{{p'}^{2}-m_{f,i}^{2}} =
i\bar u_i(p') .
\earr
\eeq
$\varepsilon(k)$, $u(p)$ and $\bar{u}(p')$ are the polarization
vectors and spinors of the external fields.
$\widetilde{\Re}$ takes the real part of the loop integrals appearing in
the self energies but not of the quark mixing matrix elements
appearing there. Since we restrict ourselves to the one-loop order we
apply it only to those quantities which depend on the quark mixing
matrix at one loop. In higher orders Re must be replaced by $\widetilde{\Re}$
everywhere. Re and $\widetilde{\Re}$ are only relevant above thresholds
and have no effect for the two-point functions of on-shell stable particles.
If the quark mixing matrix is real $\widetilde{\Re}$ can be replaced by
Re. This holds in particular for a unit quark mixing matrix which is
often used.
 
From the above equations we obtain the conditions for
the self energy functions.
\beq
\barr{ll} \label{RCG}
\disp \widetilde{\Re}\, \hat{\Sigma }^{W }_{T}(M_{W}^{2}) = 0 , \\[1em]
\disp \Re\, \hat{\Sigma }^{ZZ}_{T}(M_{Z}^{2}) = 0 , &
      \Re\, \hat{\Sigma }^{AZ}_{T}(M_{Z}^{2}) = 0 , \\[1em]
\disp       \hat{\Sigma }^{AZ}_{T}(    0    ) = 0 , &
     \hat{\Sigma }^{AA}_{T}  (    0    )   = 0, {\footnotemark} \\[1em]
\disp\left. \widetilde{\Re}\,
\frac{\partial \hat{\Sigma }^{W }_{T}(k^{2})}{\partial k^{2}}
\right\vert_{k^{2}=M_{W}^{2}}  = 0 , \\[1.2em]
\disp \left. \Re\, \disp\frac{\partial \hat{\Sigma }^{ZZ}_{T}(k^{2})}
{\partial k^{2}}
\right\vert_{k^{2}=M_{Z}^{2}}  = 0 , \quad &
\left. \Re\, \disp\frac{\partial \hat{\Sigma }^{AA}_{T}(k^{2})}{\partial k^{2}}
\right\vert_{k^{2}=0}  = 0 ,
\earr
\eeq
\beq \label{RCH}
\Re\,\hat{\Sigma }^{H }  (M_{H}^{2})   = 0 ,
\qquad\qquad\quad
\left. \Re\, \disp\frac{\partial \hat{\Sigma }^{H }(k^{2})}{\partial k^{2}}
\right\vert_{k^{2}=M_{H}^{2}}  = 0 ,
\eeq
\beq \label{RCF}
\barr{l}
\disp m_{f,j}\widetilde{\Re}\,\hat{\Sigma }_{ij}^{f,L}(m_{f,j}^{2})
+ m_{f,j}\widetilde{\Re}\,\hat{\Sigma }_{ij}^{f,S}(m_{f,j}^{2}) = 0 ,\\[1em]
\disp m_{f,j}\widetilde{\Re}\,\hat{\Sigma }_{ij}^{f,R}(m_{f,j}^{2})
+ m_{f,i}\widetilde{\Re}\,\hat{\Sigma }_{ij}^{f,S}(m_{f,j}^{2}) = 0 ,\\[1em]
\disp\widetilde{\Re}\,\hat{\Sigma }_{ii}^{f,R}(m_{f,i}^{2})
+ \widetilde{\Re}\,\hat{\Sigma }_{ii}^{f,L}(m_{f,i}^{2}) \\[1em]
\disp\mbox{}+2m_{f,i}^{2}\disp\frac{\partial}{\partial p^{2}}\left.\left(
\widetilde{\Re}\,\hat{\Sigma }_{ii}^{f,R}(p^{2})
+ \widetilde{\Re}\,\hat{\Sigma }_{ii}^{f,L}(p^{2}) +
2\widetilde{\Re}\,\hat{\Sigma }_{ii}^{f,S}(p^{2})
\right)\right\vert_{p^{2}=m_{f,i}^{2}} = 0.
\earr
\eeq
Note that the (unphysical)
longitudinal part of the gauge boson self energies drops out for on-shell
external gauge bosons. \footnotetext
{This condition is automatically fulfilled due to a Ward identity.}
 
Our choice for the renormalization condition of the quark mixing
matrix $V_{ij}$ can
be motivated as follows. To lowest order $V_{ij}$ is given by (see
eq. \ref{QMM0})
\beq
V_{0,ij}=U^{u,L}_{ik} U^{d,L\dagger}_{kj},
\eeq
where the matrices $U^{f,L}$ transform the weak interaction
eigenstates $f'_{0}$ to the lowest order mass eigenstates $f_{0}$
\beq
U^{f,L\dagger}_{ij}f^{L}_{j,0}= {f'}^{L}_{i,0}.
\eeq
In the on-shell renormalization scheme the higher order mass
eigenstates are related to the bare mass eigenstates through the field
renormalization constants of the fermions
\beq
f^{L}_{i}=Z^{1/2,f,L}_{ij}f^{L}_{j,0} .
\eeq
We define the renormalized quark mixing matrix in analogy to the
unrenormalized one through the rotation from the weak interaction
eigenstates to the renormalized mass eigenstates.
In the one-loop approximation the rotation contained in the fermion wave
function renormalizaton $1+\frac{1}{2}\delta Z^{L}$
is simply given by the anti-Hermitean part
$\delta Z^{AH}$ of $\delta Z^{L}$
\beq
\delta Z^{f,AH}_{ij}=\frac{1}{2} (\delta Z^{f,L}_{ij}-\delta
Z^{f,L\dagger}_{ij}) .
\eeq
Thus we are lead to define the renormalized quark mixing matrix as
\beq
\barr{lll}
V_{ij} & = & (\delta_{ik} + \frac{1}{2}\delta Z^{u,AH\dagger}_{ik})U^{u,L}_{km}
U^{d,L\dagger}_{mn} (\delta_{nj} + \frac{1}{2}\delta Z^{d,AH}_{nj}) \\[1em]
& = & (\delta_{ik} + \frac{1}{2}\delta Z^{u,AH\dagger}_{ik})V_{0,kn}
(\delta_{nj} + \frac{1}{2}\delta Z^{d,AH}_{nj}) .
\earr
\label{RCV}
\eeq
It has been shown that this condition correctly cancels all one-loop
divergencies and
that $V_{ij} = V_{0,ij}$ in the limit of degenerate up- or down-type
quark masses \cite{De90}.
 
\savebox{\Vr}(36,0)[bl]
{\multiput(3,0)(12,0){3}{\oval(6,4)[t]}
\multiput(9,0)(12,0){3}{\oval(6,4)[b]} }
 
\savebox{\Ftr}(36,18)[bl]
{ \put(0,0){\vector(2,1){20}} \put(18,9){\line(2,1){18}} }
\savebox{\Fbr}(36,18)[bl]
{ \put(36,0){\vector(-2,1){21}} \put(18,9){\line(-2,1){18}} }
\savebox{\Ftbr}(36,36)[bl]
{\put(00,18){\usebox{\Ftr}}
\put(00,00){\usebox{\Fbr}}}
 
Finally the electrical charge is defined as the full $ee\gamma $-coupling
for on-shell external particles in the Thomson limit.
This means that all corrections to this vertex
vanish on-shell and for zero momentum transfer\footnote{Due to the wave
function renormalization of the external particles the self energy
corrections in the external legs contribute only with a factor $1/2$ to
the S-matrix elements.}
\beq
\left.
\barr{l}
\begin{picture}(248,120)
\put(5,70){$A_{\mu}$}
\put(0,60){\usebox{\Vr}}
\put(36,46){\usebox{\blobza}}
\put(64,60){\usebox{\Vr}}
\put(100,44){\usebox{\blobdz}}
\put(130.3,67.15){\usebox{\Ftr}}
\put(164.8,77.4){\usebox{\blobza}}
\put(191.3,97.65){\usebox{\Ftr}}
\put(130.3,34.85){\usebox{\Fbr}}
\put(164.8,14.6){\usebox{\blobza}}
\put(191.3,4.35){\usebox{\Fbr}}
\put(220,97){$e^{+},p'$}
\put(220,15){$e^{-},p$}
\end{picture}
\earr
\right\vert_{p=p',\;p^{2}=p'^{2}=m_{e}^{2}}
= ie\bar{u}(p)\gamma _{\mu }u(p).
\eeq
The momenta $p$, $p'$ flow in the direction of the fermion arrows.
Due to our choice for the field renormalization the corrections
in the external legs vanish and we obtain the condition
\beq
\left.\bar{u}(p)\Gamma ^{ee\gamma}_{\mu }(p,p)u(p)\right\vert_{
p^{2}=m_{e}^{2}}
=ie \bar{u}(p)\gamma_{\mu }u(p) ,
\label{RCE}
\eeq
for the (amputated) vertex function
\beq
\hat\Gamma^{ee\gamma}_{\mu }(p,p')= \quad
\barr{l}
\begin{picture}(112,60)
\put(5,40){$A_{\mu}$}
\put(0,30){\usebox{\Vr}}
\put(36,14){\usebox{\blobdz}}
\put(66.3,37.15){\usebox{\Ftr}}
\put(66.3, 4.85){\usebox{\Fbr}}
\put(107,52){$e^{+},p'$}
\put(107,2){$e^{-},p$}
\end{picture}\qquad
\earr .
\eeq
 
\section{Explicit form of renormalization constants}
\label{secrcfix}
 
The renormalized quantities defined in Sect.\ \ref{secrcd} consist of the
unrenormalized ones and the counterterms as specified by the Feynman
rules in App.~A.
The renormalization conditions allow to express the counterterms by the
unrenormalized self energies at special external momenta. This is evident
 for all
renormalization constants apart from the one for the electrical charge. In
this case, however, we can use a Ward identity to eliminate the vertex
function.
 
From conditions (\ref{RCT}, \ref{RCG}, \ref{RCH})
we obtain for the gauge boson and Higgs sector
\beq
\barr{llllll}
\delta t & = & -T , \\[1em]
\delta M^{2}_{W} & = & \widetilde{\Re}\,\Sigma_{T}^{W}(M^{2}_{W}), \quad
&
\delta Z_{W} & = & \left.
-\Re\,\disp\frac{\partial\Sigma_{T}^{W}(k^{2})}{\partial k^{2}}
\right\vert_{k^{2}=M^{2}_{W}} , \\[1.2em]
\delta M^{2}_{Z} & = & \Re\, \Sigma_{T} ^{ZZ}(M^{2}_{Z}), &
\delta Z_{ZZ} & = &
\left.-\Re\,\disp\frac{\partial\Sigma_{T}^{ZZ}(k^{2})}{\partial k^{2}}
\right\vert_{k^{2} = M^{2}_{Z}}, \\[1.2em]
\delta Z_{AZ} & = &
-2\Re\,\disp\frac{\Sigma_{T}^{AZ}(M^{2}_{Z})}{M^{2}_{Z}}, \quad &
\delta Z_{ZA} & = & \disp2\frac{\Sigma_{T}^{AZ}(0)}{M_{Z}^{2}} ,\\[1.2em]
\delta Z_{AA} & = & \left.
-\disp\frac{\partial\Sigma_{T}^{AA}(k^{2})}{\partial k^{2}}
\right\vert_{k^{2} = 0}, \quad & \\[1.2em]
\delta M^{2}_{H} & = & \Re\,\Sigma ^{H}(M^{2}_{H}), &
\delta Z_{H} & = & \left. -\Re\,\disp\frac{\partial\Sigma^{H}(k^{2})}
{\partial k^{2}} \right\vert_{k^{2} = M^{2}_{H}} .
\earr
\eeq
In the fermion sector (\ref{RCF}) yields
\beqar
\nn \delta m_{f,i} & = & \frac{m_{f,i}}{2}\widetilde{\Re}\,\left(
\Sigma^{f,L}_{ii}(m_{f,i}^{2})+\Sigma^{f,R}_{ii}(m_{f,i}^{2})
+2\Sigma^{f,S}_{ii}({m_{f,i}}^{2})
\right), \\[1em]
\label{CTF}
\nn\delta Z_{ij}^{f,L} & = & \frac{2}{m_{f,i}^{2}-m_{f,j}^{2}}
\widetilde{\Re}\,\left[
 m_{f,j}^{2}\Sigma^{f,L}_{ij}(m_{f,j}^{2})
+m_{f,i}m_{f,j}\Sigma^{f,R}_{ij}(m_{f,j}^{2})
\right.\\[.5em]
\nn&&\phantom{\frac{2}{m_{i\sigma }^{2}-m_{j\sigma
}^{2}}\widetilde{\Re}\,\left[\right.}
\left. +(m_{f,i}^{2}+m_{f,j}^{2})\Sigma^{f,S}_{ij}(m_{f,j}^{2})
\right], \qquad i\ne j,\\[1em]
\delta Z_{ij}^{f,R} & = & \frac{2}{m_{f,i}^{2}-m_{f,j}^{2}}
\widetilde{\Re}\,\left[
 m_{f,j}^{2}\Sigma^{f,R}_{ij}(m_{f,j}^{2})
+m_{f,i}m_{f,j}\Sigma ^{f,L}_{ij}(m_{f,j}^{2})
\right.\\[.5em]
\nn &&\phantom{\frac{2}{m_{i\sigma }^{2}-m_{j\sigma
}^{2}}\widetilde{\Re}\,\left[\right.}
\left. +2m_{f,i}m_{f,j}\Sigma^{f,S}_{ij}(m_{f,j}^{2})\right],
\qquad i\ne j,\\[1em]
\nn\delta Z_{ii}^{f,L} & = &
-\widetilde{\Re}\,\Sigma _{ii}^{f ,L}(m_{f,i}^{2})
- m_{f,i}^{2}\frac{\partial}{\partial p^{2}} \widetilde{\Re}\,\left[\left.
\Sigma^{f,L}_{ii}(p^{2}) +\Sigma^{f,R}_{ii}(p^{2})
 +2\Sigma^{f,S}_{ii}(p^{2})\right]
\right\vert_{p^{2}=m_{f,i}^{2}} , \\[1em]
\nn \delta Z_{ii}^{f,R} & = &
-\widetilde{\Re}\,\Sigma _{ii}^{f,R}(m_{f,i}^{2})
- m_{f,i}^{2}\frac{\partial}{\partial p^{2}} \widetilde{\Re}\,\left[\left.
\Sigma^{f,L}_{ii}(p^{2})+\Sigma ^{f,R}_{ii}(p^{2})
 +2\Sigma^{f,S}_{ii}(p^{2})\right]
\right\vert_{p^{2}=m_{f,i}^{2}} .
\eeqar
The use of $\widetilde{\Re}$ ensures reality of the renormalized
Lagrangian. Furthermore it
yields
\beq
\delta Z^{\dagger}_{ij}=\delta Z_{ij}(m^{2}_{i}\leftrightarrow
m^{2}_{j}) ,
\eeq
and in particular
\beq
\delta Z^{\dagger}_{ii}=\delta Z_{ii}.
\eeq
 
In the lepton sector we have $V_{ij}=\delta _{ij}$. Consequently all
lepton self energies are diagonal and the off-diagonal lepton wave
function renormalization constants are zero. The same holds for the
quark sector if one replaces the quark mixing matrix by a unit matrix as
is usually done in calculations of radiative corrections for high
energy processes.
 
The renormalization constant for the quark mixing matrix $V_{ij}$ can be
directly read off from (\ref{RCV})
\beq  \label{DV}
\delta V_{ij}=\frac{1}{4} \Bigl[(\delta Z^{u,L}_{ik}-\delta
Z^{u,L\dagger}_{ik}) V_{kj} -
V_{ik}(\delta Z^{d,L}_{kj}-\delta Z^{d,L\dagger}_{kj})\Bigr] .
\eeq
Inserting the fermion field renormalization constants
(\ref{CTF}) yields
\beq
\barr{llrrl}
\delta V_{ij} & = &
\disp\frac{1}{2}\widetilde{\Re}&\biggl\{\frac{1}{m_{u,i}^{2}-
m_{u,k}^{2}} &
\biggl[m_{u,k}^{2}\Sigma ^{u,L}_{ik}(m_{u,k}^{2})+m_{u,i}^{2}\Sigma
^{u,L}_{ik}(m_{u,i}^{2})\\[1em]
&&&&+m_{u,i}m_{u,k}\left(\Sigma ^{u,R}_{ik}(m_{u,k}^{2})+\Sigma
^{u,R}_{ik}(m_{u,i}^{2})\right)\\[1em]
&&&&+(m_{u,k}^{2}+m_{u,i}^{2})\left(\Sigma^{u,S}_{ik}(m_{u,k}^{2})+\Sigma
^{u,S}_{ik}(m_{u,i}^{2})\right)\biggr] \; V_{kj} \\[1em]
&&-V_{ik}&\frac{1}{m_{d,k}^{2}-
m_{d,j}^{2}} &
\biggl[m_{d,j}^{2}\Sigma ^{d,L}_{kj}(m_{d,j}^{2})+m_{d,k}^{2}\Sigma
^{d,L}_{kj}(m_{d,k}^{2}) \\[1em]
&&&&+m_{d,k}m_{d,j}\left(\Sigma ^{d,R}_{kj}(m_{d,j}^{2})+\Sigma
^{d,R}_{kj}(m_{d,k}^{2})\right)\\[1em]
&&&&+(m_{d,k}^{2}+m_{d,j}^{2})\left(\Sigma^{d,S}_{kj}(m_{d,k}^{2})+\Sigma
^{d,S}_{kj}(m_{d,j}^{2})\right)\biggr]\biggr\}.
\earr
\eeq
It remains to fix the charge renormalization constant $\delta
Z_{e}$. This is determined from the $ee\gamma$-vertex. To be more general we
investigate the $ff\gamma$-vertex for arbitrary fermions $f$. The renormalized
vertex function reads
\beq
\hat{\Gamma }^{\gamma ff}_{ij,\mu }(p,p')=-ieQ_{f}\delta_{ij}
\gamma_{\mu } +ie\hat{\Lambda }^{\gamma ff}_{ij,\mu }(p,p').
\label{RCE1}
\eeq
For on-shell external fermions it can be decomposed as ($k=p'-p$)
\beq
\hat{\Lambda }^{\gamma ff}_{ij,\mu }(p,p')=\delta_{ij}\left(
 \gamma _{\mu }\hat{\Lambda
}^{f}_{V}(k^{2})-\gamma _{\mu }\gamma _{5}\hat{\Lambda }^{f}_{A}(k^{2})+
\frac{(p+p')_{\mu }}{2m_{f}}\hat{\Lambda }^{f}_{S}(k^{2})+\frac{(p'-
p)_{\mu }}{2m_{f}}\gamma _{5}\hat{\Lambda }^{f}_{P}(k^{2}) \right).
\label{RCE2}
\eeq
Expressing the renormalized quantities by the unrenormalized ones and
the counterterms and inserting this in the analogue of the
renormalization condition
(\ref{RCE}) for arbitrary fermions we find, using the Gordon identities,
\beq
\barr{lll}
0&=&\bar{u}(p)\hat{\Lambda}^{\gamma ff}_{ii,\mu }(p,p)u(p)\\[1em]
& = & \bar{u}(p) \gamma _{\mu }u(p)
\left[-Q_{f}(\delta Z_{e}+\delta Z^{f,V}_{ii}+\frac{1}{2}\delta Z_{AA})+
\Lambda ^{f}_{V}(0)+\Lambda ^{f}_{S}(0)+ v_{f}\frac{1}{2}\delta Z_{ZA}
\right] \\[1em]
& & -\bar{u}(p) \gamma _{\mu }\gamma _{5}u(p)
\left[-Q_{f}\delta Z^{f,A}_{ii}+ \Lambda ^{f}_{A}(0)
+a_{f}\frac{1}{2}\delta Z_{ZA}\right],
\earr
\eeq
where
\beq
\delta Z^{f,V}_{ii} = \frac{1}{2}(\delta Z^{f,L}_{ii} + \delta
Z^{f,R}_{ii}) ,\quad
\delta Z^{f,A}_{ii} = \frac{1}{2}(\delta Z^{f,L}_{ii} - \delta
Z^{f,R}_{ii}) ,\quad
\eeq
and $v_{f}$, $a_{f}$  are the vector and axialvector couplings of the
$Z$-boson to the fermion $f$, given explicitly in (\ref{vfaf}).
This yields in fact two conditions, namely
\beqar
\label{cond1}
0& = & -Q_{f}(\delta Z_{e}+\delta Z^{f,V}_{ii}+\frac{1}{2}\delta Z_{AA})
+\Lambda ^{f}_{V}(0)+\Lambda ^{f}_{S}(0)+v_{f}\frac{1}{2}\delta Z_{ZA} ,
\\[1ex]
\label{cond2}
0& = & -Q_{f}\delta Z^{f,A}_{ii} +\Lambda ^{f}_{A}(0)
+a_{f}\frac{1}{2}\delta Z_{ZA} .
\eeqar
The first one (\ref{cond1}) for $f=e$ fixes the charge renormalization
constant. The
second (\ref{cond2}) is automatically fulfilled due to a Ward identity
which can be
derived from the gauge invariance of the theory. The same Ward identity
moreover yields
\beq
\Lambda ^{f}_{V}(0)+\Lambda ^{f}_{S}(0)-Q_{f}\delta Z^{f,V}_{ii}+a_{f}
\frac{1}{2}\delta Z_{ZA}=0 .
\eeq
Inserting this in (\ref{cond1}) we finally find (using $v_{f}-a_{f}=-
Q_{f}s_{W}/c_{W}$)
\beq  \label{DZE}
\barr{lll}
\delta Z_{e}& = &\disp-\frac{1}{2}\delta Z_{AA}-
\frac{s_{W}}{c_{W}}\frac{1}{2}\delta Z_{ZA}
=\disp\left.\frac{1}{2}\frac{\partial\Sigma_{T}^{AA}(k^{2})}{\partial k^{2}}
\right\vert _{k^{2}=0}-\frac{s_{W}}{c_{W}}\frac{\Sigma_{T}^{AZ}(0)}{M^{2}_{Z}} .
\earr
\eeq
This result is independent of the fermion species, reflecting electric
charge universality. Clearly it does not
depend on a specific choice of field renormalization.
Consequently the analogue of (\ref{RCE}) holds for arbitrary fermions
$f$.
 
In the on-shell scheme the weak mixing angle is a derived quantity.
Following Sirlin \cite{Si80} we define it as
\beq \label{swren}
\sin^{2}\theta_{W} = s_{W}^{2} = 1-\frac{M_{W}^{2}}{M_{Z}^{2}},
\eeq
using the renormalized gauge boson masses. This definition is
independent of a specific process and valid to all orders of
perturbation theory.
 
Since the dependent parameters $s_{W}$ and $c_{W}$ frequently
appear, it
is useful to introduce the corresponding counterterms
\beq
c_{W,0} = c_{W} + \delta c_{W} , \quad
s_{W,0} = s_{W} + \delta s_{W} .
\eeq
Because of (\ref{swren}) these are directly related
to the counterterms to the gauge boson masses. To one-loop order we
obtain
\beq
\barr{lll}
\disp\frac{\delta c_{W}}{c_{W}} & = & \disp\frac{1}{2}
\left(\frac{\delta M_{W}^{2}}{M_{W}^{2}}
-\frac{\delta M_{Z}^{2}}{M_{Z}^{2}}\right) =
\frac{1}{2}\widetilde{\Re}\,\left(\frac{\Sigma ^{W}_{T}(M_{W}^{2})}{M_{W}^{2}}
-\frac{\Sigma ^{ZZ}_{T}(M_{Z}^{2})}{M_{Z}^{2}}\right), \\[1.3em]
\disp\frac{\delta s_{W}}{s_{W}} & = & -\disp\frac{c_{W}^{2}}{s_{W}^{2}}
\frac{\delta c_{W}}{c_{W}} =
-\disp\frac{1}{2}\disp\frac{c_{W}^{2}}{s_{W}^{2}}
\widetilde{\Re}\,\left(\frac{\Sigma ^{W}_{T}(M_{W}^{2})}{M_{W}^{2}}
-\frac{\Sigma ^{ZZ}_{T}(M_{Z}^{2})}{M_{Z}^{2}}\right) .
\earr
\eeq
 
We have now determined all renormalization constants in terms of
unrenormalized self energies. In the next sections we will describe
the methods to calculate these self energies  and more general diagrams
at the one-loop level.

\setcounter{chapter}{3}
\newcommand{\sgn}{\mbox{sgn}}
 
\savebox{\Lt}(0,36)[bl]
{ \put(0,17){\vector(0,1){3}} \put(0,0){\line(0,1){36}} }
\savebox{\Ltr}(36,18)[bl]
{ \put(17,8.5){\vector(2,1){3}} \put(0,0){\line(2,1){36}} }
\savebox{\Ltl}(36,18)[bl]
{ \put(19,8.5){\vector(-2,1){3}} \put(36,0){\line(-2,1){36}}}
\savebox{\Lb}(0,36)[bl]
{ \put(0,19){\vector(0,-1){3}} \put(0,0){\line(0,1){36}} }
\savebox{\Lbl}(36,18)[bl]
{ \put(19,9.5){\vector(-2,-1){3}} \put(0,0){\line(2,1){36}} }
\savebox{\Lbr}(36,18)[bl]
{ \put(17,9.5){\vector(2,-1){3}} \put(36,0){\line(-2,1){36}}}
 
\chapter{One-loop integrals}
\label{chatenint}
 
Perturbative calculations at one-loop order involve integrals over
the loop momentum. In this chapter we discuss their classification
and techniques for their calculation. The methods described here
are to a large
extent based on the work of Passarino and Veltman \cite{Pa79}, 't Hooft and
Veltman \cite{tH79}, and Melrose \cite{Me65}.
 
\section{Definitions}
\label{sectendef}
 
The one-loop integrals in $D$ dimensions are
classified according to the number $N$ of
propagator factors in the denominator and the number $P$ of
integration momenta in the numerator. For $P+D-2N \ge 0$
these integrals are UV-divergent. These divergencies are regularized by
calculating the integrals in general dimensions $D\ne4$ (dimensional
regularization). The UV-divergencies drop out in
renormalized quantities. For renormalizable theories we have $P \le N$
and thus a finite number of divergent integrals.
 
\begin{figure}[b]
\bma
\begin{picture}(150,144)
\put(5,100){$p_{1}$}
\put(0,25){$p_{NN-1}$}
\put(45,71){$q$}
\put(65,54){$q+p_{N-1}$}
\put(65,89){$q+p_{1}$}
\put(93,5){$p_{N-1N-2}$}
\put(93,135){$p_{21}$}
\put(106,35){$q+p_{N-2}$}
\put(106,105){$q+p_{2}$}
\put(18,90){\usebox{\Lbr}}
\put(54,36){\usebox{\Ltl}}
\put(90,90){\usebox{\Lbr}}
\put(18,36){\usebox{\Ltr}}
\put(54,90){\usebox{\Ltr}}
\put(90,36){\usebox{\Lbl}}
\put(54,54){\usebox{\Lt}}
\put(90,108){\usebox{\Lb}}
\put(90,0){\usebox{\Lt}}
\multiput(126,54)(0,6){7}{\circle*{2}}
\end{picture}
\ema
\caption{Conventions for the N-point integral.\label{fig41}}
\efi
We define the general one-loop tensor integral (see Fig.~\ref{fig41}) as
\beq \label{tenint}
T^{N}_{\mu_{1}\ldots\mu_{P}} (p_{1},\ldots,p_{N-1},m_{0},\ldots,m_{N-1})=
\frac{(2\pi\mu)^{4-D}}{i\pi^{2}}\int d^{D}\!q
\frac{q_{\mu _{1}}\cdots q_{\mu _{P}}}{D_{0}D_{1}\cdots D_{N-1}}
\eeq
with the denominator factors
\beq \label{D0Di}
D_{0}= q^{2}-m_{0}^{2}+i\varepsilon, \qquad
D_{i}= (q+p_{i})^{2}-m_{i}^{2}+i\varepsilon, \qquad i=1,\ldots,N-1 ,
\eeq
originating from the propagators in the Feynman diagram.
Furthermore we introduce
\beq
p_{i0}=p_{i} \quad \mbox{ and } \quad p_{ij}=p_{i}-p_{j}.
\eeq
 
Evidently the tensor integrals are invariant under arbitrary
permutations of the propagators $D_{i}$, $i\ne0$ and totally symmetric in
the Lorentz indices $\mu_{k}$.
$i\varepsilon$ is an infinitesimal imaginary part which is needed
to regulate singularities of the integrand. Its specific choice
ensures causality. After integration it
determines the correct imaginary parts of the logarithms and
dilogarithms.
The parameter $\mu$ has mass dimension and serves to keep the
dimension of the integrals fixed for varying $D$.
Conventionally $T^{N}$
is denoted by the $N$th character of the alphabet, i.e. $T^{1}\equiv A$,
$T^{2}\equiv B$, \ldots, and the scalar integrals carry an index $0$.
 
Lorentz covariance of the integrals allows to decompose the
tensor integrals into tensors constructed from the external momenta
$p_{i}$, and the metric tensor $g_{\mu\nu }$ with totally
symmetric coefficient functions $T^{N}_{i_{1}\ldots i_{P}}$.
We formally introduce an artificial momentum $p_{0}$ in order
to write the terms containing $g_{\mu \nu }$ in a compact way
\beq
T^{N}_{\mu_{1}\ldots \mu_{P}}(p_{1},\ldots,p_{N-1},m_{0},\ldots,m_{N-1})
=\sum_{i_{1},\ldots,i_{P}=0}^{N-1} T_{i_{1}\ldots i_{P}}^{N}
p_{i_{1}\mu_{1}}\cdots p_{i_{P}\mu_{P}}.
\label{tendec}
\eeq
From this formula the correct $g_{\mu\nu }$ terms are recovered by omitting
all terms containing an odd
number of $p_{0}$'s and replacing products of even numbers of $p_{0}$'s by the
corresponding totally symmetric tensor constructed from the $g_{\mu \nu}$, e.g.
\beq
\barr{lll}
p_{0\mu_{1}}p_{0\mu_{2}} & \to & g_{\mu_{1}\mu_{2}},  \\[1ex]
p_{0\mu_{1}}p_{0\mu_{2}}p_{0\mu_{3}}p_{0\mu_{4}} & \to &
g_{\mu_{1}\mu_{2}}g_{\mu_{3}\mu_{4}}
+g_{\mu_{1}\mu_{3}}g_{\mu_{2}\mu_{4}}
+g_{\mu_{1}\mu_{4}}g_{\mu_{2}\mu_{3}}.
\earr
\eeq
The explicit Lorentz decompositions for the lowest order integrals read
\beq
\begin{array}{lll}
\hphantom{C_{\mu \nu \rho }} &&
\hphantom{\mbox{}(g_{\mu \nu }p_{1\rho }+g_{\nu \rho }p_{1\mu }+
g_{\mu \rho }p_{1\nu }) C_{001} + (g_{\mu \nu }p_{2\rho }+g_{\nu \rho }
p_{2\mu }+ g_{\mu \rho }p_{2\nu }) C_{002}}\\[-.7em]
B_{\mu } & = & p_{1\mu } B_{1}, \\[1em]
B_{\mu \nu } & = & g_{\mu \nu } B_{00} + p_{1\mu } p_{1\nu } B_{11},
\end{array}
\label{tedecb}
\end{equation}
\begin{equation}
\begin{array}{lll}
C_{\mu } & = & p_{1\mu } C_{1} + p_{2\mu } C_{2}
= \disp\sum_{i=1}^{2} p_{i\mu} C_{i} ,\\[1.2em]
C_{\mu \nu } & = & g_{\mu \nu }C_{00} + p_{1\mu }p_{1\nu }C_{11} + p_{2\mu}
p_{2\nu }C_{22} +(p_{1\mu }p_{2\nu }+p_{2\mu }p_{1\nu }) C_{12} \\[1ex]
&=& g_{\mu \nu }C_{00} +
\disp\sum_{i,j=1}^{2} p_{i\mu} p_{j\nu }C_{ij} ,\\[1.2em]
C_{\mu \nu \rho } &=& \mbox{}(g_{\mu \nu }p_{1\rho }+g_{\nu \rho }p_{1\mu }+
g_{\mu \rho }p_{1\nu }) C_{001} + (g_{\mu \nu }p_{2\rho }+g_{\nu \rho }
p_{2\mu }+ g_{\mu \rho }p_{2\nu }) C_{002}  \\[1ex]
&&\mbox{} + p_{1\mu }p_{1\nu }p_{1\rho } C_{111}
+ p_{2\mu }p_{2\nu }p_{2\rho } C_{222} \\[1ex]
&&\mbox{} +(p_{1\mu }p_{1\nu }p_{2\rho }+
p_{1\mu }p_{2\nu }p_{1\rho }+
p_{2\mu }p_{1\nu }p_{1\rho }  ) C_{112} \\[1ex]
&&\mbox{} +(p_{2\mu }p_{2\nu }p_{1\rho } + p_{2\mu }p_{1\nu }p_{2\rho }+
p_{1\mu }p_{2\nu }p_{2\rho } ) C_{122} \\[1ex]
&=& \disp\sum_{i=1}^{2}(g_{\mu \nu }p_{i\rho }+g_{\nu \rho }p_{i\mu }
+ g_{\mu \rho }p_{i\nu }) C_{00i}
+ \disp\sum_{i,j,k=1}^{2} p_{i\mu} p_{j\nu }p_{k\rho}C_{ijk} ,
\end{array}
\label{tedecc}
\end{equation}
\begin{equation}
\begin{array}{lll}
\hphantom{C_{\mu \nu \rho }} &&
\hphantom{\mbox{}(g_{\mu \nu }p_{1\rho }+g_{\nu \rho }p_{1\mu }+
g_{\mu \rho }p_{1\nu }) C_{001} + (g_{\mu \nu }p_{2\rho }+g_{\nu \rho }
p_{2\mu }+ g_{\mu \rho }p_{2\nu }) C_{002}}\\[0em]
D_{\mu } &=& \disp\sum_{i=1}^{3} p_{i\mu} D_{i} ,\\[1em]
D_{\mu \nu } &=& g_{\mu \nu }D_{00}
+ \disp\sum_{i,j=1}^{3} p_{i\mu} p_{j\nu }D_{ij} ,\\[1em]
D_{\mu \nu \rho } &=&
\disp\sum_{i=1}^{3}(g_{\mu \nu }p_{i\rho }+g_{\nu \rho }p_{i\mu }
+ g_{\mu \rho }p_{i\nu }) D_{00i}
+ \sum_{i,j,k=1}^{3} p_{i\mu} p_{j\nu }p_{k\rho}D_{ijk} ,\\[1.4em]
D_{\mu \nu \rho \sigma } &=&
(g_{\mu\nu}g_{\rho \sigma} + g_{\mu\rho}g_{\nu \sigma }
+ g_{\mu\sigma}g_{\nu\rho }) D_{0000}\\ [.7em]
&&\mbox{}+\disp\sum_{i,j=1}^{3}(g_{\mu \nu }p_{i\rho}p_{j\sigma}
+ g_{\nu \rho }p_{i\mu }p_{j\sigma} + g_{\mu \rho }p_{i\nu}p_{j\sigma}
\\[.7em]
&& \phantom{+\disp\sum_{i,j=1}^{3}(}
+ g_{\mu \sigma}p_{i\nu}p_{j\rho }+ g_{\nu \sigma}p_{i\mu}p_{j\rho }
+ g_{\rho \sigma}p_{i\mu}p_{j\nu}) D_{00ij} \\[.7em]
&&\mbox{}+\disp\sum_{i,j,k,l=1}^{3} p_{i\mu} p_{j\nu
}p_{k\rho}p_{l\sigma}D_{ijkl} .
\end{array}
\label{tedecd}
\end{equation}
 
Since the four dimensional space is spanned by four Lorentz vectors the
terms involving $g_{\mu \nu}$ should be omitted for $N \ge 5$ and at
most four Lorentz vectors should be used in the decomposition
(\ref{tendec}). Consequently the Lorentz decomposition for a general
tensor integral with $N \ge 5$ in four dimensions can be written as
\beq
T^{N}_{\mu_{1}\ldots \mu_{P}}(p_{1},\ldots,p_{N-1},m_{0},\ldots,m_{N-1})
=\sum_{i_{1},\ldots,i_{P}=1}^{4} T_{i_{1}\ldots i_{P}}^{N}
p_{i_{1}\mu_{1}}\cdots p_{i_{P}\mu_{P}},
\eeq
where $p_{1},\ldots,p_{4}$ is any set of four linear independent Lorentz
vectors out of $p_{1},\ldots,p_{N-1}$.
The symmetry of the tensor integrals under exchange of the propagators
yields relations bet\-ween the scalar coefficient functions. Exchanging
the arguments $(p_{i},m_{i})\leftrightarrow (p_{j},m_{j})$  together
with the corresponding indices $i \leftrightarrow j$ leaves the
scalar coefficient functions invariant
\beq
\barr{l}
T^{N}_{\ldots\underbrace{\scriptstyle i\ldots i}_{n}
\ldots\underbrace{\scriptstyle j\ldots j}_{m}
\ldots}(p_{1},\ldots,p_{i},\ldots,p_{j},\ldots,p_{N-
1},m_{0},\ldots,m_{i},\ldots,m_{j},\ldots,m_{N-1}) \\
=T^{N}_{\ldots\underbrace{\scriptstyle i\ldots i}_{m}
\ldots\underbrace{\scriptstyle j\ldots j}_{n}
\ldots}(p_{1},\ldots,p_{j},\ldots,p_{i},\ldots,p_{N-
1},m_{0},\ldots,m_{j},\ldots,m_{i},\ldots,m_{N-1}),
\earr
\eeq
e.g.
\beq
\barr{lll}
C_{1}(p_{1},p_{2},m_{0},m_{1},m_{2}) & = &
C_{2}(p_{2},p_{1},m_{0},m_{2},m_{1}) , \\[1ex]
C_{00}(p_{1},p_{2},m_{0},m_{1},m_{2}) & = &
C_{00}(p_{2},p_{1},m_{0},m_{2},m_{1}) , \\[1ex]
C_{12}(p_{1},p_{2},m_{0},m_{1},m_{2}) & = &
C_{12}(p_{2},p_{1},m_{0},m_{2},m_{1}) .
\earr
\eeq
 
All one-loop tensor integrals can be reduced to the scalar ones
$T_{0}^{N}$. This is done in Sect.\ \ref{sectenred}. General analytical
results for the scalar integrals $A_{0}$, $B_{0}$, $C_{0}$ are $D_{0}$
are listed in Sect.\ \ref{sectensca}.
The scalar integrals for $N > 4$ can be expressed in terms of $D_{0}$'s
in four dimensions. The relevant formulae are given in Sect.\
\ref{sectenres}. They apply as well to the
tensor integrals with $N\le 4$
in the kinematical regions, where the usual tensor integral reduction
breaks down, because the Gram determinants appearing there are zero.
The UV-divergent parts of the one-loop integrals are explicitly given
in Sect.~\ref{sectendiv}
 
\section{Reduction of tensor integrals to scalar integrals}
\label{sectenred}
 
Using the Lorentz decomposition of the tensor integrals (\ref{tendec})
the invariant functions $T_{i_{1}\ldots i_{P}}^{N}$
can be iteratively reduced to the scalar integrals $T_{0}^{N}$ \cite{Pa79}.
We derive the relevant formulae for the general tensor integral.
 
The product of the integration momentum $q_{\mu }$ with an external
momentum can be expressed in terms of the denominators
\beq \label{qpk}
qp_{k}=\frac{1}{2}[D_{k}-D_{0}-f_{k}], \qquad f_{k}=p_{k}^{2}-
m_{k}^{2}+m_{0}^{2}.
\eeq
Multiplying (\ref{tenint}) with $p_{k}$ and substituting (\ref{qpk}) yields
\beq
\barr{lll}
R^{N,k}_{\mu_{1}\ldots \mu_{P-1}} & = &
T^{N}_{\mu_{1}\ldots \mu_{P}}p_{k}^{\mu _{P}} \\[1em]
&=& \disp \frac{1}{2} \frac{(2\pi\mu)^{4-D}}{i\pi^{2}}\int d^{D}\!q
\left[\frac{q_{\mu _{1}}\ldots q_{\mu _{P-1}}}{D_{0}\ldots D_{k-1}D_{k+1}
\ldots D_{N-1}} \right.\\[1em]
&&\disp \hspace{4em} \left.
-\frac{q_{\mu _{1}}\ldots q_{\mu _{P-1}}}{D_{1}\ldots D_{N-1}}
- f_{k}\frac{q_{\mu _{1}}\ldots q_{\mu _{P-1}}}{D_{0}\ldots D_{N-1}}
\right]\\[1em]
&=& \frac{1}{2}\left[
T^{N-1}_{\mu_{1}\ldots \mu_{P-1}}(k)
-T^{N-1}_{\mu_{1}\ldots \mu_{P-1}}(0)
-f_{k}T^{N}_{\mu_{1}\ldots \mu_{P-1}}\right] ,
\label{tens1}
\earr
\eeq
where the argument $k$ of the tensor integrals in the last line indicates
that the propagator $D_{k}$ was cancelled.
Note that $T^{N-1}_{\mu_{1}\ldots\mu_{P-1}}(0)$ has
an external momentum in its first propagator. Therefore a shift of the
integration momentum 
has to be performed in this integral in order to bring it to the form
(\ref{tenint}).
All tensor integrals on the right-hand side of eq.\ (\ref{tens1})
have one Lorentz index less than the original tensor integral. In two
of them also one propagator is eliminated.
 
For $P\ge 2$ we obtain one more relation by contracting (\ref{tenint})
with $g^{\mu\nu}$ and using
\beq
g^{\mu \nu}q_{\mu}q_{\nu }=q^{2}=D_{0}+m^{2}_{0}.
\eeq
This gives
\beq
\barr{lll}
R^{N,00}_{\mu_{1}\ldots \mu_{P-2}} & = &
T^{N}_{\mu_{1}\ldots \mu_{P}}g^{\mu_{P-1}\mu_{P}} \\ [1em]
&=& \disp \frac{(2\pi\mu)^{4-D}}{i\pi^{2}}\int d^{D}\!q
\left[\frac{q_{\mu _{1}}\ldots q_{\mu _{P-2}}}{D_{1}\ldots D_{N}}
+m_{0}^{2}\frac{q_{\mu _{1}}\ldots q_{\mu _{P-2}}}{D_{0}\ldots D_{N}}
\right]\\[1em]
&=&\left[
T^{N-1}_{\mu_{1}\ldots \mu_{P-2}}(0)
+m_{0}^{2}T^{N}_{\mu_{1}\ldots \mu_{P-2}}\right].
\label{tens2}
\earr
\eeq
Inserting the Lorentz decomposition (\ref{tendec}) for the
tensor integrals $T$ into (\ref{tens1}) and (\ref{tens2}) we obtain
set of linear equations for the corresponding coefficient functions.
This set decomposes naturally into disjoint sets of $N-1$ equations for
each tensor
integral. If the inverse of the matrix
\beq
X_{N-1} = \left(
\barr{cccc}
p_{1}^{2} & p_{1}p_{2} & \ldots & \;p_{1}p_{N-1} \\
p_{2}p_{1}& p_{2}^{2}  & \ldots & \;p_{2}p_{N-1} \\
\vdots    & \vdots     & \ddots     &\;\vdots     \\
p_{N-1}p_{1} & p_{N-1}p_{2} & \ldots &\; p_{N-1}^{2}
\earr
\right)
\eeq
exists, these can be solved for the invariant functions
$T^{N}_{i_{1}\ldots i_{p}}$ yielding them in terms of invariant
functions of tensor integrals with fewer indices (see eqs.\
(\ref{tenst}) and (\ref{tensr}) below). In this way all tensor integrals
are expressed iteratively in terms of scalar integrals $T^{L}_{0}$ with
$L\le N$.
 
If the matrix $X_{N-1}$ becomes singular, the reduction algorithm breaks
down. If this is due to the linear dependence of the momenta
we can leave out the linear dependent vectors of the set
$p_{1},\ldots,p_{N-1}$
in the Lorentz decomposition resulting in a smaller matrix
$X_{M}$. If $X_{M}$ is nonsingular the reduction algorithm works again. This
happens usually at the edge of phase space where some of the momenta $p_{i}$
become collinear.
 
If the determinant of $X_{N-1}$, the Gram determinant, is zero but the
momenta
are not linear dependent\footnote{This can happen, because of the
indefinite metric of space time.} one has to use a different reduction
algorithm \cite{Me65,St87}.
This will be discussed in Sect.\ \ref{sectenres}.
 
Here we give the results for the reduction of arbitrary N-point integrals
 depending
on $M\le N-1$ linear independent Lorentz vectors in $D$ dimensions for
nonsingular $X_{M}$.
Inserting the Lorentz decomposition of $T_{N}$ and $R^{N,k}$ as well as
$R^{N,00}$
\beq
\barr{lllll}
R^{N,k}_{\mu_{1}\ldots \mu_{P-1}} & = &
T^{N}_{\mu_{1}\ldots \mu_{P}} p_{k}^{\mu_{P}} & = &
\disp\sum_{i_{1},\ldots,i_{P-1}=0}^{M} R_{i_{1}\ldots i_{P-1}}^{N,k}
p_{i_{1}\mu_{1}}\cdots p_{i_{P-1}\mu_{P-1}} , \\[1em]
R^{N,00}_{\mu_{1}\ldots \mu_{P-2}} & = &
T^{N}_{\mu_{1}\ldots \mu_{P}} g^{\mu_{P-1}\mu_{P}} & = &
\disp\sum_{i_{1},\ldots,i_{P-2}=0}^{M} R_{i_{1}\ldots i_{P-2}}^{N,00}
p_{i_{1}\mu_{1}}\cdots p_{i_{P-2}\mu_{P-2}} ,
\label{tens4}
\earr
\eeq
into the first lines of (\ref{tens1}) and (\ref{tens2})
these equations can be solved for the $T^{N}_{i_{1}\ldots i_{p}}$:
\beq
\barr{lll}
T_{00i_{1}\ldots i_{P-2}}^{N} & = & \disp\frac{1}{D+P-2-M}
\left[R_{i_{1}\ldots i_{P-2}}^{N,00}
-\sum_{k=1}^{M} R_{ki_{1}\ldots i_{P-2}}^{N,k} \right], \\[1em]
T_{ki_{1}\ldots i_{P-1}}^{N} & = & \left(X_{M}^{-1}\right)_{kk'}
\left[R_{i_{1}\ldots i_{P-1}}^{N,k'}
-\disp\sum_{r=1}^{P-1} \delta_{i_{r}}^{k'}
T_{00i_{1}\ldots i_{r-1}i_{r+1}\ldots i_{P-1}}^{N} \right] .
\label{tenst}
\earr
\eeq
Note that the numerator of the prefactor in the first equation is
always positve in the relevant cases $P \ge 2$ and $D > M$.
Using the third lines of (\ref{tens1}) and (\ref{tens2})
the $R$'s can be expressed in terms of
$T^{N}_{i_{1}\ldots i_{P-1}}$, $T^{N}_{i_{1}\ldots i_{P-2}}$, 
and $T^{N-1}_{i_{1}\ldots i_{q}}$, with $q < P$
as follows
\beqar \label{tensr}
\nn R_{i_{1}\ldots i_{q}\underbrace{\scriptstyle M\ldots M}_{P-2-q}}^{N,00}
& = &
m_{0}^{2} T_{i_{1}\ldots i_{q}\underbrace{\scriptstyle M\ldots M}_{P-2-q}}^{N}
\\*[1em]
\nn && + \, (-1)^{P-q}\left[\tilde{T}_{i_{1}\ldots i_{q}}^{N-1}(0)
+ \disp\left({P-2-q \atop 1}\right) \disp\sum_{k_{1}=1}^{M-
1}\tilde{T}_{i_{1}\ldots i_{q}k_{1}}^{N-1}(0) \right. \\[1em]
\nn &&\qquad+ \disp\left({P-2-q \atop 2}\right) \disp\sum_{k_{1},k_{2}=1}^{M-
1}\tilde{T}_{i_{1}\ldots i_{q}k_{1}k_{2}}^{N-1}(0) + \ldots \\[1em]
\nn &&\qquad+ \disp\left. \left({P-2-q \atop P-2-q}\right)
\disp\sum_{k_{1},\ldots, k_{P-2-q}=1}^{M-
1}\tilde{T}_{i_{1}\ldots i_{q}k_{1}\ldots k_{P-2-q}}^{N-1}(0) \right],
\\[1.5em]
R_{i_{1}\ldots i_{q}\underbrace{\scriptstyle M\ldots M}_{P-1-q}}^{N,k} & = &
\frac{1}{2}\Biggl\{T_{\tilde{i}_{1}\ldots \tilde{i}_{q}
\underbrace{\scriptstyle \tilde{M}\ldots \tilde{M}}_{P-1-q}}^{N-1} (k)
\theta(k\mid i_{1},\ldots, i_{q},
\underbrace{M,\ldots ,M}_{P-1-q})
- f_{k}T_{i_{1}\ldots i_{q}\underbrace{\scriptstyle M\ldots M}_{P-1-q}}^{N}
 \\[1em]
\nn && \disp - (-1)^{P-1-q}\left[\tilde{T}_{i_{1}\ldots i_{q}}^{N-1}(0)
+ \left({P-1-q \atop 1}\right) \disp\sum_{k_{1}=1}^{M-
1}\tilde{T}_{i_{1}\ldots i_{q}k_{1}}^{N-1}(0)\right.  \\[1em]
\nn && \qquad+ \disp \left({P-1-q \atop 2}\right) \disp\sum_{k_{1},k_{2}=1}^{M-
1}\tilde{T}_{i_{1}\ldots i_{q}k_{1}k_{2}}^{N-1}(0) + \ldots \\[1em]
\nn &&\qquad+ \disp \left. \left({P-1-q \atop P-1-q}\right)
\disp\sum_{k_{1},\ldots, k_{P-1-q}=1}^{M-
1}\tilde{T}_{i_{1}\ldots i_{q}k_{1}\ldots k_{P-1-q}}^{N-1}(0) \right]
\Biggr\}
\eeqar
where $i_{1},\ldots,i_{q}\ne M$ and
\beq
\theta(k\mid i_{1},\ldots,i_{P-1}) =
\left\{\barr{ll} 1 & \quad i_{r}\ne k,\quad r=1,\ldots,P-1, \\
                0 & \quad \mbox{else}. \earr \right.
\eeq
The indices $\tilde{i}$ refer to the $i$-th momentum of the corresponding
$N$-point function $T^N$ but to the $(i-1)$-th momentum of the $N-1$-point
function $T^{N-1}(k)$ if $i>k$.
Again the arguments of the $T$'s indicate the cancelled propagators.
The tilde in $\tilde{T}(0)$ means that a shift
of the integration variable $q\to q-p_{M}$ has been performed in order to
obtain the standard form of these integrals. This shift generates the
terms in the square brackets of (\ref{tensr}). 
It is also the reason for the unsymmetric appearance of the index $M$ in
the above equations. A different shift would result in similar results.
An explicit example illustrating the use of these reduction formulae is
given in App.~C.
 
The recursion formulae above determine the coefficients $T_{i_{1}\ldots
i_{P}}$ regardless of their symmetries. Consequently coefficients whose
indices are not all equal are obtained in different ways.
This allows for checks on the analytical results as well as on numerical
stability.
 
If the number $M$ of linear independent momenta equals the dimension
$D$ of space-time then the terms containing $g_{\mu \nu }$ in the
Lorentz decomposition have to be omitted, since $g_{\mu \nu }$ can be
built up from the $D$ momenta. In this case the coefficients
$T^{N}_{i_{1}\ldots i_{P}}$ are obtained from the second equations
in (\ref{tenst}) and
(\ref{tensr}) with $T^{N}_{00i_{1}\ldots i_{P-2}}=0$.
 
\section{Scalar one-loop integrals for $N\le 4$}
\label{sectensca}
 
With the methods described in the last section all one-loop integrals
can be reduced to the scalar ones
$T_{0}^{N}$
provided the matrices $X_{M}$ are nonsingular.
General analytical results for $A_{0},B_{0},C_{0}$ and $D_{0}$ were
derived in \cite{tH79}.  Algorithms for
the numerical calculation of the scalar one-loop integrals based on
these results have been presented in \cite{Ol90}. Here we give a new formula
\cite{De91a} for $D_{0}$ involving only 16 dilogarithms compared to 24 of the
solution of \cite {tH79}. For completeness we first list the results for
$A_{0},B_{0}$ and $C_{0}$.
 
\subsection{Scalar one-point function}
 
The scalar one-point function reads
\beq
A_{0}(m)=-m^{2}(\frac{m^{2}}{4\pi\mu ^{2}})^{\frac{D-4}{2}}\Gamma
(1-\frac{D}{2})\\
=m^{2}(\Delta-\log\frac{m^{2}}{\mu ^{2}}+1)+O(D-4),
\eeq
with the UV-divergence contained in
\beq
\Delta =\frac{2}{4-D}-\gamma _{E}+\log 4\pi
\eeq
and $\gamma _{E}$ is Euler's constant.
The terms of order $O(D-4)$ are only relevant for two- or higher-loop
calculations.
 
\subsection{Scalar two-point function}
 
The two-point function is given by
\beq
\barr{lll}
B_{0}(p_{10},m_{0},m_{1})&=&\disp\Delta -
\int^{1}_{0} dx \log\frac{[p^{2}_{10}x^{2}-x(p^{2}_{10}-
m^{2}_{0}+m^{2}_{1})+m^{2}_{1}-i\varepsilon ]}{\mu^{2}}+O(D-4)\\[1em]
&=&\disp\Delta + 2 - \log\frac{m_{0}m_{1}}{\mu^{2}}
+\frac{m^{2}_{0}-m^{2}_{1}}{p^{2}_{10}}\log\frac{m_{1}}{m_{0}}
-\frac{m_{0}m_{1}}{p^{2}_{10}}(\frac{1}{r}-r)\log r \\[1em]
&&+O(D-4),
\earr
\label{B0}
\eeq
where $r$ and $\frac{1}{r}$ are determined from
\beq
x^{2}+\frac{m^{2}_{0}+m^{2}_{1}-p^{2}_{10}-i\varepsilon
}{m_{0}m_{1}}\,x+1=(x+r)(x+\frac{1}{r}) .
\eeq
The variable $r$ never crosses the negative real axis even for complex physical
masses ($m^{2}$ has a negative imaginary part!). For $r < 0$ the
$i\varepsilon $ prescription yields $\Im \,r=\varepsilon \, \sgn (r-
\frac{1}{r})$. Consequently the result (\ref {B0}) is valid for
arbitrary physical parameters.
 
For the field renormalization constants we need the
derivative of $B_{0}$ with respect to $p_{10}^{2}$. This is easily
obtained by differentiating the above result
\beq
\barr{lll}
\disp\frac{\partial}{\partial p_{10}^{2}}B_{0}(p_{10},m_{0},m_{1})&=&
\disp-\frac{m^{2}_{0}-m^{2}_{1}}{p^{4}_{10}}\log\frac{m_{1}}{m_{0}}
+\frac{m_{0}m_{1}}{p^{4}_{10}}(\frac{1}{r}-r)\log r \\[1em]
&&\disp-\frac{1}{p_{10}^{2}}\Bigl(1+\frac{r^{2}+1}{r^{2}-1}\log r\Bigr) +O(D-4).
\earr
\label{DB0}
\eeq
 
\subsection{Scalar three-point function}
 
The general result for the scalar three-point function valid for all real
momenta and physical masses was calculated by \cite{tH79}. It can be
brought into the symmetric form
\beqar \label{C0}
\nn\lefteqn{C_{0}(p_{10},p_{20},m_{0},m_{1},m_{2}) =} \\
\nn&&-\int^{1}_{0}dx\int^{x}_{0}dy[p^{2}_{21}x^{2}+p^{2}_{10}y^{2}+
(p^{2}_{20}-p^{2}_{10}-p^{2}_{21})xy \\[1ex]
&&\qquad\qquad\mbox{}
+(m^{2}_{1}-m^{2}_{2}-p^{2}_{21})x+(m^{2}_{0}-m^{2}_{1}+p^{2}_{21}-
p^{2}_{20})y+m^{2}_{2}-i\varepsilon ]^{-1} \\[1ex]
\nn&=&\frac{1}{\alpha }\sum^{2}_{i=0}\Biggl\{\sum_{\sigma =\pm}
\biggl[\Li\Bigl(\frac{y_{0i}-1}{y_{i\sigma }}\Bigr)
-\Li\Bigl(\frac{y_{0i}}{y_{i\sigma }}\Bigr)\\[1ex]
\nn&&\qquad\qquad +\eta\Bigl(1-x_{i\sigma },\frac{1}{y_{i\sigma }}\Bigr)
\log\frac{y_{0i}-1}{y_{i\sigma }}
-\eta\Bigl(-x_{i\sigma },\frac{1}{y_{i\sigma}}\Bigr)
\log\frac{y_{0i}}{y_{i\sigma}}
\biggr]\\[1ex]
\nn&& -\biggl[\eta (-x_{i+},-x_{i-}) -\eta (y_{i+},y_{i-})
-2\pi i \theta (-p_{jk}^{2})\theta \bigr(-\Im(y_{i+}y_{i-})\bigl)\biggr]
\log \frac{1-y_{i0}}{-y_{i0}}\Biggr\},
\eeqar
with ($i,j,k=0,1,2$ and cyclic)
\beqar
\nn y_{0i} & = & \frac{1}{2\alpha p_{jk}^{2}}
\bigl[ p_{jk}^{2}( p_{jk}^{2}-p_{ki}^{2}-p_{ij}^{2}
                 + 2m_{i}^{2} -m_{j}^{2} -m_{k}^{2} ) \\[1ex]
\nn && \quad -(p_{ki}^{2}-p_{ij}^{2})(m_{j}^{2} -m_{k}^{2} )
+\alpha (p_{jk}^{2}-m_{j}^{2} +m_{k}^{2})\bigr], \\[1ex]
\nn x_{i\pm} & = & \frac{1}{2 p_{jk}^{2}}
\bigl[ p_{jk}^{2}-m_{j}^{2}+m_{k}^{2}  \pm \alpha_{i}\bigr], \\[1ex]
    y_{i\pm} & = &y_{0i} - x_{i\pm}, \\[1ex]
\nn  \alpha & = & \kappa(p_{10}^{2},p_{21}^{2},p_{20}^{2})
,\\[1ex]
\nn  \alpha_{i} & = & \kappa(p_{jk}^{2},m_{j}^{2},m_{k}^{2})
\,(1+i\varepsilon p_{jk}^{2}) ,
\eeqar
and $\kappa$ is the K\"all\'en function
\beq \label{kappa}
\kappa(x,y,z) = \sqrt{x^{2}+y^{2}+z^{2}-2(xy+yz+zx)}.
\eeq
The dilogarithm or Spence function $\Li(x)$ is defined as
\beq
\Li(x) = -\int_{0}^{1}\frac{dt}{t} \log(1-xt) , \qquad |\mbox{arg}\,(1-
x)|<\pi .
\eeq
The $\eta$-function compensates for cut crossings on the Riemann-sheet
of the logarithms and dilogarithms. For $a$, $b$ on the first Riemann
sheet it is defined by
\beq
\log(ab) = \log(a) + \log(b) + \eta(a,b).
\eeq
All $\eta$-functions in (\ref{C0}) vanish if $\alpha$ and all the
masses $m_{i}$ are real. Note that $\alpha $
is real in particular for all on-shell decay and scattering processes.
 
\subsection{Scalar four-point function}
 
The scalar four-point function
$D_{0}(p_{10},p_{20},p_{30},m_{0},m_{1},m_{2},m_{3})$
can be expressed in terms of 16 dilogarithms \cite{De91a}.
 
Before we give the result we first introduce some useful variables and
functions. We define
\beq
k_{ij}=\frac{m_{i}^{2}+m_{j}^{2}-p_{ij}^{2}}{m_{i}m_{j}},\quad
i,j=0,1,2,3,
\eeq
and $r_{ij}$ and $\tilde{r}_{ij}$ by
\begin{equation}
x^{2}+k_{ij}x+1=(x+r_{ij})(x+1/r_{ij}),
\end{equation}
and
\begin{equation}
x^{2}+(k_{ij}-i\varepsilon )x+1=(x+\tilde{r}_{ij})(x+1/\tilde{r}_{ij}) .
\end{equation}
Note that for real $k_{ij}$ the $r_{ij}$'s lie either
on the real axis or on the complex unit circle.
Furthermore
\beqar
P(y_{0},y_{1},y_{2},y_{3})&=&\sum_{0\leq i<j\leq 3}k_{ij}y_{i}y_{j}+
\sum_{j=0}^{3}y_{j}^{2},\\[1em]
Q(y_{0},y_{1},0,y_{3})&=&(1/r_{02}-r_{02})y_{0}
+(k_{12}-r_{02}k_{01})y_{1}+(k_{23}-r_{02}k_{03})y_{3},\\[1em]
\overline{Q}(y_{0},0,y_{2},y_{3})&=&(1/r_{13}-r_{13})y_{3}
+(k_{12}-r_{13}k_{23})y_{2}+(k_{01}-r_{13}k_{03})y_{0}.
\eeqar
and $x_{1,2}$ is defined by
\beqar
&&\frac{r_{02}r_{13}}{x}
\left\{\Bigl[P(1,\frac{x}{r_{13}},0,0)-i\varepsilon\Bigr]
\Bigl[P(0,0,\frac{1}{r_{02}},x)- i\varepsilon \Bigr] \right. \nn\\[.8ex]
&&\qquad\quad\left.-\Bigl[P(0,\frac{x}{r_{13}},\frac{1}{r_{02}},0)
-i\varepsilon\Bigr]
\Bigl[P(1,0,0,x)- i\varepsilon \Bigr]\right\} \\[1em]
&& =a x^{2} +b x +c + i \varepsilon  d \nonumber  
= a (x-x_{1}) (x-x_{2}) ,
\eeqar
where
\beqar
a&=& k_{23}/r_{13}+r_{02}k_{01}-k_{03}r_{02}/r_{13}- k_{12}, \nn\\[1ex]
b&=& (r_{13}-1/r_{13})(r_{02}-1/r_{02})+k_{01}k_{23}- k_{03}k_{12},
\nn\\[1ex]
c&=&k_{01}/r_{02}+r_{13}k_{23}-k_{03}r_{13}/r_{02}-k_{12}  ,
\nn\\[1ex]
d&=& k_{12}-r_{02}k_{01}-r_{13}k_{23}+r_{02}r_{13}k_{03}.
\end{eqnarray}
In addition we introduce
\beq
\gamma _{kl}=\sgn\,\Re [a(x_{k}-x_{l})],\quad k,l=1,2 ,
\eeq
and
\beq
\barr{ll}
x_{k0} = x_{k}, & s_{0} = \tilde{r}_{03},\\[1ex]
x_{k1} = x_{k}/r_{13},  & s_{1} = \tilde{r}_{01},\\[1ex]
x_{k2} = x_{k} r_{02}/r_{13},\quad & s_{2} = \tilde{r}_{12},\\[1ex]
x_{k3} = x_{k} r_{02}, & s_{3} = \tilde{r}_{23}.
\earr
\eeq
as well as
\beq
x_{kj}^{(0)}= \lim_{\varepsilon  \rightarrow 0}x_{kj}
\quad\mbox{as}\quad r_{ij} = \lim_{\varepsilon  \rightarrow 0}\tilde{r}_{ij}.
\eeq
Finally we need
\begin{eqnarray}
\tilde{\eta }(a,\tilde{b})&=&
\left\{ \begin{array}{l}
        \displaystyle \eta(a,b) \hspace{1cm} \mbox{for $b$ not real,}\\
        \displaystyle 2 \pi i \left[ \theta ( - \Im\, a ) \theta (-\Im\,
         \tilde{b}) - \theta ( \Im\, a) \theta (\Im\, \tilde{b}) \right]
\quad \mbox{for $b<0$,} \\
        \displaystyle 0 \hspace{1cm} \mbox{for $b>0$}
       \end{array} \right.
\end{eqnarray}
with $b = \lim_{\varepsilon  \rightarrow 0}\tilde{b}$.
 
Then the result for real $r_{02}$ can be written as
\vspace{-.2em}
\begin{eqnarray}
\lefteqn{D_{0}(p_{10},p_{20},p_{30},m_{0},m_{1},m_{2},m_{3})=
\frac{1}{m_{1}m_{2}m_{3}m_{4}a(x_{1}-x_{2})}}\nn\\
&&\Biggl\{ \sum_{j=0}^{3} \sum_{k=1}^{2} (-1)^{j+k}
     \biggl[ \Li(1+s_{j}x_{kj}) +\eta (-x_{kj},s_{j}) \log
            (1+s_{j}x_{kj})         \nonumber\\
&&\qquad\qquad\qquad\quad\:\mbox{}+\Li(1+\frac{ x_{kj} }{ s_{j} })
   +\eta (-x_{kj},\frac{1}{s_{j}}) \log (1+\frac{x_{kj}}{s_{j}})
 \biggr]   \nonumber\\
&& +\mbox{} \sum_{k=1}^{2} (-1)^{k+1}
\Biggl[ \tilde{\eta}(-x_{k},\tilde{r}_{02})
  \biggl[  \log ( r_{02} x_{k}) +
      \log \Bigl( Q(\frac{1}{x_{k}^{(0)}},0,0,1) - i\varepsilon \Bigr)
 \nonumber \\
&&\qquad\qquad\qquad\mbox{}
+ \log \Bigl( \frac{\overline{Q}(0,0,1,r_{02} x_{k}^{(0)})}{d} +
               i \varepsilon \gamma _{k,3-k}
              \sgn (r_{02} \Im\, \tilde{r}_{13}) \Bigr) \biggr]
              \nonumber\\
&& \qquad\qquad\mbox{}+\tilde{\eta}(-x_{k},\frac{1}{\tilde{r}_{13}})
   \biggl[\log \Bigl( \frac{x_{k}}{r_{13}} \Bigr) +
      \log \Bigl( Q(\frac{r_{13}}{x_{k}^{(0)}},1,0,0) -i \varepsilon
      \Bigr)
     \nonumber\\
&& \qquad\qquad\qquad\mbox{}
+ \log \Bigl( \frac{\overline{Q}(1,0,0,x_{k}^{(0)})}{d} +
               i \varepsilon \gamma _{k,3-k}
              \sgn (\Im\, \tilde{r}_{13}) \Bigr) \biggr] \nonumber \\
&& \qquad\qquad\mbox{}
- \left[ \tilde{\eta}(-x_{k} ,\frac{\tilde{r}_{02}}{\tilde{r}_{13}})
          + \eta (\tilde{r}_{02},\frac{1}{\tilde{r}_{13}})  \right]
  \biggl[ \log \Bigl( \frac{r_{02} x_{k}}{r_{13}}  \Bigr) +
         \log \Bigl( Q(\frac{r_{13}}{x_{k}^{(0)}},1,0,0)
         -i \varepsilon \Bigr)
    \nonumber \\
&& \qquad\qquad\qquad \mbox{}
+ \log \Bigl( \frac{\overline{Q}(0,0,1,r_{02} x_{k}^{(0)})}{d} +
               i \varepsilon \gamma _{k,3-k}
              \sgn (r_{02} \Im\, \tilde{r}_{13}) \Bigr)
    \biggr]   \nonumber \\
&& \qquad\qquad\mbox{}+\eta (\tilde{r}_{02},\frac{1}{\tilde{r}_{13}})
    \tilde{\eta} (-x_{k},-\frac{\tilde{r}_{02}}{\tilde{r}_{13}})  \Biggr]
    \Biggr\}\;.
\end{eqnarray}
 
In the case that $|r_{ij}|=1$ for all $r_{ij}$, the result reads:
\vspace{-.2em}
\begin{eqnarray}
\lefteqn{D_{0}(p_{10},p_{20},p_{30},m_{0},m_{1},m_{2},m_{3})=
\frac{1}{m_{1}m_{2}m_{3}m_{4}a(x_{1}-x_{2})}}\nn\\*
&& \Biggl\{\sum_{j=0}^{3} \sum_{k=1}^{2} (-1)^{j+k}
     \Bigl[ \Li(1+s_{j}x_{kj}) +\eta (-x_{kj},s_{j}) \log
            (1+s_{j}x_{kj})         \nonumber\\
&&\qquad\qquad\qquad\quad\mbox{}+\Li(1+\frac{ x_{kj} }{ s_{j} })
   +\eta (-x_{kj},\frac{1}{s_{j}}) \log (1+\frac{x_{kj}}{s_{j}})
  \Bigr]   \nonumber\\
&&+\mbox{} \sum_{k=1}^{2} (-1)^{k+1} \Biggl[ \eta (-x_{k},\frac{1}{r_{13}})
   \biggl[\log \Bigl( \frac{r_{13}}{x_{k}^{(0)}} P(1,\frac{x_{k}^{(0)}}
{r_{13}},0,0)
           - \frac{x_{k}^{(0)}}{r_{13}} \varepsilon b \gamma _{k,3-k} \Bigr)
          + \log \Bigl( \frac{x_{k}^{(0)}}{r_{13}} \Bigr) \biggr] \nonumber\\
&&\qquad\qquad\mbox{}  + \eta (-x_{k},r_{02})   \left[
        \log \Bigl(\frac{1}{r_{02}x_{k}^{(0)}}P(0,0,1,r_{02}x_{k}^{(0)})
              - r_{02} x_{k}^{(0)} \varepsilon b \gamma _{k,3-k}  \Bigr)
         + \log (r_{02}x_{k}^{(0)})   \right]       \nonumber\\
&&\qquad\qquad\mbox{} -\left[ \eta (-x_{k}, \frac{r_{02}}{r_{13}})
      +\eta (r_{02},\frac{1}{r_{13}}) \right]  \left[
   \log \Bigl( \frac{r_{13}}{r_{02}x_{k}^{(0)}}P(0,1,\frac{r_{02}x_{k}^{(0)}}
{r_{13}},0)
       - \frac{r_{02}x_{k}^{(0)}}{r_{13}}  \varepsilon b \gamma _{k,3-k} \Bigl)
     \right.    \nonumber\\
&&\qquad\qquad\mbox{} \left.
+ \log \Bigl( \frac{r_{02}x_{k}^{(0)}}{r_{13}}\Bigr) \right]
    + \Bigl(1 -  \gamma _{k,3-k} \sgn (b) \Bigr)
    \eta (-x_{k},-\frac{r_{02}}{r_{13}}) \eta (r_{02},\frac{1}{r_{13}})
       \Biggr]\Biggr\}. \nonumber
\end{eqnarray}
$\varepsilon $ is understood as infinitesimally small.

\section{Reduction of scalar and tensor integrals for vanishing Gram
determinant}
\label{sectenres}
 
Using the four-dimensionality of space-time the scalar five-point function
can be reduced to five scalar four-point functions \cite{Me65,Ol90}.
Furthermore if the Gram determinant of the external momenta of a tensor
integral $T^{N}$ vanishes,
\beq
\left\vert X_{N-1} \right\vert =
\left\vert
\barr{cccc}
p_{1}^{2} & p_{1}p_{2} & \ldots & \;p_{1}p_{N-1} \\
p_{2}p_{1}& p_{2}^{2}  & \ldots & \;p_{2}p_{N-1} \\
\vdots    & \vdots     & \ddots     &\;\vdots     \\
p_{N-1}p_{1} & p_{N-1}p_{2} & \ldots &\; p_{N-1}^{2}
\earr
\right\vert = 0,
\eeq
this tensor integral can be expressed in terms of $N$ integrals
$T^{N-1}$. This is always the case for $N\ge6$, because any five
momenta are linear dependent in four dimensions.
 
\subsection{Reduction of scalar five-point functions}
 
Here we assume that the four external momenta
appearing in the five-point function span the whole four-dimensional
space\footnote{The exceptional case, when they are linear dependent will
be covered in the next section.}.
Then the integration momentum $q$ depends linearly on these four
external momenta and the following equations holds
\beq
0=\left\vert
\barr{cccc}
2q^{2}  & 2qp_{1}    & \ldots & \;2qp_{4} \\
2p_{1}q & 2p_{1}^{2} & \ldots & \;2p_{1}p_{4} \\
\vdots    & \vdots     & \ddots     &\;\vdots     \\
2p_{4}q & 2p_{4}p_{1} & \ldots &\; 2p_{4}^{2}
\earr
\right\vert =
\left\vert\barr{cccc}
2D_{0}+Y_{00}  & 2qp_{1}    & \ldots & \;2qp_{4} \\
D_{1}-D_{0}+Y_{10}-Y_{00} & 2p_{1}^{2} & \ldots & \;2p_{1}p_{4} \\
\vdots    & \vdots     & \ddots     &\;\vdots     \\
D_{4}-D_{0}+Y_{40}-Y_{00} & 2p_{4}p_{1} & \ldots &\; 2p_{4}^{2}
\earr\right\vert
\eeq
with
\beq
Y_{ij} = m_{i}^{2} + m_{j}^{2} - (p_{i}-p_{j})^{2}.
\eeq
and $D_{i}$ as defined in (\ref{D0Di}). Thus we have
\beq
\frac{1}{i\pi^{2}}\int d^{D}\!q
\frac{1}{D_{0}D_{1}\cdots D_{4}}
\left\vert
\barr{cccc}
2D_{0}+Y_{00}  & 2qp_{1}    & \ldots & \;2qp_{4} \\
D_{1}-D_{0}+Y_{10}-Y_{00} & 2p_{1}^{2} & \ldots & \;2p_{1}p_{4} \\
\vdots    & \vdots     & \ddots     &\;\vdots     \\
D_{4}-D_{0}+Y_{40}-Y_{00} & 2p_{4}p_{1} & \ldots &\; 2p_{4}^{2}
\earr
\right\vert = 0.
\eeq
 
Expanding the determinant along the first column we obtain
\beq \label{E0red1}
\barr{lll}
0&=&\Bigl[2T^{4}_{0}(0)+Y_{00}T^{5}_{0}
\Bigr]
\left\vert\barr{ccc}
 2p_{1}p_{1}    & \ldots & \;2p_{1}p_{4} \\
 \vdots     & \ddots     &\;\vdots     \\
 2p_{4}p_{1} & \ldots &\; 2p_{4}p_{4}
\earr\right\vert \\[1em]
&&+\disp\sum_{k=1}^{4}(-1)^{k}\Bigl[T^{4}_{\mu}(k)
-T^{4}_{\mu}(0)
-p_{4\mu}T^{4}_{0}(0)  \\[1em]
&&\qquad\qquad +p_{4\mu}T^{4}_{0}(0)
+(Y_{k0}-Y_{00})T^{5}_{\mu}\Bigr]\\[1em]
&&\qquad\qquad\times \left\vert\barr{ccc}
 2p_{1\mu }    & \ldots & \;2p_{4\mu } \\
 2p_{1}p_{1} & \ldots &\; 2p_{1}p_{4}\\
 \vdots     & \ddots     &\;\vdots     \\
 2p_{k-1}p_{1} & \ldots &\; 2p_{k-1}p_{4}\\
 2p_{k+1}p_{1} & \ldots &\; 2p_{k+1}p_{4}\\
 \vdots     & \ddots     &\;\vdots     \\
 2p_{4}p_{1} & \ldots &\; 2p_{4}^{2}
\earr\right\vert
\earr
\eeq
where the arguments of the functions $T^{4}$ denote again the cancelled
propagators.
 
The Lorentz decomposition of the vector integrals in (\ref{E0red1})
involves only the momenta $p_{1},\ldots,p_{4}$. 
$T^{4}_{\mu\mu_{1}\ldots\mu_{P}}(k)$ does not depend on
$p_{k}$, consequently each term in its Lorentz decomposition contains
a factor $p_{i\mu}$, $i\ne k$ and its contraction with the
corresponding determinant in (\ref{E0red1}) vanishes. Similarly
all terms in the tensor integral decomposition
of $T^{4}_{\mu}(0)+p_{4\mu}T^{4}_{0}(0)$  involve a factor $p_{i\mu }-
p_{4\mu }$, $i=1,2,3$, if one performs the shift $q \to q-p_{4}$ to
bring the tensor integral to the standard form.
Multiplying with the determinants and performing
the sum in (\ref{E0red1}) these terms drop out. Finally in the term
$p_{4\mu}T^{4}_{0}(0)$ the determinant is nonzero 
only for $k=4$
where it can be combined with the first term in (\ref{E0red1}).
Rewriting the resulting equation as a determinant and reinserting
the explicit form of the tensor integrals we find
\beq \label{eq46}
\frac{1}{i\pi^{2}}\int d^{D}\!q
\frac{1}{D_{0}D_{1}\cdots D_{4}}
\left\vert
\barr{cccc}
D_{0}+Y_{00}  & 2qp_{1}    & \ldots & \;2qp_{4} \\
Y_{10}-Y_{00} & 2p_{1}^{2} & \ldots & \;2p_{1}p_{4} \\
\vdots    & \vdots     & \ddots     &\;\vdots     \\
Y_{40}-Y_{00} & 2p_{4}p_{1} & \ldots &\; 2p_{4}^{2}
\earr
\right\vert = 0.
\eeq
Using
\beq
\barr{l}
2p_{i}p_{j}= Y_{ij}-Y_{i0}-Y_{0j}+Y_{00},\\
2qp_{j}= D_{j}-D_{0}+Y_{0j}-Y_{00} ,
\earr
\eeq
adding the first column to each of the other columns and then
enlarging the determinant by one column and one row
this can be written as
\beq
\left\vert\barr{cccc}
1 & Y_{00}    & \ldots & \;Y_{04} \\
0 & \quad T^{4}_{0}(0)+Y_{00}T^{5}_{0}
\quad & \ldots & \quad
T^{4}_{0}(4)+Y_{40}T^{5}_{0} \\
0 & Y_{10}-Y_{00} & \ldots &\; Y_{14}-Y_{04}  \\
\vdots    & \vdots     & \ddots     &\;\vdots     \\
0 & Y_{40}-Y_{00} & \ldots &\; Y_{44}-Y_{04}
\earr\right\vert
=0
\eeq
This is equivalent to
\beq \label{E0red}
\left\vert \barr{cccccc}
T^{5}_{0} &\:-T^{4}_{0}(0)&\:-T^{4}_{0}(1)&\:-T^{4}_{0}(2)&
\:-T^{4}_{0}(3)&\:-T^{4}_{0}(4)\\
  1   &  Y_{00}   &  Y_{01}   &  Y_{02}   &  Y_{03}   &  Y_{04}   \\
  1   &  Y_{10}   &  Y_{11}   &  Y_{12}   &  Y_{13}   &  Y_{14}   \\
  1   &  Y_{20}   &  Y_{21}   &  Y_{22}   &  Y_{23}   &  Y_{24}   \\
  1   &  Y_{30}   &  Y_{31}   &  Y_{32}   &  Y_{33}   &  Y_{34}   \\
  1   &  Y_{40}   &  Y_{41}   &  Y_{42}   &  Y_{43}   &  Y_{44}
\earr \right\vert = 0,
\eeq
which can be solved for 
$T^{5}_{0}$
if the determinant of the
matrix $Y_{ij}$, $i,j = 0,\ldots, 4$ is nonzero. Note that in the 
integral 
$T_{0}^{4}(0)$
the momenta have not been shifted. In particular
(\ref{E0red}) yields the scalar five-point function $T_{0}^{5}$ in terms of
five scalar four-point functions.
 
\subsection{Reduction of N-point functions for
zero Gram determinant} 
 
For vanishing Gram determinant $\vert X_{N-1}\vert$ the following
relation holds, if the Lorentz decomposition of the appearing tensor
integrals contains only momenta and no metric tensors, which is the
case for $N\ge6$ or $P=0$ (scalar integrals)
\beq
\frac{(2\pi\mu)^{4-D}}{i\pi^{2}}\int d^{D}\!q
\frac{q_{\mu _{1}}\cdots q_{\mu _{P}}}{D_{0}D_{1}\cdots D_{N-1}}
\left\vert
\barr{cccc}
D_{0}+Y_{00}  & 2qp_{1}    & \ldots & \;2qp_{N-1} \\
Y_{10}-Y_{00} & 2p_{1}^{2} & \ldots & \;2p_{1}p_{N-1} \\
\vdots    & \vdots     & \ddots     &\;\vdots     \\
Y_{N-10}-Y_{00} & 2p_{N-1}p_{1} & \ldots &\; 2p_{N-1}^{2}
\earr
\right\vert = 0.
\eeq
Performing the same manipulations
of the determinant as in (\ref{eq46}) to (\ref{E0red}) above
this results in
\beq\label{TN}
\left\vert \barr{ccccc}
T_{\mu_{1}\ldots\mu_{P}}^{N} & -T_{\mu_{1}\ldots\mu_{P}}^{N-1}(0) &
-T_{\mu_{1}\ldots\mu_{P}}^{N-1}(1) & \ldots & -T_{\mu_{1}\ldots\mu_{P}}^{N-1}
(N-1)
\\
  1   &  Y_{00}   &  Y_{01}   &  \ldots   &  Y_{0N-1}   \\
  1   &  Y_{10}   &  Y_{11}   &  \ldots   &  Y_{1N-1}   \\
\vdots&  \vdots   &  \vdots   &  \ddots   &  \vdots   \\
  1   &  Y_{N-10}   &  Y_{N-11}   &  \ldots   &  Y_{N-1N-1}
\earr \right\vert = 0 ,
\label{TNred}
\eeq
valid for $|X_{N-1}|=0$ and $N\ge6$ or $P=0$.
We stress again that in the tensor
integral $T^{N}_{\mu _{1}\ldots\mu _{P}}(0)$ appearing in
(\ref{TNred}) the momenta have not been shifted.
Eq.~(\ref{TNred}) determines $T^{N}_{\mu_{1}\ldots\mu_{P}}$ in terms of
$T_{\mu_{1}\ldots\mu_{P}}^{N-1}(i)$, $i=0,\ldots,N-1$, if the
determinant of the matrix
$Y_{ij}$ is nonzero. The vanishing of this determinant
corresponds to the leading Landau singularity of $T^{N}$ which is
clearly not contained in $T^{N-1}$. In this case one has to calculate
$T^{N}$ directly \cite{St87}.
 
Eq.\ (\ref{TNred}) in particular expresses $T^{N}_{0}$ by $T^{N-1}_{0}$.
For $N=5$ and $P=0$ (\ref{TNred}) coincides with (\ref{E0red}), which is thus valid for
arbitrary momenta.
For $N>6$ one can choose any six out of the N denominator factors
resulting in different reductions. For $N\le 4$,
where (\ref{TNred}) is only valid for scalar integrals,
the Gram determinant is
singular at the edge of phase space where some of the momenta $p_{i}$
become collinear, i.e.\ for forward or backward scattering or at the
threshold of a certain process. Because in this special situations all
integrals can be reduced to lower ones one can obtain considerably
simpler formulae than in the general case (see e.g.\ \cite{Gr87a}).
 
With the methods described in this section all tensor integrals
with $N\ge 6$ can be reduced directly to
tensor integrals with smaller $N$. Note that this may yield tensor
integrals with $P > N$ because $P$ is not reduced simultaneously as in
the reduction method described in Sect.~\ref{sectenred}. These tensor
integrals are not directly present in renormalizable
theories. Nevertheless their
reduction to scalar integrals can be done with the
formulae given in Sect.~\ref{sectenred}.
 
\section{UV-divergent parts of tensor integrals}
\label{sectendiv}
 
For practical calculations it is useful to know the UV-divergent parts
of the tensor integrals explicitly. We give directly the products of
$D-4$ with all divergent one-loop tensor coefficient integrals appearing in
renormalizable theories up to terms of the order $O(D-4)$
\beq \label{TNdiv}
\barr{lll}
(D-4)\,A_{0}(m) &=& -2m^{2} ,\\[1ex]
(D-4)\,B_{0}(p_{10},m_{0},m_{1}) &=& -2 ,\\[1ex]
(D-4)\,B_{1}(p_{10},m_{0},m_{1}) &=& \phantom{-} 1 ,\\[1ex]
(D-4)\,B_{00}(p_{10},m_{0},m_{1}) &=&
\phantom{-}\frac{1}{6}(p_{10}^{2}-3m_{0}^{2}-3m_{1}^{2}) ,\\[1ex]
(D-4)\,B_{11}(p_{10},m_{0},m_{1}) &=&  -\frac{2}{3} ,\\[1ex]
(D-4)\,C_{00}(p_{10},p_{20},m_{0},m_{1},m_{2}) &=&  -\frac{1}{2} ,\\[1ex]
(D-4)\,C_{00i}(p_{10},p_{20},m_{0},m_{1},m_{2}) &=&  \phantom{-}\frac{1}{6} ,
\\[1ex]
(D-4)\,D_{0000}(p_{10},p_{20},p_{30},m_{0},m_{1},m_{2},m_{3}) &=&
-\frac{1}{12} .
\earr
\eeq
All other scalar coefficients defined in (\ref{tedecc}) and
(\ref{tedecd}) are UV-finite.

\chapter{Standard matrix elements}
\label{chaSME}
 
\section{Definition}
 
The invariant matrix elements for scattering and decay processes involving
external fermions and/or vector bosons depend on the polarizations
$\sigma_{i}$,
$\sigma_{i}'$ and $\lambda_{i}$ of these
particles. This dependence is completely contained in the polarization
vectors $\varepsilon_{\mu_{i}}(k_{i},\lambda_{i})$ and spinors
$\bar{v}_{\alpha _{i}}(p_{i}',\sigma_{i}')$ and $u_{\alpha
_{i}}(p_{i},\sigma_{i})$.
$k_{i}$, $p_{i}'$ and $p_{i}$ denote the incoming momenta of the vector
bosons, antifermions and fermions, respectively.
For outgoing fermions one has to replace $p$, $p'$ by $-p$, $-p'$ and
one must use $u(-p)=v(p)$.
If we split off the polarization vectors and spinors from the invariant
matrix element ${\cal M}$ we are left with a tensor involving Lorentz and
Dirac indices in the general case
\beq
\barr{lll}
{\cal M}&=&\bar{v}_{\alpha _{1}}(p'_{1},\sigma_{1}')\ldots
\bar{v}_{\alpha_{n}}(p'_{n},\sigma_{n}')
{\cal M}^{\mu_{1}\ldots\mu_{m}}_{\alpha _{1}\ldots\alpha_{n}\beta
_{1}\ldots\beta _{n}}
u_{\beta_{1}}(p_{1},\sigma_{1})\ldots u_{\beta _{n}}(p_{n},\sigma_{n})\\
&&\times\varepsilon_{\mu _{1}}(k_{1},\lambda_{1})\ldots
\varepsilon_{\mu _{m}}(k_{m},\lambda_{m}).
\earr
\eeq
To be definite we choose $m$ external vector bosons and $n$ external
fermion-antifermion pairs. The tensor ${\cal M}^{\mu _{1}\ldots\mu_{m}}
_{\alpha _{1}\ldots\beta _{n}}$ can be decomposed into a set of
covariant operators together with the corresponding scalar formfactors $F_{i}$
\beq
{\cal M}^{\mu _{1}\ldots\mu _{m}}_{\alpha _{1}\ldots\beta _{n}}=
\sum_{i}{\cal M}^{\mu _{1}\ldots\mu _{m}}_{i,\alpha
_{1}\ldots\beta _{n}} F_{i}.
\eeq
We call the covariant operators ${\cal M}^{\mu _{1}\ldots\mu _{n}}_{i,\alpha
_{1}\ldots\beta _{n}}$ multiplied by the corresponding polarization
vectors and spinors standard matrix elements ${\cal M}_{i}$
\beq
\barr{lll}
{\cal M}_{i}&=&\bar{v}_{\alpha _{1}}(p'_{1},\sigma_{1}')\ldots
\bar{v}_{\alpha_{n}}(p'_{n},\sigma_{n}')
{\cal M}^{\mu_{1}\ldots\mu_{m}}_{i,\alpha _{1}\ldots\beta _{n}}
u_{\beta_{1}}(p_{1},\sigma_{1})\ldots u_{\beta _{n}}(p_{n},\sigma_{n})\\
&&\times\varepsilon_{\mu _{1}}(k_{1},\lambda_{1})\ldots
\varepsilon_{\mu _{m}}(k_{m},\lambda_{m}).
\earr
\eeq
In this way the invariant amplitude ${\cal M}$ is decomposed into polarization
independent formfactors $F_{i}$ and the standard matrix elements ${\cal
M}_{i}$
\beq
{\cal M}=\sum_{i}{\cal M}_{i}F_{i}.
\eeq
The formfactors $F_{i}$ are complicated model dependent functions
involving in general the invariant integrals $T^{N}$ and the counterterms.
The standard matrix elements in contrast are simple model
independent expressions which depend on the external particles only but
contain the whole information on their polarization.
They are purely kinematical objects. All of the dynamical information
is contained in the formfactors.
 
The covariant tensor operators forming the standard matrix elements can be
constructed from the external four-momenta $p_{i}$,  $p'_{i}$ and $k_{i}$,
the Lorentz tensors
$g^{\mu \nu }$ and $\varepsilon ^{\mu \nu \rho \sigma }$ and the Dirac
matrices $\gamma ^{\mu },\gamma _{5}$. In general one thus obtains an
overcomplete set. Dirac algebra and momentum conservation are used to
eliminate superfluous operators. Since the external particles are
on-shell, the Dirac equation for the fermion spinors
\beq
\barr{l}
\ps_{i} u (p_{i},\sigma_{i}) = m_{i} u(p_{i},\sigma_{i}) ,\\
\bar{v} (p'_{i},\sigma'_{i})\ps'_{i} =-m'_{i}\bar{v} (p'_{i},\sigma'_{i})
\earr
\eeq
and the transversality condition for the polarization vectors
\beq
k_{i}^{\mu _{i}}\varepsilon_{\mu _{i}}(k_{i},\lambda_{i}) =0
\eeq
reduce the number of independent standard matrix elements further.
 
The number of independent standard matrix elements cannot be larger
than the number of independent polarization combinations of the external
particles. In four dimensions there are only four linear independent
four-vectors.
Expressing all four-vectors in a definite basis allows to derive the
missing relations between the remaining standard matrix elements.
Thus a minimal set of \SME\ can be constructed.
 
If there are only few external particles there may be less
independent standard matrix elements than different polarization
combinations, since there are
only few momenta available for their construction.
In this case some of the polarized amplitudes are related.
 
The number of standard matrix elements can be reduced further
if the model under consideration exhibits certain symmetries. These
evidently also apply to the relevant standard matrix elements.
 
For many applications it is not essential to minimize the number of
\SME. All one needs is a complete set.
 
Furthermore the choice of the \SME\ is not unique.
This allows to arrange for the most convenient set according to
simplicity, the structure of the lowest
order amplitudes and, if present, symmetries.
At least some of the formfactors
can be chosen as generalizations of the lowest order couplings. This is
useful in establishing improved Born approximations.
 
The concept of \SME\ is not indispensable for the calculation of
amplitudes in higher orders. It is, however, extremely helpful in
organizing lengthy calculations, which often are inevitable. All
complicated expressions are cast into the formfactors which are
polarization independent and thus have to be evaluated only once.
 
An alternative method would be to calculate directly the polarized
amplitudes. This requires a definite representation for the spinors
and/or polarization vectors from the start. The whole calculation
has to be done for each polarization separately. A closer look shows
that this method can be represented as a particular case of the standard
matrix element approach.
The corresponding covariant operators are constructed from the polarization
vectors and spinors instead of the momenta, Lorentz tensors and Dirac
matrices. Their explicit form is
\beq \label{SMEpol}
\barr{lll}
{\cal M}^{\mu_{1}\ldots\mu_{m}}_{i,\alpha _{1}\ldots\alpha_{n}\beta
_{1}\ldots\beta _{n}}
&=&(-1)^{n}\frac{v_{\alpha_{1}}(p'_{1},\sigma_{1}')}{2m'_{1}}\ldots
\frac{v_{\alpha_{n}}(p'_{n},\sigma_{n}')}{2m'_{n}}
\frac{\bar{u}_{\beta_{1}}(p_{1},\sigma_{1})}{2m_{1}}\ldots
\frac{\bar{u}_{\beta _{n}}(p_{n},\sigma_{n})}{2m_{n}}\\
&&\times\varepsilon^{*}_{\mu _{1}}(k_{1},\lambda_{1})\ldots
\varepsilon^{*}_{\mu _{m}}(k_{m},\lambda_{m}),
\earr
\eeq
where $m$, $m'$ are the masses of the external spinors.
The indices $i$ correspond to different polarization combinations.
Consequently the number of different \SME\ equals the number of
polarizations of the external particles.
For each polarization only one standard matrix element is nonzero.
In this sense the set of \SME\ (\ref{SMEpol}) is orthogonal. The
formfactors equal the polarization amplitudes and are directly obtained
by inserting explicitly the polarization vectors and spinors in the
invariant matrix element.  Unlike in the approach outlined above these
formfactors are no direct generalizations of the lowest order couplings.
 
In the following we list complete sets of \SME\ relevant for
the production of bosons in fermion-antifermion annihilation.
 
\section{Standard matrix elements for processes with two external fermions}
 
In this section we will give the \SME\ for processes involving two external
fermions ($F\bar{F}$) and one [or two] scalar ($S$) or vector ($V$) bosons.
The momenta and spinors of the fermions are denoted by $p_{1}$, $p_{2}$ and
$\bar{v}(p_{1})=\bar{v}(p_{1},\sigma_{1})$, $u(p_{2})=u(p_{2},\sigma_{2})$,
the momenta and polarization vectors of the bosons
by $k_{1}$, $\varepsilon_{1}=\varepsilon(k_{1},\lambda_{1})$
[and $k_{2}$, $\varepsilon_{2}=\varepsilon(k_{2},\lambda_{2})$]. The
numbers of different polarizations for each scalar, fermion and vector
boson are 1, 2 and 3, respectively. If we use momentum conservation to
eliminate
$k_{1}$ [or $k_{1}+k_{2}$]  the \SME\ are constructed from
the momenta $p_{2}$ and $p_{1}$ [and $k_{1}-k_{2}$], the polarization
vectors of the
vector bosons, the totally antisymmetric tensor $\varepsilon ^{\mu \nu
\rho \sigma }$ and Dirac matrices between the spinors. If there are
products of $\varepsilon $-tensors, pairs of them can be eliminated using
\beq \label{Shouten}
\varepsilon ^{\mu \nu \rho \sigma }\varepsilon ^{\alpha \beta \gamma
\delta }=-\left\vert
\barr{llll}
g^{\mu \alpha }&g^{\mu \beta }&g^{\mu \gamma }&g^{\mu \delta }\\
g^{\nu \alpha }&g^{\nu \beta }&g^{\nu \gamma }&g^{\nu \delta }\\
g^{\rho\alpha }&g^{\rho\beta }&g^{\rho\gamma }&g^{\rho\delta }\\
g^{\sigma\alpha}&g^{\sigma\beta}&g^{\sigma\gamma}&g^{\sigma\delta}
\earr
\right\vert .
\eeq
If any of the left over $\varepsilon $-tensors are contracted with
four four-momenta, we write for one of these momenta
$p^{\alpha }=\frac{1}{2}\{\ps,\gamma^{\alpha }\}$ between the spinors.
Now all remaining $\varepsilon$-tensors are contracted with
one $\gamma$-matrix at least and can be eliminated using the Chisholm
identity\footnote{Eq.\ (\protect\ref{Shouten}) and (\protect\ref{Chisholm})
can be applied because the \SME\ involve only
external vectors and spinors which remain four-dimensional also
in dimensional regularization.}
\beq \label{Chisholm}
\varepsilon _{\mu \nu \rho \sigma }\gamma ^{\sigma}=-i[\gamma _{\mu }\gamma
_{\nu }\gamma _{\rho }-g_{\mu \nu }\gamma _{\rho }+g_{\mu \rho }\gamma
_{\nu }-g_{\nu \rho }\gamma _{\mu }] \gamma _{5}.
\eeq
All Dirac matrices contracted with $p_{1}$ or $p_{2}$ can be eliminated using
Dirac algebra and the Dirac equation. Consequently the only remaining
Dirac matrices are contracted with polarization vectors [and $\ks_{1}-
\ks_{2}$] and there is at most one of each type. Finally in the scalar
products involving the polarization vectors only one [or two] independent
momenta may appear because of transversality and momentum conservation.
 
Thus we arrive at the following sets of \SME\ (we suppress polarization
indices in the following):
 
\subsection{$S \to F\bar{F}$}
 
There are $2\times2=4$ different polarization combinations but only
two \SME\
\beq
{\cal M}^{\sigma } = \bar{u}(p_{1})\;\omega_{\sigma }\;v(p_{2}),
\label{MSFF}
\eeq
where $\sigma = \pm$ and $\omega_{\pm}=\frac{1\pm\gamma_{5}}{2}$
and the fermions are outgoing.
 
\subsection{$V \to F\bar{F}$}
 
Replacing the scalar by a vector results in $3\times2\times2=12$
different polarizations and yet only four \SME\
\beq
\barr{lll}  \label{MVFF}
{\cal M}^{\sigma }_{1}&=&\bar{u}(p_{1})\;\es_{1}\;\omega_{\sigma }\;v(p_{2}),\\
{\cal M}^{\sigma }_{2}&=&\bar{u}(p_{1})\;\omega_{\sigma }\;v(p_{2})
\;\varepsilon _{1}p_{1}.
\earr
\eeq
 
\subsection{$F\bar{F}\to SS$}
\newcommand{\br}{{\textstyle\frac{1}{2}}}
 
Here the number of independent polarizations four equals the number of
\SME\
\beq
\barr{lll}
{\cal M}^{\sigma }_{1} & = & \bar{v}(p_{1})\;\br(\ks_{1}-\ks_{2})\;
\omega_{\sigma}\;u(p_{2}),\\
{\cal M}^{\sigma }_{2} & = & \bar{v}(p_{1})\;\omega_{\sigma }\;u(p_{2}).
\earr
\label{MFFSS}
\eeq
 
\subsection{$F\bar{F}\to SV$}
 
In this case we find twelve \SME\ for $2\times2\times3=12$ different
polarizations
\beq
\barr{lll}
{\cal M}^{\sigma }_{1} & = & \bar{v}(p_{1})\;\es_{2}\;\omega_{\sigma
}\;u(p_{2}),\\
{\cal M}^{\sigma }_{2} & = & \bar{v}(p_{1})\;\br(\ks_{1}-\ks_{2})\;
\omega_{\sigma }\;u(p_{2})\varepsilon_{2}p_{1},\\
{\cal M}^{\sigma }_{3} & = & \bar{v}(p_{1})\;\br(\ks_{1}-\ks_{2})\;
\omega_{\sigma }\;u(p_{2})\varepsilon_{2}p_{2},\\
{\cal M}^{\sigma }_{4} & = & \bar{v}(p_{1})\;\es_{2}\ks_{2}\;\omega_{\sigma
}\;u(p_{2}),\\
{\cal M}^{\sigma }_{5} & = & \bar{v}(p_{1})\;\omega_{\sigma }\;u(p_{2})
\;\varepsilon_{2}p_{1},\\
{\cal M}^{\sigma }_{6} & = & \bar{v}(p_{1})\;\omega_{\sigma }\;u(p_{2})
\;\varepsilon _{2}p_{2}.
\earr
\label{MFFSV}
\eeq
 
\subsection{$F\bar{F}\to VV$}
 
There are $2\times2\times3\times3=36$ different polarization
combinations, however, we can construct 40 \SME
\beqar \label{MFFVV}
{\cal{M}}^{\sigma }_{1} & = & \bar{v}(p_{1})\;
\es _{1}\;(\ks _{1} - \ps_{1})\; \es _{2}\;\omega _{\sigma }\;u(p_{2}),
\nn \\[1ex]
{\cal{M}}^{\sigma }_{2} & = & \bar{v}(p_{1})\;\br(\ks _{1}-\ks_{2})
\;(\varepsilon _{1}\varepsilon _{2})\;\omega _{\sigma }\; u(p_{2}),
\nn \\[1ex]
{\cal{M}}^{\sigma }_{3,1} & = & \phantom{-}\bar{v}(p_{1})\;
\es _{1}\;\omega _{\sigma }\;u(p_{2})\;(\varepsilon _{2}k_{1}),
\nn \\[1ex]
{\cal{M}}^{\sigma }_{3,2} & = & -\bar{v}(p_{1})\;
\es _{2}\;\omega _{\sigma }\;u(p_{2})\;(\varepsilon _{1}k_{2}),
\nn \\[1ex]
{\cal{M}}^{\sigma }_{4,1} & = & \phantom{-}\bar{v}(p_{1})\;
\es _{1}\;\omega _{\sigma }\;u(p_{2})\;(\varepsilon _{2}p_{2}),
\nn \\[1ex]
{\cal{M}}^{\sigma }_{4,2} & = & -\bar{v}(p_{1})\;
\es _{2}\;\omega _{\sigma }\;u(p_{2})\;(\varepsilon _{1}p_{1}),
\nn \\[1ex]
{\cal{M}}^{\sigma }_{5} & = & \bar{v}(p_{1})\;\br(\ks _{1}-\ks_{2})
\;\omega _{\sigma }\;u(p_{2})\;(\varepsilon _{1}k_{2})\;(\varepsilon
_{2}k_{1}),
\nn \\[1ex]
{\cal{M}}^{\sigma }_{6} & = & \bar{v}(p_{1})\;\br(\ks _{1}-\ks_{2})
\;\omega _{\sigma }\;u(p_{2})\;(\varepsilon _{1}p_{1})\;(\varepsilon
_{2}p_{2}),
\nn \\[1ex]
{\cal{M}}^{\sigma }_{7,1} & = & \bar{v}(p_{1})\;
\br(\ks _{1}-\ks_{2})\;\omega _{\sigma }\;u(p_{2})\;
(\varepsilon _{1}k_{2})\;(\varepsilon _{2}p_{2}) ,
\nn \\[1ex]
{\cal{M}}^{\sigma }_{7,2} & = & \bar{v}(p_{1})\;
\br(\ks _{1}-\ks_{2})\;\omega _{\sigma }\;u(p_{2})\;
(\varepsilon _{1}p_{1})\;(\varepsilon _{2}k_{1}) ,
\\[1ex]
{\cal{M}}^{\sigma }_{11} & = & \bar{v}(p_{1})\;
\es _{1}\; \es _{2}\;\omega _{\sigma }\;u(p_{2}),
\nn \\[1ex]
{\cal{M}}^{\sigma }_{12} & = & \bar{v}(p_{1})\;\omega _{\sigma }\; u(p_{2})\;
(\varepsilon _{1}\varepsilon _{2}),
\nn \\[1ex]
{\cal{M}}^{\sigma }_{13,1} & = & \bar{v}(p_{1})\;
\es _{1}\;\ks_{1}\;\omega _{\sigma }\;u(p_{2})\;(\varepsilon _{2}k_{1}),
\nn \\[1ex]
{\cal{M}}^{\sigma }_{13,2} & = & \bar{v}(p_{1})\;
\ks_{2}\;\es _{2}\;\omega _{\sigma }\;u(p_{2})\;(\varepsilon _{1}k_{2}),
\nn \\[1ex]
{\cal{M}}^{\sigma }_{14,1} & = & \bar{v}(p_{1})\;
\es _{1}\;\ks_{1}\;\omega _{\sigma }\;u(p_{2})\;(\varepsilon _{2}p_{2}),
\nn \\[1ex]
{\cal{M}}^{\sigma }_{14,2} & = & \bar{v}(p_{1})\;
\ks_{2}\;\es _{2}\;\omega _{\sigma }\;u(p_{2})\;(\varepsilon _{1}p_{1}),
\nn \\[1ex]
{\cal{M}}^{\sigma }_{15} & = & \bar{v}(p_{1})\;
\omega _{\sigma }\;u(p_{2})\;(\varepsilon _{1}k_{2})\;(\varepsilon
_{2}k_{1}),
\nn \\[1ex]
{\cal{M}}^{\sigma }_{16} & = & \bar{v}(p_{1})\;
\omega _{\sigma }\;u(p_{2})\;(\varepsilon _{1}p_{1})\;(\varepsilon
_{2}p_{2}),
\nn \\[1ex]
{\cal{M}}^{\sigma }_{17,1} & = & \bar{v}(p_{1})\;
\omega _{\sigma }\;u(p_{2})\;
(\varepsilon _{1}k_{2})\;(\varepsilon _{2}p_{2}) ,
\nn \\[1ex]
{\cal{M}}^{\sigma }_{17,2} & = & \bar{v}(p_{1})\;
\omega _{\sigma }\;u(p_{2})\;
(\varepsilon _{1}p_{1})\;(\varepsilon _{2}k_{1}) .
\nn 
\eeqar
We have obtained more than 36 \SME\
because we have not yet used the four dimensionality of space time,
i.e.\ the fact that the five vectors
$p_{1}$, $p_{2}$, $k_{1}-k_{2}$, $\varepsilon _{1}$ and $\varepsilon
_{2}$ are linear dependent. The relations between
the 40 \SME\ can be found by representing these vectors in a certain
basis using for example $v_{1}=p_{1}+p_{2}$, $v_{2}=p_{1}-p_{2}$,
$v_{3}=k_{1}-k_{2}$, $v_{4,\mu }=\varepsilon _{\mu \nu \rho \sigma
}v_{1}^{\nu }v_{2}^{\rho }v_{3}^{\sigma }$. In this way one can derive
the relation
\beqar \label{MFFVVrel}
0&=&
\nn2(p_{1}p_{2}-m_{1}m_{2})\;({\cal M}_{1}^{\sigma} + {\cal M}_{2}^{\sigma})\\
\nn&&\mbox{} - 2\,(p_{2}k_{2}-m_{2}^{2}-m_{1}m_{2})\;{\cal M}_{3,1}^{\sigma}
- 2\,(p_{1}k_{1}-m_{1}^{2}-m_{1}m_{2})\;{\cal M}_{3,2}^{\sigma} \\*
\nn&& \nn \mbox{} - 2\,(k_{1}^{2}+(k_{1}k_{2}))\;{\cal M}_{4,1}^{\sigma}
- 2\,(k_{2}^{2}+(k_{1}k_{2}))\;{\cal M}_{4,2}^{\sigma}
- 2{\cal M}_{5}^{\sigma} + 2({\cal M}_{7,1}^{\sigma} + {\cal
M}_{7,2}^{\sigma}) \\
&&\mbox{}-2(m_{1}(m_{2}^{2}-p_{2}k_{2})+m_{2}(m_{1}^{2}-p_{1}k_{1}))
({\cal M}_{11}^{\sigma} - {\cal M}_{12}^{\sigma})\\
\nn&&\mbox{} +(m_{1}+m_{2})(p_{1}p_{2}-m_{1}m_{2})\;{\cal M}_{12}^{\sigma}
- 2\,m_{2}\;{\cal M}_{13,1}^{\sigma} - 2\,m_{1}\;{\cal M}_{13,2}^{\sigma} \\
\nn&& \mbox{} +(m_{1}+m_{2})\,(2{\cal M}_{14,1}^{\sigma}
+2{\cal M}_{14,2}^{\sigma} -{\cal M}_{15}^{\sigma} -4{\cal M}_{16}^{\sigma})
\\
\nn&& \mbox{}+(3m_{1}+m_{2})\,{\cal M}_{17,1}^{\sigma}
 +(3m_{2}+m_{1})\,{\cal M}_{17,2}^{\sigma}
\eeqar
and a similar independent one allowing to eliminate four of the 40
\SME\ (\ref{MFFVV}).
 
The construction of complete sets of \SME\
described above is straightforward. The
reduction of general structures to these \SME\ is therefore easy to
implement into computer algebra programs.
In practical
applications some of the \SME\ may not contribute due to the presence of
symmetries and/or the neglection of fermion masses. These aspects will be
discussed together with the applications in the following chapters.
 
\section{Calculation of \SME}
\label{secmatcal}
 
For the calculation of the \SME\ one has to choose a certain
representation for the polarization vectors and spinors. This has to be
done only once for each process and not in the calculation of
individual Feynman diagrams.
If there are at
least four external particles the polarization vectors can be constructed
from their four-momenta respecting
\beq \label{conpol}
\barr{l}
k_{i}\cdot\varepsilon_{i}(k_{i},\lambda _{i})=0,\\
\varepsilon _{i}(k_{i},\lambda _{i})\varepsilon
_{i}(k_{i},\lambda'_{i})=-\delta_{\lambda _{i},\lambda'_{i}} .
\earr
\eeq
We thus obtain for $\varepsilon_{2}$
\beqar
\nn \varepsilon _{2}^{\mu }(k_{2},\parallel ) &=&\disp
\frac{1}{\sqrt{[p_{1}p_{2}(2p_{1}k_{2}\,p_{2}k_{2}-
k_{2}^{2}\,p_{1}p_{2})+p_{1}^{2}p_{2}^{2}k_{2}^{2}-
p_{2}^{2}(p_{1}k_{2})^{2}-p_{1}^{2}(p_{2}k_{2})^{2}]}}\\[1.2em]
\nn&&\times\frac{1}{\sqrt{[(p_{2}k_{2}+p_{1}k_{2})^{2}-
k_{2}^{2}(p_{1}+p_{2})^{2}]}} \\ [1em]
\nn&&\times\biggl[ p_{2}^{\mu } \Bigl(p_{1}(p_{1}+p_{2})\,k_{2}^{2} -
p_{1}k_{2}\,(p_{2}k_{2}+p_{1}k_{2})\Bigr) \\[1em]
\nn&&\mbox{}- p_{1}^{\mu } \Bigl(p_{2}(p_{1}+p_{2})\,k_{2}^{2} -
p_{2}k_{2}\,(p_{2}k_{2}+p_{1}k_{2})\Bigr)
\\[1em]\nn&&\mbox{}
+ k_{2}^{\mu } \Bigl(p_{1}p_{2}\,(p_{1}k_{2}-p_{2}k_{2})
+p_{2}^{2}\,p_{1}k_{2}-p_{1}^{2}\,p_{2}k_{2}\Bigr)
\biggr]\\[1em]
&=&\disp (0,\cos \vartheta ,0,-\sin \vartheta ) \\[2em]
\nn\varepsilon _{2}^{\mu }(k_{2},\perp) &=& \disp
\frac{1}{\sqrt{p_{1}p_{2}(2p_{1}k_{2}\,p_{2}k_{2}-
k_{2}^{2}p_{1}p_{2})+p_{1}^{2}p_{2}^{2}k_{2}^{2}-
p_{2}^{2}(p_{1}k_{2})^{2}-p_{1}^{2}(p_{2}k_{2})^{2} }}
\;\epsilon ^{\mu}_{\;\nu \rho \sigma } p_{2}^{\nu }
p_{1}^{\rho } k_{2}^{\sigma } \\[1em]
&=&\disp (0,0,1,0) \\[2em]
\nn\varepsilon _{2}^{\mu }(k_{2},L) &=& \disp
\frac{1}{\sqrt{k_{2}^{2}[(p_{2}k_{2}+p_{1}k_{2})^{2}-
k_{2}^{2}(p_{2}+p_{1})^{2}]}} 
\biggl[ k_{2}^{\mu } (p_{2}k_{2}+p_{1}k_{2}) -
(p_{1}+p_{2})^{\mu }k_{2}^{2} \biggr] \\[1ex]
&=&\disp (k, E_{2}\sin \vartheta ,0, E_{2}\cos \vartheta )/\sqrt{k_{2}^{2}} \; .
\label{polar}
\eeqar
where we have also given the simple expressions in the CMS-system of the
fermions and bosons. In this system the four-momenta of the external
particles read
\beq
\barr{lll}
p_{1,2} &=& (\tilde{E}_{1,2},0,0,\mp |{\bf p}|), \\
k_{1,2} &=& (E_{1,2},\mp |{\bf k}| \sin \vartheta ,0,\mp|{\bf k}|
\cos \vartheta ).
\earr
\eeq
$\tilde{E}_{1,2}$ are the energies and ${\bf p}$ the three-momentum
of the fermions and $E_{1,2}$ the energies and ${\bf k}$
the three-momentum of the bosons. $\vartheta $ is the angle between
the spatial vectors ${\bf p}$ and ${\bf k}$,
 
From the polarization vectors given above the ones for helicity states
are obtained as
\beq
\varepsilon_{2}^{\mu}(k_{2},\pm) = \frac{1}{\sqrt{2}}\Bigl[
\varepsilon_{2}^{\mu}(k_{2},\parallel) \pm
i\varepsilon_{2}^{\mu}(k_{2},\perp)\Bigr], \quad
\varepsilon_{2}^{\mu}(k_{2},0) =
\varepsilon_{2}^{\mu}(k_{2},L).
\eeq
 
The polarization vector $\varepsilon_{1}$ can be
obtained by interchanging $1 \leftrightarrow 2$.
 
For the case of only three external particles one needs a further independent
vector. It can be chosen freely but linear independent of the momenta.
Using this additional vector as one of the polarization vectors the
others can be constructed using (\ref{conpol}).
 
Inserting the polarization vectors (\ref{polar}) into the \SME\
these can be reduced to the ones for
external scalars, i.e.\ to (\ref{MSFF}) for the decay $V\to F\bar{F}$, and
to (\ref{MFFSS}) for the annihilation processes $F\bar{F}\to VV$, $VS$.
To calculate these remaining Dirac matrix elements one either inserts a
definite representation for the spinors or evaluates the quantities
${\cal M}^{\sigma *}_{i}{\cal M}^{\sigma'}_{j}$ via traces and
reconstructs ${\cal M}_{i}^{\sigma }$ from those if needed.
Note that for the calculation of $|{\cal M}|^{2}$ to one-loop order
one only has to evalulate the products
${\cal M}^{\sigma *}_{i}{\cal M}^{\sigma'}_{j}$ for those values of $i$,
where $F_{i}^{\sigma}$ is nonzero in lowest order
\beq
\barr{lll}
|{\cal M}|^{2} &=&
|{\cal M}_{0}+\delta{\cal M}_{1}|^{2} \approx
|{\cal M}_{0}|^{2}
+2\Re\,\{{\cal M}_{0}^{*}\delta{\cal M}_{1}\} \\[1ex]
&=&\disp\sum_{i,j} F_{i,0}^{\sigma *}(F_{j,0}^{\sigma}+2\delta F_{j,1}^{\sigma})
\Re\,\{{\cal M}_{i}^{\sigma *}{\cal M}_{j}^{\sigma}\}.
\earr
\eeq
Here ${\cal M}_{0}$, $F_{i,0}^{\sigma}$ denote the lowest order quantities and
${\delta \cal M}_{1}$, $\delta F_{i,1}^{\sigma}$  the one-loop quantities.
 
For massless external fermions the Dirac matrix elements (\ref{MSFF})
and (\ref{MFFSS}) are equivalent to the helicity matrix elements. They do
not interfere and can thus
easily be obtained from $|{\cal M}^{\sigma }_{i}|^{2}$ as
\beq
\barr{lll}
\bar{v}(p_{1})\;\omega_{\sigma }\;u(p_{2}) & = &
\sqrt{2p_{1}p_{2}}, \\[1ex]
\bar{v}(p_{1})\;(\ks_{1}-\ks_{2})\;\omega_{\sigma}\;u(p_{2}) & = &
\sqrt{4p_{1}(k_{1}-k_{2})\,p_{2}(k_{1}-k_{2})
- 2p_{1}p_{2} (k_{1}-k_{2})^{2}}.
\earr
\eeq
 
If one is only interested in unpolarized quantities it suffices to
calculate $\sum _{pol}{\cal M} ^{\sigma* }_{i}{\cal
M}^{\sigma}_{j}$ using the polarization sums for vector bosons and
spinors.

\chapter{Calculation of one-loop amplitudes} 
 
We have described all the ingredients necessary for the calculation
of one-loop radiative corrections. This chapter shows how one-loop
amplitudes are evaluated in practice.
 
First one has to specify a Lagrangian and to derive the
corresponding Feynman rules. Then renormalization has to be carried
out and the counterterms have to be determined. Both were done
in Chap.\ \ref{chaSM} and \ref{charen} for the SM.
Once the Feynman
rules and the counterterms are fixed, the following steps apply to any
renormalizable model.
 
To calculate the amplitude of a certain process at the one-loop level one
has to construct all tree and one-loop Feynman diagrams with the given
external particles allowed
by the specified model. Next each Feynman diagram has to be reduced
algebraically to a form suitable for numerical evaluation. This procedure
is explained in more detail in Sect.\ \ref{secrfd}. Finally the expressions for
all diagrams have to be put together into a numerical program which
calculates the amplitude and the corresponding cross section or decay
rate.
 
\section{Algebraic reduction of Feynman diagrams}
\label{secrfd}
 
The algorithm for the reduction of one-loop diagrams is the following.
The loop integral obtained from the Feynman rules contains a product of
propagators as denominator and a numerator composed of Lorentz vectors
and tensors, Dirac matrices and spinors and polarization
vectors of the external particles. The numerator is simplified using
tensor and Dirac algebra, the mass shell conditions for the external
particles and momentum conservation. One can also try to separate terms
proportional to one or more of the denominators. Cancelling these yields
N-point functions of lower degree. Next the loop integral is organized
into the tensor integrals defined in Sect.\ \ref{sectendef}. The
Lorentz decomposition of these integrals is inserted and the
whole Dirac and Lorentz structure is separated off from the
integrals. Using again Dirac
algebra and mass shell conditions it can be reduced to the appropriate
standard matrix elements as discussed in Chap.\ \ref{chaSME}.
 
We thus arrive at an expression of the form
\beq \label{Mdec}
\delta {\cal M}=\sum_{i}{\cal M}_{i}\delta F_{i}
\eeq
for each one-loop Feynman diagram. The formfactors are linear combinations of
the invariant coefficient functions of the tensor integrals with coefficients
being functions of the kinematical invariants.
 
The formfactors can be further evaluated by applying the reduction
scheme for the invariant integrals described in Sects.\ \ref{sectenred}
and \ref{sectenres}.
Finally they are obtained as linear combinations of the scalar one-loop
integrals $A_{0}$, $B_{0}$, $C_{0}$ and $D_{0}$
which are given explicitly in Sect.\ \ref{sectensca}. This last step may
lead to
very lengthy expressions. Their algebraic evaluation needs a lot of time
and space. This can be avoided by performing the reduction to scalar
integrals numerically.
 
The evaluation of the counterterm diagrams and the Born diagrams is done
in a similar way. Since no integrations have to be performed
their calculation is much
easier. In most cases the counterterm diagrams can be obtained from
the Born diagrams by replacing the lowest order couplings by the
corresponding counterterm.
 
\savebox{\Vr}(36,0)[bl]
{\multiput(3,0)(12,0){3}{\oval(6,4)[t]}
\multiput(9,0)(12,0){3}{\oval(6,4)[b]} }
\savebox{\Vt}(0,36)[bl]
{\multiput(0,3)(0,12){3}{\oval(4,6)[l]}
\multiput(0,9)(0,12){3}{\oval(4,6)[r]} }
 
\savebox{\Fr}(36,0)[bl]
{ \put(0,0){\vector(1,0){20}} \put(18,0){\line(1,0){18}} }
\savebox{\Fl}(36,0)[bl]
{ \put(36,0){\vector(-1,0){20}} \put(18,0){\line(-1,0){18}} }
\savebox{\Ft}(0,36)[bl]
{ \put(0,0){\vector(0,1){20}} \put(0,18){\line(0,1){18}} }
 
As an illustration of the reduction method we present the explicit
calculation of a box
diagram contributing to $e^{+}e^{-} \to W^{+}W^{-}$ (Fig.\ \ref{fig611}).
\bfi
\bma
\begin{picture}(150,72)
\put(0,14){$e^{-}$}
\put(0,51){$e^{+}$}
\put(131,14){$W^{-}$}
\put(131,51){$W^{+}$}
\put(68,5){$W$}
\put(68,60){$W$}
\put(36,32){$\nu_{e}$}
\put(100,32){$Z$}
\put(18,54){\usebox{\Fl}}
\put(18,18){\usebox{\Fr}}
\put(54,18){\usebox{\Ft}}
\put(54,18){\usebox{\Vr}}
\put(54,54){\usebox{\Vr}}
\put(90,18){\usebox{\Vr}}
\put(90,54){\usebox{\Vr}}
\put(90,18){\usebox{\Vt}}
\end{picture}
\ema
\caption{Box diagram contributing to $e^{+}e^{-} \to W^{+}W^{-}$.}
\label{fig611}
\efi
According to
the Feynman rules the corresponding contribution to the invariant matrix
element $\delta {\cal M}$ is given by
(we include a global factor $i$ in the Feynman rules in order to obtain
real amplitudes)
\begin{equation}
\barr{lll}
\delta {\cal M} &=&
\disp -i\frac{e^{4}c_{W}^{2}}{2s_{W}^{4}}\mu^{4-D} \\[1em]&&
\disp\int \frac{d^{D}q}{(2\pi)^{D}}\frac{\bar{v}(p_{1}) \gamma ^{\mu }
(\qs +\ks_{1} - \ps_{1})\gamma ^{\nu } \omega _{-} u(p_{2})
\Gamma _{\lambda \mu \rho }
\Gamma ^{\lambda }_{\;\sigma \nu } \varepsilon _{1}^{\rho } \varepsilon
_{2}^{\sigma }
}{\left[q^{2}-M_{Z}^{2}\right]
\left[ (q+k_{1})^{2}-M_{W}^{2}\right]  (q+k_{1}-p_{1})^{2}
\left[ (q-k_{2})^{2}-M_{W}^{2}\right]},
\earr
\end{equation}
with
\begin{equation}
\begin{array}{l}
\Gamma _{\lambda \mu \rho } = g_{\lambda \mu }(2q+k_{1})_{\rho } + g_{\mu
\rho } (-q-2k_{1})_{\lambda }+g_{\rho \lambda }(k_{1}-q)_{\mu } ,\\[1em]
\Gamma _{\lambda \sigma \nu }= g_{\lambda \sigma }(-k_{2}-q)_{\nu }+
g_{\sigma \nu }(2k_{2}-q)_{\lambda }+g_{\nu \lambda }(2q-k_{2})_{\sigma}
\; .
\end{array}
\end{equation}
 
Evaluating the numerator and introducing the tensor integrals $D$
and $C$ we arrive at
\begin{eqnarray}
\nn \delta {\cal M} &=& \frac{\alpha^{2} c_{W}^{2}}{2s_{W}^{4}}
\bar{v}(p_{1})
\Bigl\{ -8 \gamma ^{\mu } \;\varepsilon_1^{\nu } \;\varepsilon_2^{\rho}
\;
D_{\mu \nu \rho } \\[1.5em]
\nn && + D_{\mu \nu } \;\Bigl[ \es _{1} \gamma ^{\mu }
\es _{2} \;
2 (k_{1}-k_{2})^{\nu } \\[1em]
\nn && \quad + \es _{1} \Bigl(-2\varepsilon _{2}^{\mu }\;
k_{2}^{\nu } -8 p_{1}^{\mu }\;\varepsilon _{2}^{\nu }+2p_{2}^{\mu }\;
\varepsilon _{2}^{\nu } \Bigr)
+ \es _{1}\gamma ^{\mu } \ks_{2} \; 8 \varepsilon _{2}^{\nu } \\[1em]
\nn && \quad - \es _{2} \Bigl(-2\varepsilon _{1}^{\mu }\;
k_{1}^{\nu } -8 p_{2}^{\mu }\;\varepsilon _{1}^{\nu }+2p_{1}^{\mu }\;
\varepsilon _{1}^{\nu } \Bigr)
- \ks_{1}\gamma ^{\mu } \es _{2} \; 8 \varepsilon _{1}
^{\nu }  \\[1em]
\nn && \quad + \ks_{1}\; 8\varepsilon _{1}^{\mu }\;
\varepsilon _{2}^{\nu } + \gamma ^{\mu } \Bigl(-16
\varepsilon _{1}^{\nu }\; p_{2}\varepsilon _{2}+16\varepsilon _{2}^{\nu }\;
p_{1} \varepsilon _{1} +2(k_{1}-p_{1})^{\nu }\; \varepsilon _{1}
\varepsilon _{2} \Bigr) \Bigr] \\[1.5em]
\nn && + D_{\mu }\; \Bigl[ \es _{1} (\ks_{1} - \ps_{1}) \es_{2}
\; 2(k_{1}-k_{2})^{\mu } + \es_{1}\gamma ^{\mu }
\es _{2} (M_{Z}^{2} -4k_{1}k_{2} ) \\[1em]
\nn && \quad + \es _{1} \Bigl(-\varepsilon _{2}^{\mu }(3 M_{Z}^{2}
+3M_{W}^{2} +2p_{1}k_{1}) -4 k_{2}^{\mu}\;\varepsilon_{2}k_{1}+4 p_2^{\mu} \;
\varepsilon _{2}k_{1} \Bigr) \\[1em]
\nn && \quad - \es _{2} \Bigl( \varepsilon _{1}^{\mu }(3 M_{Z}^{2}
+3M_{W}^{2} +2p_{2}k_{2}) +4 k_{1}^{\mu }\; \varepsilon _{1}k_{2}
-4 p_{1}^{\mu }
\; \varepsilon _{1}k_{2} \Bigr) \\[1em]
 && \quad + \ks_{1} \Bigl(2(k_{1}-p_{1})^{\mu } \; \varepsilon _{1}
\varepsilon _{2}
- 8 \varepsilon _{1}^{\mu }\; \varepsilon _{2}p_{2}
+8 \varepsilon _{2}^{\mu }
\; \varepsilon _{1} p_{1} \Bigr) \\[1em]
\nn && \quad + \gamma ^{\mu }  \varepsilon _{1}\varepsilon _{2}
(M_{Z}^{2}-3M_{W}^{2}+6p_{1}k_{1})
+ \ks_{1}\gamma ^{\mu } \es _{2} \; 4 \varepsilon _{1}k_{2} + 4
\es_{1}\gamma ^{\mu } \ks_{2} \; 4 \varepsilon _{2}k_{1}
\Bigr] \\[1.5em]
\nn && +D_{0} \Bigl[ \es _{1} (\ks_{1}- \ps_{1})
\es_{2} (M_{Z}^{2}-4k_{1}k_{2}) +\ks_{1} \; \varepsilon _{1}
\varepsilon _{2} ( M_{W}^{2}-2 p_{1}k_{1} -3 M_{Z}^{2} ) \\[1em]
\nn && \qquad+ \es _{1} \varepsilon _{2}k_{1} ( 4
p_{2}k_{2} -2M_{W}^{2}+2 M_{Z}^{2} ) -\es _{2} \; \varepsilon _{1}k_{2} (
4 p_{1}k_{1} -2M_{W}^{2}+2 M_{Z}^{2} )
\Bigr]  \\[1.5em]
\nn && + C_{\mu } \;
\Bigl[ \es _{1} \gamma ^{\mu }
\es _{2}
-\es _{2}\;3 \varepsilon _{1}^{\mu } -
\es_{1}\; 3 \varepsilon _{2}^{\mu } + \gamma ^{\mu }\varepsilon _{1}
\varepsilon _{2}  \Bigr]  \\[1em]
\nn && + C_{0} \Bigl[ \es _{1}\;
2\varepsilon _{2}k_{1} -
\es_{2}\; 2 \varepsilon _{1}k_{2}
+ \es _{1}\; 3\varepsilon _{2}p_{2} -
\es_{2}\; 3 \varepsilon _{1}p_{1}
- \ks_{1} \; 4\varepsilon
_{1}\varepsilon _{2} \Bigr] \Bigl\} \omega _{-}u(p_{2}) .
\end{eqnarray}
The three-point integrals arise from $q^{2}$ terms in the numerator by
writing $q^{2}$=($q^{2}-M^{2}_{Z}$)+$M^{2}_{Z}$ and cancelling the
first denominator factor. After that the shift $q\to q-k_{1}+p_{1}$
was performed in the three-point integrals (this shift conserves the
manifest CP symmetry of the diagram).
The arguments of the $C$ and $D$ functions are as follows
\begin{equation}
\begin{array}{l}
D = D(k_{1},k_{1}-p_{1},-k_{2},M_{Z},M_{W},0,M_{W}) ,\\[1em]
C = C(p_{1},-p_{2},0,M_{W},M_{W}) \; .
\end{array}
\end{equation}
Inserting the tensor integral decomposition eqs.\ (\ref{tedecc}, \ref{tedecd})
yields the final expression 
\begin{eqnarray} \label{eq66}
\nonumber
\delta {\cal M}&=&
\begin{array}[t]{l}
 \displaystyle \frac{\alpha^{2} c_{W}^{2}}{2s_{W}^{4}}
\left\{ {\cal M}_{1}^{-} \Bigl[ 20 \;D_{00}
+2(4M_{W}^{2}-s)D_{33} +2 (M_{W}^{2}+t)D_{22} \right. \\[1em]
\hspace{5mm} \mbox{}+(12M_{W}^{2}+4t-2s)D_{23} +2(4M_{W}^{2}-s) D_{13}
+(16M_{W}^{2}-6s+2M_{Z}^{2}) D_{3}  \\[1em]
\hspace{5mm} \mbox{} +(2t-2s+6M_{W}^{2}+M_{Z}^{2}) D_{2}
+(4M_{W}^{2}-2s+M_{Z}^{2}) D_{0}
\Bigl]   \\[1em]
\end{array}\\
\nonumber
&&\begin{array}{l}
\mbox{} +{\cal M}_{2}^{-} \Bigl[-4 C_{0}-16\,D_{003} -8\, D_{002} +10 \,
D_{00} +2t D_{22} +2(M_{W}^{2}+t) (D_{33} + D_{13})
\\[1em]
\hspace{5mm} \mbox{} +2(M_{W}^{2}+3t)D_{23} +2(M_{W}^{2}-2t +M_{Z}^{2}) D_{3}
+(M_{Z}^{2}-t) D_{2} +(t-3M_{Z}^{2}) D_{0} \Bigl] \\[1em]
\end{array}  \\
&&\begin{array}{l}
\mbox{}+\Bigl[{\cal M}_{3,1}^{-}+{\cal M}_{3,2}^{-}\Bigr]
 \Bigl[ -3C_{2} +2C_{0} -8 D_{003} -8 D_{00}
+(4s-11M_{W}^{2}+5t) D_{13}  \\[1em]
\hspace{5mm} \mbox{}-3(M_{W}^{2}+t)D_{33}-2(2M_{W}^{2}+t)D_{23}
+(t-4M_{W}^{2}-3M_{Z}^{2})D_{3}+2(M_{Z}^{2}-t) D_{0} \Bigl] \\[1em]
\end{array}    \\
\nonumber
&&\begin{array}{l}
\mbox{}+\Bigl[{\cal M}_{4,1}^{-}+{\cal M}_{4,2}^{-}\Bigr]
\Bigl[4 C_{2} +3 C_{0} -8D_{002} -26 D_{00}
+2(s-4M_{W}^{2})(D_{33} +D_{13})  \\[1em]
\hspace{5mm}  - 2(t+2M_{W}^{2})D_{22}+(4s-2t-18M_{W}^{2}) D_{23}
 \\[1em]
\hspace{5mm} \mbox{} +(4s-8M_{W}^{2}-2M_{Z}^{2}) D_{3}
+(t-4M_{W}^{2}-3M_{Z}^{2})D_{2} \Bigl] \\[1em]
\end{array}      \\
\nonumber
&&\mbox{}+{\cal M}_{5}^{-} \Bigl[ 16 D_{113}+8 D_{123} -8 D_{13}
\Bigl] \\[1em]
\nonumber
&&\mbox{}+{\cal M}_{6}^{-} \Bigl[ 8 D_{222} +16 D_{223} +24 D_{22}
+32 D_{23}+16 D_{2} \Bigl] \\[1em]
\nonumber
&&\left.
\mbox{}+\Bigl[{\cal M}_{7,1}^{-}+{\cal M}_{7,2}^{-}\Bigr]
\Bigl[ 8 D_{332} +8 D_{223} + 8 D_{123}
 +8 D_{23} +16 D_{13} \Bigl]  \right\}  ,
\end{eqnarray}
where we introduced the standard matrix elements (\ref{MFFVV}),
the Mandelstam variables
\beq
s = (p_{1}+p_{2})^{2}, \quad t=(p_{1}-k_{1})^{2},
\eeq
and put
\beq
k_{1}^{2}=k_{2}^{2}= M_{W}^{2}, \quad p_{1}^{2} = p_{2}^{2} = 0 .
\eeq
Furthermore we made use of the relations which follow from the symmetry
of the diagram under the exchange $e^{+} \leftrightarrow e^{-},W^{+}
\leftrightarrow W^{-}$ (CP invariance)
\begin{equation}
\begin{array}{lll}
D_{1}=D_{3}, \hspace{1cm}& D_{11}= D_{33}, \hspace{1cm}
& D_{12}=D_{23}, \\[.7em]
D_{001}=D_{003}, & D_{112}=D_{332}, & D_{122}=D_{223}, \\[.7em]
D_{111}=D_{333}, & D_{113}=D_{133}, & C_{1} = C_{2}.
\end{array}
\end{equation}
These reduce the number of independent invariant integrals considerably.
Note that not all of the 40 \SME\ of (\ref{MFFVV}) appear in
(\ref{eq66}). This is due to the neglection of fermion masses and CP
invariance of the box diagram.
 
This example shows that the reduction method is straightforward and universally
applicable to one-loop Feynman diagrams, since they all have a similar
structure.
 
\section{Generic Feynman diagrams}
 
The huge number of algebraic calculations makes the evaluation of each
Feynman diagram very lengthy and tedious. Furthermore there are a large
number of diagrams contributing to each process. Fortunately many of the
diagrams resemble each other in their algebraic structure and can be
considered as special cases of
generic diagrams. These are the Feynman diagrams of a theory with only
one generic scalar, fermion, vector boson and Faddeev-Popov ghost
each and arbitrary
renormalizable couplings between those fields (for more details see
\cite{Ec90}). It suffices to do the
algebraic calculations for these generic diagrams only. All actual
diagrams are obtained from those by substituting the actual fields
together with their coupling constants and masses. This saves a lot of
work especially if there are many fields in the theory.
 
Clearly the generic diagrams can be calculated with the methods
described above. The efficiency of generic diagrams is illustrated in the next
section using the decay of the $W$-boson into massless fermions as
example.
 
\section{The decay $W\to f_{i}\bar{f}'_{j}$ for massless fermions}
\label{secWff0}
We will now apply the methods described above by calculating the
one-loop amplitude for the decay of the $W$-boson into massless fermions
\beq
W^{+}(k) \to f_{i}(p_{1}) \bar{f}'_{j}(p_{2}).
\eeq
In lowest order there is only one Feynman diagram (Fig.\ref{fig621})
leading to the amplitude
\beq
{\cal M} _{0}=-\frac{eV_{ij}}{\sqrt{2}s_{W}} \bar{u}(p_{1})\es
(k)\omega _{-}v(p_{2})
=-\frac{eV_{ij}}{\sqrt{2}s_{W}}{\cal M}^{-}_{1}.
\label{M0Wff}
\eeq
\savebox{\Vr}(48,0)[bl]
{\multiput(3,0)(12,0){4}{\oval(6,4)[t]}
\multiput(9,0)(12,0){4}{\oval(6,4)[b]} }
\savebox{\Ftr}(32,24)[bl]
{ \put(0,0){\vector(4,3){18}} \put(16,12){\line(4,3){16}} }
\savebox{\Fbr}(32,24)[bl]
{ \put(32,0){\vector(-4,3){19}} \put(16,12){\line(-4,3){16}} }
\savebox{\Ftbr}(32,48)[bl]
{\put(00,24){\usebox{\Ftr}}
\put(00,00){\usebox{\Fbr}}}
\bfi
\bma
\barr{l}
\begin{picture}(96,72)
\put(87,55){\makebox(10,18)[bl]{$f_{i}$}}
\put(0,42){\makebox(10,20)[bl]{$W^{+}$}}
\put(87,10){\makebox(10,18)[bl]{$\bar{f}'_{j}$}}
\put(48,36){\circle*{4}}
\put(0,36){\usebox{\Vr}}
\put(48,12){\usebox{\Ftbr}}
\end{picture}
\earr
\ema
\caption{Born diagram to $W \to f_{i}\bar{f}'_{j}$.}
\label{fig621}
\efi
Neglecting the fermion masses the amplitude
(\ref{M0Wff}) leads to the following lowest order decay width
\beq
\Gamma _{0}=\frac{\alpha }{6}\frac{M_{W}}{2s^{2}_{W}}\left\vert
V_{ij}\right\vert^{2}.
\eeq
At one-loop order there are six loop diagrams and one counterterm diagram
(Fig.\ \ref{fig622}; the counterterm is indicated by a cross).
\bfi
\savebox{\Vr}(36,0)[bl]
{\multiput(3,0)(12,0){3}{\oval(6,4)[t]}
\multiput(9,0)(12,0){3}{\oval(6,4)[b]} }
\savebox{\Vt}(0,36)[bl]
{\multiput(0,3)(0,12){3}{\oval(4,6)[l]}
\multiput(0,9)(0,12){3}{\oval(4,6)[r]} }
\savebox{\Vtr}(36,18)[bl]
{\multiput(4,0)(8,4){5}{\oval(8,4)[tl]}
\multiput(4,4)(8,4){4}{\oval(8,4)[br]} }
\savebox{\Vbr}(36,18)[bl]
{\multiput(4,18)(8,-4){5}{\oval(8,4)[bl]}
\multiput(4,14)(8,-4){4}{\oval(8,4)[tr]}}
\savebox{\Vtbr}(36,36)[bl]
{\put(00,18){\usebox{\Vtr}}
\put(00,00){\usebox{\Vbr}}}
\savebox{\Sr}(36,0)[bl]
{ \multiput(0,0)(13,0){3}{\line(4,0){10}} }
\savebox{\St}(0,36)[bl]
{ \multiput(0,0)(0,13){3}{\line(0,1){10}} }
\savebox{\Str}(36,18)[bl]
{ \multiput(0,0)(13,6.5){3}{\line(2,1){10}} }
\savebox{\Sbr}(36,18)[bl]
{\multiput(0,18)(13,-6.5){3}{\line(2,-1){10}} }
\savebox{\Stbr}(36,36)[bl]
{\put(00,18){\usebox{\Str}}
\put(00,00){\usebox{\Sbr}}}
\savebox{\Fr}(36,0)[bl]
{ \put(0,0){\vector(1,0){20}} \put(18,0){\line(1,0){18}} }
\savebox{\Ft}(0,36)[bl]
{ \put(0,0){\vector(0,1){21}} \put(0,18){\line(0,1){18}} }
\savebox{\Ftr}(36,18)[bl]
{ \put(0,0){\vector(2,1){20}} \put(18,9){\line(2,1){18}} }
\savebox{\Fbr}(36,18)[bl]
{ \put(36,0){\vector(-2,1){21}} \put(18,9){\line(-2,1){18}} }
\savebox{\Ftbr}(72,72)[bl]
{\put(00,36){\usebox{\Ftr}}\put(36,54){\usebox{\Ftr}}
 \put(00,18){\usebox{\Fbr}}\put(36,00){\usebox{\Fbr}}}
\savebox{\Vp}(36,36)[bl]
{ \put(0,18){\circle*{4}} \put(36,0){\circle*{4}}
\put(36,36){\circle*{4}} }
\bma
\begin{picture}(108,72)
\put(5,40){\makebox(10,10){$W$}}
\put(77,32){\makebox(20,10){$\gamma,Z$}}
\put(43,54){\makebox(20,10){$f_{i}$}}
\put(43,8){\makebox(20,10){$f'_{j}$}}
\put(100,54){\makebox(10,10){$f_{i}$}}
\put(100,8){\makebox(10,10){$\bar{f}'_{j}$}}
\put(36,18){\usebox{\Vp}}
\put(0,36){\usebox{\Vr}}
\put(36,00){\usebox{\Ftbr}}
\put(72,18){\usebox{\Vt}}
\end{picture}
\ema
\savebox{\Ftbr}(72,72)[bl]
{\put(36,18){\usebox{\Ft}}
\put(36,54){\usebox{\Ftr}} \put(36,00){\usebox{\Fbr}}}
\bma
\barr{cc}
\begin{picture}(108,72)
\put(5,40){\makebox(10,10){$W$}}
\put(77,32){\makebox(20,10){$f_{i}$}}
\put(43,54){\makebox(20,10){$\gamma,Z$}}
\put(43,8){\makebox(20,10){$W$}}
\put(100,54){\makebox(10,10){$f_{i}$}}
\put(100,8){\makebox(10,10){$\bar{f}'_{j}$}}
\put(36,18){\usebox{\Vp}}
\put(0,36){\usebox{\Vr}}
\put(36,00){\usebox{\Ftbr}}
\put(36,18){\usebox{\Vtbr}}
\end{picture}  \qquad & \qquad
\begin{picture}(108,72)
\put(5,40){\makebox(10,10){$W$}}
\put(77,32){\makebox(20,10){$f'_{j}$}}
\put(43,54){\makebox(20,10){$W$}}
\put(43,8){\makebox(20,10){$\gamma,Z$}}
\put(100,54){\makebox(10,10){$f_{i}$}}
\put(100,8){\makebox(10,10){$\bar{f}'_{j}$}}
\put(36,18){\usebox{\Vp}}
\put(0,36){\usebox{\Vr}}
\put(36,00){\usebox{\Ftbr}}
\put(36,18){\usebox{\Vtbr}}
\end{picture}
\earr
\ema
\savebox{\Vr}(48,0)[bl]
{\multiput(3,0)(12,0){4}{\oval(6,4)[t]}
\multiput(9,0)(12,0){4}{\oval(6,4)[b]} }
\savebox{\Ftr}(32,24)[bl]
{ \put(0,0){\vector(4,3){18}} \put(16,12){\line(4,3){16}} }
\savebox{\Fbr}(32,24)[bl]
{ \put(32,0){\vector(-4,3){19}} \put(16,12){\line(-4,3){16}} }
\savebox{\Ftbr}(32,48)[bl]
{\put(00,24){\usebox{\Ftr}}
\put(00,00){\usebox{\Fbr}}}
\bma
\barr{l}
\begin{picture}(96,72)
\put(87,55){\makebox(10,18)[bl]{$f_{i}$}}
\put(0,42){\makebox(10,20)[bl]{$W^{+}$}}
\put(87,10){\makebox(10,18)[bl]{$\bar{f}'_{j}$}}
\put(43.5,30){\line(3,4){9}}
\put(43.5,42){\line(3,-4){9}}
\put(0,36){\usebox{\Vr}}
\put(48,12){\usebox{\Ftbr}}
\end{picture}
\earr
\ema
\caption{One-loop diagrams and corresponding counterterm diagram
to $W \to f_{i}\bar{f}'_{j}$.}
\label{fig622}
\efi
 
The first two loop diagrams correspond to one generic diagram and the
other four to another generic one. We first calculate the two generic
diagrams.
The expression for the first reads
\beq
\barr{lll}
\delta {\cal M}_{1}&=&\disp i\mu^{4-D}\int \frac{d^{D}q}{(2\pi)^{D}}
\disp\frac{1}{(q^{2}-M^{2})(q+p_{1})^{2}(q-p_{2})^{2}}\\[1em]
&&\bar{u}(p_{1})\gamma ^{\nu }(g^{-}_{1}\omega _{-}+g^{+}_{1}\omega_{+})(\qs
+\ps_{1})\es(g^{-}_{3}\omega_{-}+g^{+}_{3}\omega_{+})
(\qs -\ps_{2})\gamma _{\nu }(g^{-}_{2}\omega _{-}+g^{+}_{2}\omega
_{+})v(p_{2}),
\earr
\eeq
where $g^{\pm}$ denote the generic left- and right-handed
fermion-fermion-vector couplings.
Simplification and decomposition into tensor integrals yields
\beq
\barr{lll}
\delta {\cal M}_{1}&=&\disp i\mu^{4-D}\int \frac{d^{D}q}{(2\pi)^{D}}
\disp\frac{1} {(q^{2}-M^{2})(q+p_{1})^{2}(q-p_{2})^{2}}\\[1em]
&&\bar{u}(p_{1})\bigl[-2(\qs-\ps_{2})\es (\qs+\ps_{1})+(4-D)(\qs \es \qs)\bigr]
(g^{-}_{1}g^{-}_{3}g^{-}_{2}\omega _{-}
+g^{+}_{1}g^{+}_{2}g^{+}_{3}\omega_{+}\bigr)v(p_{2})\\[1em]
& = & -\frac{1}{16\pi^{2}}
\bar{u}(p_{1})\bigl[(2-D)C_{\mu \nu} \gamma^{\mu}\es\gamma^{\nu}
+2 C_{\mu } (\ps_{2} \es \gamma ^{\mu }-
\gamma ^{\mu }\es \ps_{1})+ 2 C_{0}\ps_{2} \es \ps_{1}\bigr]\\[1em]
&&\hspace{3em}
\bigl(g^{-}_{1}g^{-}_{2}g^{-}_{3}\omega _{-
}+g^{+}_{1}g^{+}_{2}g^{+}_{3}\omega _{+})v(p_{2}).
\earr
\eeq
Insertion of the Lorentz decomposition and further simplification gives
\beq
\barr{lll}
\delta {\cal M}_{1}&=&-\frac{1}{16\pi^{2}}
(g^{-}_{1}g^{-}_{3}g^{-}_{2}{\cal M}^{-}_{1}
+g^{+}_{1}g^{+}_{3}g^{+}_{2}{\cal M}^{+}_{1})\\[1ex]
&&\bigl[(2-D)^{2}C_{00}-2k^{2}(C_{12}+C_{1}+C_{2}+C_{0})\bigr].
\earr
\eeq
Finally the reduction of the invariant integrals and the use of
(\ref{TNdiv})  leads to
\beq
\barr{lll}
\delta {\cal M}_{1}&=&-\frac{1}{16\pi^{2}}
(g^{-}_{1}g^{-}_{3}g^{-}_{2}{\cal M}^{-}_{1}
+g^{+}_{1}g^{+}_{3}g^{+}_{2}{\cal M}^{+}_{1})\\[1ex]
&&\big[-2k^{2}C_{0}(0,k^{2},0,M,0,0)(1+\frac{M^{2}}{k^{2}})^{2}\\[1ex]
&&\quad-B_{0}(k^{2},0,0)(3+2\frac{M^{2}}{k^{2}})
+2B_{0}(0,M,0)(2+\frac{M^{2}}{k^{2}})-2\big] \\[1ex]
&=&-\frac{1}{16\pi^{2}}(g^{-}_{1}g^{-}_{3}g^{-}_{2}{\cal M}^{-}_{1}
+g^{+}_{1}g^{+}_{3}g^{+}_{2}{\cal M}^{+}_{1}){\cal
V}_{a}(0,k^{2},0,M,0,0),
\earr
\eeq
where we introduced the generic vertex function ${\cal V}_{a}$ which is
defined in the general case in App.~C.
 
Similarly we obtain for the second generic diagram
\beq
\barr{lll}
\delta {\cal M}_{2}&=&-\disp i\mu^{4-D}\int \frac{d^{D}q}{(2\pi)^{D}}
\frac{\bar{u}(p_{1})\gamma ^{\nu }(g^{-}_{1}\omega _{-}+g^{+}_{1}\omega_{+})
(-\qs)\gamma_{\rho} (g^{-}_{2}\omega _{-}+g^{+}_{2}\omega_{+})v(p_{2})}
{q^{2}[(q+p_{1})^{2}-M^{2}_{1}][(q-p_{2})^{2}-M^{2}_{2}]}\\[1em]
&&g_{3}\Bigr[g_{\rho \mu }(p_{1}+2p_{2}-q)_{\nu }-g_{\mu \nu
}(2p_{1}+p_{2}+q)_{\rho}
+g_{\nu\rho }(2q+p_{1}-p_{2})_{\mu}]\varepsilon ^{\mu }\\[1em]
&=&\frac{1}{16\pi^{2}}g_{3}(g^{-}_{1}g^{-}_{2}{\cal M}^{-}_{1}
+g^{+}_{1}g^{+}_{2}{\cal M}^{+}_{1})
\Bigl[4(D-1)C_{00}-2k^{2}(C_{12}+C_{1}+C_{2})\Bigr]\\[1em]
&=&\frac{1}{16\pi^{2}}g_{3}
(g^{-}_{1}g^{-}_{2}{\cal M}^{-}_{1}+g^{+}_{1}g^{+}_{2}{\cal M}^{+}_{1})\\[1ex]
&&\Bigl[2(M^{2}_{1}+M^{2}_{2}+\frac{M^{2}_{1}M^{2}_{2}}{k^{2}})
C_{0}(0,k^{2},0,0,M_{1},M_{2})
-(1+\frac{M^{2}_{1}+M^{2}_{2}}{k^{2}})B_{0}(k^{2},M_{1},M_{2})\\[1ex]
&&+(2+\frac{M^{2}_{1}}{k^{2}})B_{0}(0,0,M_{1})
  +(2+\frac{M^{2}_{2}}{k^{2}})B_{0}(0,0,M_{2})\Bigr]\\[1em]
&=&\frac{1}{16\pi^{2}}g_{3}
(g^{-}_{1}g^{-}_{2}{\cal M}^{-}_{1}+g^{+}_{1}g^{+}_{2}{\cal M}^{+}_{1})
{\cal V}_{b}^{-}(0,k^{2},0,0,M_{1},M_{2}).
\earr
\eeq
The general definition of ${\cal V}_{b}^{-}$ can again be found in
App.~C.
 
Inserting the actual couplings and masses of the six actual diagrams
into the generic diagrams and
adding the counterterm diagram, which can be easily obtained from the
Feynman rules or the Born diagram,
we find for the virtual one-loop corrections to the
invariant amplitude for $W\to f_{i}\bar{f}'_{j}$
\beq
\barr{lll}
\delta {\cal M}&=&-\disp\frac{e}{\sqrt{2}s_{W}}\frac{\alpha}{4\pi }V_{ij}
{\cal M}^{-}_{1}\\[1ex]
&&\{Q_{f}Q_{f'}{\cal V}_{a}(0,M^{2}_{W},0,\lambda ,0,0)\\[1ex]
&&+g^{-}_{f}g^{-}_{f'}{\cal V}_{a}(0,M^{2}_{W},0,M_{Z},0,0)\\[1ex]
&&+Q_{f}{\cal V}^{-}_{b}(0,M^{2}_{W},0,0,\lambda ,M_{W})\\[1ex]
&&-Q_{f'}{\cal V}^{-}_{b}(0,M^{2}_{W},0,0,M_{W},\lambda )\\[1ex]
&&+\frac{c_{W}}{s_{W}}g^{-}_{f}{\cal V}^{-}_{b}(0,M^{2}_{W},0,0,M_{Z},M_{W})
\\[1ex]
&&-\frac{c_{W}}{s_{W}}g^{-}_{f'}{\cal V}^{-}_{b}(0,M^{2}_{W},0,0,M_{W},M_{Z})
\\[1ex]
&&+\frac{1}{2}\delta Z^{f,L\dagger}_{ii}+\frac{1}{2}\delta Z^{f',L}_{jj}
+\frac{1}{2}\delta Z_{W}+\delta Z_{e}-\frac{\delta s_{W}}{s_{W}}\} .
\earr
\eeq
The left- and right-handed couplings $g_{f}^{\sigma}$ of the fermions
to the $Z$-boson are defined in (\ref{geZ}).
Note that only one out of the four \SME\ (\ref{MVFF}) is contributing
there
and that we need no counterterm to the quark mixing matrix for massless
fermions.
The counterterms are expressed in terms of the self energies in Sect.\
\ref{secrcfix}. These have
to be calculated to one-loop order to determine $\delta {\cal M}$
completely.
 
$\delta {\cal M}$ contains infrared divergencies. These are
regularized with a photon mass $\lambda $. They drop out in the decay
width if the contribution from the decay $W\to f_{i}\bar{f}'_{j}\gamma$
is added. This will be discussed in more detail in Chap.~\ref{chasopho}.
 
The example above was rather simple. If we keep the fermion masses
finite or consider processes with more external particles the number and
complexity of Feynman diagrams raises considerably.
 
\section{Computeralgebraic calculation of one-loop diagrams}
 
The procedure of generation and algebraic reduction of Feynman
diagrams as described above
is algorithmic and can be implemented in symbolic computation
systems. There are several attempts to create such systems for high
energy physics calculations \cite{comal}. In addition there exist
special packages written in general purpose languages
\cite{Ye90,Il90,Ku90,Me91}. In particular
the {\it MATHEMATICA\/} packages \FA\/ \cite{Ku90} and \FC\/ \cite{Me91}
have been developed for the automatic calculation of one-loop
diagrams following the approach outlined in this paper.
 
\FA\/ generates all graphs to a given process in a specified model
together with their combinatorial factors (weights). It yields both
analytical expressions and drawings of the graphs. There is a
version under development which uses the concept of generic diagrams.
It creates all
relevant generic graphs together with a list of all possible
substitutions yielding the actual graphs.
 
\FC\/ performs the algebraic evaluation of Feynman diagrams. It
starts from the output of \FA\/ and uses exactly the
reduction algorithm described above. It can deal
with generic diagrams. The \FC\/ output can easily be translated into
{\it FORTRAN\/} code.

\chapter{Soft photon bremsstrahlung}
\label{chasopho}
 
As mentioned in the last chapter the virtual one-loop corrections to
the decay matrix element $W\to f_{i}\bar{f}'_{j}$ are infrared divergent.
These divergencies originate from photonic corrections and show up
in any process with charged external particles.
 
However, these processes are not of direct
physical relevance since they cannot be distinguished experimentally
from those involving additional soft external photons.
Since the photons are massless their energies can be arbitrarily small
and thus less than the resolution of any detector. Therefore in
observable processes in addition to the basic
process those with arbitrary numbers of soft photons are included.
 
For these observable processes one obtains theoretically satisfactory results.
Adding incoherently the cross sections of all the different processes with
arbitrary
numbers of photons, all infrared divergencies cancel \cite{Bl37}. This
cancellation takes place between the virtual photonic corrections and the real
bremsstrahlung corrections order by order in perturbation theory.
To one-loop order one only needs to consider single photon radiation.
For the cancellations only the soft
photons, i.e.\ photons with energy $k_{0}\le \Delta E$, are relevant,
where $\Delta E$ is a cutoff parameter, which should be small compared to all
relevant energy scales. Photons with energies $k_{0} > \Delta E$ are
called hard. They can also yield sizable contributions especially
arising from photon emission collinear to the external charged particles.
 
In Sect.~\ref{secspa} we introduce the soft
photon approximation and show that in this approximation the
bremsstrahlung diagrams are
proportional to the Born diagrams. The corresponding soft
photon cross section for arbitrary Born diagrams is given in Sect.\
\ref{secspc}.
 
\section{Soft photon approximation}
\label{secspa}
 
Attaching soft photons to a charged external particle line of an arbitrary
Feynman diagram yields diagrams which become singular for vanishing
momentum of the soft photon. This divergence arises from the propagator
of the charged particle
generated by the inclusion of the radiated photon line.
In the soft photon approximation the momenta of the radiated photons
are neglected everywhere but in this singular propagator. This
approximation is valid if the matrix element of the basic process does
not change much if a photon with energy $\Delta E$ is emitted, i.e.\
the basic matrix element
is a slowly varying function of the photon energy for $k_{0}<\Delta E$.
This is not the case if the basic process contains a narrow resonance as
e.g.\ in
$e^{+}e^{-}\to \mu ^{+}\mu ^{-}$. Then one must either choose $\Delta E$
small compared to the width of the resonance or take into account the
strong variation exactly \cite{Gr80,Bo82}.
 
We now extract the soft photon matrix elements for external fermions,
scalars and vector bosons. The general renormalizable
couplings of these particles to
the photon allowed by electromagnetic gauge invariance are (momenta are
considered as incoming):
\savebox{\Vr}(48,0)[bl]
{\multiput(3,0)(12,0){4}{\oval(6,4)[t]}
\multiput(9,0)(12,0){4}{\oval(6,4)[b]} }
\savebox{\Vtr}(32,24)[bl]
{\multiput(4,0)(8,6){4}{\oval(8,6)[tl]}
\multiput(4,6)(8,6){4}{\oval(8,6)[br]} }
\savebox{\Vbr}(32,24)[bl]
{\multiput(4,24)(8,-6){4}{\oval(8,6)[bl]}
\multiput(4,18)(8,-6){4}{\oval(8,6)[tr]}}
\savebox{\Vtbr}(32,48)[bl]
{\put(00,24){\usebox{\Vtr}}
\put(00,00){\usebox{\Vbr}}}
 
\savebox{\Sr}(48,0)[bl]
{ \multiput(0,0)(12.5,0){4}{\line(4,0){10}} }
\savebox{\Str}(32,24)[bl]
{ \multiput(-2,-1.5)(12,9){3}{\line(4,3){10}} }
\savebox{\Sbr}(32,24)[bl]
{\multiput(-2,25.5)(12,-9){3}{\line(4,-3){10}} }
\savebox{\Stbr}(32,48)[bl]
{\put(00,24){\usebox{\Str}}
\put(00,00){\usebox{\Sbr}}}
 
\savebox{\Fr}(48,0)[bl]
{ \put(0,0){\vector(1,0){26}} \put(24,0){\line(1,0){24}} }
\savebox{\Ftr}(32,24)[bl]
{ \put(0,0){\vector(4,3){18}} \put(16,12){\line(4,3){16}} }
\savebox{\Fbr}(32,24)[bl]
{ \put(32,0){\vector(-4,3){19}} \put(16,12){\line(-4,3){16}} }
\savebox{\Ftbr}(32,48)[bl]
{\put(00,24){\usebox{\Ftr}}
\put(00,00){\usebox{\Fbr}}}
\beqar
\raisebox{-36pt}[36pt][42pt]{
\begin{picture}(108,72)
\put(83,56){\makebox(10,20)[bl]{$\bar{F}$}}
\put(0,43){\makebox(10,20)[bl]{$A_{\mu}$}}
\put(83,9){\makebox(10,20)[bl]{$F$}}
\put(48,36){\circle*{4}}
\put(0,36){\usebox{\Vr}}
\put(48,12){\usebox{\Ftbr}}
\end{picture} } \quad
&=&
-ie Q_{F}\gamma _{\mu }
, \\[1ex]
\raisebox{-36pt}[42pt][42pt]{
\begin{picture}(108,72)
\put(82,56){\makebox(10,20)[bl]{$S^{*},p'$}}
\put(0,43){\makebox(10,20)[bl]{$A_{\mu}$}}
\put(82,9){\makebox(10,20)[bl]{$S,p$}}
\put(48,36){\circle*{4}}
\put(0,36){\usebox{\Vr}}
\put(48,12){\usebox{\Stbr}}
\put(64,24){\vector(-4,3){3}} \put(62,46.5){\vector(4,3){3}}
\end{picture}  } \quad
&=&
-ie Q_{S}(p-p')_{\mu },\\[1ex]
\raisebox{-36pt}[42pt][42pt]{
\begin{picture}(108,72)
\put(82,56){\makebox(10,20)[bl]{$V^{*}_{\nu},p'$}}
\put(0,43){\makebox(10,20)[bl] {$A_{\mu},k$}}
\put(82,9){\makebox(10,20)[bl] {$V_{\rho},p$}}
\put(48,36){\circle*{4}}
\put(0,36){\usebox{\Vr}}
\put(48,12){\usebox{\Vtbr}}
\put(64,26){\vector(-4,3){3}} \put(64,50){\vector(4,3){3}}
\end{picture}  } \quad
&=&
+ie Q_{V}g_{\nu \rho }(p-p')_{\mu }
-ie \kappa _{V}[k_{\rho }g_{\mu \nu }-k_{\nu }g_{\mu \rho }].
\eeqar
Quartic boson couplings do not give rise to IR-singularities
and are thus not relevant in the soft photon approximation.
The terms involving the charges $Q$ are obtained directly from the
covariant derivative with respect to QED. The term proportional to
$\kappa_{V}$, which contributes only to the magnetic moment,
is gauge invariant by
itself. Further terms present in the $\gamma WW$ coupling in
the SM do not contribute for physical vector bosons. Since we will use
the unitary gauge in this section they drop out. In renormalizable
gauges their contributions are cancelled by those of the corresponding
unphysical Higgs bosons.
 
Consider first radiation from a fermion line. Let the basic
matrix element without soft photons be
\savebox{\blobdz}(32,32)[lb]{
\put(16,16){\circle{32}}
\put(5,5){\line(1,1){22}}
\put(16,0){\line(1,1){16}}
\put(0,16){\line(1,1){16}}
\put(9.5,1.5){\line(1,1){21}}
\put(1.5,9.5){\line(1,1){21}}}
\beq
\barr{l}
{\cal M} _{0}= A(p) u(p) = \quad
\earr
\barr{l}
\begin{picture}(80,32)(0,3.9)
\put(5,20){$F$}
\put(0,16){\usebox{\Fr}}
\put(48,00){\usebox{\blobdz}}
\end{picture}\quad
\earr ,
\eeq
where $u(p)$ is the fermion spinor with momentum $p$, $p^{2}=m^{2}$
and $A(p)$ the remaining part
of the matrix element. Inserting one photon (polarization vector
$\varepsilon$, momentum $k$) into the fermion line yields
\beq
\barr{l}
\barr{l}
{\cal M}_{1}=\quad
\earr
\barr{l}
\begin{picture}(128,50)(0,-5)
\put(6,22){$F,p$}
\put(85,40){$\varepsilon ,k$}
\put(0,16){\usebox{\Fr}}
\put(48,16){\usebox{\Fr}}
\put(48,16){\usebox{\Vtr}}
\put(96,00){\usebox{\blobdz}}
\end{picture}
\earr\\[1ex]
\phantom{{\cal M}_{1}}
=\disp A(p-k)\frac{i(\ps- \ks+m)}{(p-k)^{2}-m^{2}}(-ie Q_{F}\es)u(p) .
\earr
\eeq
Anticommuting $\ps -\frac{1}{2}\ks$ with $\es$ and using the Dirac
equation this can be written as
\beq \label{M1}
{\cal M} _{1}=\frac{eQ_{F}}{-2pk}A(p-k)[2\varepsilon p-i\varepsilon
^{\mu }\sigma_{\mu\nu}k^{\nu }]u(p) ,
\eeq
where $\sigma_{\mu \nu }= \frac{i}{2}[\gamma _{\mu },\gamma _{\nu }]$.
The denominator $\frac{1}{2pk}$ contains the IR-singularity.
Neglecting all terms proportional $k$ in the numerator
we obtain the soft photon approximation
\beq
{\cal M}_{1,s}=-eQ_{F}\frac{\varepsilon
p}{kp}A(p)u(p)=-eQ_{F}\frac{\varepsilon p}{kp}{\cal M}_{0} .
\label{MBRF}
\eeq
Note that the contributions of the magnetic moment term, the second
term in the square bracket in (\ref{M1}), are neglected in the soft
photon approximation and that the
soft photon matrix element is proportional to the Born matrix element.
For an outgoing fermion ($\bar{u}(p)$) one finds in the same way
\beq
{\cal M}_{1,s} = eQ_{F}\frac{\varepsilon p}{kp}{\cal M}_{0}.
\eeq
This is equivalent to (\ref{MBRF}) apart from a minus sign originating
from the different charge flow.
 
For an external vector line with polarization vector $\varepsilon
_{V}(p)$ the basic matrix element is
\beq
\barr{ll}
{\cal M}_{0} = A_{\sigma}(p) \varepsilon ^{\sigma}_{V}(p) = \quad &
\earr
\barr{l}
\begin{picture}(80,32)(0,3.5)
\put(5,23){$V_{\sigma}$}
\put(0,16){\usebox{\Vr}}
\put(27,18){\vector(1,0){3}}
\put(48,00){\usebox{\blobdz}}
\end{picture}
\earr\quad .
\eeq
The corresponding soft photon contribution is obtained from
\beq
\barr{lll}
{\cal M} _{1}&=&\disp A_{\sigma }(p-k)\frac{-i}{(p-k)^{2}-M^{2}}
(g^{\sigma \nu }-\frac{(p-k)^{\sigma }(p-k)^{\nu }}{M^{2}})
\varepsilon^{\rho }_{V}(p) \varepsilon ^{\mu }(k) \\[1em]
&&ie[Q_{V}g_{\nu \rho }(2p-k)_{\mu }-\kappa _{V}(k_{\rho }g_{\mu \nu
}-k_{\nu}g_{\mu \rho })],
\earr
\eeq
using
\beq
\varepsilon _{V}\cdot p=0
\eeq
as
\beq
{\cal M}_{1,s}=-eQ_{V}\frac{p\varepsilon }{pk}{\cal M}_{0}.
\eeq
It is proportional to the Born matrix element and independent of the
contribution of the magnetic moment $\kappa_{V}$ of the boson $V$.
Again an outgoing vector yields an extra minus sign.
 
The soft photon matrix element for an external scalar line is derived
analogously.
 
Radiation from internal charged lines or quartic vertices does not lead to
IR-singularities and is neglected in the soft photon approximation.
 
Summarizing, the $O(\alpha)$ soft photon matrix element corresponding to
an arbitrary matrix element ${\cal M}_{0}$ can be written as
\beq \label{MSP}
{\cal M}_{1,s}=-e {\cal M}_{0}\sum_{i}\,(\pm Q_{i})
\frac{\varepsilon p_{i}}{kp_{i}},
\eeq
where $p_{i},Q_{i}$ are the momentum and the charge of the $i$-th external
particle and $k$ is the outgoing photon momentum. The $\pm$ sign refers
to charges flowing into or out of the diagram, respectively.
The soft photon matrix element is
always proportional to the Born matrix element. The proportionality
factor depends only on the charges and momenta of the external particles.
 
\section{Soft photon cross section}
\label{secspc}
 
The soft photon cross section is obtained by squaring the soft photon
matrix element (\ref{MSP}), summing over the photon polarizations and
integrating
over the photon phase space with $|{\bf k}|\le \Delta E$
\beq \label{dssoft}
\left(\frac{d\sigma }{d\Omega }\right)_{s}=
-\left(\frac{d\sigma }{d\Omega }\right)_{0}
\frac{e^{2}}{(2\pi)^{3}}\int_{|{\bf k}|\le \Delta E}
\frac{d^{3}k}{2\omega_{k} }\sum_{ij}
\frac{\pm p_{i}p_{j}\,Q_{i}Q_{j}}{p_{i}k\,p_{j}k},
\label{dsbrem}
\eeq
where
\beq
\omega _{k}=\sqrt{{\bf k}^{2}+\lambda ^{2}}
\eeq
and $\pm$ refers to the relative sign of the $i$-th and $j$-th term in
(\ref{MSP}). As in the virtual corrections the IR-singularities are
regularized by
the photon mass $\lambda $.
Note that these integrals are not Lorentz-invariant due to the
integration region.
The basic integrals
\beq
I_{ij}= \int_{|{\bf k}|\le\Delta E}\frac{d^{3}k}{2\omega _{k}}
\frac{2p_{i}p_{j}}{p_{i}k\,p_{j}k}
\eeq
have been worked out e.g. by 't Hooft and Veltman \cite{tH79}.
 
The result is
\beqar
I_{ij}&=&4\pi  \frac{\alpha p_{i}p_{j}}{(\alpha p_{i})^{2}-
p_{j}^{2}}\left\{\frac{1}{2}\log \frac{(\alpha p_{i})^{2}}{p_{j}^{2}}\log
\frac{4\Delta E^{2}}{\lambda ^{2}} \right.\\[1em]
\nn&&+\biggl[\frac{1}{4}\log ^{2}\frac{u_{0}-|{\bf u}|}{u_{0}+|{\bf u}|}+
\Li\biggl(1 -\frac{u_{0}+\vert{\bf u}\vert}{v}\biggr)
\left.+\Li\biggl(1-\frac{u_{0}-|{\bf u}|}{v}\biggr)\biggr]^{u=\alpha
p_{i}}_{u=p_{j}}\right\},
\eeqar
with
\beq
v=\frac{(\alpha p_{i})^{2}-p_{j}^{2}}{2(\alpha p_{i0}-p_{j0})},
\eeq
and $\alpha $ defined through
\beq
\alpha ^{2}p_{i}^{2}-2\alpha p_{i}p_{j}+p_{j}^{2}=0,
\qquad\frac{\alpha p_{i0}-p_{j0}}{p_{j0}}>0.
\eeq
For $p_{i}=p_{j}$ this simplifies to
\beq
I_{ii}=2\pi \left\{\log \frac{4\Delta E^{2}}{\lambda ^{2}}
+\frac{p_{0}}{|{\bf p}|}
\log\frac{p_{0}-|{\bf p}|}{p_{0}+|{\bf p}|}\right\},
\eeq
and for ${\bf p}_{i}=-{\bf p}_{j}={\bf p}$
\beqar
I_{ij}&=&2\pi
\frac{pq}{(p_{0}+q_{0})|{\bf p}|}
\left\{\frac{1}{2}\log \frac{p_{0}+|{\bf p}|}{p_{0}-|{\bf p}|}
\log \frac{4\Delta E^{2}}{\lambda ^{2}}
-\Li\biggl(\frac{2|{\bf p}|}{p_{0}+|{\bf p}|}\biggr)
-\frac{1}{4}\log^{2}\frac{p_{0}+|{\bf p}|}{p_{0}-|{\bf p}|}
\right.\quad \\[1em]
\nn &&\phantom{2\pi\frac{pq}{(p_{0}+q_{0})|{\bf p}|}}
\left.+\frac{1}{2}\log \frac{q_{0}+|{\bf p}|}{q_{0}-|{\bf p}|}
\log \frac{4\Delta E^{2}}{\lambda ^{2}}
-\Li\biggl(\frac{2|{\bf p}|}{q_{0}+|{\bf p}|}\biggr)
-\frac{1}{4}\log^{2}\frac{q_{0}+|{\bf p}|}{q_{0}-|{\bf p}|}
\right\}.\quad
\eeqar
 
Inserting the results for $I_{ij}$ into (\ref{dsbrem}) yields the soft
photon cross section. Adding it to the one-loop corrected cross section
for the corresponding basic process
the IR-divergencies cancel and the limit $\lambda \to 0$ can be taken.
 
Although the inclusion of the real soft photon emission is sufficient
to obtain IR-finite results, it is often not adequate for real
experiments, because realistic detectors do not provide a sufficiently
small resolution $\Delta E/E$ necessary for the validity of the soft
photon approximation. Therefore also hard photons (with
$k_{0}>\Delta E$) are important.
Their contribution is UV-and IR-finite and can be treated separately. One
merely has to make sure that the soft and hard part are properly adapted
to each other.
 
Hard photon corrections are treated with methods different from the ones
presented in this work. Their contribution depends sensitively on the
experimental setup. They are usually incorporated by Monte Carlo
simulations \cite{Kl89}.

\chapter{Input parameters and leading higher order contributions}
 
In order to complete all ingredients necessary for the calculation of
radiative corrections we have to specify the input parameters. This is
done in Sect.~\ref{secpar}. The leading higher order corrections which
become important for precision experiments are discussed in
Sect.~\ref{sechicor}.
 
\section{Input parameters}
\label{secpar}
 
In its original symmetric version the SM depends on the parameters (\ref
{sympar}), which are essentially the couplings allowed by the
$SU(2)_{W}\times U(1)_{Y}$ symmetry.
These were replaced by the physical parameters (\ref{phypar}), i.e.\
the particle masses, the electromagnetic coupling constant, and the
quark mixing matrix.
In the on-shell
renormalization scheme the renormalized parameters are equal to these
physical parameters in all orders of perturbation theory.
 
The numerical values of the physical parameters must be fixed
through experimental input. However, this input may not necessarily
consist of direct measurements of the renormalized parameters;
it may be obtained from any suitable set of experimental results.
In practice one uses those experiments which have the highest
experimental accuracy and theoretical reliability.
This criterion is certainly fulfilled for the following set of parameters
whose numerical values are taken from \cite{PDG90}:
\bit
\item the fine structure constant
\bma    \alpha = 1/137.0359895 (61)  \ema
      corresponding to the classical electron charge
$e=\sqrt{4\pi\alpha}$,
\item the masses of the charged leptons
\bma
\barr{ll}
      m_{e}= 0.51099906(15) \:\mbox{MeV}, \qquad &
      m_{\mu}= 105.658387(34) \:\mbox{MeV},  \qquad \\[1ex]
      m_{\tau}= 1784.1 {+2.7 \atop -3.6} \:\mbox{MeV},
\earr
\ema
\item the mass of the $Z$-boson \cite{Dy90}
\bma      M_{Z}=91.177 (21) \:\mbox{GeV},     \ema
\item  and the Fermi constant
\bma      G_{F}=1.16637(2) 10^{-5} \: \mbox{GeV}^{-2},    \ema
      which is directly related to the muon lifetime.
\eit
We do not use the $W$-mass as input parameter because it is
experimentally not known with comparable accuracy.
 
Besides the above listed well known parameters the still unknown masses of
the top quark and the Higgs scalar are kept as free parameters.
If the minimal SM is correct, the present experimental data restrict the
top quark mass to the region $80\:\mbox{GeV} < m_{t} < 200\:\mbox{GeV}$
\cite{Dy90,El90}.
For the Higgs mass we use $40\:\mbox{GeV} < M_{H} < 1\:\mbox{TeV}$, where
the lower bound is experimental \cite{Dy90} and the upper bound is favored
by theoretical consistency arguments. If not stated otherwise we will
use the values $m_{t}=140 \:\mbox{GeV}$ and $M_{H}=100 \:\mbox{GeV}$.
 
The quark mixing matrix elements $V_{ij}$ are directly taken
from experiment. We use the parametrization of Harari and Leurer
\cite{Ha86} as advocated by the Particle Data Group and choose the
following numerical values for the parameters in agreement with \cite{PDG90}
\beq \label{sqmm}
s_{12} = 0.220 ,\qquad s_{23} = 0.046 ,\qquad s_{13} = 0.007
\eeq
and $\delta = 0$ for simplicity. This yields approximately the
following numbers for the quark mixing matrix elements:
\beq
\barr{llrllrllr}
V_{ud} &=&  0.975, & \qquad V_{us} &=&  0.220, & \qquad V_{ub} &= &0.007, \\
V_{cd} &=& -0.220, & \qquad V_{cs} &=&  0.974, & \qquad V_{cb} &= &0.046, \\
V_{td} &=&  0.003, & \qquad V_{ts} &=& -0.046, & \qquad V_{tb} &= &0.999.
\earr
\eeq
 
It remains to discuss the masses $m_{q}$ of the light quarks
($q=d,\,u,\,s,\,c,\,b$). In the
electroweak Lagrangian the quarks are treated as free particles with
appropriate masses. This is not correct due to the presence of the
strong interaction. Therefore the quark masses can at best be considered as
somewhat effective parameters. Fortunately in typical high energy
experiments ($s\gg m^{2}_{q}$) theoretical predictions depend on the
quark masses only through universal quantities such as the hadronic vacuum
polarization or the quark structure functions. These can be directly
determined from experiment. Nonuniversal contributions are
suppressed as $m^{2}_{q}/s$ and thus negligible for sufficiently
high energies.
 
For processes without external quarks only the hadronic contribution to
the vacuum polarization
\beq
\Pi^{AA}(s) = \frac{\Sigma_{T}^{AA}(s)}{s}
\eeq
is relevant. In perturbation theory the contribution of light quarks is given by
\beq  \label{Pihad}
\hat{\Pi}^{AA}_{had}(s)=3\frac{\alpha }{3\pi}\sum_{d,u,s,c,b}Q^{2}_{q}
\left(\frac{5}{3} - \log\frac{-s-i\varepsilon}{m^{2}_{q}}\right).
\eeq
 
The large logarithmic terms contained in (\ref{Pihad}) constitute a
dominant contribution to the radiative corrections. They originate from
the charge renormalization in the on-shell scheme at zero momentum
transfer (see eq.~\ref {DZE}) involving
\beq
\Pi ^{AA}(0)= \left.\frac{\partial \Sigma_{T}^{AA}(k^{2})}{\partial k^{2}}
\right|_{k^{2}=0}.
\eeq
In this quantity nonperturbative strong interaction
effects cannot be neglected. Since no reliable theoretical predictions
are available one has to extract $\Pi ^{AA}_{had}(0)$ from experimental data.
Writing
\beq
\barr{lll}
\Pi ^{AA}_{had}(0)&=&\Pi ^{AA}_{had}(0)-\Re\,\Pi^{AA}_{had}(s)
+\Re\,\Pi^{AA}_{had}(s)\\[1ex]
&=&-\Re\,\hat{\Pi}^{AA}_{had}(s)+\Re\,\Pi ^{AA}_{had}(s),
\earr
\eeq
the unrenormalized hadronic vacuum polarization $\Pi^{AA}_{had}(s)$
can be evaluated
perturbatively for $s\gg m^{2}_{q}$ and the renormalized one
$\Re\, {\hat \Pi}^{AA}_{had}(s)$ is given by the dispersion relation
\beq
\Re\, \hat{ \Pi }^{AA}_{had}(s)=\frac{\alpha }{3\pi }s \,
\Re\int^{\infty}_{4m^{2}_{\pi }}
ds' \frac{R^{AA}(s')}{s'(s'-s-i\varepsilon)}
\eeq
with
\beq
R^{AA}(s')=\frac{\sigma (e^{+}e^{-}\to\gamma ^{*}\to \mbox{hadrons})}{\sigma
(e^{+}e^{-}\to\gamma ^{*}\to\mu ^{+}\mu ^{-})}.
\eeq
$R^{AA}(s)$ can be taken from experiment up to some scale $s$,
for larger $s$ perturbative QCD is used.
A recent analysis \cite{Bu89} involving data for the energy range
below $40\:\mbox{GeV}$ yields for the contribution of the 5 light quarks
in the energy region $50\:\mbox{GeV} < s < 200\:\mbox{GeV}$
\beqar
\Re\, \hat{\Pi}^{AA(5)}_{had}(s) &=&-0.0288 \pm 0.0009 \\[1em]
\nn&& - 0.002980\left[
\log\left(\frac{s}{(92\:\mbox{GeV})^{2}}\right)
+ 0.006307 \left(\frac{s}{(92\:\mbox{GeV})^{2}} - 1 \right) \right].
\eeqar
 
For energies around the Z-boson mass this can be approximated by
(\ref{Pihad}) using the effective quark masses
\beq
\barr{lll}
m_{u}= 0.041 \:\mbox{GeV}, \qquad & m_{c}= 1.5   \:\mbox{GeV}, \\
m_{d}= 0.041 \:\mbox{GeV}, \qquad & m_{s}= 0.15  \:\mbox{GeV}, \qquad &
 m_{b}= 4.5   \:\mbox{GeV}.
\earr
\eeq
These quark masses, in particular the ones for the three lightest
quarks, are effective parameters adjusted to fit the
dispersion integral and have no further significance.
 
In addition to the above parameters we need the
strong coupling constant $\alpha_{s}$ for the QCD corrections.
Its value at the scale of the
$Z$-boson mass is given by \cite{Dy90}
\beq
\alpha_{s}= 0.115 \pm 0.010\:.
\eeq
For the numerical evaluation we use in the following
\beq
\alpha_{s}= 0.12 \:.
\eeq
 
The $W$-mass is determined from the parameters given above through
the relation
\beq \label{GF1loop}
M^{2}_{W}(1-\frac{M^{2}_{W}}{M^{2}_{Z}})=\frac{\pi \alpha
}{\sqrt{2}G_{F}} [1+\Delta r] .
\eeq
$\Delta r$ summarizes the radiative corrections to muon decay \cite{Si80}
apart from the QED corrections which coincide with those of the
Fermi model. It depends on all parameters of the SM and
is given by
\beqar
\nn\Delta r &=& \Pi^{AA}(0) -\frac{c_{W}^{2}}{s_{W}^{2}}
\left(\frac{\Sigma^{ZZ}_{T}(M_{Z}^{2})}{M_{Z}^{2}}
-\frac{\Sigma^{W}_{T}(M_{W}^{2})}{M_{W}^{2}}\right)
+\frac{\Sigma^{W}_{T}(0)-\Sigma^{W}_{T}(M_{W}^{2})}{M_{W}^{2}} \\[1em]
&&+2\frac{c_{W}}{s_{W}}\frac{\Sigma^{AZ}_{T}(0)}{M_{Z}^{2}}
+\frac{\alpha }{4\pi s_{W}^{2}}
\left(6 +  \frac{7-4s_{W}^{2}}{2s_{W}^{2}}\log c_{W}^{2}\right).
\eeqar
The relation (\ref{GF1loop}) can be improved by summing the leading
higher order reducible corrections. This is done in the next section
(eq.~\ref{GFloop}). For the set of parameters specified above we obtain
for the $W$-boson mass
\beq
M_{W}= 80.23\:\mbox{GeV}.
\eeq
 
\section {Leading higher order contributions}
\label{sechicor}
 
The natural order of magnitude of one-loop radiative corrections is set
by the loop expansion parameter $\frac{\alpha}{\pi}\sim 0.0023$.
Consequently typical one-loop corrections are of the order of one
percent. There are, however, two important types of radiative corrections
which are enhanced by large mass ratios.
The first type is associated with light fermions the second with a
heavy top quark. These corrections can reach several percent.
Consequently the corresponding higher loop corrections may become as big
as several permill and thus have to be taken into account in
predictions for precision experiments.
 
\subsection {Leading logarithms from light fermion masses}
 
The first type of enhanced corrections originates from the
renormalization of $\alpha$ at zero momentum transfer where the
relevant scale is set by the fermion masses. These are much
smaller than the relevant scales in high energy experiments. The large
ratio of these different scales leads to large logarithms which can be
summarized in the universal quantity
\beq
\Delta \alpha (s)=-\Re\hat{\Pi}^{AA}(s)  =
\frac{\alpha }{3\pi}\sum_{f}N_{C}^{f}Q^{2}_{f}
\log\frac{|s|}{m^{2}_{f}} + \ldots = (\Delta\alpha)_{LL} + \ldots  ,
\eeq
where $N_{C}^{f}$ is the colour factor of the fermions and the dots
indicate nonleading contributions.
Renormalization group arguments can be used to show \cite{Ma79} that
the leading logarithms $(\Delta \alpha)_{LL}$ are correctly summed to
all orders in
perturbation theory by the replacement
\beq
1+(\Delta \alpha )_{LL}\to \frac{1}{1-(\Delta \alpha )} _{LL}.
\eeq
Since not only the leading logarithms but the whole fermionic
contribution to $\Delta \alpha $ are gauge invariant we will sum the
latter. The gauge dependent bosonic contribution can not be summed, however,
because this would violate gauge invariance in higher orders. Thus we
arrive at
\beq
1+\Delta \alpha =1+(\Delta \alpha )_{ferm}+(\Delta \alpha )_{bos}\to
\frac{1}{1-(\Delta \alpha )_{ferm}}+(\Delta \alpha )_{bos}.
\eeq
This corresponds to a resummation of the iterated one-loop fermionic
vacuum polarization to all orders.
 
Since the leading logarithms contained in $\Delta \alpha $ originate
from the charge renormalization constant $\delta Z_{e}$ they are universal,
i.e.\ they appear everywhere where $\alpha $ appears in lowest order. They can
be taken into acount by replacing the lowest order $\alpha $ by a
running $\alpha (s)$ defined as
\beq \label{a(s)}
\alpha = \alpha(0)\to \alpha(s)=\frac{\alpha}{1-
(\Delta\alpha(s))_{ferm}}.
\eeq
$\Delta \alpha (0)=0$ due to the on-shell renormalization condition
for $\alpha $. Using $\alpha(s)$ instead of $\alpha $ effectively
corresponds to renormalize $\alpha $ not at zero momentum transfer but
at momentum transfer $s$. Then the light fermion masses can be neglected
everywhere and no large logarithms appear.
 
There are similar large logarithms associated with external fermion
lines. These are related to collinear singularities, arising from
the radiation of photons collinear to external particles. They can be
consistently treated with the structure function method \cite{Be89}.
 
\subsection {Leading $m^{2}_{t}$ contributions}
 
The second type of important corrections
is also connected to the fermionic sector.
The top quark gives rise to corrections $\propto  m^{2}_{t}/M^{2}_{W}$,
which become large if the top quark mass
is large compared to the $W$-boson mass.
For $m_{t}=200\:\mbox{GeV}$ they reach several percent.
These terms arise from fermion loop
contributions to the boson self energies and from the Yukawa couplings
of the physical and unphysical Higgs fields, which show up in vertex and
fermionic self energy corrections. The latter are process dependent and
are therefore not discussed here. The former, however, are universal,
they can be traced back to the renormalization of $s_{W}$ in the
on-shell scheme
\beqar
\nn \frac{\delta s_{W}}{s_{W}} & = & -\frac{1}{2}\frac{c^{2}_{W}}{s^{2}_{W}}
\widetilde{\Re}\,\left(\frac{\Sigma^{W}_{T}(M^{2}_{W})}{M^{2}_{W}}
-\frac{\Sigma^{ZZ}_{T}(M^{2}_{Z})}{M^{2}_{Z}}\right)\\[1em]
& = & \frac{1}{2}\frac{c^{2}_{W}}{s^{2}_{W}}\frac{\alpha }{4\pi }
\frac{3}{4s^{2}_{W}}\frac{m^{2}_{t}}{M^{2}_{W}}+\ldots
=\frac{1}{2}\frac{c^{2}_{W}}{s^{2}_{W}}\Delta \rho +\ldots
\eeqar
where the dots indicate again nonleading contributions.
 
There is no general principle that determines the resummation of these
corrections. The following recipe has been shown to yield the correct
leading terms to $O(\alpha^{2})$ \cite{Co89}
\beq \label{sweff}
\barr{lll}
s^{2}_{W} &\to& s^{2}_{W}+c^{2}_{W}\Delta \bar{\rho}=\bar{s}_{W}^{2},\\[1ex]
c^{2}_{W} &\to& c^{2}_{W}(1-\Delta \bar{\rho})=\bar{c}_{W}^{2},
\earr
\eeq
where
\beq
\Delta \bar{\rho }=\frac{3G_{F}m^{2}_{t}}{8\sqrt{2}\pi^{2}}
\left[1+\frac{G_{F}m_{t}^{2}}{8\sqrt{2}\pi^{2}}(19-2\pi^{2})\right]
\eeq
incorporates the result \cite{Bi87a} from two-loop irreducible diagrams.
Note that $\alpha $ has been replaced by $G_{F}$ in order to obtain the
correct leading $O(\alpha ^{2})$ terms.
 
In particular for the relation between $M_{W}$ and $G_{F}$ the correct
resummation of the leading corrections is given by
\beq \label{GFloop}
G_{F}=\frac{\pi }{M^{2}_{W}\sqrt{2}}\frac{\alpha}{s^{2}_{W}}
\frac{1}{1-(\Delta \alpha)_{ferm}}
\frac{1}{1+\frac{c^{2}_{W}}{s^{2}_{W}}\Delta \bar{\rho}}
\left[1+\Delta r-(\Delta \alpha )_{ferm}+\frac{c^{2}_{W}}{s^{2}_{W}}\Delta
\rho\right].
\eeq
Note that the two types of leading corrections are summed separately.
Inserting $s_{W}^{2}=1-M_{W}^{2}/M_{Z}^{2}$ this relation can be used to
determine the $W$-boson mass $M_{W}$ including leading higher order
contributions.
 
Neglecting the nonleading contributions and using (\ref{a(s)})
and (\ref{sweff}) eq.\ (\ref{GFloop}) can be written as
\beq \label{GFappr}
\frac{\pi \alpha (M^{2}_{W})}{\bar{s}_{W}^{2}}\approx
\sqrt{2}G_{F}M^{2}_{W}.
\eeq
With this relation the appearance of $G_{F}$ in $\Delta \bar{\rho }$ can
be easily understood. All leading universal corrections arise from the
renormalization constants of $\alpha $ and $s_{W}$. Consequently they
can be absorbed by incorporating the leading finite parts of these
renormalization constants into the effective parameters
$\alpha (s)$ and $\bar{s}_{W}^{2}$ (including the
leading higher order corrections). Thus one obtains from the
lowest order result directly the corresponding result including the
leading universal corrections. In particular (\ref{GFappr}) can be obtained
in this way. Applying this recipe to $\Delta \rho$ and using
(\ref{GFappr}) introduces naturally $G_{F}$.
 
\subsection {Recipes for leading universal corrections}
 
From the discussion above we infer that the universal leading higher order
contributions can be taken into account by the following replacements
\beqar \label{reclco}
\nn \alpha &\to& \alpha (s),\\[1ex]
\nn s^{2}_{W} &\to& \bar{s}_{W}^{2}, \qquad \qquad
c^{2}_{W}\to \bar{c}_{W}^{2},\\[1em]
\nn \frac{e^{2}}{2s^{2}_{W}}& = & \frac{2\pi \alpha }{s^{2}_{W}} \quad \to
\; \frac{2\pi \alpha (s)}{\bar{s}_{W}^{2}}\approx
2\sqrt{2}G_{F}M^{2}_{W},\\[1em]
\frac{e^{2}}{4s^{2}_{W}c^{2}_{W}}& = & \frac{\pi \alpha
}{s^{2}_{W}c^{2}_{W}} \; \to \;\; \frac{\pi \alpha
(s)}{\bar{s}_{W}^{2}\bar{c}_{W}^{2}}\approx
\sqrt{2}G_{F}M^{2}_{Z}\frac{1}{1-\Delta \bar{\rho }}.
\eeqar
Note that this does not include the nonuniversal corrections $\propto
\alpha\,m_{t}^{2}/M_{W}^{2}$ arising from enhanced Yukawa couplings.
These have to be evaluated for each process directly.

\chapter{The width of the $W$-boson}
\label{chaWff}
 
In the coming years the upgrade of the LEP electron positron storage ring
to $180-190\:\mbox{GeV}$ CM-energy will allow to study the properties of the
$W$-boson in detail in a model independent way. Besides its mass $M_{W}$
also its total and partial decay widths are of interest. While the
leptonic partial widths allow to test lepton universality the hadronic
partial widths can serve to determine the quark mixing matrix elements
\cite{Lo87}. The expected accuracy is $\delta M_{W} \approx 100\:\mbox{MeV}$ and
$\delta \Gamma_{W}\approx 200\:\mbox{MeV}$ \cite{Bi87}.
 
An accurate comparison between these experiments and theoretical
predictions of the SM requires at least the inclusion of one-loop radiative
corrections both for the production and the decay of $W$-bosons.
The $W$-bosons decay dominantly into fermion-antifermion pairs.
The total and partial widths of the $W$-boson and the corresponding
one-loop radiative corrections are discussed in this chapter.
 
The electroweak and QCD corrections for decays into massless fermions
($m_{f}\ll M_{W}$) have been calculated in \cite{Al80,Ba86,Je86}.
The hard bremsstrahlung contribution has been investigated in
\cite{Be85,Je85}. The QCD corrections for the decay into a massive top quark
and a massless bottom quark were given in \cite {Al88}. The full
one-loop electroweak and QCD corrections together with the complete
photonic and gluonic bremsstrahlung were evaluated for arbitrary finite
fermion masses in \cite {De90a}.
 
Since the top quark is probably heavier than the $W$-boson and since all other
quark masses are small compared to the $W$-boson mass, the fermion mass
effects are not of great importance for the $W$-decay. However, the
matrix element for the $W$-decay into two fermions is directly related
via crossing to the one for the decay of the top quark into a $W$-boson
and a bottom quark. In this case the fermion masses are crucial. Since
we will discuss top decay in Chap.~\ref{chatWb}
we give the results for $W$-decay for finite fermion masses.
 
\section{Lowest order}
The Born amplitude for the decay $W^{+}\to f_{i}\bar{f}'_{j}$ was already
given in (\ref {M0Wff}). For finite fermion masses the following result
is obtained for the corresponding partial decay width
\beq \label{G0Wff}
\Gamma ^{Wf_{i}f'_{j}}_{0}(M_{W},m_{f,i},m_{f',j})=N_{C}^{f}\frac{\alpha}{12}
\frac{1}{2s_{W}^{2}}|V_{ij}|^{2}
\frac{\kappa(M_{W}^{2},m_{f,i}^{2},m_{f',j}^{2})}{M^{3}_{W}}G_{1}^{-},
\eeq
where $\kappa/M^{3}_{W}$ originates from phase space and
\beq
G^{-}_{1}=\sum_{pol}{\cal M}^{-{\dagger}}_{1}{\cal M} ^{-}
_{1}=2M_{W}^{2}-m_{f,i}^{2}-m_{f',j}^{2}-\frac{(m_{f,i}^{2}-
m_{f',j}^{2})^{2}}{M_{W}^{2}}
\eeq
from the polarization sum of the matrix element squared.
The K\"all\'en function $\kappa$ was
defined in (\ref{kappa}).
The colour factor $N_{C}^{f}$ is given by
\beq
N_{C}^{f}=\left\{ \begin{tabular}{l} 3 \quad for quarks,\\
       1 \quad for leptons. \end{tabular}  \right.
\eeq
For leptonic decays we have $V_{ij}=\delta _{ij}$. The total width is obtained
as a sum over the partial fermionic decay widths with
$m_{f,i}+m_{f',j}< M_{W}$
\beq
\Gamma ^{W}_{0}=\sum_{ij}\Gamma_{0}^{Wu_{i}d_{j}}
+\sum _{i}\Gamma_{0}^{W\nu_{i} l_{i}}.
\eeq
We can write down another possible tree level representation for the
partial decay width by eliminating $\alpha/s^{2}_{W}$ in
favour of $G_F$ 
\beq  \label{GGWff}
\bar{\Gamma}^{Wf_{i}f'_{j}}_{0}(M_{W},m_{f,i},m_{f',j})=N_{C}^{f}
\frac{G_F}{12\pi
\sqrt{2}}|{V_{ij}}|^2\frac{\kappa(M_W^2,m_{f,i}^2,m_{f',j}^{2})}{M_{W}}
G_{1}^{-}.
\eeq
 
\section{Electroweak virtual corrections}
 
In our formulation of the on-shell renormalization scheme the invariant
matrix element to one-loop order has the following form
\beqar \label{M1Wff}
\nn{\cal M}^{Wf_{i}f'_{j}}=-\frac{e}{\sqrt{2}s_{W}}
&&\Bigl\{V_{ij}{\cal M}^{-}_{1}
[1+ \delta Z_{e}-\frac{\delta s_{W}}{s_{W}}]
+{\cal M}^{-}_{1}\delta V_{ij}\\[.9ex]
\nn&&\mbox{}+V_{ij}{\cal M}^{-}_{1}\frac{1}{2}\delta Z_{W}
+{\cal M}^{-}_{1}\frac{1}{2}\sum_{k}[\delta Z^{f,L\dagger}_{ik}
V_{kj}+V_{ik}\delta Z^{f',L}_{kj}]\\[.9ex]
&&\mbox{}+\sum_{a=1}^{2}\sum_{\sigma =\pm}
{\cal M}^{\sigma}_{a}\delta F^{\sigma}_{a}(M_{W},m_{f,i},m_{f',j})\Bigr\}.
\eeqar
The standard matrix elements $M_{a}^{\sigma}$ were defined in
(\ref{MVFF}). The functions $\delta F^{\sigma}_{a}$ summarize the loop
corrections to
the $Wf_{i}f'_{j}$-vertex. There are no explicit self energy corrections
from the external lines. These are all absorbed into the field
renormalization constants $\delta Z_{W}$, $\delta Z^{f,L}$ and $\delta
Z^{f',L}$. These and the parameter renormalization constants were given
in terms of self energies in Sect.~\ref{secrcfix}.
The explicit forms of the self energies can be found in App.~B.
 
\begin{figure}[p]
\savebox{\Vr}(36,0)[bl] 
{\multiput(3,0)(12,0){3}{\oval(6,4)[t]}
\multiput(9,0)(12,0){3}{\oval(6,4)[b]} }
\savebox{\Vt}(0,36)[bl] 
{\multiput(0,3)(0,12){3}{\oval(4,6)[l]}
\multiput(0,9)(0,12){3}{\oval(4,6)[r]} }
\savebox{\Vtr}(36,18)[bl] 
{\multiput(4,0)(8,4){5}{\oval(8,4)[tl]}
\multiput(4,4)(8,4){4}{\oval(8,4)[br]} }
\savebox{\Vbr}(36,18)[bl] 
{\multiput(4,18)(8,-4){5}{\oval(8,4)[bl]}
\multiput(4,14)(8,-4){4}{\oval(8,4)[tr]}}
\savebox{\Vtbr}(36,36)[bl] 
{\put(00,18){\usebox{\Vtr}}
\put(00,00){\usebox{\Vbr}}}
 
\savebox{\Sr}(36,0)[bl] 
{ \multiput(0,0)(13,0){3}{\line(4,0){10}} }
\savebox{\St}(0,36)[bl] 
{ \multiput(0,0)(0,13){3}{\line(0,1){10}} }
\savebox{\Str}(36,18)[bl] 
{ \multiput(0,0)(13,6.5){3}{\line(2,1){10}} }
\savebox{\Sbr}(36,18)[bl] 
{\multiput(0,18)(13,-6.5){3}{\line(2,-1){10}} }
\savebox{\Stbr}(36,36)[bl] 
{\put(00,18){\usebox{\Str}}
\put(00,00){\usebox{\Sbr}}}
 
\savebox{\Fr}(36,0)[bl] 
{ \put(0,0){\vector(1,0){20}} \put(18,0){\line(1,0){18}} }
\savebox{\Ft}(0,36)[bl] 
{ \put(0,0){\vector(0,1){21}} \put(0,18){\line(0,1){18}} }
\savebox{\Ftr}(36,18)[bl] 
{ \put(0,0){\vector(2,1){20}} \put(18,9){\line(2,1){18}} }
\savebox{\Fbr}(36,18)[bl] 
{ \put(36,0){\vector(-2,1){21}} \put(18,9){\line(-2,1){18}} }
\savebox{\Ftbr}(72,72)[bl] 
{\put(00,36){\usebox{\Ftr}}\put(36,54){\usebox{\Ftr}}
 \put(00,18){\usebox{\Fbr}}\put(36,00){\usebox{\Fbr}}}
\savebox{\Vp}(36,36)[bl] 
{ \put(0,18){\circle*{4}} \put(36,0){\circle*{4}} 
\put(36,36){\circle*{4}} }
 
\bma
\barr{l}
\makebox{
\begin{picture}(135,90)
\put(-40,60){\makebox(10,10){$a)$}}
\put(5,40){\makebox(10,10){$W$}}
\put(77,32){\makebox(20,10){$\gamma,Z$}}
\put(43,54){\makebox(20,10){$f_{i}$}}
\put(43,8){\makebox(20,10){$f'_{j}$}}
\put(100,8){\makebox(10,10){$\bar{f}'_{j}$}}
\put(100,54){\makebox(10,10){$f_{i}$}}
\put(36,18){\usebox{\Vp}}
\put(0,36){\usebox{\Vr}}
\put(36,00){\usebox{\Ftbr}}
\put(72,18){\usebox{\Vt}}
\end{picture}}  \\
 
\savebox{\Ftbr}(72,72)[bl] 
{\put(36,18){\usebox{\Ft}}
\put(36,54){\usebox{\Ftr}} \put(36,00){\usebox{\Fbr}}}
 
\makebox{
\begin{picture}(135,90)
\put(-40,60){\makebox(10,10){$b)$}}
\put(5,40){\makebox(10,10){$W$}}
\put(77,32){\makebox(20,10){$f_{i}$}}
\put(43,54){\makebox(20,10){$\gamma,Z$}}
\put(43,8){\makebox(20,10){$W$}}
\put(100,8){\makebox(10,10){$\bar{f}'_{j}$}}
\put(100,54){\makebox(10,10){$f_{i}$}}
\put(36,18){\usebox{\Vp}}
\put(0,36){\usebox{\Vr}}
\put(36,00){\usebox{\Ftbr}}
\put(36,18){\usebox{\Vtbr}}
\end{picture}}  
\makebox{
\begin{picture}(135,90)
\put(5,40){\makebox(10,10){$W$}}
\put(77,32){\makebox(20,10){$f'_{j}$}}
\put(43,54){\makebox(20,10){$W$}}
\put(43,8){\makebox(20,10){$\gamma,Z$}}
\put(100,8){\makebox(10,10){$\bar{f}'_{j}$}}
\put(100,54){\makebox(10,10){$f_{i}$}}
\put(36,18){\usebox{\Vp}}
\put(0,36){\usebox{\Vr}}
\put(36,00){\usebox{\Ftbr}}
\put(36,18){\usebox{\Vtbr}}
\end{picture}}  \\
 
\savebox{\Ftbr}(72,72)[bl] 
{\put(00,36){\usebox{\Ftr}}\put(36,54){\usebox{\Ftr}}
 \put(00,18){\usebox{\Fbr}}\put(36,00){\usebox{\Fbr}}}
\makebox{
\begin{picture}(135,90)
\put(-40,60){\makebox(10,10){$c)$}}
\put(5,40){\makebox(10,10){$W$}}
\put(77,32){\makebox(20,10){$H,\chi$}}
\put(43,8){\makebox(20,10){$f'_{j}$}}
\put(43,54){\makebox(20,10){$f_{i}$}}
\put(100,8){\makebox(10,10){$\bar{f}'_{j}$}}
\put(100,54){\makebox(10,10){$f_{i}$}}
\put(36,18){\usebox{\Vp}}
\put(0,36){\usebox{\Vr}}
\put(36,00){\usebox{\Ftbr}}
\put(72,18){\usebox{\St}}
\end{picture}}  
\\
\savebox{\Ftbr}(72,72)[bl] 
{\put(36,18){\usebox{\Ft}}
\put(36,54){\usebox{\Ftr}} \put(36,00){\usebox{\Fbr}}}
\makebox{
\begin{picture}(135,90)
\put(-40,60){\makebox(10,10){$d)$}}
\put(5,40){\makebox(10,10){$W$}}
\put(77,32){\makebox(20,10){$f_{i}$}}
\put(43,54){\makebox(20,10){$H$}}
\put(43,8){\makebox(20,10){$W$}}
\put(100,8){\makebox(10,10){$\bar{f}'_{j}$}}
\put(100,54){\makebox(10,10){$f_{i}$}}
\put(36,18){\usebox{\Vp}}
\put(0,36){\usebox{\Vr}}
\put(36,00){\usebox{\Ftbr}}
\put(36,18){\usebox{\Vbr}}
\put(36,36){\usebox{\Str}}
\end{picture}} 
\makebox{
\begin{picture}(135,90)
\put(5,40){\makebox(10,10){$W$}}
\put(77,32){\makebox(20,10){$f'_{j}$}}
\put(43,54){\makebox(20,10){$W$}}
\put(43,8){\makebox(20,10){$H$}}
\put(100,8){\makebox(10,10){$\bar{f}'_{j}$}}
\put(100,54){\makebox(10,10){$f_{i}$}}
\put(36,18){\usebox{\Vp}}
\put(0,36){\usebox{\Vr}}
\put(36,00){\usebox{\Ftbr}}
\put(36,18){\usebox{\Sbr}}
\put(36,36){\usebox{\Vtr}}
\end{picture}} 
\\
\savebox{\Ftbr}(72,72)[bl] 
{\put(36,18){\usebox{\Ft}}
\put(36,54){\usebox{\Ftr}} \put(36,00){\usebox{\Fbr}}}
\makebox{
\begin{picture}(135,90)
\put(-40,60){\makebox(10,10){$e)$}}
\put(5,40){\makebox(10,10){$W$}}
\put(77,32){\makebox(20,10){$f_{i}$}}
\put(43,54){\makebox(20,10){$H,\chi$}}
\put(43,8){\makebox(20,10){$\phi$}}
\put(100,8){\makebox(10,10){$\bar{f}'_{j}$}}
\put(100,54){\makebox(10,10){$f_{i}$}}
\put(36,18){\usebox{\Vp}}
\put(0,36){\usebox{\Vr}}
\put(36,00){\usebox{\Ftbr}}
\put(36,18){\usebox{\Stbr}}
\end{picture}} 
\makebox{
\begin{picture}(135,90)
\put(5,40){\makebox(10,10){$W$}}
\put(77,32){\makebox(20,10){$f'_{j}$}}
\put(43,54){\makebox(20,10){$\phi$}}
\put(43,8){\makebox(20,10){$H,\chi$}}
\put(100,8){\makebox(10,10){$\bar{f}'_{j}$}}
\put(100,54){\makebox(10,10){$f_{i}$}}
\put(36,18){\usebox{\Vp}}
\put(0,36){\usebox{\Vr}}
\put(36,00){\usebox{\Ftbr}}
\put(36,18){\usebox{\Stbr}}
\end{picture}}
\\
\savebox{\Ftbr}(72,72)[bl] 
{\put(36,18){\usebox{\Ft}}
\put(36,54){\usebox{\Ftr}} \put(36,00){\usebox{\Fbr}}}
\makebox{
\begin{picture}(135,90)
\put(-40,60){\makebox(10,10){$f)$}}
\put(5,40){\makebox(10,10){$W$}}
\put(77,32){\makebox(20,10){$f_{i}$}}
\put(43,54){\makebox(20,10){$\gamma,Z$}}
\put(43,8){\makebox(20,10){$\phi$}}
\put(100,8){\makebox(10,10){$\bar{f}'_{j}$}}
\put(100,54){\makebox(10,10){$f_{i}$}}
\put(36,18){\usebox{\Vp}}
\put(0,36){\usebox{\Vr}}
\put(36,00){\usebox{\Ftbr}}
\put(36,18){\usebox{\Sbr}}
\put(36,36){\usebox{\Vtr}}
\end{picture}} 
\makebox{
\begin{picture}(135,90)
\put(5,40){\makebox(10,10){$W$}}
\put(77,32){\makebox(20,10){$f'_{j}$}}
\put(43,54){\makebox(20,10){$\phi$}}
\put(43,8){\makebox(20,10){$\gamma,Z$}}
\put(100,8){\makebox(10,10){$\bar{f}'_{j}$}}
\put(100,54){\makebox(10,10){$f_{i}$}}
\put(36,18){\usebox{\Vp}}
\put(0,36){\usebox{\Vr}}
\put(36,00){\usebox{\Ftbr}}
\put(36,18){\usebox{\Vbr}}
\put(36,36){\usebox{\Str}}
\end{picture}}
\earr
\ema
\caption{One-loop diagrams for $W \to f_{i}\bar{f}'_{j}$.}
\label{9.1}
\efi
 
\savebox{\Vr}{}
\savebox{\Vt}{}
\savebox{\Vtr}{}
\savebox{\Vbr}{}
\savebox{\Vtbr}{}
\savebox{\Sr}{}
\savebox{\St}{}
\savebox{\Str}{}
\savebox{\Sbr}{}
\savebox{\Stbr}{}
\savebox{\Fr}{}
\savebox{\Ft}{}
\savebox{\Ftr}{}
\savebox{\Fbr}{}
\savebox{\Ftbr}{}
\savebox{\Vp}{}
 
Fig.~\reff{9.1} shows the Feynman diagrams contributing to the
vertex corrections for massive fermions. They yield the following
vertex form factors
\beqar
\nn\lefteqn{ \delta F_1^{-}(M_{W},m_{1},m_{2}) =
\frac{\alpha}{4\pi} \times} \\
\nn&&\Biggl\{ Q_{f} Q_{f'}
{\cal V}_{a} (m_{1}^2,M_W^2,m_{2}^2,\lambda,m_{1},m_{2})
+ g_{f}^- g_{f'}^{-} {\cal V}_{a} (m_{1}^2,M_W^2,m_{2}^2,M_Z,m_{1},m_{2})
 \\[1em]
\nn && \mbox{}+ \sum_{\sigma =\pm}
 \biggl[ Q_{f} {\cal V}^{\sigma}_{b}(m_{1}^2,M_W^2,m_{2}^2,m_{1},\lambda,M_W)
- Q_{f'}  {\cal V}^{\sigma}_{b}(m_{2}^2,M_W^2,m_{1}^2,m_{2},\lambda,M_W)
\\[1em]
\nn &&\qquad \mbox{}+ \frac{c_{W}}{s_{W}}  g_{f}^{\sigma} {\cal V}^{\sigma}_{b}
(m_{1}^2,M_W^2,m_{2}^2,m_{1},M_Z,M_W)
- \frac{c_{W}}{s_{W}} g_{f'}^{\sigma}
{\cal V}^{\sigma}_{b} (m_{2}^2,M_W^2,m_{1}^2,m_{2},M_Z,M_W)
        \biggl] \\[1em]
\nn&& \mbox{}+ \frac{1}{4 s_{W}^2} 
{\cal V}_c (m_{1}^2,M_W^2,m_{2}^2,M_H,m_{1},m_{2}) \\[1em]
&& \mbox{}+ \frac{1}{2s_{W}^2}  \biggl[
{\cal V}_{d} (m_{1}^2,M_W^2,m_{2}^2,m_{1},M_H,M_W) + {\cal V}_{d}
(m_{2}^2,M_W^2,m_{1}^2,m_{2},M_H,M_W)
\biggl]  \\[1em]
\nn&& \mbox{}+ \frac{1}{2s_{W}^2} 
\biggl[ {\cal V}_{e} (m_{1}^2,M_W^2,m_{2}^2,m_{1},M_H,M_W)
+ {\cal V}_{e} (m_{2}^2,M_W^2,m_{1}^2,m_{2},M_H,M_W)
\\[1em]
\nn && \qquad
\mbox{}+ {\cal V}_{e} (m_{1}^2,M_W^2,m_{2}^2,m_{1},M_Z,M_W)
+ {\cal V}_{e} (m_{2}^2,M_W^2,m_{1}^2,m_{2},M_Z,M_W) \biggl]
\\[1em]
\nn&& \mbox{}+ \sum_{\sigma =\pm} \biggl[Q_{f} {\cal V}^{\sigma}_{f}
(m_{1}^2,M_W^2,m_{2}^2,m_{1},\lambda,M_W)
- Q_{f'} {\cal V}^{\sigma}_{f} (m_{2}^2,M_W^2,m_{1}^2,m_{2},\lambda,M_W)
\\[1em]
\nn&&\qquad \mbox{}- \frac{s_{W}}{c_{W}}  g_{f}^{\sigma} {\cal V}^{\sigma}_{f}
(m_{1}^2,M_W^2,m_{2}^2,m_{1},M_Z,M_W)
+ \frac{s_{W}}{c_{W}}  g_{f'}^{\sigma}
{\cal V}^{\sigma}_{f} (m_{2}^2,M_W^2,m_{1}^2,m_{2},M_Z,M_W)
\biggl]  \Biggl\} \: ,
\eeqar
\beqar
\nn\lefteqn{\delta F_1^{+}(M_{W},m_{1},m_{2})  =
\frac{\alpha}{4\pi}  \; m_{1}m_{2} \;\times} \\
\nn&& \Biggl\{ \sum_{\sigma =\pm} \biggl[ Q_{f} Q_{f'} {\cal W}^{\sigma}_{a}
(m_{1}^2,M_W^2,m_{2}^2,\lambda,m_{1},m_{2})
+  g_{f}^{\sigma} g_{f'}^{\sigma} {\cal W}^{\sigma}_{a}
(m_{1}^2,M_W^2,m_{2}^2,M_Z,m_{1},m_{2})
\\[1em]
\nn&&\qquad \mbox{}
+  Q_{f}{\cal W}^{\sigma}_{b}(m_{1}^2,M_W^2,m_{2}^2,m_{1},\lambda,M_W)
- Q_{f'}  {\cal W}^{\sigma}_{b}(m_{2}^2,M_W^2,m_{1}^2,m_{2},\lambda,M_W)
     \\[1em]
\nn&&\qquad \mbox{}+ \frac{c_{W}}{s_{W}} g_{f}^{\sigma} {\cal W}^{\sigma}_{b}
(m_{1}^2,M_W^2,m_{2}^2,m_{1},M_Z,M_W)
- \frac{c_{W}}{s_{W}}  g_{f'}^{\sigma} {\cal W}^{\sigma}_{b}
(m_{2}^2,M_W^2,m_{1}^2,m_{2},M_Z,M_W)
    \biggl] \\[1em]
\nn && \mbox{}+ \frac{1}{4 s_{W}^2} \biggl[ \sum_{\sigma =\pm}
{\cal W}^{\sigma}_c (m_{1}^2,M_W^2,m_{2}^2,M_H,m_{1},m_{2})
- {\cal W}_c^- (m_{1}^2,M_W^2,m_{2}^2,M_Z,m_{1},m_{2}) \biggl] \\[1em]
&& \mbox{}+ \frac{1}{2s_{W}^2}  \biggl[
{\cal W}_{d} (m_{1}^2,M_W^2,m_{2}^2,m_{1},M_H,M_W)
+ {\cal W}_{d} (m_{2}^2,M_W^2,m_{1}^2,m_{2},M_H,M_W) \biggl]  \\[1em]
\nn&& \mbox{}+ \frac{1}{2s_{W}^2} \biggl[
{\cal W}_{e} (m_{1}^2,M_W^2,m_{2}^2,m_{1},M_H,M_W)
+ {\cal W}_{e} (m_{2}^2,M_W^2,m_{1}^2,m_{2},M_H,M_W)
    \\[1em]
\nn &&\qquad
\mbox{}- {\cal W}_{e} (m_{1}^2,M_W^2,m_{2}^2,m_{1},M_Z,M_W)
- {\cal W}_{e} (m_{2}^2,M_W^2,m_{1}^2,m_{2},M_Z,M_W)
\biggl] \\[1em]
\nn && \mbox{}+ \sum_{\sigma =\pm} \biggl[ Q_{f} {\cal W}^{\sigma}_{f}
(m_{1}^2,M_W^2,m_{2}^2,m_{1},\lambda,M_W)
- Q_{f'} {\cal W}^{\sigma}_{f}(m_{2}^2,M_W^2,m_{1}^2,m_{2},\lambda,M_W)
\\[1em]
\nn&&\qquad \mbox{}- \frac{s_{W}}{c_{W}} g_{f}^{\sigma} {\cal W}^{\sigma}_{f}
(m_{1}^2,M_W^2,m_{2}^2,m_{1},M_Z,M_W)
+ \frac{s_{W}}{c_{W}} g_{f'}^{\sigma} {\cal W}^{\sigma}_{f}
(m_{2}^2,M_W^2,m_{1}^2,m_{2},M_Z,M_W)
 \biggl]   \Biggl\} ,
\eeqar
\beqar
\nn\lefteqn{\delta F_2^{-}(M_{W},m_{1},m_{2}) =
\frac{\alpha}{4\pi} \; m_{1} \times}\\
\nn&&\Biggl\{ \sum_{\sigma =\pm} \biggl[ Q_{f} Q_{f'} {\cal X}^{\sigma}_{a}
(m_{1}^2,M_W^2,m_{2}^2,\lambda,m_{1},m_{2})
+  g_{f}^{\sigma} g_{f'}^{-} {\cal X}^{\sigma}_{a}
(m_{1}^2,M_W^2,m_{2}^2,M_Z,m_{1},m_{2})\\[1em]
\nn&&\qquad \mbox{}
+ Q_{f} {\cal X}^{\sigma}_{b}(m_{1}^2,M_W^2,m_{2}^2,m_{1},\lambda,M_W)
+ \frac{c_{W}}{s_{W}} g_{f}^{\sigma} {\cal X}^{\sigma}_{b}
(m_{1}^2,M_W^2,m_{2}^2,m_{1},M_Z,M_W)\biggl] \\[1em]
\nn&& \mbox{}- Q_{f'} {\cal X}^{-}_{b}(m_{1}^2,M_W^2,m_{2}^2,m_{2},M_W,\lambda)
- \frac{c_{W}}{s_{W}} g_{f'}^{-}
  {\cal X}^-_{b} (m_{1}^2,M_W^2,m_{2}^2,m_{2},M_W,M_Z)   \\[1em]
\nn&&\mbox{}\mbox{}+ \frac{1}{4 s_{W}^2} \biggl[ \sum_{\sigma =\pm}
{\cal X}^-_c (m_{1}^2,M_W^2,m_{2}^2,M_H,m_{1},m_{2})
 -{\cal X}^-_c (m_{1}^2,M_W^2,m_{2}^2,M_Z,m_{1},m_{2}) \biggl]
\\[1em]
&& \mbox{}
+ \frac{1}{2s_{W}^2}  {\cal X}_{d} (m_{1}^2,M_W^2,m_{2}^2,m_{1},M_H,M_W)
  \\[1em]
\nn&& \mbox{}- \frac{1}{2s_{W}^2} \sum_{\sigma = \pm} \biggl[
 \sigma \Bigl( {\cal X}^{\sigma}_{e} (m_{1}^2,M_W^2,m_{2}^2,m_{1},M_H,M_W)
 - {\cal X}^{\sigma}_{e} (m_{2}^2,M_W^2,m_{1}^2,m_{2},M_H,M_W) \Bigl) \\[1em]
\nn && \qquad\mbox{}
-{\cal X}^{\sigma}_{e} (m_{1}^2,M_W^2,m_{2}^2,m_{1},M_Z,M_W)
 - {\cal X}^{\sigma}_{e} (m_{2}^2,M_W^2,m_{1}^2,m_{2},M_Z,M_W) \biggl] \\[1em]
\nn && \mbox{}- \frac{1}{2s_{W}^2} \frac{m_{2}^2}{M_W^2} \biggl[
 {\cal X}^0_{e} (m_{1}^2,M_W^2,m_{2}^2,m_{1},M_H,M_W)
 - {\cal X}^0_{e} (m_{2}^2,M_W^2,m_{1}^2,m_{2},M_H,M_W)  \\[1em]
\nn && \qquad \mbox{}- {\cal X}^0_{e} (m_{1}^2,M_W^2,m_{2}^2,m_{1},M_Z,M_W)
 - {\cal X}^0_{e} (m_{2}^2,M_W^2,m_{1}^2,m_{2},M_Z,M_W) 
\biggl] \\[1em]
\nn && \mbox{}+ Q_{f} {\cal X}_{f} (m_{1}^2,M_W^2,m_{2}^2,m_{1},\lambda,M_W)
 + Q_{f'}  {\cal X}_{f} (m_{2}^2,M_W^2,m_{1}^2,m_{2},\lambda,M_W)  \\[1em]
\nn&& \mbox{}- \frac{s_{W}}{c_{W}} \biggl[ g_{f}^{+} {\cal X}_{f}
(m_{1}^2,M_W^2,m_{2}^2,m_{1},M_Z,M_W) + g_{f'}^{-} {\cal X}_{f}
(m_{2}^2,M_W^2,m_{1}^2,m_{2},M_Z,M_W)
  \biggl]  \Biggl\}.
\eeqar
To obtain these results we had to evaluate six generic diagrams labeled
by a, b, c, d, e and f in Fig.~\ref{9.1}. The
corresponding invariant functions ${\cal V}$, ${\cal W}$ and
${\cal X}$ are listed in App.~C.
The formfactor $\delta F_2^{+}$ can be obtained
from $\delta F_2^{-}$ by the substitutions
$m_{1} \leftrightarrow m_{2}$, $Q_{f} \leftrightarrow -Q_{f'}$,
$g_{f} \leftrightarrow -g_{f'}$.
 
Squaring
the matrix element (\ref{M1Wff}), summing over the polarizations of the
external particles and multiplying the phase space factor yields the
one-loop corrected width
\beqar \label{G1Wff}
\nn \Gamma ^{Wf_{i}f'_{j}}_{1}&=&N_{C}^{f}\frac{\alpha}{12}\frac{1}{2s_{W}^{2}}
\frac{\kappa(M_{W}^{2},m_{f,i}^{2},m_{f',j}^{2})}{M^{3}_{W}}\\[1em]
\nn &&\Bigl\{|{V_{ij}}|^{2}G^{-}_{1}[1
+2 \delta Z_{e}-2\frac{\delta
s_{W}}{s_{W}}+\delta Z_{W}]\\[1em]
\nn &&\mbox{}+\frac{1}{2}G^{-}_{1}\sum_{k}[(\delta Z^{f,L\dagger}_{ik}
+\delta Z^{f,L}_{ik})
V_{kj}+V_{ik}(\delta Z^{f',L{\dagger}}_{kj}+\delta
Z^{f',L}_{kj})]V^\dagger_{ij}\\[1em]
&&\mbox{}+2|{V_{ij}}|^{2}\sum_{a=1}^{2}\sum_{\sigma =\pm}
G^{\sigma}_{a}\delta F^{\sigma}_{a}(M_{W},m_{f,i},m_{f',j})\\[1em]
\nn&=& \Gamma ^{Wf_{i}f'_{j}}_{0}(1+\delta _{virt}^{ew}),
\eeqar
where
\beqar
G^{+}_{1}&=&\sum_{pol}{\cal M}^{-{\dagger}}_{1}{\cal M}
^{+}_{1}=\phantom{-}6\,m_{f,i}m_{f',j},\\
\nn G^{-}_{2}&=&\sum_{pol}{\cal M}^{-{\dagger}}_{1}{\cal M}^{-}_{2}=-
\frac{m_{f,i}}{2}
\frac{\kappa^{2}(M_{W}^{2},m_{f,i}^{2},m_{f',j}^{2})}{M_{W}^{2}}
,\\
\nn G^{+}_{2}&=&\sum_{pol}{\cal M}^{-{\dagger}}_{1}{\cal M}^{+}_{2}=-
\frac{m_{f',j}}{2}
\frac{\kappa^{2}(M_{W}^{2},m_{f,i}^{2},m_{f',j}^{2})}{M_{W}^{2}}
,
\eeqar
and we have inserted $\delta V_{ij}$ from (\ref{DV}).
 
\section{Photon bremsstrahlung}
 
\bfi
\savebox{\Vr}(36,0)[bl]
{\multiput(3,0)(12,0){3}{\oval(6,4)[t]}
\multiput(9,0)(12,0){3}{\oval(6,4)[b]} }
\savebox{\Vtr}(36,18)[bl]
{\multiput(4,0)(8,4){5}{\oval(8,4)[tl]}
\multiput(4,4)(8,4){4}{\oval(8,4)[br]} }
\savebox{\Vbr}(36,18)[bl]
{\multiput(4,18)(8,-4){5}{\oval(8,4)[bl]}
\multiput(4,14)(8,-4){4}{\oval(8,4)[tr]}}
\savebox{\Sr}(36,0)[bl]
{ \multiput(0,0)(13,0){3}{\line(4,0){10}} }
\savebox{\Ftr}(36,18)[bl]
{ \put(0,0){\vector(2,1){20}} \put(18,9){\line(2,1){18}} }
\savebox{\Fbr}(36,18)[bl]
{ \put(36,0){\vector(-2,1){21}} \put(18,9){\line(-2,1){18}} }
\savebox{\Vtr}(32,24)[bl]
{\multiput(4,0)(8,6){4}{\oval(8,6)[tl]}
\multiput(4,6)(8,6){4}{\oval(8,6)[br]} }
\savebox{\Ftbr}(72,72)[bl]
{\put(00,36){\usebox{\Ftr}} \put(00,18){\usebox{\Fbr}}}
\savebox{\Vp}(36,36)[bl]
{ \put(0,18){\circle*{4}} \put(36,18){\circle*{4}} }
\bma
\barr{l}
\makebox{
\begin{picture}(160,100)
\put(5,40){\makebox(10,10){$W$}}
\put(65,60){\makebox(20,10){$\gamma $}}
\put(45,20){\makebox(20,10){$W$}}
\put(100,5){\makebox(10,10){$f'_{j}$}}
\put(100,59){\makebox(10,10){$f_{i}$}}
\put(36,18){\usebox{\Vp}}
\put(0,36){\usebox{\Vr}}
\put(72,00){\usebox{\Ftbr}}
\put(36,36){\usebox{\Vtr}}
\put(36,36){\usebox{\Vr}}
\end{picture}}
\makebox{
\begin{picture}(160,100)
\put(5,40){\makebox(10,10){$W$}}
\put(65,60){\makebox(20,10){$\gamma $}}
\put(45,20){\makebox(20,10){$\phi$}}
\put(100,5){\makebox(10,10){$f'_{j}$}}
\put(100,59){\makebox(10,10){$f_{i}$}}
\put(36,18){\usebox{\Vp}}
\put(0,36){\usebox{\Vr}}
\put(72,00){\usebox{\Ftbr}}
\put(36,36){\usebox{\Vtr}}
\put(36,36){\usebox{\Sr}}
\end{picture}}
\\
\savebox{\Vtr}(36,18)[bl]
{\multiput(4,0)(8,4){5}{\oval(8,4)[tl]}
\multiput(4,4)(8,4){4}{\oval(8,4)[br]} }
\savebox{\Ftbr}(72,72)[bl]
{ \put(0,36){\vector(2,1){38}} \put(36,54){\line(2,1){36}}
 \put(72,0){\vector(-2,1){39}} \put(36,18){\line(-2,1){36}} }
\savebox{\Vp}(36,36)[bl]
{ \put(0,18){\circle*{4}} \put(36,0){\circle*{4}} }
\makebox{
\begin{picture}(160,100)
\put(5,40){\makebox(10,10){$W$}}
\put(43,8){\makebox(20,10){$f'_{j}$}}
\put(77,32){\makebox(20,10){$\gamma$}}
\put(100,10){\makebox(10,10){$\bar{f}'_{j}$}}
\put(100,54){\makebox(10,10){$f_{i}$}}
\put(36,36){\circle*{4}} \put(72,18){\circle*{4}}
\put(0,36){\usebox{\Vr}}
\put(36,00){\usebox{\Ftbr}}
\put(72,18){\usebox{\Vtr}}
\put(52,28){\vector(-2,1){3}} \put(88,10){\vector(-2,1){3}}
\end{picture}}
\makebox{
\begin{picture}(160,100)
\put(5,40){\makebox(10,10){$W$}}
\put(43,54){\makebox(20,10){$f_{i}$}}
\put(77,32){\makebox(20,10){$\gamma$}}
\put(100,10){\makebox(10,10){$\bar{f}'_{j}$}}
\put(100,54){\makebox(10,10){$f_{i}$}}
\put(36,36){\circle*{4}} \put(72,54){\circle*{4}}
\put(0,36){\usebox{\Vr}}
\put(36,00){\usebox{\Ftbr}}
\put(72,36){\usebox{\Vbr}}
\put(56,46){\vector(2,1){3}} \put(92,64){\vector(2,1){3}}
\end{picture}}
\earr
\ema
\caption{Bremsstrahlung Feynman diagrams for $W \to f_{i}\bar{f}'_{j}\gamma$.}
\label{9.2}
\efi
\savebox{\Vr}{}
\savebox{\Vt}{}
\savebox{\Vtr}{}
\savebox{\Vbr}{}
\savebox{\Vtbr}{}
\savebox{\Sr}{}
\savebox{\St}{}
\savebox{\Str}{}
\savebox{\Sbr}{}
\savebox{\Stbr}{}
\savebox{\Fr}{}
\savebox{\Ft}{}
\savebox{\Ftr}{}
\savebox{\Fbr}{}
\savebox{\Ftbr}{}
\savebox{\Vp}{}
Like any one-loop amplitude with external charged particles (\ref{M1Wff})
and consequently (\ref {G1Wff}) are IR-divergent due to virtual
photonic corrections. These singularities are compensated by the
the real bremsstrahlung corrections, i.e.\ the three-body decay
\beq
W^{+}(k)\to f_{i}(p_{1})\bar{f}'_{j}(p_{2})\gamma (q) .
\eeq
The corresponding matrix element as given by the Feynman diagrams
(Fig.~\reff{9.2}) is
\beq
\begin{array}{ll}
{\cal M}_b = & \disp V_{ij}
\frac{e^{2}}{\sqrt{2}s_{W}} \bar{u}(p_1)
\biggl\{ \frac{-Q_{f}}{2p_{1}q} \Bigl[ 2 p_1\eta \;
\rlap{/} \epsilon + \rlap{/} \eta \rlap{/}q \rlap{/}\epsilon \Bigl]
+ \frac{Q_{f'}}{2p_{2}q} \Bigl[ 2 p_2\eta \;
\rlap{/} \epsilon + \rlap{/} \epsilon \rlap{/}q \rlap{/}\eta \Bigl] \\[1em]
& \displaystyle \hspace{5mm}
+ \frac{(Q_{f}-Q_{f'})}{-2kq} \Bigl[ (q\eta-2k\eta) \;
\rlap{/} \epsilon + 2 \epsilon\eta \; \rlap{/} q
-2 q \epsilon \; \rlap{/} \eta \Bigl] \biggl\} \omega_- v(p_2),
\end{array}
\label{matb}
\eeq
where $\eta$ denotes the polarization vector of the photon. Performing the
polarization sum over the square of the amplitude gives
\begin{eqnarray} \label{M2Wffg}
\sum_{pol} \bigl| {\cal M}_b \bigl|^2  & = &
\frac{\alpha^{2}}{2s_{W}^2}  (64 \pi^{2}) \left|V_{ij} \right|^2
 \Biggl\{ \frac{Q_{f} Q_{f'}}{(2p_{1}q)\,(2p_{2}q)}
 \biggl[  (M_W^2-m_{f,i}^2-m_{f',j}^2)  G_{1}^{-}
 \biggl]  \nonumber  \\[1em]
 &  & - \frac{Q_{f}^2}{(2p_{1}q)^2}\biggl[ (m_{f,i}^2 +2p_{1}q) G_{1}^{-}
+(1+\frac{m_{f,i}^2+m_{f',j}^2}{2M_W^2})
(2p_{1}q)\,(-2kq) + (2p_{1}q)^2  \biggl]
\nonumber \\[1em]
\nn &  & - \frac{Q_{f'}^2}{(2p_{2}q)^2}
\biggl[  (m_{f',j}^2 +2p_{2}q) G_{1}^{-}
+(1+\frac{m_{f,i}^2+m_{f',j}^2}{2M_W^2}) (2p_{2}q)\,(-2kq)
+ (2p_{2}q)^2  \biggl]
 \label{matsqare}  \\[1em]
 &  & - \frac{(Q_{f}-Q_{f'})^2}{(-2kq)^2} \biggl[  (M_W^2 -2kq) G_{1}^{-}
+\frac{m_{f,i}^2+m_{f',j}^2}{2M_W^2} (-2kq)^2 -2(2p_{1}q)(2p_{2}q)\biggl]
 \nonumber \\[1em]
 &  & - \frac{Q_{f}(Q_{f}-Q_{f'})}{-2kq\,2p_{1}q}
\biggl[ (M_W^2+m_{f,i}^2-m_{f',j}^2) G_{1}^{-} - 2(2p_{1}q)(2p_{2}q)
  \biggl]   \\[1em]
 &  & + \frac{Q_{f'}(Q_{f}-Q_{f'})}{-2kq\,2p_{2}q}
\biggl[ (M_W^2-m_{f,i}^2+m_{f',j}^2) G_{1}^{-} - 2(2p_{1}q)(2p_{2}q)
  \biggl]   \Biggl\} \; .  \nonumber
\end{eqnarray}
From this the complete bremsstrahlung cross section (including soft and
hard photons) is obtained by integrating
over the phase space of the photon and the two fermions as
\beqar
\nn \Gamma_b^{Wf_{i}f'_{j}}(M_{W},m_{f,i},m_{f',j})&=&
\frac{1}{(2\pi)^5} \frac{N_C^{f}}{2M_W} \int
\frac{d^3 q}{2 q_{0}} \frac{d^3 p_1}{2 p_{10}} \frac{d^3 p_2}{2 p_{20}}
\delta^{(4)} (p_1+p_2+q-k) \frac{1}{3} \sum_{pol} \bigl| {\cal M}_b
\bigl|^2  \\[1em]
&=&\Gamma_{0}^{Wf_{i}f'_{j}} \delta _{b}^{ew}(M_{W},m_{f,i},m_{f',j})
\eeqar
with
\begin{eqnarray}
\nn\lefteqn{\delta_b ^{ew}(M_{W},m_{f,i},m_{f',j}) =} \\[1em]
& & \left( -\frac{\alpha}{\pi} \right)
\frac{4M_W^2}{\kappa(M_{W}^{2},m_{f,i}^{2},m_{f',j}^{2})}
\Biggl\{ - Q_{f} Q_{f'}  \biggl[  (M_W^2-m_{f,i}^2-m_{f',j}^2)  I_{12}
 \biggl]  \nonumber  \\[1em]
 &  & \displaystyle + Q_{f}^2 \biggl[ (m_{f,i}^2 I_{11}+I_1)
+(1+\frac{m_{f,i}^2+m_{f',j}^2}{2M_W^2}) \frac{I_1^0}{G_{1}^{-}}
+\frac{I}{G_{1}^{-}}  \biggl]
\nonumber \\[1em]
\nn  &  & \displaystyle
+ Q_{f'}^2 \biggl[  (m_{f',j}^2 I_{22}+I_2)
+(1+\frac{m_{f,i}^2+m_{f',j}^2}{2M_W^2}) \frac{I_2^0}{G_{1}^{-}}
+\frac{I}{G_{1}^{-}}  \biggl]
 \label{delb}  \\[1em]
 &  & \displaystyle
+ (Q_{f}-Q_{f'})^2 \biggl[  (M_W^2 I_{00}+I_0)
+\frac{m_{f,i}^2+m_{f',j}^2}{2M_W^2} \frac{I}{G_{1}^{-}}
-2\frac{I_{00}^{12}}{G_{1}^{-}}  \biggl]
 \nonumber \\[1em]
 &  & \displaystyle
+ Q_{f}(Q_{f}-Q_{f'}) \biggl[  (M_W^2+m_{f,i}^2-m_{f',j}^2) I_{01}
- 2 \frac{I_0^2}{G_{1}^{-}}
  \biggl]   \nonumber  \\[1em]
 &  & \displaystyle
- Q_{f'}(Q_{f}-Q_{f'}) \biggl[ (M_W^2-m_{f,i}^2+m_{f',j}^2) I_{02}
- 2 \frac{I_0^1}{G_{1}^{-}}
  \biggl]   \Biggl\}.
\end{eqnarray}
The bremsstrahlung phase space integrals $I_{\cdots}^{\cdots} =
I_{\cdots}^{\cdots}(M_{W},m_{f,i},m_{f',j})$
are given in App.~D. The IR-singularities contained in the $I_{kl}$
are again regularized by a photon mass $\lambda $.
 
\section{QCD corrections}
\label{secGWffQCD}
 
Like the electroweak corrections also the QCD corrections consist of
virtual and real contributions
\begin{equation}
\delta ^{QCD} (M_{W},m_{f,i},m_{f',j}) =  \delta _{virt}^{QCD} + \delta
_b^{QCD},
\end{equation}
which are individually IR-divergent.
They are obtained from the electroweak results of the previous sections
by keeping only the terms containing $Q_{f}^{2}$, $Q_{f'}^{2}$ or
$Q_{f}Q_{f'}$, setting $Q_{f}=Q_{f'}=1$,
replacing $\alpha $ by the strong coupling constant $\alpha _s$
and multiplying an overall colour factor $C_F = \frac{4}{3}$.
In particular the virtual QCD-corrections arise only from the diagram of
Fig.~\reff{9.1}a with the photon replaced by a gluon and the
corresponding corrections to the fermion wave function renormalization
constants. Since this allows many simplifications we give the explicit results
\beqar
\nn\delta ^{QCD}_{virt} &= & \left( -\frac{\alpha_s}{4\pi} \right)
\; 2 \, C_F \; \Biggl\{
6 - B_{0}( M_W^2,m_{f,i},m_{f',j} ) \\[1em]
\nn&& + \frac{1}{2}B_{0}(0,m_{f,i},m_{f,i})
+\frac{1}{2}B_{0}(0,m_{f',j},m_{f',j})    \\[1em]
&& + 2 (M_W^2 -m_{f,i}^2 - m_{f',j}^2 ) ( C_0 +C_{1} + C_{2} )
-2 m_{f,i}^2 C_{1} -2 m_{f',j}^2 C_{2}   \\[1em]
\nn&& -2 \log ( \frac{m_{f,i}m_{f',j}}{\lambda^2} )
 - 2 \frac{G_1^{+}}{G_1^{-}} m_{f,i}m_{f',j}\Bigl[ C_{1} +C_{2} \Bigl] \\[1em]
\nn&& +4 \frac{G_2^{-}}{G_1^{-}} m_{f,i}\Bigl[ C_{11}+C_{12} +C_{1} \Bigl]
+ 4 \frac{G_2^{+}}{G_1^{-}} m_{f',j}\Bigl[ C_{22}+C_{12} +C_{2} \Bigl]
 \Biggl\}  \; .
\eeqar
The arguments of the three-point functions are
$C = C(m_{f,i}^2,M_W^2,m_{f',j}^2,\lambda,m_{f,i},m_{f',j})$.
To the gluonic bremsstrahlung $\delta^{QCD}_b$ only the first three terms
of (\ref{delb}) contribute.
For zero fermion masses the total QCD-correction reduces to
\begin{equation}
\delta^{QCD} (M_{W},0,0) = \frac{\alpha_s}{\pi}.
\end{equation}
 
\section{Results and Discussion}
\label{secWffres}
 
For numerical evaluation of the previous results we use the parameters
listed in Sect.~\ref{secpar} including the values for the quark mixing
matrix as given by (\ref{sqmm}). The $W$-mass is determined from the
relation (\ref{GFloop}).
 
In the on-shell scheme the lowest order width is parametrized by
$\alpha$ and the particle masses (\ref{G0Wff}).
In this scheme large electroweak corrections arise due to fermion
loop contributions to the renormalization of $\alpha $ and $s_{W}$.
We can improve the results
in this scheme by resumming the corresponding one-loop contributions to
all orders as discussed in Sect.~\ref{sechicor}.
Thus we obtain for the corrected width
\begin{equation} \label{Gsummed}
\barr{lll}
\Gamma &=& \disp\Gamma_0 \Bigl[ 1+ \delta_1
- (\Delta \alpha)_{ferm} + \frac{c_{W}^{2}}{s_{W}^{2}}\Delta \rho \Bigl]
 \; \frac{1}{1- (\Delta \alpha)_{ferm} }
\frac{1}{1+\frac{c_{W}^{2}}{s_{W}^{2}}\Delta \bar{\rho} } \\[1em]
& =& \Gamma_0 \Bigl[ 1+ \delta \Bigl],
\earr
\end{equation}
where
\beq
\delta _{1}= \delta_{virt}^{ew} + \delta_{b}^{ew} +
\delta_{virt}^{QCD} + \delta_{b}^{QCD}
\eeq
is the proper one-loop correction without resummation.
As in any charged current process
we can avoid the large corrections by parametrizing the lowest order decay
width with $G_F$ and $M_W$ instead of $\alpha $ and $s_{W}^2$ (\ref{GGWff}).
Using (\ref{GFloop}) we find the relation between the decay width in both
parametrizations
\beqar \label{GFpar}
\bar{\Gamma}_0 &=& \Gamma_0
\frac{1}{1- (\Delta \alpha)_{ferm}}
\frac{1}{1+\frac{c_{W}^{2}}{s_{W}^{2}}\Delta \bar{\rho} }
\left[1+\Delta r - (\Delta \alpha)_{ferm}
+\frac{c_{W}^{2}}{s_{W}^{2}}\Delta \rho \right], \\[1em]
\nn \bar{\Gamma} &=& \bar{\Gamma}_0 \Bigl[ 1 + \delta_1-\Delta r\Bigl] \;
= \bar{\Gamma}_{0}(1+\bar{\delta}).
\eeqar
The large fermionic contributions contained in $\delta _{1}$ are exactly
cancelled by equal contributions in $\Delta r$ and consequently the
remaining corrections $\bar{\delta}$ are small.
 
\bfi  \label{9.3}
\includegraphics[bb=20 20 592 402, width=16.cm]{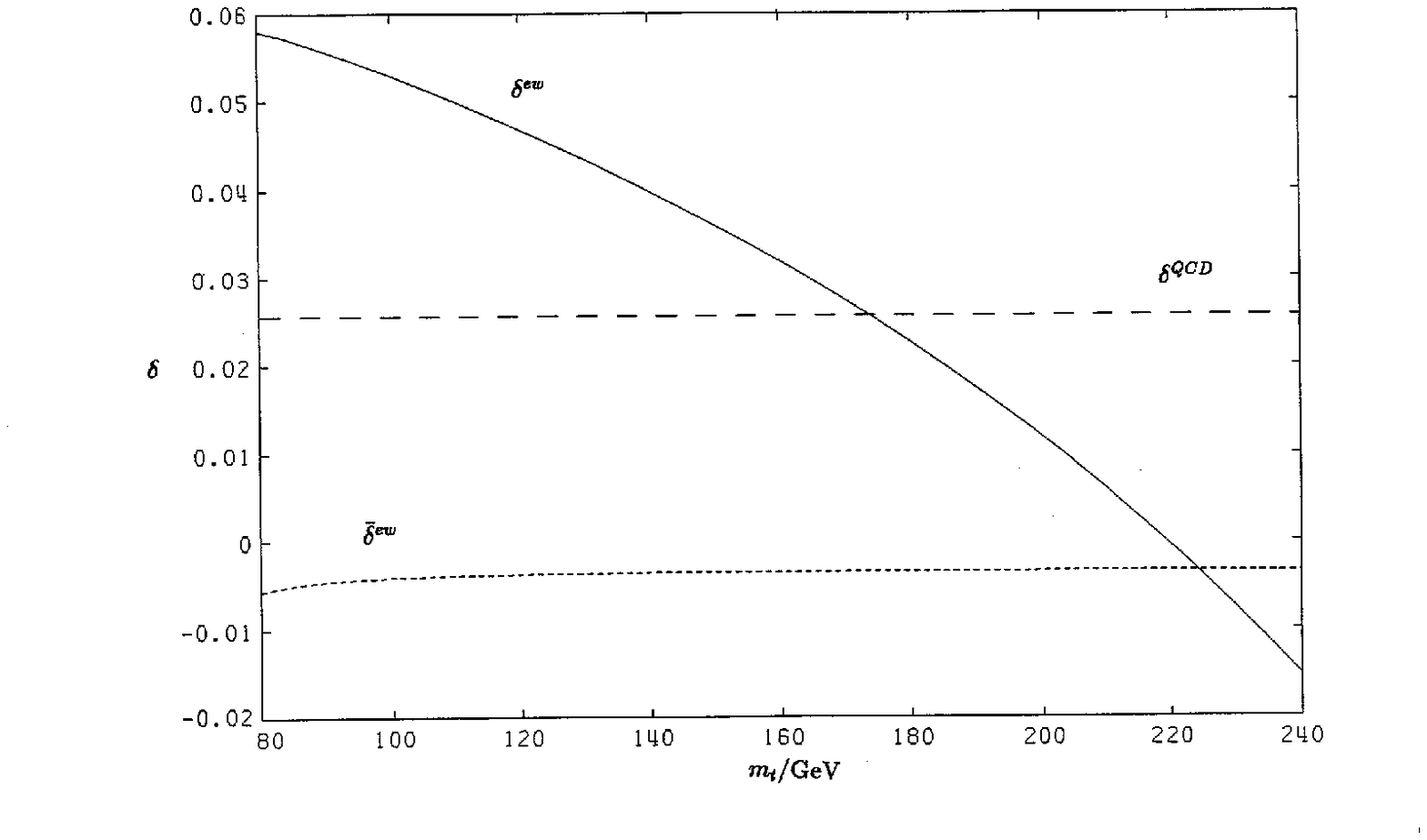}
\caption{Electroweak radiative corrections $\delta^{ew} $ and $\bar{\delta
}^{ew}$ and QCD-corrections $\delta^{QCD}$ to the total $W$-width
$\Gamma^{W}$ versus the top mass.}
\efi
The relative corrections to the total $W$-boson decay width
are shown in Fig.~\reff{9.3}.
They are large and strongly $m_{t}$-dependent in the on-shell scheme.
This behaviour arises from the fermionic contributions to the
renormalization of the weak mixing angle in (\ref{M1Wff}) ($\delta
s_{W}^{2}/s_{W}^{2}$) which contain terms $\propto \alpha\,
m_{t}^{2}/M_{W}^{2}$.
In contrast to this in the parametrization with $G_F$ the corrections
depend only weakly on $m_{t}$ and remain below 0.6\%.
The QCD corrections are practically constant and equal to
$2\alpha_{s}/(3\pi)$, their value for zero fermion
masses.
 
\begin{table}
\bma
\begin{array}{||c|c|c|c|c|c||} \hline\hline
m_t & M_W & \Gamma_0^{W} & \bar{\Gamma}_0^{W} & \Gamma_{ew}^{W} &
\bar{\Gamma}_{ew}^{W}\\
\hline
\quad 80.0 \quad &  \quad 79.87 \quad & \quad 1.8778 \quad &
\quad 2.0056 \quad & \quad 1.9866 \quad & \quad 1.9940 \quad \\
 \hline
100.0 &  79.99 & 1.8996 & 2.0147 & 2.0002 & 2.0063 \\
 \hline
120.0 &  80.10 & 1.9211 & 2.0235 & 2.0110 & 2.0158 \\
 \hline
140.0 &  80.23 & 1.9447 & 2.0330 & 2.0221 & 2.0256 \\
 \hline
160.0 &  80.37 & 1.9713 & 2.0435 & 2.0340 & 2.0363 \\
 \hline
180.0 &  80.52 & 2.0018 & 2.0553 & 2.0469 & 2.0481 \\
 \hline
200.0 &  80.69 & 2.0368 & 2.0684 & 2.0608 & 2.0612 \\
 \hline
 \hline
\end{array}
\ema
\caption{W-decay width $\Gamma^{W}$ parametrized by $\alpha$ and
$\bar{\Gamma}^{W}$ parametrized by $G_F$ in lowest order and including
electroweak corrections. All values are given in $\mbox{GeV}$. \label{tab91}}
\end{table}
Tab.~\ref{tab91} shows the lowest order width $\Gamma_{0}$ and the
width including electroweak corrections $\Gamma_{ew}$
in both parametrizations for various values of the top mass.
The $W$-mass obtained from (\ref{GFloop}) is also listed there.
While the results for the lowest order width in the two parametrizations
differ by several percent, the deviation of the first order expressions
is always less than 0.4\%.
 
The analytical results were presented above for finite external fermion
masses
and with correct renormalization of the quark mixing matrix. However,
since the top quark is presumably heavier than the $W$-boson
\cite{CDF90} all
relevant actual fermion masses are small compared to the $W$-boson
mass. Therefore in addition to the  completely corrected
numerical results $\Gamma(M_{W},m_{f,i},m_{f',j})$ for finite fermion
masses we also give those for vanishing fermion masses
$\Gamma(M_{W},0,0)$ in Tab.~\ref{9.2}.
The analytical results for vanishing fermion masses were listed in
Sect.~\ref{secWff0}.
Finally we include an improved Born approximation consisting of
the Born widths with zero fermion masses parametrized by $G_{F}$
and multiplied by the QCD correction factor for zero quark masses
\beqar \label{GWapp}
\nn \Gamma^{W\nu l}_{imp}
&=&\frac{G_{F}M_{W}^{3}}{6\sqrt{2}\pi},\\[1em]
\Gamma^{Wud}_{imp} &=&\frac{G_{F}M_{W}^{3}}{2\sqrt{2}\pi}
\left\vert V_{ij}\right\vert^{2}
\left(1+\frac{\alpha_{s}}{\pi}\right),\\[1em]
\nn \Gamma^{W}_{imp} &=&\frac{3G_{F}M_{W}^{3}}{2\sqrt{2}\pi}
\left(1+\frac{2}{3}\frac{\alpha_{s}}{\pi}\right),
\eeqar
for the leptonic partial widths, the hadronic partial widhts and
for the total width, respectively.
 
The numerical values for the partial and total $W$-widths
in these different approximations are given in Tab.~\ref{tab92} assuming
a top quark
mass of $140\:\mbox{GeV}$ and a Higgs boson mass of $100\:\mbox{GeV}$.
The improved Born approximation (\ref{GWapp}) reproduces the exact results up
to 0.4\% (0.6\% for the decays into a b-quark).
The effects of the fermion masses are below 0.3\%.
They are suppressed by $m_{q}^{2}/s$.
There are no mass singularities since the width is obtained by integrating over
the full phase space of the final state particles
\cite{Ki62}. Consequently the exact numerical values of the masses of
the external fermions
masses are irrelevant\footnote{The values given in
Sect.~\ref{secpar} are not appropriate for the external quarks.
We only use them here
to demonstrate the numerical irrelevance of the fermion mass effects.}.
The branching ratios derived from (\ref{GWapp})
\beqar
\nn BR(W \to l\nu ) &=& \frac{1}{9(1+2\alpha_{s}/3\pi)}, \\[1em]
\nn BR(W \to \mbox{leptons}) &=& \frac{1}{3(1+2\alpha_{s}/3\pi)}, \\[1em]
\nn BR(W \to u_{i}d_{j}) &=& \frac{\vert
V_{ij}\vert^{2}(1+\alpha_{s}/\pi)}{3(1+2\alpha_{s}/3\pi)}, \\[1em]
BR(W \to \mbox{hadrons}) &=&
\frac{2(1+\alpha_{s}/\pi)}{3(1+2\alpha_{s}/3\pi)}
\eeqar
agree numerically within 0.1\% with those obtained from the full one
loop results. They depend only on $\alpha_{s}$ and $V_{ij}$.
\begin{table}
\begin{center}
\begin{tabular}{||l|l|l|l|l||l||} \hline\hline
& \multicolumn{1}{|c|}{Born}      & \multicolumn{1}{c|}{complete}
& \multicolumn{1}{c|}{$m_{f}=0$}
& \multicolumn{1}{c||}{improved}
& \multicolumn{1}{c||}{Branching} \\
& \multicolumn{1}{|c|}{width}     & \multicolumn{1}{c|}{one-loop}
& \multicolumn{1}{c|}{in $\Gamma$}
& \multicolumn{1}{c||}{Born}
& \multicolumn{1}{c||}{ratio} \\
 \hline
 $\Gamma(W \rightarrow e \nu_{e})$
 \ &\  0.2260   \ &\  0.2252   \ &\  0.2252  \ &\  0.2260 \ &\  0.1084 \\
 \hline
 $\Gamma(W \rightarrow \mu\nu_{\mu})$
 \ &\  0.2259   \ &\  0.2252   \ &\  0.2252  \ &\  0.2260 \ &\  0.1084 \\
 \hline
 $\Gamma(W \rightarrow \tau \nu_{\tau})$
 \ &\  0.2258   \ &\  0.2250   \ &\  0.2252  \ &\  0.2260 \ &\  0.1083 \\
 \hline
 $\Gamma(W \rightarrow lep.)$
 \ &\  0.6777   \ &\  0.6754   \ &\  0.6756  \ &\  0.6778 \ &\  0.3251 \\
 \hline
 $\Gamma(W \rightarrow u d)$
 \ &\  0.6450   \ &\  0.6672   \ &\  0.6672  \ &\  0.6696 \ &\  0.3211 \\
 \hline
 $\Gamma(W \rightarrow u s)\times 10$
 \ &\  0.3281   \ &\  0.3393   \ &\  0.3393  \ &\  0.3406 \ &\  0.0163 \\
 \hline
 $\Gamma(W \rightarrow u b)\times 10^{4}$
 \ &\  0.3306   \ &\  0.3428  \ &\   0.3436  \ &\  0.3448 \ &\  0.00002 \\
 \hline
 $\Gamma(W \rightarrow c d)\times 10$
 \ &\  0.3281   \ &\  0.3396  \ &\   0.3396  \ &\  0.3409 \ &\  0.0163 \\
 \hline
 $\Gamma(W \rightarrow c s)$
 \ &\  0.6432   \ &\  0.6656   \ &\  0.6657  \ &\  0.6682 \ &\  0.3204 \\
 \hline
 $\Gamma(W \rightarrow c b)\times 10^{2}$
 \ &\  0.1427   \ &\  0.1480   \ &\  0.1484  \ &\  0.1489 \ &\  0.0007 \\
 \hline
 $\Gamma(W \rightarrow had.)$
 \ &\  1.3553   \ &\  1.4022   \ &   1.4023  \ &\  1.4075 \ &\  0.6749 \\
 \hline
 $\Gamma(W \rightarrow all)$
 \ &\  2.0330   \ &\  2.0776   \ &\  2.0779  \ &\  2.0853  & \\
 \hline \hline
\end{tabular}
\end{center}
\caption{Partial and total W-decay widths $\bar{\Gamma}$ in different
approximations
for $m_{t}= 140\:\mbox{GeV}$, $M_{H}=100\:\mbox{GeV}$
and the corresponding W-mass $M_{W}= 80.23\:\mbox{GeV}$.
\label{tab92}}
\end{table}
 
The dependence of the $W$-width on the unknown top and Higgs masses is shown in
Tab.~\ref{tab93} for $\Gamma ^{Wud}$ and in
Tab.~\ref{tab94} for $\Gamma ^{We\nu}$. A variation of $m_{t}$
between 80 and $200\:\mbox{GeV}$ affects the partial widths by $\sim
4\%$, a
variation of $M_{H}$ between 50 and $1000\;\mbox{GeV}$ by $\sim 1\%$.
This holds as well for the total width as shown in Tab.~\ref{tab95}.
All this is valid for constant $\alpha $, $G_{F}$ and $M_{Z}$.
In this case  the top mass dependence is mainly due to the variation of
$M_{W}$ with $m_{t}$. Keeping instead $M_{W}$,  $G_{F}$ and $M_{Z}$
fixed the dependence on $m_{t}$ is considerably smaller.
Remember, however, that the prediction for the decay
width has the same uncertainty in this parametrization due to the
uncertainty of the experimental value for the $W$-boson mass.
 
We have compared our results for the partial leptonic width for zero
fermion masses
to those of Jegerlehner \cite{Je86} and Bardin et al.\ \cite{Ba86},
who both use the parametrization with $G_F$.
Furthermore Jegerlehner includes two-loop QCD
corrections into the boson self energies.
If these are switched off the difference between
his and our results is less than $0.1\:\mbox{MeV}$.
Performing the same comparison with Bardin et al., our values for the
partial widths are $0.7\:\mbox{MeV}$ to $0.8\:\mbox{MeV}$ larger than theirs.
Our results for the QCD corrections agree with those obtained by Alvarez
et al.\ \cite{Al88}.
 
\begin{table}
\begin{center}
\begin{tabular}{||r|c|c|c|c||} \hline\hline
\multicolumn{1}{||c|}{$ m_t $} &  $M_H=50 $ &
 $ M_H=100 $ & $ M_H=300 $  & $ M_H=1000 $      \\
 \hline
 80.0 & 0.2212 & 0.2209 & 0.2202 & 0.2193 \\
 \hline
100.0 & 0.2227 & 0.2224 & 0.2217 & 0.2208 \\
 \hline
120.0 & 0.2239 & 0.2236 & 0.2229 & 0.2220 \\
 \hline
140.0 & 0.2251 & 0.2248 & 0.2241 & 0.2232 \\
 \hline
160.0 & 0.2264 & 0.2261 & 0.2254 & 0.2245 \\
 \hline
180.0 & 0.2278 & 0.2276 & 0.2269 & 0.2260 \\
 \hline
200.0 & 0.2294 & 0.2291 & 0.2284 & 0.2275 \\
 \hline\hline
\end{tabular}
\end{center}
\caption{Partial W-decay width $\Gamma^{We\nu}$ including first
order QCD and electroweak corrections for different values of the top 
and Higgs masses.                                                 
All values are given in $\mbox{GeV}$.\label{tab93}}
\end{table}
\begin{table} 
\begin{center}
\begin{tabular}{||r|c|c|c|c||} \hline\hline
\multicolumn{1}{||c|}{$ m_t $} &  $M_H=50 $ &
 $ M_H=100 $ & $ M_H=300 $  & $ M_H=1000 $      \\
 \hline
 80.0 & 0.6539 & 0.6530 & 0.6509 & 0.6481 \\
 \hline
100.0 & 0.6585 & 0.6575 & 0.6554 & 0.6526 \\
 \hline
120.0 & 0.6622 & 0.6612 & 0.6591 & 0.6563 \\
 \hline
140.0 & 0.6659 & 0.6650 & 0.6629 & 0.6601 \\
 \hline
160.0 & 0.6700 & 0.6691 & 0.6670 & 0.6642 \\
 \hline
180.0 & 0.6745 & 0.6736 & 0.6715 & 0.6686 \\
 \hline
200.0 & 0.6793 & 0.6784 & 0.6763 & 0.6735 \\
 \hline\hline
\end{tabular}
\end{center}
\caption{Partial W-decay width $\Gamma^{Wud}$ including first
order QCD and electroweak corrections for different values of the 
top and Higgs masses. All values are given in $\mbox{GeV}$.\label{tab94}}
\end{table}
\begin{table}
\begin{center}
\begin{tabular}{||r|c|c|c|c||} \hline\hline
\multicolumn{1}{||c|}{$ m_t $} &  $M_H=50 $ &
 $ M_H=100 $ & $ M_H=300 $  & $ M_H=1000 $      \\
 \hline
 80.0 & 2.0375 & 2.0347 & 2.0283 & 2.0196 \\
 \hline
100.0 & 2.0517 & 2.0488 & 2.0423 & 2.0336 \\
 \hline
120.0 & 2.0630 & 2.0602 & 2.0537 & 2.0449 \\
 \hline
140.0 & 2.0747 & 2.0719 & 2.0654 & 2.0566 \\
 \hline
160.0 & 2.0872 & 2.0844 & 2.0780 & 2.0692 \\
 \hline
180.0 & 2.1009 & 2.0981 & 2.0917 & 2.0830 \\
 \hline
200.0 & 2.1157 & 2.1129 & 2.1066 & 2.0980 \\
 \hline\hline
\end{tabular}
\end{center}
\caption{Total W-decay width $\Gamma^{W}$ including first
order QCD and electroweak corrections for different values of the 
top and Higgs masses. All values are given in $\mbox{GeV}$.\label{tab95}}
\end{table}

\chapter{The top width}
\label{chatWb}
 
The present lower limit from CDF data indicates
that the top mass is at least $89\:\mbox{GeV}$ \cite{CDF90}. Moreover,
LEP data in  combination with
radiative correction calculations require $m_t = 137 \pm 40 \:\mbox{GeV}$
within the minimal standard model \cite{Dy90,El90} at the $1\sigma$
level.
Therefore the top mass lies presumably above the $Wb$ threshold and
the dominant decay of the top quark is the one
into a $W$-boson and a bottom quark ($t\rightarrow Wb$) and
the total width of the top quark can be well
described by the partial width $\Gamma ^{tWb} = \Gamma(t\rightarrow Wb)$.
 
While the measurement of the top mass will provide a long missing input
parameter, the measurement of its width will serve
as a consistency check on the standard model. With the operation of
LHC, SSC and/or a high energy $e^{+}e^{-}$ collider
one expects to obtain a sufficiently large number of tops
so that both the mass and the width can be measured
with good accuracy.
 
The QCD corrections to the top decay $t \rightarrow Wb$
were already evaluated in \cite{Je89,Li90}.
The first order electroweak corrections have been
calculated by \cite{De91,Ei90}.
 
The electroweak corrections to this decay involve
particularly interesting
contributions of $\displaystyle O(\alpha\,m_t^2/M_W^2)$
which are potentially large for large top masses.
Those terms arise not only from fermion loop contributions to the
boson self energies but also from the Yukawa couplings of the
Higgs fields, which show
up in vertex and fermionic self energy corrections.
As discussed in Sect.~\ref{sechicor} contributions of the first type
can be eliminated if the
Born approximation is expressed by $G_F$ and $M_W$.
Surprisingly the effects from strong Yukawa couplings
turn out to be small, as will be demonstrated in the following.
 
We will only consider the decay of free top quarks and
sum over the polarizations of the $W$-bosons.
The results are obtained via crossing from the ones for $W\to t\bar{b}$.
 
\section{Notation and lowest order decay width}
 
Because we want to use our results for the decay $W^{+}\to t\bar{b}$ we
consider the decay of an anti-top quark. The corresponding decay width is
identical to the one of the top quark because of the CPT theorem.
 
The lowest order decay of an anti-top quark
\begin{equation}
\bar{t} (p_1)\rightarrow W^-(k)  \bar{b} (p_2)
\end{equation}
is described by the
Feynman diagram of Fig.~\reff{10.1}
\begin{figure}[b]
\bce \savebox{\Vtr}(32,24)[bl]
{\multiput(4,0)(8,6){4}{\oval(8,6)[tl]}
\multiput(4,6)(8,6){4}{\oval(8,6)[br]} }
\savebox{\Fr}(48,0)[bl] {
\put(48,0){\vector(-1,0){26}} \put(24,0){\line(-1,0){24}} }
\savebox{\Fbr}(32,24)[bl] {
\put(32,0){\vector(-4,3){19}} \put(16,12){\line(-4,3){16}} }
\begin{picture}(96,72)
\put(87,55){\makebox(10,18)[bl]{$W^{-}$}}
\put(0,42){\makebox(10,20)[bl]{$\bar{t}$}}
\put(87,10){\makebox(10,18)[bl]{$\bar{b}$}}
\put(48,36){\circle*{4}}
\put(0,36){\usebox{\Fr}}
\put(48,36){\usebox{\Vtr}}
\put(48,12){\usebox{\Fbr}}
\end{picture}
\ece
\caption{Born diagram for the decay $\bar{t}\to W^{-}\bar{b}$} \label{10.1}
\efi
yielding the amplitude
\begin{equation}  \label{M0tWb}
{\cal M}_{0} = \frac{-e}{\sqrt{2} s_W} \; V_{tb}\;
\bar{v} (p_1)  \rlap{/}\epsilon \omega_- v(p_2).
\end{equation}
It can be obtained from the Born amplitude (\ref{M0Wff})
for the decay $W^{+}\to t\bar{b}$ by crossing. This amounts to
change the signs of $p_{1}$ and
$k$ and use $u(-p_{1})=v(p_{1})$ and $\varepsilon(-k)=\varepsilon(k)$.
From (\ref{M0tWb}) we get the lowest order
width
\begin{equation} \label{G0tWb}
\Gamma_0^{tWb} (m_t,M_W,m_b) = \frac{\alpha}{8}\;
\frac{1}{2s_W^2}\; |V_{tb}|^2
\frac{\kappa(m_{t}^{2},M_{W}^{2},m_{b}^{2})}{m_t^3} G_{1}^{-},
\eeq
with
\beq \label{G1mtWb}
G_{1}^{-}=\left[m^2_t + m^2_b
- 2M_W^2+ \frac{(m^2_t - m^2_b)^2}{M_W^2}\right].
\end{equation}
Eq.~(\ref{G0tWb}) can directly be derived from (\ref{G0Wff}) by
substituting $m_{i}\to m_{t}$ and $m_{j}\to m_{b}$ in $G_{1}^{-}$
coming from the matrix element squared, exchanging $M_{W}$
and $m_{t}$ in the phase space factors, changing the spin average from
1/3 to 1/2 and supplying a minus sign originating from the different
signs of the momenta entering the matrix element squared.
This minus sign has been incorporated into the definition of $G_{1}^{-}$
which differs from the one in Chap.~\ref{chaWff}.
 
Introducing $G_{F}$ instead of $\alpha $ the lowest order width reads
\begin{equation}
\bar{\Gamma}_0^{tWb} (m_t,M_W,m_b) = \frac{G_{F}M_{W}^{2}}{8\pi \sqrt{2}}\;
|V_{tb}|^2 \frac{\kappa(m_{t}^{2},M_{W}^{2},m_{b}^{2})}{m_t^3}
G_{1}^{-}.
\label{GGtWb}
\end{equation}
 
\section{Virtual corrections}
 
With the four Dirac matrix elements
\begin{equation}
\begin{array}{l}
{\cal M}_1^{-} = \bar{v} (p_1)  \rlap{/}\epsilon \omega_- v(p_2) ,\\[1ex]
{\cal M}_1^{+} = \bar{v} (p_1)  \rlap{/}\epsilon \omega_+ v(p_2) ,\\[1ex]
{\cal M}_2^{-} = \bar{v} (p_1)  \omega_- v(p_2) \; \epsilon \cdot p_1 ,\\[1ex]
{\cal M}_2^{+} = \bar{v} (p_1)  \omega_+ v(p_2) \; \epsilon \cdot p_1
\end{array}
\label{Mi}
\end{equation}
obtained from (\ref{MVFF}) by setting $p_{1} \to -p_{1}$
the virtual electroweak one-loop corrections take the form of
(\ref{M1Wff}) with $t$ and $b$ instead of $i$ and $j$.
The corresponding decay width $\Gamma_{1}^{tWb}$
follows from (\ref{G1Wff}) using the substitutions specified after
(\ref{G1mtWb}).
 
The QCD corrections can be extracted from the
electroweak ones in the same way as in Sect.~\ref{secGWffQCD}.

\section{Bremsstrahlung}
 
The real photonic contributions of $O(\alpha)$ to the top width arise from the
radiative decay
\begin{equation}
\bar{t}(p_1) \rightarrow W^{-}(k) \bar{b}(p_2) \gamma(q) .
\end{equation}
The corresponding amplitude squared,
summed over all polarizations, can be derived
from (\ref{M2Wffg}) by replacing the momenta $k \to -k$,
$p_1 \to -p_1$ and multiplying an overall factor $(-1)$.
From this we get the bremsstrahlung contribution to the top width by
integrating over the appropriate phase space
\beqar
\nn\Gamma^{tWb}_b (m_t,M_W,m_b) &=& \frac{1}{(2\pi)^5} \frac{1}{2m_t} \int
\frac{d^3 q}{2 q_{0}} \frac{d^3 k}{2 k_{0}} \frac{d^3 p_2}{2 p_{20}}
\delta^{(4)} (p_1-p_2-q-k) \frac{1}{2} \sum_{pol} \bigl| {\cal M}_b
\bigl|^2  \\[1em]
&=& \Gamma _{0}^{tWb} \delta _{b}^{ew}(m_{t},M_{W},m_{b}) .
\eeqar
The correction factor reads
\begin{eqnarray}
\delta_b^{ew} (m_t,M_W,m_b) & = & \displaystyle  \left( -
\frac{\alpha}{\pi} \right)
\frac{4m_t^2}{\kappa(m_{t}^{2},M_{W}^{2},m_{b}^{2})}
\Biggl\{ - Q_t Q_b  \biggl[  (M_W^2-m_t^2-m_b^2)  I_{02}
 \biggl]  \nonumber  \\[1em]
 &  & \displaystyle + Q_t^2 \biggl[ (m_t^2 I_{00}+I_0)
-(1+\frac{m_t^2+m_b^2}{2M_W^2}) \frac{I_0^1}{G_{1}^{-}}
-\frac{I}{G_{1}^{-}}  \biggl]
\nonumber \\[1em]
 &  & \displaystyle
+ Q_b^2 \biggl[  (m_b^2 I_{22}+I_2)
-(1+\frac{m_t^2+m_b^2}{2M_W^2}) \frac{I_2^1}{G_{1}^{-}}
-\frac{I}{G_{1}^{-}}  \biggl]
 \label{brems}  \\[1em]
 &  & \displaystyle
+ (Q_t-Q_b)^2 \biggl[  (M_W^2 I_{11}+I_1)
-\frac{m_t^2+m_b^2}{2M_W^2} \frac{I}{G_{1}^{-}}
+2\frac{I_{11}^{02}}{G_{1}^{-}}  \biggl]
 \nonumber \\[1em]
 &  & \displaystyle
+ Q_t(Q_t-Q_b) \biggl[  (M_W^2+m_t^2-m_b^2) I_{01} + 2 \frac{I_1^2}{G_{1}^{-}}
  \biggl]   \nonumber  \\[1em]
 &  & \displaystyle
- Q_b(Q_t-Q_b) \biggl[ (M_W^2-m_t^2+m_b^2) I_{12} + 2 \frac{I_1^0}{G_{1}^{-}}
  \biggl]   \Biggl\} . \nonumber
\end{eqnarray}
The bremsstrahlung phase space integrals carry the arguments
$I_{\cdots}^{\cdots}=I_{\cdots}^{\cdots}(m_{t},M_{W},m_{b})$. They are
given in App.~D.
 
From eq.~(\ref{brems}) the gluonic bremsstrahlung corrections can be
obtained by setting $Q_t=Q_b=1$, replacing $\alpha $ by the strong coupling
constant $\alpha_s $ and multiplying with the colour
factor $C_F = \frac{4}{3}$.
 
\section{Results and discussion}
 
We again use the parameters listed in Sect.~\ref{secpar} as numerical
input and calculate the $W$-mass from the relation (\ref{GFloop}).
Unless stated otherwise, we choose for the Higgs mass $M_H = 100 \: \mbox{GeV}$.
 
We perform the same summation of the leading higher order corrections
as discussed in Sect.\ \ref{secWffres}, eq.~(\ref{Gsummed}), and introduce the
parametrization with $G_{F}$ and $M_{W}$ as in (\ref{GFpar}).
In this parametrization the large corrections arising from the
renormalization of $\alpha$ and $s_{W}^{2}$ are absorbed into the lowest
order expression.
This is not the case for large contributions proportional to
$\alpha\,m_t^2/M_W^2$
arising from vertex and fermion self energy diagrams with enhanced
Yukawa couplings.
 
In Tab.~\ref{tab101} we give the lowest order width as well as the
width including electroweak corrections
in both parametrizations for various values of the top
mass together with the $W$-mass obtained from (\ref{GFloop}).
The results for the first order expressions of both parametrizations
agree within 0.05\%.
\begin{table}
\begin{center}
\begin{tabular}{||c|c|c|c|c|c||} \hline\hline
\multicolumn{1}{||c|}{$m_t$} & $M_W$ & $\Gamma_0^{tWb}$ & $\bar{\Gamma}_0^{tWb}$
 & $\Gamma_{ew}^{tWb}$ & $\bar{\Gamma}_{ew}^{tWb}$ \\
\hline
100.0 &  79.99 &  0.0887 &  0.0940 &  0.0951 &  0.0951 \\
 \hline
120.0 &  80.10 &  0.3095 &  0.3260 &  0.3306 &  0.3305 \\
 \hline
140.0 &  80.23 &  0.6393 &  0.6684 &  0.6788 &  0.6786 \\
 \hline
160.0 &  80.37 &  1.0850 &  1.1248 &  1.1435 &  1.1433 \\
 \hline
180.0 &  80.52 &  1.6627 &  1.7071 &  1.7365 &  1.7362 \\
 \hline
200.0 &  80.69 &  2.3927 &  2.4299 &  2.4724 &  2.4720 \\
 \hline
220.0 &  80.88 &  3.2989 &  3.3083 &  3.3665 &  3.3661 \\
\hline\hline
\end{tabular}
\end{center}
\caption{Top decay width $\Gamma^{tWb}$ parametrized by $\alpha$
and $\bar{\Gamma}^{tWb}$ by $G_F$ in lowest order and including electroweak
corrections. All values are given in $\mbox{GeV}$.\label{tab101}}
\end{table}
 
\bfi
\includegraphics[width=15.cm]{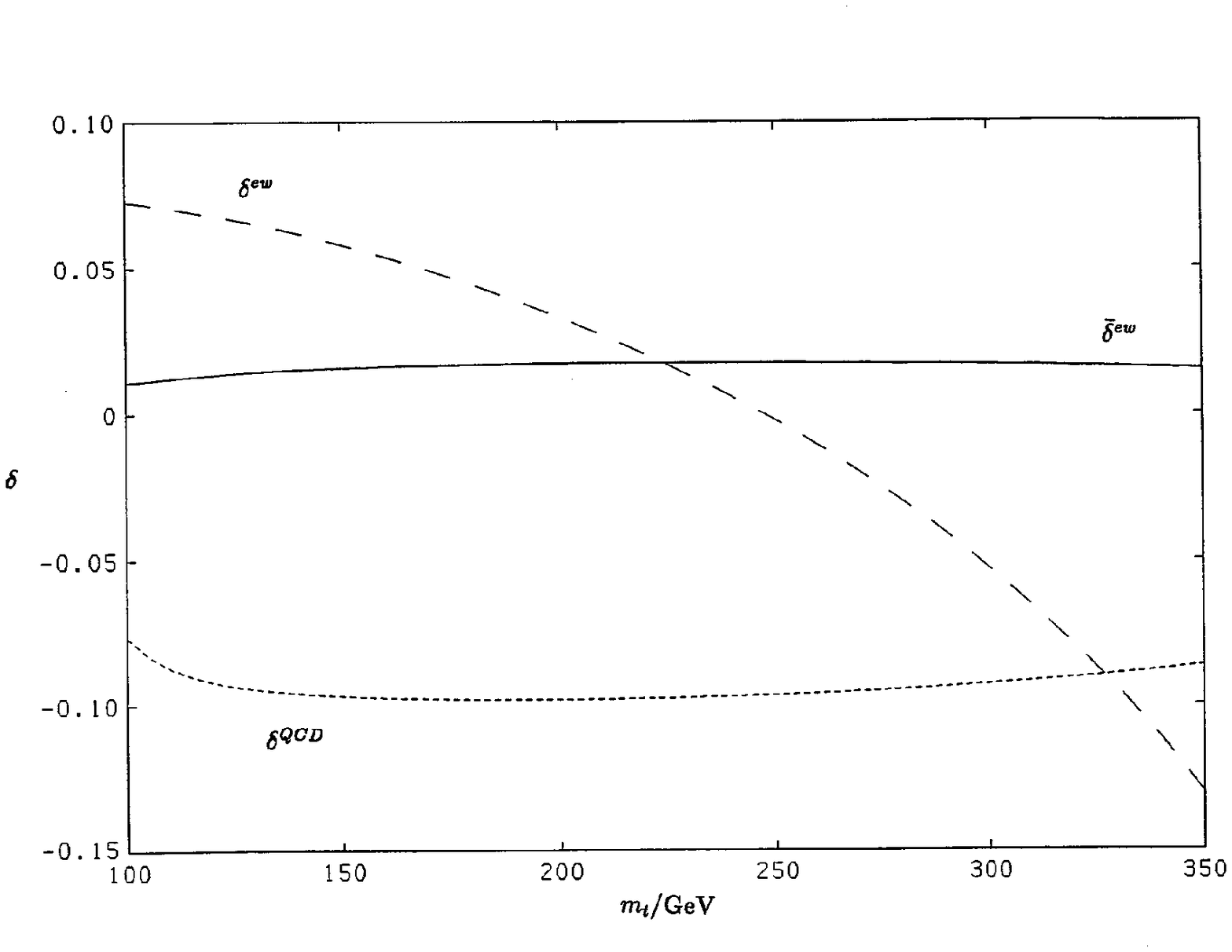}
\caption{Electroweak radiative corrections $\delta ^{ew}$ and
$\bar{\delta}^{ew}$ and QCD corrections $\delta^{QCD}$ to the top decay
width versus the top mass.}
\label{10.2}
\efi
According to (\ref{G0tWb}) the width increases
with the top mass approximately like $m_t^3/M_W^2$.
The corresponding relative corrections are shown in Fig.~\reff{10.2}.
The QCD corrections yield about $-10\%$ with only
a weak dependence on the top mass. 
In the on-shell scheme we find sizable electroweak corrections
which range from
$+7 \%$ at $m_t = 100 \: \mbox{GeV}$ to $-13\%$ at $m_t = 350 \:\mbox{GeV}$.
The large variation arises from terms $\propto \alpha\,m_t^2/M_{W}^{2}$
in the first order corrections.
Contrarily in the
parametrization with $G_F$ the corrections are only $\approx +1\%$ for
$m_t = 100 \:\mbox{GeV}$ and remain almost constant at $\approx +1.7\%$ for
$m_t \geq 160 \:\mbox{GeV}$. This feature is independent of the Higgs
boson mass.
However, we expected large corrections $\propto m_t^2$
to arise from the vertex diagrams containing large Yukawa couplings;
in the similar case of $Z^0 \rightarrow b \bar{b}$ the corresponding
contributions are noticeable.
The slow variation of the relative correction in this parametrization
indicates that the strong Yukawa couplings have no
sizable effect on the top width. In order to demonstrate the absence of
large corrections the plot in Fig.~\reff{10.2}
has been extended up to $m_t = 350\:\mbox{GeV}$, although this
is well above the present upper limits on the top quark mass.
 
Some understanding of this surprising feature can be obtained from the expansion
of the relative correction factor in the parametrization with $G_{F}$
and $M_{W}$, $\bar{\delta}^{ew}$, for large top quark masses
$(m_t \gg M_W,M_Z,M_H)$
\beq \label{expan}
\bar{\delta}^{ew}=\delta^{ew}_1 - \disp \Delta r
\sim  \displaystyle \frac{\alpha}{4\pi}  \, \frac{1}{2s_W^2}
\, \frac{m_t^2}{M_W^2} \biggl\{ \Bigl[ \frac{17}{4}
+ \log (\frac{M_H^2}{m_t^2}) \Bigl]
+\Bigl[ -\frac{7}{2} \pi \, \frac{M_H}{m_t}
 \Bigl] \; + \; O\Bigl( \log^2 (\frac{m_t^2}{M_i^2})\Bigl)
\biggl\}
\eeq
with $M_i=M_W,M_Z,M_H$. The two leading terms are evaluated in
Tab.~\ref{tab102} for a wide range of values of the top quark mass
and $M_{W}=80\:\mbox{GeV}$, $M_{H}=100\:\mbox{GeV}$. For $m_t \,<
\, 1\:\mbox{TeV}$ we find the leading term
of the expansion to be smaller than the next to leading term.
Consequently the expansion is only asymptotic and the contributions
$\propto \,\alpha \,m_t^2/M_W^2$ are not dominant
unless the top mass has a value of several $\mbox{TeV}$.
In the physically acceptable range of top quark masses,
the quadratic terms are numerically compensated by logarithmic contributions.
This remains true also for large Higgs masses.
The small corrections result from intricate cancellations between
leading and nonleading terms.
\begin{table}
\begin{center}
\begin{tabular}{||l|r|r|r|r||} \hline\hline
 $ m_t/\mbox{GeV} $ &  $ 200 $ &  $300 $ &
 $ 500 $ & $ 1000 $ \\[1ex] \hline
 leading term & 0.023 &  0.036   &  0.051     & -0.070  \\ \hline
 next to  & -0.043 & -0.065   &  -0.108     & -0.217  \\[-.5em]
 leading term & & & & \\ \hline\hline
\end{tabular}
\end{center}
\caption{The two leading terms of the expansion in
eq.~(\protect{\ref{expan}})
for various values of the top quark mass.\label{tab102}}
\end{table}
 
In Fig.~\reff{10.3} we show the lowest order width in both parametrizatons
$\Gamma_{0}$, $\hat{\Gamma}_{0}$,
the electroweak  corrections $\delta\Gamma_{ew}$, the QCD corrections
$\delta\Gamma_{QCD}$ as well as the
fully corrected width $\Gamma$ as a function of the top quark mass.
\bfi
\includegraphics[width=15.cm]{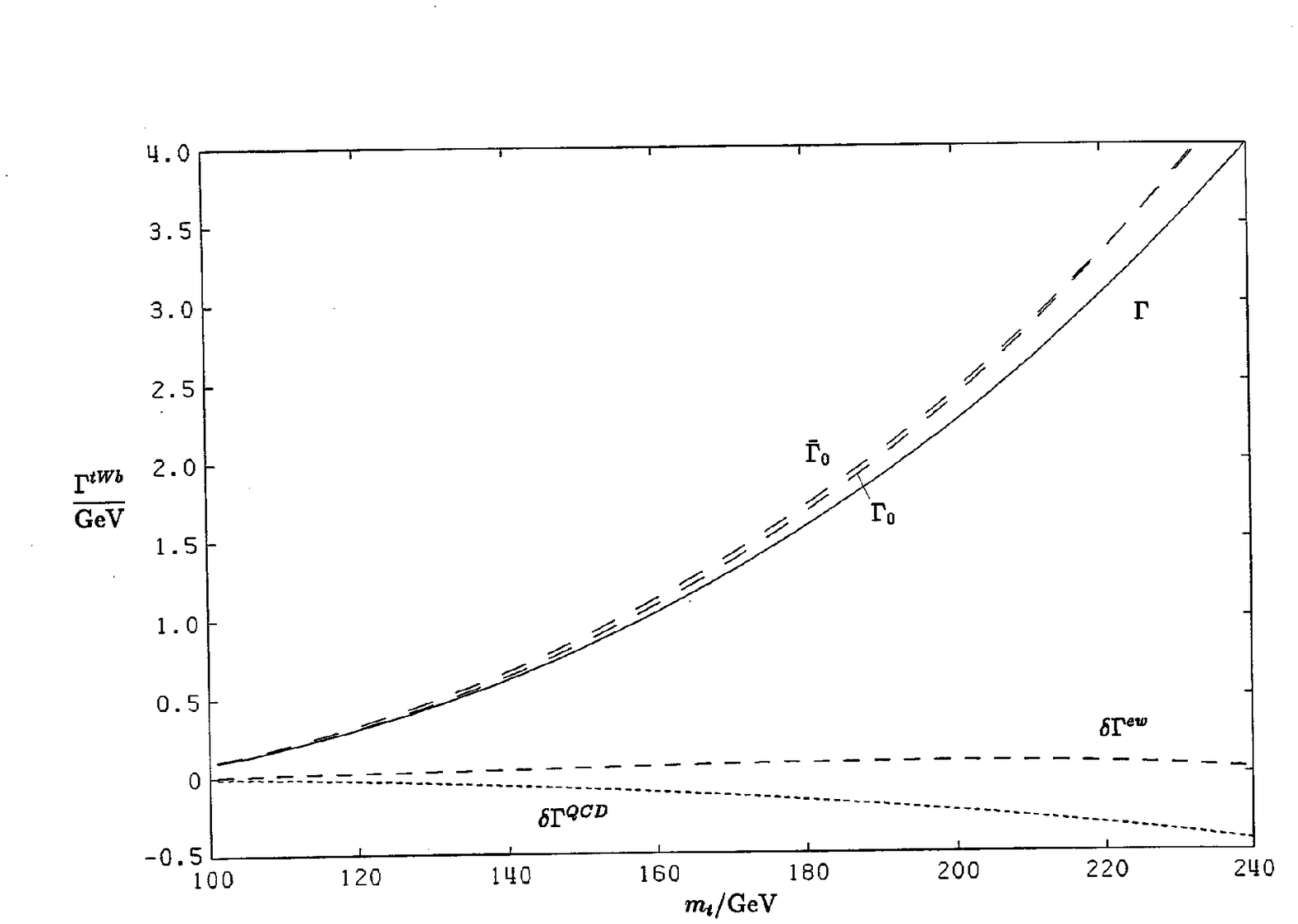}
\caption{Top decay width $\Gamma^{tWb}$ versus the top mass in
lowest order $\Gamma_{0}$, $\bar{\Gamma}_{0}$ including all corrections
$\Gamma $ and the contribution of the electroweak $\delta\Gamma_{ew}$
and QCD corrections $\delta\Gamma_{QCD}$.}
\label{10.3}
\efi
 
The dependence of the total width on the Higgs mass is displayed in
Tab.~\ref{tab103} where this parameter is varied from $50$ to
$1000 \:\mbox{GeV}$.
Although the influence of $M_H$ becomes stronger for large top masses
it never exceeds $1\%$.
\begin{table}
\begin{center}
\begin{tabular}{||r|c|c|c|c||} \hline\hline
 \multicolumn{1}{||c|}{$m_t$} & $M_H$ =   50 & $M_H$ =  100 &
 $M_H =  300$ & $M_H = 1000$ \\
 \hline
100.0 &  0.0882 &  0.0883 &  0.0886 &  0.0891 \\
 \hline
120.0 &  0.3022 &  0.3021 &  0.3020 &  0.3022 \\
 \hline
140.0 &  0.6179 &  0.6174 &  0.6165 &  0.6157 \\
 \hline
160.0 &  1.0385 &  1.0377 &  1.0356 &  1.0329 \\
 \hline
180.0 &  1.5742 &  1.5734 &  1.5697 &  1.5641 \\
 \hline
200.0 &  2.2373 &  2.2376 &  2.2321 &  2.2221 \\
 \hline
220.0 &  3.0411 &  3.0438 &  3.0369 &  3.0207 \\
 \hline\hline
\end{tabular}
\end{center}
\caption{Top decay width $\Gamma^{tWb}$ including first
order QCD and electroweak corrections for different values of the Higgs mass.
All values are given in $\mbox{GeV}$.\label{tab103}}
\label{higgs}
\end{table}
 
We have compared the pure QCD corrections
to those obtained in \cite{Je89} and found complete agreement.
Our results for the electroweak corrections agree with those of
\cite{Ei90}.
 
 
 

\chapter{$W$-pair production}
 
One of the predominant aims of LEP200 is the high precision
investigation of the properties of the $W$-boson, i.e.\ its mass, its
total and partial widths and its couplings. Probably the most
interesting aspect will be the study of the nonabelian gauge interaction
which has no direct experimental evidence so far.
 
The general properties
of $W$-pair production at LEP200 have been studied in \cite{El86,Bo87}.
While the total cross
section of $e^{+}e^{-}\to W^{+}W^{-}$ is extremely sensitive to
deviations from  the triple gauge boson couplings of
the SM at energies
high above the production threshold, the sensitivity to variations of
this coupling in the LEP200 energy range is only at the percent level.
Consequently theoretical predictions should be better than 1\% to obtain
reasonable limits on the structure of the gauge boson self interaction.
 
The $W$-pair production process allows an independent direct measurement
of the
$W$-boson mass with an expected accuracy of about $100\:\mbox{MeV}$
\cite{Bi87}.
Again this requires the knowledge of the total cross section with a
precision better than 1\%.
 
Much effort has been made in recent years to obtain such precise
theoretical predictions for $W$-pair production. The virtual electroweak
and soft photonic corrections were calculated by several authors
\cite{Le80,Ph82,Bo88,Fl89}. The complete analytical
results for arbitrary polarizations of the external particles were
published in \cite{Bo88}. These will be used for our evaluations. For
the unpolarized case they numerically agree with those of \cite{Fl89}
better than 0.3\% and
essentially also with \cite{Le80}. The hard photon bremsstrahlung
corrections have been evaluated by \cite{Be90,Be91,Ko91,Ta90}
for definite initial and final state polarizations.
The effects arising from the finite width of the W-bosons have been
studied in the Born approximation in \cite{Mu86}
and including the leading weak corrections in \cite{Gr87}.
Recently also the hard bremsstrahlung to the process
$e^{+}e^{-}\to W^{+}W^{-}\to 4 \; \mbox{fermions}$ has been evaluated
\cite{Ae91}.
 
In the following we will review and update existing results for the
virtual and real electroweak corrections and the finite width effects.
We will add some new results on approximate formulae for the $W$-pair
production cross section.
 
\section {Notation and amplitudes}
 
We discuss the process
\beq
e^{+}(p_{1},\sigma_{1})+e^{-}(p_{2},\sigma_{2})\to
W^{+}(k_{1},\lambda_{1})+W^{-}(k_{2},\lambda_{2}).
\eeq
The arguments indicate the momenta and helicities of the incoming
fermions and outgoing bosons ($\sigma_{i}=\pm \frac{1}{2},\:\lambda_{i}=
1,0,-1$).
We introduce the usual Mandelstam variables
\beq
\barr{lll}
s&=&(p_{1}+p_{2})^{2}=(k_{1}+k_{2})^{2}=4E^{2}, \\[1ex]
t&=&(p_{1}-k_{1})^{2}=(p_{2}-k_{2})^{2}=M^{2}_{W}-
2E^{2}+2E^{2}\beta\cos\vartheta  .
\earr
\eeq
Here $E$ is the beam energy, $\vartheta$ the scattering
angle between the $e^{-}$ and the $W^{-}$ and $\beta =\sqrt{1-
M^{2}_{W}/E^{2}}$ the velocity of the $W$-bosons in the center
of mass frame. The electron
mass has been consistently neglected.
 
In the approximation of zero electron mass the invariant
matrix element vanishes due to chiral symmetry
for equal helicities of the $e^{+}$ and $e^{-}$.
Consequently we can write
\beq
{\cal M}(\sigma_{1},\sigma_{2},\lambda_{1},\lambda_{2},s,t)={\cal M}(\sigma
,\lambda_{1},\lambda_{2},s,t)
\eeq
with $\sigma =\sigma_{2}=-\sigma_{1}$.
If we neglect the CP-violating phase in the quark mixing matrix, CP is a
symmetry of the process leading to the relation
\beq
{\cal M} (\sigma ,\lambda_{1}, \lambda_{2},s,t)=
{\cal M} (\sigma,-\lambda_{2},-\lambda_{1},s,t).
\eeq
Consequently there are only 12 independent helicity matrix elements
instead of 36.
 
As discussed in Chap.~\ref{chaSME} the general matrix element
\beq
{\cal M}(\sigma ,\lambda_{1},\lambda_{2},s,t)=
\bar {v}(p_{1},-\sigma ){\cal M}^{\mu \nu }u(p_{2},\sigma )
\varepsilon_{\mu} (k_{1},\lambda_{1})\varepsilon_{\nu}(k_{2},\lambda_{2})
\eeq
can be decomposed into formfactors and standard matrix elements. Due to
the above-mentioned symmetries we do not need all of the (overcomplete)
40 \SME\ given in
(\ref{MFFVV}) for a general process of vector pair production in
fermion-antifermion annihilation, but only seven for each fermion helicity
\beq \label{SMEWW}
{\cal M}(\sigma ,\lambda_{1},\lambda_{2},s,t)=
\sum^{7}_{i=1} {\cal M}^{\sigma }_{i} F^{\sigma }_{i}(s,t).
\eeq
The standard matrix elements ${\cal M}^{\sigma }_{i}$ are defined in
(\ref{MFFVV}) together with
\beq
\barr{lll}
{\cal M}^{\sigma}_{3}&=&{\cal M}^{\sigma}_{3,1}+{\cal M}^{\sigma}_{3,2},\\
{\cal M}^{\sigma}_{4}&=&{\cal M}^{\sigma}_{4,1}+{\cal M}^{\sigma}_{4,2},\\
{\cal M}^{\sigma}_{7}&=&{\cal M}^{\sigma}_{7,1}+{\cal M}^{\sigma}_{7,2}.
\earr
\eeq
Only six are linear independent. Using (\ref{MFFVVrel}) we can express
${\cal M}_{7}$ by the others
\beq
{\cal M}^{\sigma}_{7}=-\frac{s}{2}({\cal M}^{\sigma }_{1}
+{\cal M}^{\sigma}_{2})+\frac{M^{2}_{W}-t}{2}{\cal M}^{\sigma }_{3}
+\frac{s}{2}{\cal M}^{\sigma }_{4}
+{\cal M}^{\sigma }_{5}.
\eeq
Therefore there are only twelve independent formfactors. The \SME\ can be
calculated with the methods described in Sect.~\ref{secmatcal}. The
explicit results can be found in \cite{Bo88,Sa87}. 
 
\section {Born cross section}
 
At the Born level three diagrams contribute to $W$-pair production
(Fig.~\reff{11.1}). We omitted a Higgs-exchange diagram, which is
suppressed by a factor $m_{e}/M_{W}$ and thus completely
negligible. The $t$-channel $\nu_{e}$-exchange diagram contributes
only for left-handed electrons whereas the $s$-channel diagrams
containing the nonabelian gauge coupling contribute also for right
handed electrons.
The analytical expressions read
\beq \label{M0eeWW}
\barr{lll}{\cal M}_{0}(-,\lambda_{1},\lambda_{2},s,t)&=&
\disp {\cal M}^{-}_{1}\frac{e^{2}}{2s^{2}_{W}}\frac{1}{t}
+2({\cal M}^{-}_{3}-{\cal M}^{-}_{2})e^{2}
\biggl[\frac{1}{s}-\frac{c_{W}}{s_{W}}g^{-}_{e}\frac{1}{s-
M^{2}_{Z}}\biggr] \\[1.2em]
&=&\disp \frac{e^{2}}{2s^{2}_{W}}\biggl[\frac{1}{t}{\cal M}^{-}_{1}
+\frac{2}{s-M^{2}_{Z}}({\cal M}^{-}_{3}-{\cal M}^{-}_{2})\biggr] \\[1em]
&&\disp +e^{2}\biggl[\frac{1}{s}-\frac{1}{s-M^{2}_{Z}}\biggr]
2({\cal M}^{-}_{3}-{\cal M}^{-}_{2}), \\[1.2em]
{\cal M}_{0}(+,\lambda_{1},\lambda_{2},s,t)&=&
\disp2({\cal M}^{+}_{3}-{\cal M}^{+}_{2})e^{2}
\biggl[\frac{1}{s}-\frac{c_{W}}{s_{W}}g^{+}_{e}\frac{1}{s-
M^{2}_{Z}}\biggr] \\[1.2em]
&=&\disp e^{2}\biggl[\frac{1}{s}-\frac{1}{s-M^{2}_{Z}}\biggr]2({\cal
M}^{+}_{3}-{\cal M}^{+}_{2}),
\earr
\eeq
where we have inserted the explicit form of the $Z$-boson fermion
couplings $g_{e}^{-},\:g_{e}^{+}$ (\ref{geZ}).
The corresponding cross section for arbitrary longitudinal polarizations
of the leptons and bosons is given by
\beq
\left(\frac{d\sigma }{d\Omega }\right)_{0}=
\frac{\beta}{64\pi^{2}s}\sum_{\lambda_{1},\lambda_{2}}
\frac{1}{4}(1-2\sigma P^{+})(1+2\sigma P^{-})
\left|{\cal M}_{0}(\sigma,\lambda_{1},\lambda_{2},s,t)\right|^{2},
\eeq
and $P^{\pm}$ are the polarization degrees of the leptons
($P^-=\pm 1$ corresponds to purely right- and  left-handed
electrons, respectively).
\bfi
\savebox{\Vr}(48,0)[bl]
{\multiput(3,0)(12,0){4}{\oval(6,4)[t]}
\multiput(9,0)(12,0){4}{\oval(6,4)[b]} }
\savebox{\Vtr}(32,24)[bl]
{\multiput(4,0)(8,6){4}{\oval(8,6)[tl]}
\multiput(4,6)(8,6){4}{\oval(8,6)[br]} }
\savebox{\Vbr}(32,24)[bl]
{\multiput(4,24)(8,-6){4}{\oval(8,6)[bl]}
\multiput(4,18)(8,-6){4}{\oval(8,6)[tr]}}
\savebox{\Vtbr}(32,48)[bl]
{\put(00,24){\usebox{\Vtr}}
\put(00,00){\usebox{\Vbr}}}
\savebox{\Ft}(0,48)[bl]
{ \put(0,0){\vector(0,1){27}} \put(0,24){\line(0,1){24}} }
\savebox{\Fr}(48,0)[bl]
{ \put(0,0){\vector(1,0){26}} \put(24,0){\line(1,0){24}} }
\savebox{\Fl}(48,0)[bl]
{ \put(48,0){\vector(-1,0){26}} \put(24,0){\line(-1,0){24}} }
\savebox{\Ftr}(32,24)[bl]
{ \put(0,0){\vector(4,3){18}} \put(16,12){\line(4,3){16}} }
\savebox{\Fbr}(32,24)[bl]
{ \put(32,0){\vector(-4,3){19}} \put(16,12){\line(-4,3){16}} }
\savebox{\Ftbr}(32,48)[bl]
{\put(00,24){\usebox{\Ftr}}
\put(00,00){\usebox{\Fbr}}}
\bma
\barr{lll}
\begin{picture}(176,72)
\put(00,60){\makebox(10,10)[bl]{$e^{+}$}}
\put(00,5){\makebox(10,10)[bl] {$e^{-}$}}
\put(116,60){\makebox(10,10)[bl]{$W^{+}$}}
\put(116,5){\makebox(10,10)[bl] {$W^{-}$}}
\put(70,33){\makebox(20,10)[bl] {$\nu$}}
\put(16,60){\usebox{\Fl}}
\put(16,12){\usebox{\Fr}}
\put(64,12){\usebox{\Ft}}
\put(64,60){\circle*{4}}
\put(64,12){\circle*{4}}
\put(64,12){\usebox{\Vr}}
\put(64,60){\usebox{\Vr}}
\end{picture}  &
\begin{picture}(148,72)
\put(00,60){\makebox(10,10)[bl]{$e^{+}$}}
\put(00,5){\makebox(10,10)[bl] {$e^{-}$}}
\put(134,60){\makebox(10,10)[bl]{$W^{+}$}}
\put(134,5){\makebox(10,10)[bl] {$W^{-}$}}
\put(62,43){\makebox(20,10)[bl] {$\gamma,Z$}}
\put(16,36){\usebox{\Fbr}}
\put(16,12){\usebox{\Ftr}}
\put(48,36){\circle*{4}}
\put(96,36){\circle*{4}}
\put(48,36){\usebox{\Vr}}
\put(96,36){\usebox{\Vtr}}
\put(96,12){\usebox{\Vbr}}
\end{picture}
\earr
\ema
\caption{Born diagrams for $e^{+}e^{-}\to W^{+}W^{-}$.}
\label{11.1}
\efi
 
The Born cross section determines the main features of $W$-pair
production.
We first study the threshold behaviour \cite{Ha87,Gr88}.
For small $\beta $ the matrix elements behave as
\beq
{\cal M}^{\sigma }_{2},{\cal M}^{\sigma }_{3}\propto\beta, \qquad
{\cal M}^{\sigma }_{1}\propto 1.
\eeq
Consequently the $s$-channel diagrams vanish at threshold and the
$t$-channel graph dominates in the threshold region. For $\beta \ll 1$
the total cross section is given by
\beq \label{sigthr}
\sigma_{0}(s)\approx \frac{\pi \alpha ^{2}}{s}\frac{1}{4s^{4}_{W}}4\beta
+O(\beta ^{3}).
\eeq
All terms $\propto\beta ^{2}$ which are present in the differential
cross section drop out in the total cross section.
$s$-channel diagrams yield contributions
$\propto \beta ^{3}$. In the SM the coefficient of the $\beta^{3}$ term
in (\ref{sigthr}) is roughly equal to the one of the leading $\beta$
term. As long as that coefficient is not enhanced drastically the
$\beta^{3}$ term is negligible in the threshold region, i.e.\ for
$2M_{W} < \sqrt{s} < 2M_{W}+10\:\mbox{GeV}$ (for
$\sqrt{s}=2M_{W}+10\:\mbox{GeV}$ we have $\beta =0.33$ and $\beta
^{3}=0.04$). Consequently the shape of the total cross section close to
threshold is completely governed by the linear rise in $\beta$ and hence
by kinematics alone. Any change in the couplings will only affect the
coefficient of the $\beta $ term and thus the normalization of
the cross section. Moreover many new physics effects such as
anomalous gauge couplings contribute to the s-channel only
and thus do not affect
the leading term. The inclusion of the finite width of the
$W$-boson smears the threshold considerably (see Sect.~\ref{secWfiwi}).
However, since the
next to leading $\beta^{3}$ term becomes only sizable
several $\Gamma^{W}$ above threshold
only the leading term is relevant for the cross section in the region of
the nominal threshold.
 
This fact allows a model independent determination of the $W$-mass from
the $W$-pair production threshold \cite{Ha87}. The measured cross
section up to about 10 GeV above threshold is fitted with a
three-parameter curve
\beq \label{sifit}
\sigma (s)=\frac{a}{s}+b\,\sigma _{SM}(M_{W},s)
\eeq
where $a/s$ accounts
for the background, $b$ is a model dependent normalization factor and
$\sigma _{SM}$ the $W$-pair production cross section in the SM
depending on the $W$-mass. Eq.~(\ref{sifit}) is valid including
radiative corrections and finite width effects.
 
At high energies the $W$-pair production cross section is subject to
large gauge cancellations arising from the contributions of
longitudinally polarized $W$-bosons. For $s\gg M^{2}_{W}$ the matrix
elements behave as
\beq
\frac{{\cal M}^{\sigma }_{1}}{t},\frac{{\cal M}^{\sigma }_{2,3}}{s}\sim
\frac{s}{M^{2}_{W}},
\eeq
but
\beq
\barr{lll}
\disp \frac{1}{t}{\cal M}^{\sigma }_{1}
+\frac{1}{s-M_{Z}^{2}}2({\cal M}^{\sigma }_{3}-
{\cal M}^{\sigma }_{2})&\sim& \disp \frac{M^{2}_{W}}{t}+O(1),\\[1em]
\disp \left(\frac{1}{s}-\frac{1}{s-M_{Z}^{2}}\right)2({\cal M}^{\sigma }_{3}-
{\cal M}^{\sigma }_{2})&\sim& O(1).
\earr
\eeq
Consequently the Born matrix elements (\ref{M0eeWW}) have a good high
energy behaviour. While the cross sections corresponding to the $s$-channel
or $t$-channel diagrams alone violate unitarity at high energies
\beq
\sigma_{0,t}(s)\approx
\sigma_{0,s}(s)\approx
\frac{\pi \alpha^{2}}{s}\frac{1}{4s^{4}_{W}} \frac{s^{2}}{24M^{4}_{W}},
\eeq
the SM cross section respects it
\beq
\sigma_{0}\approx\frac{\pi \alpha^{2}}{s}\frac{1}{4s^{4}_{W}}
\biggl[2\Bigl(\log\frac{s}{M^{2}_{W}}-1\Bigr)-
\frac{1}{2}-\frac{1}{3c_{W}^{2}}+\frac{5}{24c^{4}_{W}}\biggr].
\eeq
\bfi
\includegraphics[bb=20 40 592 402, width=15.cm]{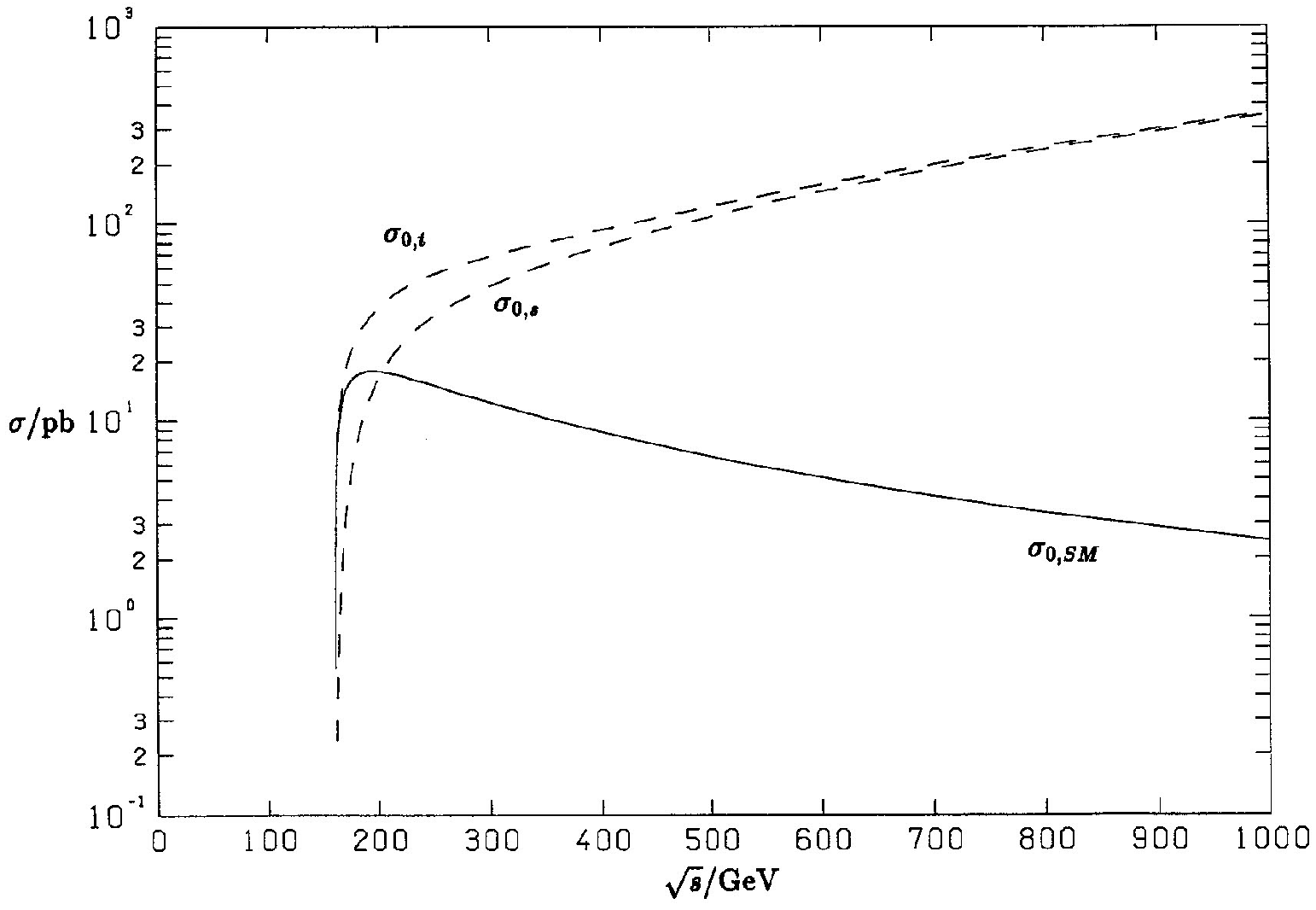}
\caption{Gauge cancellations in the total cross section for W-pair
production. Shown are the Born cross sections arising from the
$s$-channel $\sigma_{0,s}$ and $t$-channel $\sigma_{0,t}$ diagrams alone
and the SM cross section $\sigma_{0}$.}
\label{11.2}
\efi
The gauge cancellations are illustrated in Fig.~\reff{11.2}. They
reach one order of magnitude at 400 GeV and two orders at 1 TeV. They
only occur for longitudinal $W$-bosons.
After the gauge cancellations the $t$-channel again dominates the
SM cross section.
Compared to the $t$-channel contribution all
other contributions to the total cross section
are suppressed by $\sim 50 \log(s/M^{2}_{W})$. For example the cross section for
right handed electrons is
\beq
\sigma_{0}(e_{R}^{-}e^{+}\to W^{+}W^{-}) \approx
\frac{\pi \alpha^{2}}{s}\frac{1}{12c^{4}_{W}},
\eeq
and the cross section for longitudinal $W$-bosons
\beq
\sigma_{0}(e^{-}e^{+}\to W_{L}^{+}W_{L}^{-})
\approx\frac{\pi \alpha^{2}}{s}\frac{1}{24c^{4}_{W}}
\left(\frac{1}{4s^{4}_{W}}+1\right).
\eeq
 
We now consider the complete Born expressions for the $W$-pair production
cross section.
The differential cross section for the unpolarized case and for
longitudinally polarized $W$-bosons is shown in Fig.~\ref{11.3} and
\ref{11.4}, respectively.
Due to the $t$-channel pole the unpolarized cross section is
strongly peaked in forward direction at high energies and drops
smoothly with increasing scattering angle.
In contrast the differential cross section for longitudinal $W$-bosons
has a minimum for a certain energy dependent finite scattering angle
$\ne \pi $.
\bfi
\includegraphics[bb=20 40 592 422,width=15.cm]{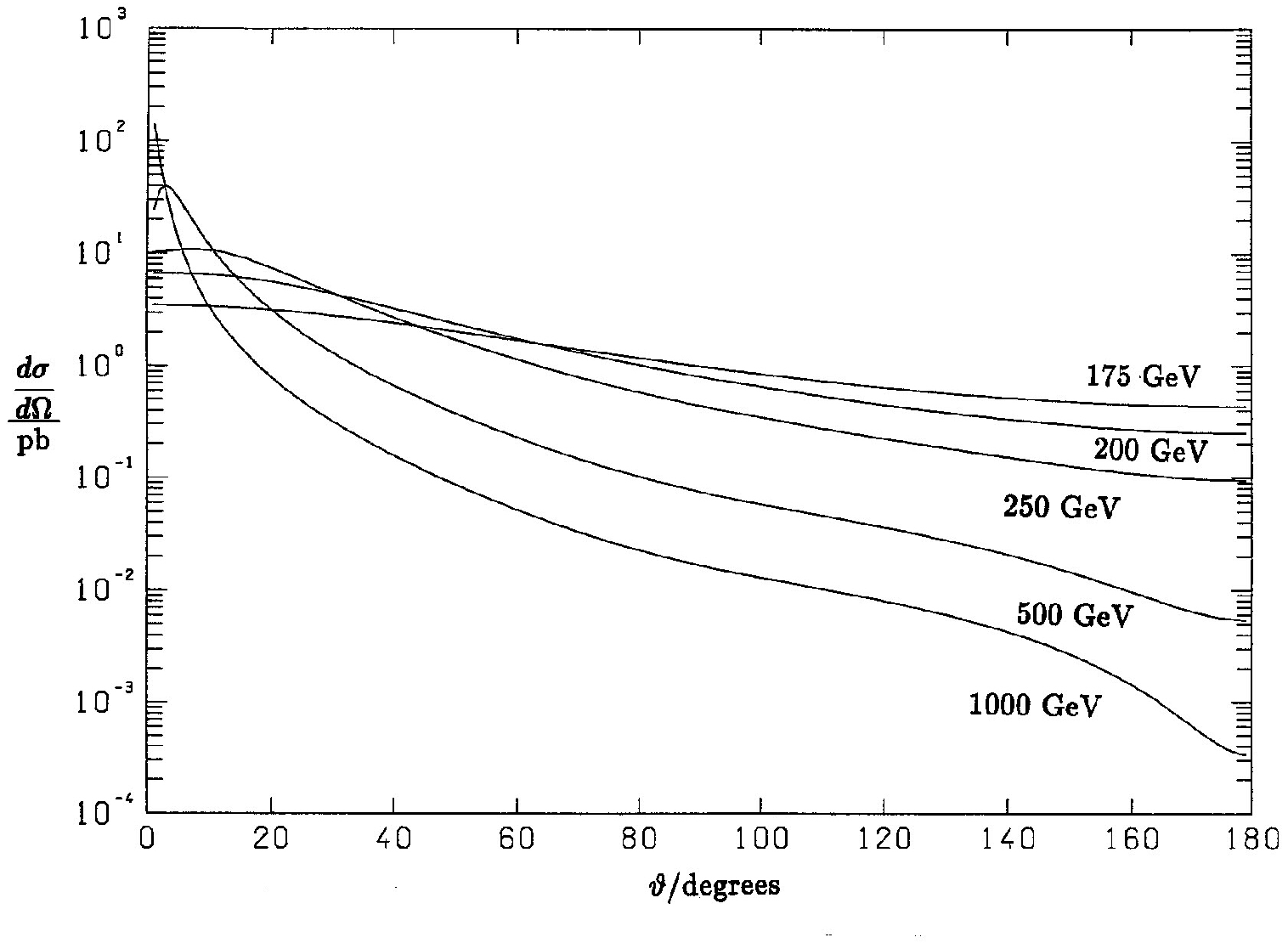}
\caption{Lowest order differential cross section for the production
of unpolarized $W$-pairs at different center of mass energies.}
\label{11.3}
\efi
\bfi
\includegraphics[bb=20 40 592 422,width=15.cm]{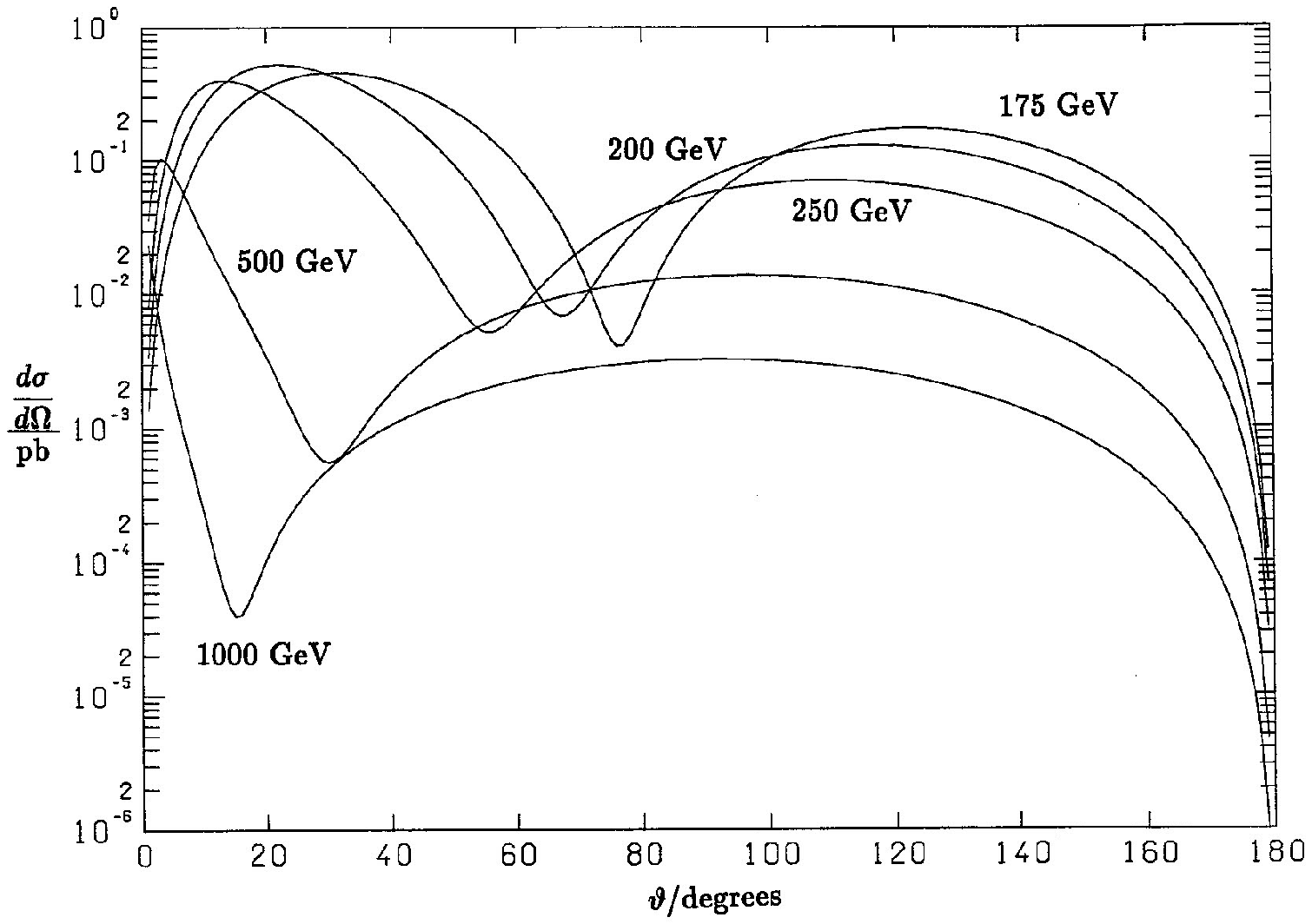}
\caption{Lowest order differential cross section for the production
of longitudinal $W$-pairs at different center of mass energies.}
\label{11.4}
\efi
 
In order to show the importance of the separate contributions we give
in Tab.~\ref{tab111} the integrated cross section for different
center of mass energies and different polarizations
of the leptons and $W$-bosons.
\btab
\bce
\begin{tabular}{||r|c|c|c|c|c|c||} \hline\hline
&\multicolumn{6}{|c||}{$\sigma_{0}/\mbox{pb}$} \\
\hline
$\sqrt{s}/\mbox{GeV}$ &  unpolarized & $e_{L}^{-}$ & $e_{R}^{-}$ &
 $W_{T}^{+}W_{T}^{-}$ & $W_{L}^{+}W_{L}^{-}$
 & $W_{T}^{+}W_{L}^{-}+ W_{L}^{+}W_{T}^{-}$
 \\ \hline
 165.0 & 10.761 & 21.466 &  0.056 &  5.254  &  1.349 & 4.158 \\
 \hline
 180.0 & 16.928 & 33.600 &  0.256 &  9.397  &  1.743 & 5.788 \\
 \hline
 200.0 & 17.709 & 35.070  & 0.348 & 11.189\phantom{1}  &  1.428 & 5.092 \\
 \hline
 250.0 & 15.011 & 29.745 &  0.277 & 11.487\phantom{1}  &  0.731 & 2.793 \\
 \hline
 500.0 &  6.534 & 13.019 &  0.050 &  6.178  &  0.113 & 0.244 \\
 \hline
1000.0 &  2.439 & \phantom{1}4.867 &  0.010 &  2.395  &  0.027 & 0.017 \\
 \hline\hline
\end{tabular}
\ece
\caption{Integrated lowest order cross section for different
polarizations and different center of mass energies and $M_{W} =
80.23\:\mbox{GeV}$. \label{tab111}}
\etab
 
The gauge cancellations depend crucially on the values of the SM
couplings. Any deviations from these values can lead to sizable effects
at higher energies since they are enhanced by a factor
$\beta s/M^{2}_{W}$.
This fact concerns especially anomalous three gauge boson couplings which
have been studied by many authors \cite{Ne87}. The sensitivity
to these effects is best at high energies and large scattering angles
where the $t$-channel pole is not dominant. Nevertheless one hopes to
determine the anomalous couplings up to 20\% at LEP200 \cite{Da87}.
 
Using right-handed electrons one could study a pure triple gauge
coupling process, but this would require longitudinally
polarized electron beams. Furthermore the right-handed cross section is
suppressed by two orders of magnitude compared to the dominant left-handed
mode, mainly because there is no $t$-channel contribution.
On the other hand, nonstandard couplings or other new physics can enhance it
drastically exactly for this reason.

\section {Virtual and soft photonic corrections}
 
The radiative corrections can be naturally divided into three classes, the
virtual corrections, the soft photonic, and the hard photonic corrections.
Since the process $e^{+}e^{-}\to W^{+}W^{-}$ involves the charged
current, the radiative corrections cannot be separated into
electromagnetic and weak ones in a gauge invariant way. We first discuss
the virtual and soft photonic corrections.
 
\subsection {Virtual corrections}
 
The virtual corrections get
contributions from the $\nu_{e}$-, $\gamma$- and $Z$-self energies, the
$\gamma$-$Z$-mixing energy, from the vertex corrections to the $ee\gamma$-,
$eeZ$-, $e\nu_{e} W$-, $WW\gamma$- and $WWZ$-vertices and from box diagrams.
The necessary counterterms involve in addition the $e$- and $W$-self
energies. Altogether one has to calculate more than 200 individual
diagrams. These can be
treated using the methods described in the first part of this review.
The number of generic diagrams to be evaluated is about 30. The results
can be expressed in terms of the formfactors defined in (\ref{SMEWW})
\beq
\delta {\cal M}_{1}(\sigma,\lambda_{1},\lambda_{2},s,t) =
\sum^{7}_{i=1}{\cal M}^{\sigma }_{i}\delta F^{\sigma }_{i} .
\eeq
The formfactors $\delta F^{\sigma }_{i}$ can be evaluated for every
CP-invariant set of diagrams separately.
For CP-violating diagrams we need in addition the standard matrix elements
${\cal M}_{3,1}^{\sigma}-{\cal M}_{3,2}^{\sigma}$, ${\cal M}_{4,1}^{\sigma}
-{\cal M}_{4,2}^{\sigma}$ and ${\cal M}_{7,1}^{\sigma}-{\cal
M}_{7,2}^{\sigma}$.
These drop out in CP-invariant combinations. The explicit analytical
results for the formfactors are given in terms of the scalar
coefficients of tensor integrals in \cite{Bo88,Sa87}. The reduction
to scalar integrals and their evaluation is done numerically using
the formulae given in Chap.~\ref{chatenint}. The contribution of the
virtual corrections to the cross section is given by
\beq \label{dsigmaV}
\delta \left(\frac{d\sigma }{d\Omega }\right)_{V}=
\frac{\beta}{64\pi^{2}s}\sum_{\lambda_{1},\lambda_{2}}
\frac{1}{4}(1-2\sigma P^{+})(1+2\sigma P^{-})
2\mbox{Re}\left({{\cal M}^{*}_{0}\delta {\cal M} _{1}}\right).
\eeq
 
The cancellations already present at the Born level occur as well
at the level of radiative corrections. These cancellations only work
for gauge invariant quantities. Consequently the
inclusion of the leading higher order contributions must be done such
that gauge invariance is respected. Otherwise one may introduce sizable
unphysical corrections. This will be discussed in more detail in
Sect.~\ref{secWWleaco}.
 
The presence of these cancellations enforces very careful
tests of the numerical stability of the computer programs. The
reliability of the results is founded on agreement between
independent calculations \cite{Bo88,Fl89}.
 
\subsection {Soft photonic corrections}
 
The soft photonic corrections can be easily obtained using the results of
Chap.~\ref{chasopho}.
The soft photon matrix element reads ($k$ is the photon momentum)
\beq
{\cal M}_{s}=e{\cal M}_{0}\left[\frac{\varepsilon p_{2}}{kp_{2}}-
\frac{\varepsilon p_{1}}{kp_{1}}+\frac{\varepsilon k_{1}}{kk_{1}}-
\frac{\varepsilon k_{2}}{kk_{2}}\right].
\eeq
This yields the soft photon cross section as
\beq
\left(\frac{d\sigma }{d\Omega }\right)_{s}=\left(\frac{d\sigma}
{d\Omega }\right)_{0}\delta _{s}
\eeq
with
\beqar
\nn \delta_{s} &=& -\frac{\alpha }{2\pi ^{2}}\int_{|{\bf k}|<\Delta E}
\frac{d^{3}k}{2\omega_{k}} \biggl\{
\frac{p_{1}^{2}}{(p_{1}k)^{2}} + \frac{p_{2}^{2}}{(p_{2}k)^{2}}
- \frac{2p_{1}p_{2}}{(p_{1}k)(p_{2}k)} \\[1em]
\nn && \qquad\mbox{}+\frac{k_{1}^{2}}{(k_{1}k)^{2}}
+ \frac{k_{2}^{2}}{(k_{2}k)^{2}}
- \frac{2k_{1}k_{2}}{(k_{1}k)(k_{2}k)} \\[1em]
\nn && \qquad\mbox{}- \frac{2p_{1}k_{1}}{(p_{1}k)(k_{1}k)}
- \frac{2p_{2}k_{2}}{(p_{2}k)(k_{2}k)}
+ \frac{2p_{1}k_{2}}{(p_{1}k)(k_{2}k)}
+ \frac{2p_{2}k_{1}}{(p_{2}k)(k_{1}k)} \biggr\} \\[1em]
\nn &=& - \frac{\alpha }{\pi }
\left\{ 4 \log\frac{2\Delta E}{\lambda } - 2 \log \frac{2\Delta E}{\lambda }
\log \frac{s}{m_{e}^{2}} +4 \log\frac{2\Delta E}{\lambda } \log
\frac{M_{W}^{2}-u}{M_{W}^{2}-t}  \right. \\[1em]
\nn && \mbox{}+ \frac{1+\beta^{2}}{\beta} \log \frac{2\Delta
E}{\lambda } \log \left( \frac{1-\beta }{1+\beta } \right) \\[1em]
\nn && \mbox{}+ \log \frac{m_{e}^{2}}{s} + \frac{1}{\beta } \log\left(
\frac{1-\beta }{1+\beta } \right) +\frac{\pi ^{2}}{3} +\frac{1}{2}
\log^{2} \frac{m_{e}^{2}}{s} \\[1em]
&& + \frac{1+\beta^{2}}{\beta} \left[\Li\left(
\frac{2\beta }{1+\beta } \right) + \frac{1}{4} \log^{2} \left(
\frac{1-\beta }{1+\beta } \right) \right] \\[1em]
\nn && \mbox{}+2  \left[ \Li \left(1-\frac{s(1-\beta )}{2(M_{W}^{2}-t)}
\right) +  \Li \left(1-\frac{s(1+\beta )}{2(M_{W}^{2}-t)} \right) \right.
\\[1em]
\nn && \mbox{}\hspace{8mm} - \left. \left. \Li \left(1-
\frac{s(1-\beta )}{2(M_{W}^{2}-u)} \right)
- \Li \left(1-\frac{s(1+\beta )}{2(M_{W}^{2}-u)} \right) \right]
\right\}.
\eeqar
 
Adding the soft photon cross section to the contribution of the virtual
corrections (\ref{dsigmaV}) the IR-singularities cancel.
Moreover also the large Sudakov
double logarithms $\log^{2}(m_{e}^{2}/s)$ drop out.
 
\subsection {Leading weak corrections}
\label{secWWleaco}
 
In order to set up improved Born approximations which are often very
handy the first step is to extract the leading corrections \cite{De91b}.
The universal corrections involving $\Delta \alpha$ and $\Delta \rho$ can be
easily obtained from (\ref{reclco})
including the leading $O(\alpha^{2})$ contributions. There are no
nonuniversal corrections $\propto \alpha \,m_{t}^{2}/M_{W}^{2}$ to the
$W$-pair production cross section for not too high energies, i.e.\ as
long as the unitarity cancellations are not sizeable.
In the LEP200 energy region also terms involving $\log
m_{t}^{2}$ or $\log M_{H}^{2}$ may become important.
These have been evaluated in the limit $M^{2}_{H},\,m^{2}_{t}\gg s$.
In addition close to threshold apart from the large bremsstrahlung
corrections which will be discussed in the next section there is a
sizable effect of the Coulomb singularity. This can be simply obtained
from general considerations or to $O(\alpha )$ directly from the loop
diagrams involving photons exchanged between the final state $W$-bosons.
Altogether this yields the following approximation
\beqar \label{WWapp1}
\nn {\cal M}_{a}^{-} & = & \frac{e^{2}}{2s^{2}_{W}}\left[\frac{1}{t}{\cal M}^{-}
_{1}+\frac{1}{s-M^{2}_{Z}}2({\cal M}^{-}_{3}-{\cal M}^{-
}_{2})\right] 
\left[\frac{1}{1-\Delta \alpha}\,
\frac{1}{1+\frac{c^{2}_{W}}{s^{2}_{W}}\Delta \rho }
\right. \\[1em]
\nn &&\qquad \mbox{}+\frac{\alpha }{4\pi }\frac{1}{2s^{2}_{W}}\left(\frac{1}{3}-
\frac{c^{2}_{W}}{s^{2}_{W}}\right)\log\frac{m^{2}_{t}}{M^{2}_{W}}
\left.+\frac{\alpha }{4\pi }
\frac{11}{6}\frac{1}{2s^{2}_{W}}\log\frac{M^{2}_{H}}{M_{W}^{2}}
+\frac{\alpha \pi }{4\beta}\right]\\[1em]
&&\mbox{}+e^{2}\left(\frac{1}{s}-\frac{1}{s-M^{2}_{Z}}\right)2\left({\cal M}
^{-}_{3}-{\cal M}^{-}_{2}\right)
\left[\frac{1}{1-\Delta \alpha}+\frac{\alpha \pi }{4\beta}\right]\\[1em]
\nn &&\mbox{}+e^{2}\frac{\alpha }{4\pi }\frac{1}{s-M^{2}_{Z}}2({\cal M}^{-}_{3}-
{\cal M}^{-}_{2})\left[\frac{4s^{2}_{W}-3}{12c^{2}_{W}s^{4}_{W}}\log
\frac{m^{2}_{t}}{M^{2}_{W}}-\frac{1}{48c^{2}_{W}s^{4}_{W}}\log
\frac{M^{2}_{H}}{M_{W}^{2}}\right],\\[1em]
\nn{\cal M}_{a}^{+} & = & e^{2}\left(\frac{1}{s}-\frac{1}{s-M^{2}_{Z}}\right)
2({\cal M}^{+}_{3}-{\cal M}^{+}_{2})
\left[\frac{1}{1-\Delta \alpha} +\frac{\alpha\pi}{4\beta}\right]\\[1em]
\nn &&\mbox{}+e^{2}\frac{\alpha }{4\pi }\frac{1}{s-M^{2}_{Z}}2({\cal M}^{+}_{3}-
{\cal M}^{+}_{2})\frac{1}{6s^{2}_{W}c^{2}_{W}}
\left[\log\frac{m^{2}_{t}}{M^{2}_{W}}-\frac{1}{4}\log
\frac{M^{2}_{H}}{M_{W}^{2}}\right].
\eeqar
All terms in (\ref{WWapp1}) respect the high
energy cancellations apart from those involving $\log m_{t}^{2}$ and
$\log M_{H}^{2}$. However, these were obtained for
$M^{2}_{H},m^{2}_{t}\gg s$, whereas
the unitarity cancellations work for $s\gg M^{2}_{H},m^{2}_{t}$.
In this limit the terms containing $\log m_{t}^{2}$ and $\log M_{H}^{2}$
are absent. These may, however, cause large effects for
small energies and large top quark or Higgs boson masses. This
phenomenon was called delayed unitarity cancellation in \cite{Ah88}.
Introducing $G_{F}$ instead of $e^{2}/s^{2}_{W}$ and the running
$\alpha(s) $ we obtain
\beqar \label{WWapp2}
\nn {\cal M}_{a}^{-} & = & 2\sqrt{2}G_{F}M^{2}_{W}\left[\frac{1}{t}{\cal M}^{-}
_{1}+\frac{1}{s-M^{2}_{Z}}2({\cal M}^{-}_{3}-{\cal M}^{-}_{2})\right]
\left[1+\frac{\alpha\pi}{4\beta}+C_{1}^{-}(s,t)\right]\\[1em]
\nn&&\mbox{}+4\pi \alpha (s)\left(\frac{1}{s}-\frac{1}{s-M^{2}_{Z}}\right)
2\left({\cal M}^{-}_{3}-{\cal M}^{-}_{2}\right)
\left[1+\frac{\alpha\pi}{4\beta}+C_{2}^{-}(s,t)\right]\\[1em]
\nn &&\mbox{}+e^{2}\frac{\alpha }{4\pi }\frac{1}{s-M^{2}_{Z}}2({\cal M}^{-}_{3}-
{\cal M}^{-}_{2})\left[\frac{4s^{2}_{W}-3}{12c^{2}_{W}s^{4}_{W}}\log
\frac{m^{2}_{t}}{M^{2}_{Z}}-\frac{1}{48c^{2}_{W}s^{4}_{W}}\log
\frac{M^{2}_{H}}{s}\right],\\[1em]
{\cal M}_{a}^{+} & = & 4\pi \alpha (s)\left(
\frac{1}{s}-\frac{1}{s-M^{2}_{Z}}\right) 2({\cal M}^{+}_{3}-{\cal M}^{+}_{2})
\left[1+\frac{\alpha\pi}{4\beta}+C_{2}^{+}(s,t)\right]\\[1em]
\nn &&\mbox{} + 2\sqrt{2}G_{F}M^{2}_{W}\left[\frac{1}{t}{\cal M}^{+}
_{1}+\frac{1}{s-M^{2}_{Z}}2({\cal M}^{+}_{3}-{\cal M}^{+}_{2})\right]
C_{1}^{+}(s,t)\\[1em]
\nn &&\mbox{}+e^{2}\frac{\alpha }{4\pi }\frac{1}{s-M^{2}_{Z}}2({\cal M}^{+}_{3}-
{\cal M}^{+}_{2})\frac{1}{6s^{2}_{W}c^{2}_{W}}
\left[\log\frac{m^{2}_{t}}{M^{2}_{Z}}-\frac{1}{4}\log
\frac{M^{2}_{H}}{s}\right].
\eeqar
We have included four functions $C_{i}(s,t),\: i=1,2,\:\sigma =\pm$ in
this approximation. These are necessary to describe the angular
dependence of the differential cross section.
The complete one-loop invariant matrix element for $W$-pair production
involves 12 formfactors $F_{i}^{\sigma}$. It turns out, however,
that only four of them namely the $C_{i}^{\sigma}$
are relevant in the LEP200 energy region. For
higher energies even $C_{1}^{+}$ can be omitted. The functions
$C_{i}^{\sigma}$ have been determined such that they reproduce the
corresponding exact one-loop formfactors sufficiently well in the
LEP200 energy region \cite{De91b}.
 
We want to stress that the naive summation of the Dyson series of the
self energies may lead to incorrect results, i.e.\ a wrong high energy
behaviour. This happens because the leading corrections are not only
contained in the self energies but also in the vertex corrections. The
actual place of their appearance depends on the choice of the field
renormalization.
 
\subsection {Numerical results for the virtual and soft photonic
corrections}
 
We now present some numerical results for the radiative corrections in
the soft photon approximation. The numerical input parameters are
defined in Sect.~\ref{secpar}. The soft photon cutoff is chosen as
$\Delta E/E =0.1$. Different choices of $\Delta E/E$ uniformly shift the
absolute value of the corrections but do not influence their angular
dependence very much.
Fig.~\reff{11.5} shows the relative correction factor $\delta $ defined
through
\beq
\frac{d\sigma }{d\Omega }=\left(\frac{d\sigma }{d\Omega
}\right)_{0}(1+\delta_{s} )  +\delta \left(\frac{d\sigma }{d\Omega
}\right)_{V}
=\left(\frac{d\sigma }{d\Omega }\right)_{0}(1+\delta )
\eeq
for the unpolarized case. While the variation with the scattering angle
is relatively flat for LEP200 energies it becomes stronger with
increasing energy. In the forward direction where the Born cross section
is dominated by the $t$-channel pole the energy dependence is very weak.
In the backward direction,
however, the percentage correction varies strongly with energy
and reaches large negative values up to $-50\%$ at $1\:\mbox{TeV}$.
Nevertheless since the
absolute cross section is small for large scattering angles (see
Fig.~\ref{11.3}), the relative corrections to the integrated cross section
stay below $20\%$ up to $1\:\mbox{TeV}$. Note that the one-loop
corrections are large exactly in that region where the sensitivity to
new physics is highest.
\bfi
\includegraphics[bb=20 40 592 422, width=15.cm]{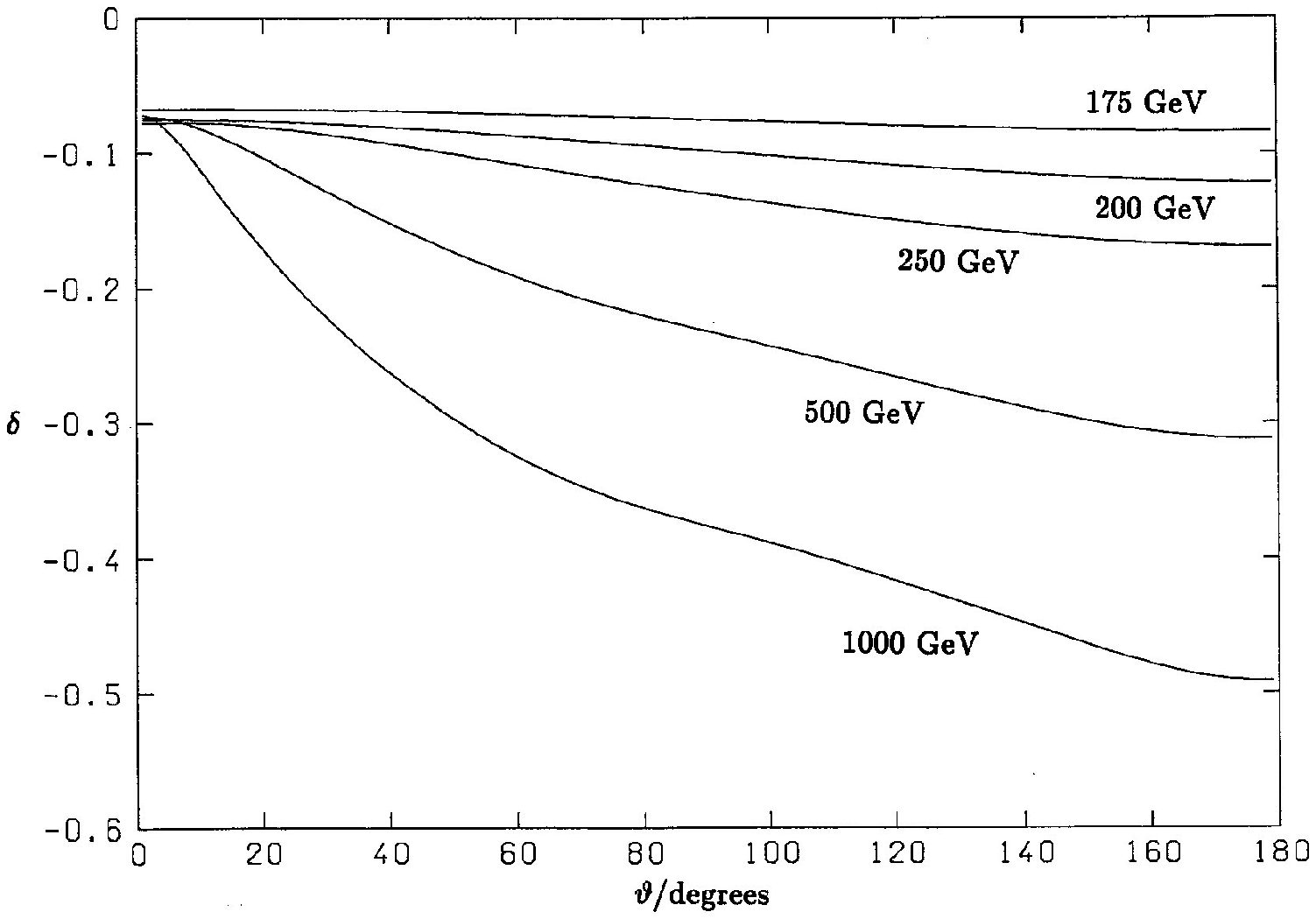}
\caption{Radiative corrections to the differential cross section
relative to lowest order for the unpolarized case at various center
of mass energies.}
\label{11.5}
\efi
 
The behaviour of the corrections for purely transverse $W$-bosons is
similar to the unpolarized case. In contrast to this the
corrections for purely longitudinal bosons (Fig.~\reff{11.6})
 exhibit a strong angular dependence arising from the minima in the
lowest order cross section (Fig.~\reff{11.4}).
\bfi
\includegraphics[bb=20 40 592 422,width=15.cm]{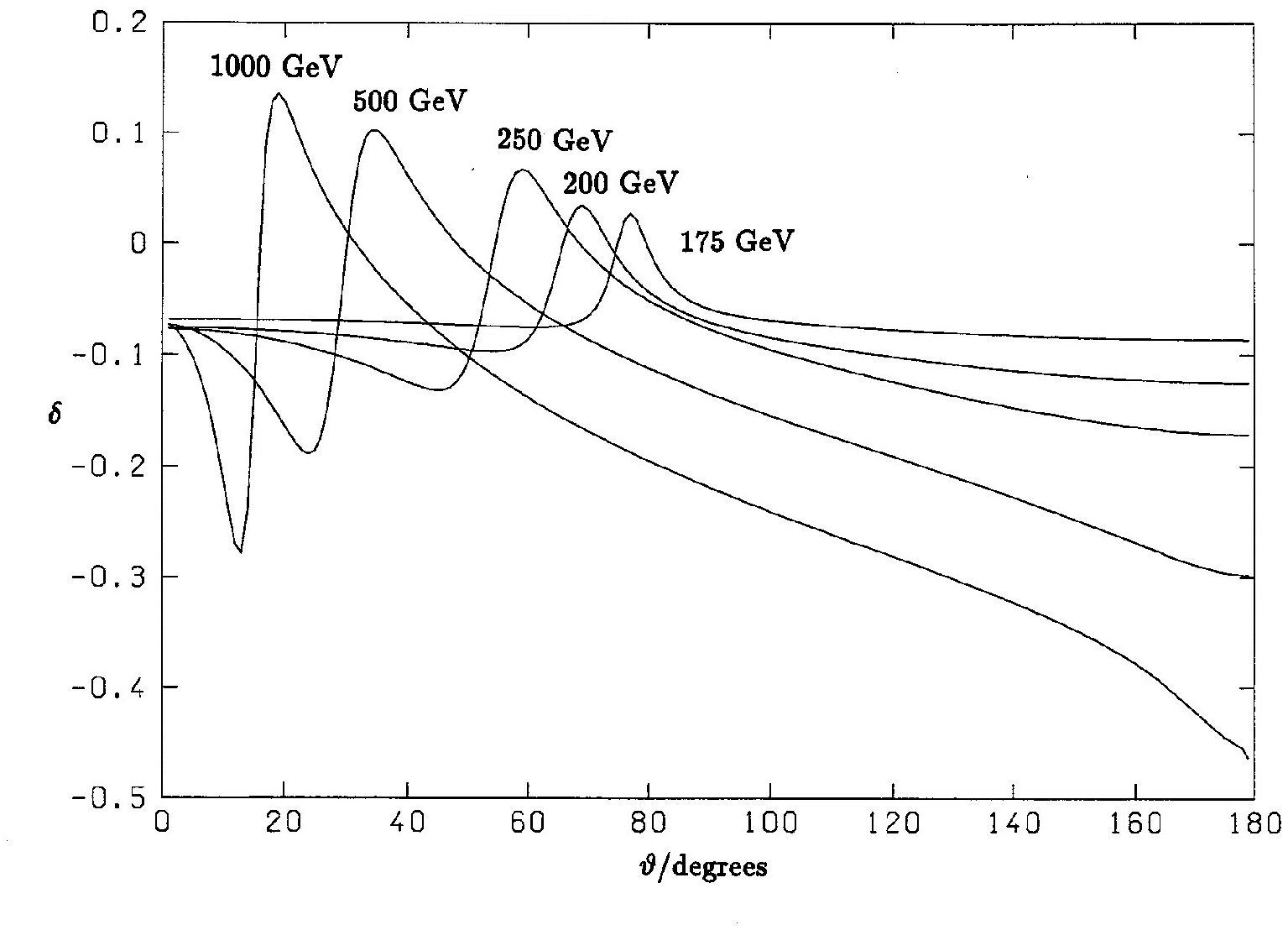}
\caption{Radiative corrections to the differential cross section
relative to lowest order for purely longitudinal $W$-bosons
 at various center of mass energies.}
\label{11.6}
\efi
 
The sensitivity of the total unpolarized cross
section on the unknown masses of the Higgs boson and top quark is
illustrated in Tab.~\ref{tab112} and \ref{tab113}.
A change of $m_{t}$ from 80
to 200 GeV affects the cross section by less than 3\% apart from the
region very close to threshold. The large effect close to threshold is due to
the variation of $M_{W}$ and thus the variation of the threshold
with $m_{t}$. A variation of $M_{H}$
between 50 and $1000\:\mbox{GeV}$ influences the total cross section by less
than 0.5\%, again with the exception of the threshold region.
Note that this is valid for
constant $\alpha$, $G_{F}$ and $M_{Z}$. Fixing instead $G_{F}$, $M_{W}$
and $M_{Z}$
the dependence on $m_{t}$ is much weaker. This allows to determine
$M_{W}$ from the cross section practically independently on $m_{t}$ and
$M_{H}$ as pointed out by Jegerlehner \cite{Je89a}.
These results for the top and Higgs mass dependence remain valid if we
include hard photonic corrections.
\btab
\bce
\begin{tabular}{||r|r|r|r|r||} \hline\hline
 $m_t/\mbox{GeV}$ =  &   80 &  120 &  160 &  200 \\
 \hline
 $M_W/\mbox{GeV}$ =  &    79.87 &    80.10 &    80.37 &    80.69 \\
 \hline
 $\sqrt{s}/\mbox{GeV}$ & \multicolumn{4}{|c||}{$\sigma/\mbox{pb}$} \\
 \hline
165.0 & 10.120 &  9.743 &  9.218 &  8.461 \\
 \hline
 180.0 & 15.521 & 15.620 & 15.646 & 15.626 \\
 \hline
 200.0 & 15.944 & 16.112 & 16.220 & 16.301 \\
 \hline
 500.0 &  5.689 &  5.760 &  5.807 &  5.847 \\
 \hline
1000.0 &  2.064 &  2.088 &  2.103 &  2.113 \\
\hline \hline
\end{tabular}
\ece
\caption{Total unpolarized cross section for $e^{+}e^{-}\to W^{+}W^{-}$
including
virtual and soft photonic corrections for different top quark masses
 at various center of mass energies.
\label{tab112}}
\etab
\btab
\bce
\begin{tabular}{||r|r|r|r|r||} \hline\hline
 $M_H/\mbox{GeV}$ =  &     50 &    100 &    300 &   1000 \\
 \hline
 $M_W$/\mbox{GeV} =  &    80.26 &    80.23 &    80.16 &    80.06 \\
 \hline
 $\sqrt{s}/\mbox{GeV}$ & \multicolumn{4}{|c||}{$\sigma/\mbox{pb}$} \\
 \hline
  165.0 &  9.488 &  9.503 &  9.614 &  9.793 \\
 \hline
  180.0 & 15.654 & 15.638 & 15.612 & 15.598 \\
 \hline
  200.0 & 16.168 & 16.168 & 16.143 & 16.105 \\
 \hline
  500.0 &  5.785 &  5.785 &  5.776 &  5.764 \\
 \hline
 1000.0 &  2.097 &  2.096 &  2.093 &  2.088 \\
 \hline \hline
\end{tabular}
\ece
\caption{Total unpolarized
cross section for $e^{+}e^{-}\to W^{+}W^{-}$ including
virtual and soft photonic corrections for different Higgs boson masses
 at various center of mass energies.
\label{tab113}}
\etab
 
Using the functions $C_{i}^{\sigma} $ given in \cite{De91b}
the improved Born approximation (\ref{WWapp2}) agrees with the
full one-loop order result within
0.5\% for $\sqrt{s}<220\:\mbox{GeV}$ and within 1\%
for $\sqrt{s}<270\:\mbox{GeV}$ in the case of the total cross section.
For the differential cross section the
deviation is at most 1\% for $\sqrt{s}<210\:\mbox{GeV}$.
The largest difference occurs for large scattering angles, i.e.\ where
the cross section is small.
 
\section {Hard photon bremsstrahlung}
 
\subsection {Complete calculations}
 
The complete hard photonic bremsstrahlung to $e^{+}e^{-}\to W^{+}W^{-}$
was determined in \cite{Be90,Ta90}. The polarized amplitudes for the process
$e^{+}e^{-}\to W^{+}W^{-}\gamma $ were calculated using three different
methods. The first one, described in detail in \cite{Ko91} uses the
Weyl representation for Dirac matrices and spinors and results in
expressions for the amplitudes in terms of the components of
momentum and polarization vectors
in the center of mass frame of the incoming leptons.
The second method used in \cite{Be91} is based on the Weyl-van der
Waerden formalism. It yields concise analytical formulae for the
amplitude which are manifestly Lorentz invariant. The relative numerical
difference between both results was found to be less than $10^{-6}$ for
the amplitude squared.  In \cite{Ta90} the amplitudes were calculated
numerically.
 
From this the total cross section is obtained as
\beq \label{DSWWg}
\barr{lll}
\sigma (s)&=&\disp \frac{1}{(2\pi)^{5}}\frac{1}{2s}\int
\frac{d^{3}k_{1}}{2k_{10}}\frac{d^{3}k_{2}}{2k_{20}}\frac{d^{3}k}{2k_{0}}
\frac{1}{4}\sum_{pol} \vert {\cal M}\vert^{2}\delta^{(4)}
(p_{1}+p_{2}-k_{1}-k_{2}-k)\\[1em]
&=&\disp \frac{1}{2s}\frac{1}{8(2\pi )^{4}}\int \,d\cos\vartheta_{2}\,
d\cos\vartheta \, d\phi \, dk_{0}
\left\vert \frac{k_{0}|{\bf k}_{2}|^{2}}{|{\bf k}_{2}|(\sqrt{s}-
k_{0})+k_{0}k_{20}c_{20}}\right\vert\frac{1}{4}\sum_{pol}
\left|{\cal M}\right|^{2},
\earr
\eeq
where $\vartheta_{2}$, $\vartheta $ are the polar angles of the
$W^{-}$-boson and the photon, $\phi $ is the azimuthal angle of the photon
with respect to the incoming electron and
\beq
c_{20}=\sin\vartheta_{2}\sin\vartheta\cos\phi
+\cos\vartheta_{2}\cos\vartheta .
\eeq
The nontrivial phase space integrations are performed using Monte Carlo
routines. Thus experimental cuts can be easily implemented.
 
Eq.~(\ref{DSWWg}) contains the soft photon poles. These are eliminated by
a cut $k_{0}>\Delta E$ on the photon energy. After combining soft and
hard photonic
corrections the cut dependence drops out. This has been checked
numerically for $\Delta E/E$ between $10^{-3}$ and $10^{-6}$.
 
\subsection {Leading logarithmic approximation}
 
The leading logarithmic (LL) QED corrections to the $W$-pair production cross
section were already calculated in \cite{Be89}.
The resulting cross section is given by \cite{Be90}
\beq \label{LLint}
\sigma_{LL}(s)=\int^{1}_{4M^{2}_{W}/s}dz\phi(z)\hat{\sigma}_{0}(zs),
\eeq
where $\hat{\sigma}_{0}(zs)$ denotes the (improved) Born cross section
at the reduced CMS energy squared $zs$. The flux $\phi(z)$ reads
\beqar
\nn \phi (z) & = &\delta (1-z)\\[1ex]
\nn &&\disp \mbox{}+\frac{\alpha }{\pi }(L-1)
\biggl[\delta (1-z)2\log\varepsilon +\theta (1-
\varepsilon -z)\frac{2}{1-z}\biggr]\\[1ex]
&&\disp \mbox{}+\frac{\alpha }{\pi }L
\biggl[\delta (1-z)\frac{3}{2}-\theta (1-\varepsilon -
z)(1+z)\biggr]\\[1ex]
\nn &&\disp \mbox{}+\left(\frac{\alpha }{\pi}L\right)^{2}\biggl\{\delta (1-
z)\left(2\log^{2}\varepsilon +3\log\varepsilon +\frac{9}{8}-\frac{\pi
^{2}}{3}\right)\\[1ex]
\nn &&\disp \mbox{}+\theta (1-\varepsilon -z)
\biggl[\frac{1+z^{2}}{1-z}(2\log(1-z)-
\log z+\frac{3}{2})
\disp+\frac{1}{2}(1+z)\log z-(1-z)\biggr]\biggr\}
\eeqar
where $L=\log (Q^{2}/m_{e}^{2})$ is the leading logarithm and
$\varepsilon = \Delta E/E$ the soft photon cutoff. $\phi(z)$ is given including
$O(\alpha ^{2})$ $LL$-contributions. Furthermore some nonleading terms
are incorporated taking into account the fact that the residue of the
soft photon pole is proportional to $L-1$ rather than $L$ for the
initial state radiation. The scale $Q^{2}$ is a free parameter. It can
only be determined through higher order calculations. In \cite{Be90}
the integral in (\ref{LLint}) was performed numerically. Neglecting the
$O(\alpha ^{2})$ leading logarithms and the nonlogarithmic terms it has
been evaluated analytically \cite{De91b}.
 
\subsection {Numerical results}
 
The results presented in this section were calculated by \cite{Be90,Sa91}.
The parameters are the same as in Sect.~\ref{secpar}. In the Born cross
section $\hat{\sigma}_{0}$, entering the $LL$ approximation, $\alpha $
has been replaced by $G_{F}$ everywhere.
Thus the large fermionic corrections are absorbed
at least in the dominant $t$-channel contributions.
 
The total cross section is plotted in Fig.~\reff{11.7} in the LEP200
energy range. Shown are the Born cross section with $\alpha $ replaced
by $G_{F}$, the cross section 
including the full $O(\alpha)$ corrections and including in addition
the $O(\alpha^{2})$ LL corrections.
The corresponding numbers are given in Tab.~\ref{tab114}
for various CMS energies. The uncertainty of the full $O(\alpha )$
result is due to the Monte Carlo integration of the hard bremsstrahlung
corrections. This error refers also to
the last column of Tab.~\ref{tab114}. The $O(\alpha )$ $LL$ results
were evaluated for two scale choices
\beq
\barr{lll}
Q^{2}&=&s\\
Q^{2}&=&\disp -t_{min} =-M^{2}_{W}+\frac{s}{2}(1-\beta ).
\earr
\eeq
The second one is motivated by the fact that the total cross section is
dominated by the $t$-channel pole. It reproduces the exact
$O(\alpha )$ results better.
The difference is found to be less than 2\% for $\sqrt{s} >
165\:\mbox{GeV}$. Choosing the scale $Q^{2}=s$ the deviation from the
exact $O(\alpha )$ result is about 5\% at $165\:\mbox{GeV}$.
Also at higher energies the scale choice $Q^{2}=-t_{min}$ turns out to
be preferable.
It reproduces the complete $O(\alpha)$ result including hard photon
bremsstrahlung within 1\% for $170\:\mbox{GeV} < \sqrt{s} <
500\:\mbox{GeV}$.
The effect of the $O(\alpha ^{2})$ $LL$ contribution is demonstrated in
the last column of Tab.~\ref{tab114}. It reaches about 5\% at
$165\:\mbox{GeV}$, decreases with increasing energy and is small for
$\sqrt{s} > 190\:\mbox{GeV}$. A practically identical result is obtained
by soft photon exponentiation.
\bfi
\includegraphics[bb=20 40 592 422,width=15.cm]{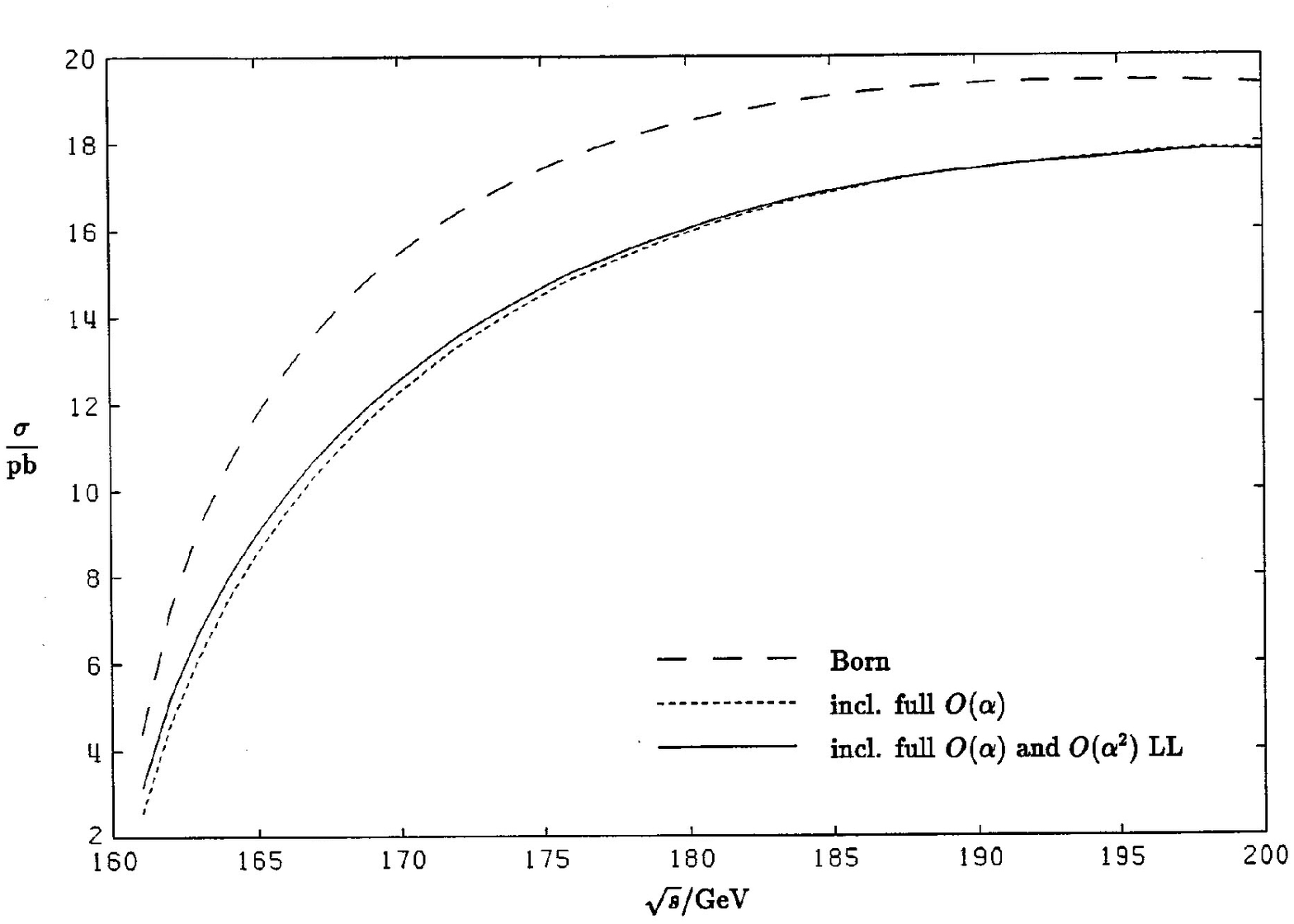}
\caption{Total cross section for $W$-pair production including hard
photonic corrections.}
\label{11.7}
\efi
 
The large deviation between $O(\alpha)$ LL and the exact result close
to threshold is due
to the Coulomb singularity (see Sect.~\ref{secWWleaco}).
It amounts to about 10\% at $1\:\mbox{GeV}$
above threshold and is not included in the $O(\alpha)$ LL result.
Note also the large $O(\alpha^{2})$ correction close to threshold
(28\% at $161\:\mbox{GeV}$).
\begin{table}
\begin{center}
\begin{tabular}{||c|c|c|c|r|c||} \hline\hline
 &  & incl. &  incl. & \mbox{incl.\qquad }  &incl. \\
 $ \sqrt{s}/GeV $ &  Born$(G_F) $ & $ O(\alpha) $ LL
 & $ O(\alpha) $ LL & $O(\alpha) $ exact & $O(\alpha) $ exact  \\
 &   & $Q^2=s$  &  $Q^2=-t_{min} $  &  & $+O(\alpha^{2}) $ LL \\[1ex] \hline
 161.0 &  4.411 &  2.003 &  2.158 &  2.556 $\pm$  0.002 &  3.255 \\ \hline
 165.0 & 11.761 &  8.141 &  8.429 &  8.553 $\pm$  0.006 &  9.089 \\ \hline
 170.0 & 15.465 & 11.967 & 12.285 & 12.264 $\pm$  0.010 & 12.606 \\ \hline
 175.0 & 17.413 & 14.264 & 14.578 & 14.484 $\pm$  0.013 & 14.690 \\ \hline
 180.0 & 18.501 & 15.730 & 16.028 & 15.920 $\pm$  0.014 & 16.033 \\ \hline
 190.0 & 19.361 & 17.272 & 17.525 & 17.373 $\pm$  0.016 & 17.375 \\ \hline
 200.0 & 19.354 & 17.810 & 18.015 & 17.796 $\pm$  0.017 & 17.742 \\ \hline
 250.0 & 16.406 & 16.223 & 16.257 & 16.033 $\pm$  0.023 & 15.937 \\ \hline
 300.0 & 13.473 & 13.734 & 13.682 & 13.543 $\pm$  0.026 & 13.470 \\ \hline
 500.0 &  7.142 &  7.664 &  7.519 &  7.449 $\pm$  0.019 &  7.430 \\ \hline\hline
\end{tabular}
\end{center}
\caption{Total cross section for $e^+e^- \rightarrow W^+W^-$ in $pb$
including hard photonic corrections.}
\label{tab114}
\end{table}
 
\section {Finite widths effects}
\label{secWfiwi}
 
Realistic calculations for $W$-pair production must include the decays
of the $W$-bosons into fermions. These are especially important around
threshold.
 
In real experiments one observes the reaction
\beq
e^{+}e^{-}\to W^{+}W^{-}\to \mbox{final states}.
\eeq
The $W$-bosons give rise to peaks in the invariant mass
distributions of the final state particles.
Therefore one has to calculate the cross section for
$e^{+}e^{-}\to f_{1}\bar{f}_{2}f_{3}\bar{f}_{4}(\gamma ,g)$. This
task has been attacked but not completed so far \cite{Ae91a}.
Above the $W$-pair production threshold the dominant contributions come
from Feynman diagrams containing resonant $W$-propagators.
The cross section for $W$-pair production is obtained by calculating all
diagrams containing two resonant $W$-propagators.
Diagrams contributing to the same final state but without two resonant
propagators
are considered as background. They are suppressed by a factor
$M_{W}/\Gamma ^{W}\approx 40$ if the full range of invariant masses
$\sqrt{s_{i}}$ of the final state particles is included.
If $\sqrt{s_{i}}$ is restricted by a cut $\Delta $
\beq
M_{W}-\Delta < \sqrt{s_{i}} < M_{W}+\Delta ,
\eeq
the suppression is even
$M^{2}_{W}/(\Gamma^{W}\Delta)\approx 300$ for $\Delta \approx 10\:\mbox{GeV}$.
Explicit calculations show that the background contributions are below
1\% for $\sqrt{s}\ge2M_{W}$ \cite{Ae91a}.
It becomes, however, more relevant below the nominal threshold.
But even above threshold nonresonant Born
contributions to $e^{+}e^{-}\to f_{1}\bar{f}_{2}f_{3}\bar{f}_{4}
(\gamma,g)$ must be taken into account to obtain an accuracy of better
than 1\%.
 
\bfi
\savebox{\Vtr}(36,12)[bl]
{\multiput(6,0)(12,4){3}{\oval(12,4)[tl]}
\multiput(6,4)(12,4){3}{\oval(12,4)[br]} }
\savebox{\Vbr}(36,12)[bl]
{\multiput(6,12)(12,-4){3}{\oval(12,4)[bl]}
\multiput(6,8)(12,-4){3}{\oval(12,4)[tr]}}
\savebox{\Ftr}(36,18)[bl]
{ \put(0,0){\vector(2,1){20}} \put(18,9){\line(2,1){18}} }
\savebox{\Fbr}(36,18)[bl]
{ \put(36,0){\vector(-2,1){21}} \put(18,9){\line(-2,1){18}} }
\savebox{\Lr}(36,0)[bl]
{ \put(0,0){\line(1,0){20}} \put(18,0){\line(1,0){18}} }
\savebox{\Ltr}(36,12)[bl]
{ \put(0,0){\line(3,1){20}} \put(18,6){\line(3,1){18}} }
\savebox{\Lbr}(36,12)[bl]
{ \put(36,0){\line(-3,1){21}} \put(18,6){\line(-3,1){18}} }
\bma
\begin{picture}(280,120)
\put(5,84){$e^{+}$}
\put(5,30){$e^{-}$}
\put(101,80){$W^{+}$}
\put(101,34){$W^{-}$}
\put(170,80){$W^{+}$}
\put(170,34){$W^{-}$}
\put(20.11,33.06){\usebox{\Ftr}}
\put(20.11,68.94){\usebox{\Fbr}}
\put(74,60){\circle{40}}
\put(92.974,66.325){\usebox{\Vtr}}
\put(92.974,41.675){\usebox{\Vbr}}
\put(142.255,82.752){\circle{28}}
\put(142.255,37.248){\circle{28}}
\put(155.538,87.75){\usebox{\Vtr}}
\put(155.538,20.25){\usebox{\Vbr}}
\put(206.715,104.239){\circle{32}}
\put(206.715,15.761){\circle{32}}
\put(221.894,109.298){\usebox{\Ltr}}
\put(221.894,88.444){\usebox{\Lbr}}
\put(221.894,19.556){\usebox{\Ltr}}
\put(221.894,-1.298){\usebox{\Lbr}}
\put(240,100.5){$\vdots$}
\put(240,10.6){$\vdots$}
\end{picture}
\ema
\caption{General structure of diagrams containing two resonant
$W$-propagators.}
\label{11.8}
\efi
There are three types of diagrams which may give resonant contributions.
The most important ones are factorizable diagrams
with the structure shown in Fig.~\ref{11.8}
which evidently contain two resonant $W$-propagators. The corresponding
cross section is given by
\beq \label{SiWWFW}
\sigma (s)=\int^{(M_{W}+\Delta)^2}_{(M_{W}-\Delta)^2}ds_{1}ds_{2}\sigma^{*}
(s,s_{1},s_{2})\rho(s_{1})\rho(s_{2})\theta (\sqrt{s}
-\sqrt{s_{1}}-\sqrt{s_{2}}),
\eeq
where $\sigma^{*}(s,s_{1},s_{2})$ is the 'cross section' for the
production of two off-shell $W$-bosons and
\beq
\rho(s)=\frac{1}{\pi }\frac{\sqrt{s}\Gamma ^{W}(s)}
{(s-M^{2}_{W})^{2}+s(\Gamma ^{W}(s))^{2}},
\eeq
with the 'decay width' $\Gamma ^{W}(s)$ for an off-shell $W$-boson.
Note that
\beq
\rho (s)\to \delta (s-M^{2}_{W}) \mbox{\quad for \quad} \Gamma ^{W}\to 0.
\eeq
The off-shell quantities $\sigma^{*}(s,s_{1},s_{2})$ and $\Gamma
^{W}(s)$ are not gauge invariant. However, the leading resonant
contributions to $\sigma (s)$ are.
Eq.~(\ref{SiWWFW}) closely resembles a Breit-Wigner
approximation for the unstable $W$-bosons.
In the threshold region $\sigma^{*}(s,s_{1},s_{2})$ depends strongly
on $s_{1}$ and $s_{2}$. Consequently $\sigma(s)$ deviates
considerably from $\sigma^{*}(s,M_{W}^{2},M_{W}^{2})$, the cross section for
on-shell stable $W$'s. Fig.~\reff{11.9} \cite{Sa91} shows this effect in
lowest order and with the full $O(\alpha )$ corrections to $\sigma
^{*}$ and $\Gamma^{W}(s)$ included.
This dependence is mainly due to the threshold factor $\kappa ^{1/2}
(s,s_{1},s_{2})$ contained in $\sigma^{*}(s,s_{1},s_{2})$. Extracting
this factor, $\sigma^{*}/\kappa^{1/2}$ depends only weakly on
$s_{1}$ and $s_{2}$. This is also the case for $\Gamma ^{W}(s)/\sqrt{s}$
with respect to $s$.
Replacing these quantities by their on-shell values we find the
following approximation
\beq
\sigma(s) \approx \frac{\sigma^{*}(s,M_{W}^{2},M_{W}^{2})}
{\kappa ^{1/2}(s,M_{W}^{2},M_{W}^{2})}\\
\int^{(M_{W}+\Delta)^2 }_{(M_{W}-\Delta)^2 }
 ds_{1}ds_{2}\kappa ^{1/2}(s,s_{1},s_{2})\tilde{\rho}(s_{1})
\tilde{\rho}(s_{2}) \theta
(\sqrt{s}-\sqrt{s_{1}}-\sqrt{s_{2}})
\eeq
with
\beq
\tilde{\rho}(s)=\frac{1}{\pi }\frac{\frac{s}{M_{W}}\Gamma ^{W}(M_{W}^{2})}
{(s-M^{2}_{W})^{2}+\frac{s^{2}}{M_{W}^{2}}\bigl(\Gamma ^{W}(M_{W}^{2})\bigr)^{2}} .
\eeq
This approximation is gauge invariant because
$\sigma^{*}(s,M_{W}^{2},M_{W}^{2})$ and $\Gamma ^{W}(M_{W}^{2})$ are
physical on-shell quantities. It is particulary useful above the nominal
threshold, whereas it gets worse below threshold because there at least one of
the $W$-bosons has to be off-shell.
 
\bfi
\includegraphics[bb=20 40 592 402,width=15.cm]{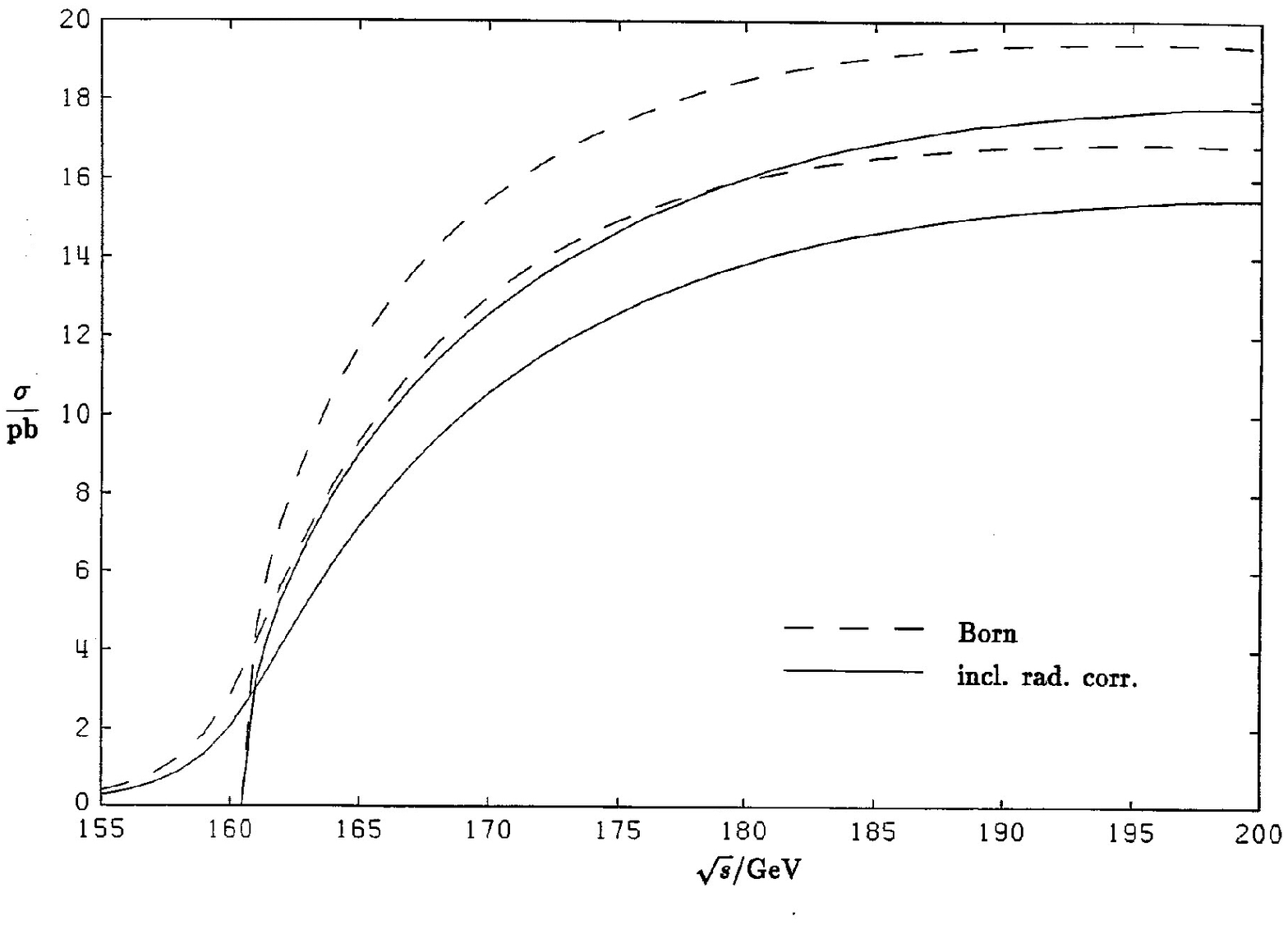}
\caption{Total cross section for $W$-pair production in lowest order
and including the full $O(\alpha )$ corrections with and
without finite width effects.}
\label{11.9}
\efi
For high energies ($s\gg M_{W}^{2}$) also $\kappa (s,s_{1},s_{2})$ varies only
weakly with $s_{1}$ and $s_{2}$ in the resonance region $s_{1}
\approx s_{2}\approx M^{2}_{W}$.
Replacing it by its on-shell value we can perform the integrations and
obtain
\beq
\sigma(s)\approx \sigma^{*}(s,M_{W}^{2},M_{W}^{2}),\quad \mbox{for}\quad
s\gg M_{W}\quad \mbox{and}\quad \Delta \gg \Gamma ^{W}.
\eeq
Eq.~(\ref{SiWWFW}) incorporates all resonant lowest order contributions and all
one-loop corrections associated either with the production of $W$-pairs
or the decay of the $W$-bosons (given in Chap.~\ref{chaWff}). This includes
in particular all self energy and vertex corrections and thus all
leading corrections. A more thorough analysis of this kind
has been carried through for the case of $Z$-pair production \cite
{De90b}.
 
Feynman diagrams which do not fit into the structure shown in
Fig.~\ref{11.8} give nonresonant contributions and can
thus be neglected with the exception of two classes.
Both originate from photonic corrections. The first type results from
virtual photons exchanged between the external lines connected to different
blobs in Fig.~\reff{11.8}. An example is shown in Fig.~\reff{11.10}a. These
diagrams give rise to resonant contributions coming from photons which are
nearly on-shell. From similar cases in $\mu $-pair production we know
that these are cancelled by the corresponding bremsstrahlung diagrams if
one integrates over the whole photon phase space. For stringent cuts,
however, resonant contributions survive.
 
The second type of diagrams consists of those where a real photon is
emitted from the internal $W$-boson line (Fig.~\reff{11.10}b).
There are three $W$-propagators in the diagram.
If the photon is hard the corresponding resonances appear in three
different regions of phase space.
Therefore these diagrams seem not to fit into the
simple Breit-Wigner-like picture discussed above.
\bfi
\savebox{\Vtr}(36,24)[bl]
{\multiput(3,0)(6,4){6}{\oval(6,4)[tl]}
\multiput(3,4)(6,4){6}{\oval(6,4)[br]} }
\savebox{\Vr}(36,0)[bl]
{\multiput(3,0)(12,0){3}{\oval(6,4)[t]}
\multiput(9,0)(12,0){3}{\oval(6,4)[b]} }
\savebox{\Vt}(0,36)[bl]
{\multiput(0,3)(0,12){3}{\oval(4,6)[r]}
\multiput(0,9)(0,12){3}{\oval(4,6)[l]} }
\savebox{\Ft}(0,48)[bl]
{ \put(0,0){\vector(0,1){27}} \put(0,24){\line(0,1){24}} }
\savebox{\Fr}(36,0)[bl]
{ \put(0,0){\vector(1,0){20}} \put(18,0){\line(1,0){18}} }
\savebox{\Fl}(36,0)[bl]
{ \put(36,0){\vector(-1,0){20}} \put(18,0){\line(-1,0){18}} }
\savebox{\Ftr}(36,12)[bl]
{ \put(0,0){\vector(3,1){20}} \put(18,6){\line(3,1){18}} }
\savebox{\Fbr}(36,12)[bl]
{ \put(36,0){\vector(-3,1){21}} \put(18,6){\line(-3,1){18}} }
\savebox{\Ltr}(36,12)[bl]
{ \put(0,0){\line(3,1){20}} \put(18,6){\line(3,1){18}} }
\savebox{\Lbr}(36,12)[bl]
{ \put(36,0){\line(-3,1){21}} \put(18,6){\line(-3,1){18}} }
\bma
\barr{ll}
\begin{picture}(168,72)
\put(-20,65){a)}
\put(90,33){$\gamma$}
\put(48,12){\circle*{4}}
\put(48,60){\circle*{4}}
\put(84,12){\circle*{4}}
\put(84,60){\circle*{4}}
\put(102,18){\circle*{4}}
\put(102,54){\circle*{4}}
\put(12,12){\usebox{\Fr}}
\put(12,60){\usebox{\Fl}}
\put(48,12){\usebox{\Ft}}
\put(48,12){\usebox{\Vr}}
\put(48,60){\usebox{\Vr}}
\put(84,60){\usebox{\Ftr}}
\put(84,48){\usebox{\Lbr}}
\put(93,57){\vector(-3,1){3}}
\put(111,51){\vector(-3,1){3}}
\put(84,12){\usebox{\Ltr}}
\put(93,15){\vector(3,1){3}}
\put(111,21){\vector(3,1){3}}
\put(84,00){\usebox{\Fbr}}
\put(102,18){\usebox{\Vt}}
\end{picture}
&
\begin{picture}(132,72)
\put(-20,65){b)}
\put(110,33){$\gamma$}
\put(48,12){\circle*{4}}
\put(48,60){\circle*{4}}
\put(84,12){\circle*{4}}
\put(84,60){\circle*{4}}
\put(66,12){\circle*{4}}
\put(12,12){\usebox{\Fr}}
\put(12,60){\usebox{\Fl}}
\put(48,12){\usebox{\Ft}}
\put(48,12){\usebox{\Vr}}
\put(48,60){\usebox{\Vr}}
\put(84,60){\usebox{\Ftr}}
\put(84,48){\usebox{\Fbr}}
\put(84,12){\usebox{\Ftr}}
\put(84,00){\usebox{\Fbr}}
\put(66,12){\usebox{\Vtr}}
\end{picture}
\earr
\ema
\caption{Examples for additional diagrams leading to resonant contributions.}
\label{11.10}
\efi
 
In order to take into account these resonant contributions properly one
has to calculate the virtual and real photonic corrections to $e^{+}e^{-}
\to 4 \:\mbox{fermions}$. This has been done for real photon radiation
\cite{Ae91}. Evaluation of the virtual photonic corrections is under way
\cite{Ae91a}.

\chapter{Conclusion}
 
With the electroweak standard model (SM) we have a theory that describes
all known
experimental facts about the electroweak interaction. It has succesfully
survived all precision experiments at low energies and at LEP100. The
upcoming experiments at LEP200, HERA and the planned hadron colliders
will allow to investigate sectors of the SM which were not directly
accessible so far. For a conclusive confrontation of future
experimental results with the SM precise predictions are mandatory.
 
For the adequate theoretical description of the experiments at LEP100
the calculation of radiative corrections was inevitable.
Although experiments outside the $Z$-region will not profit from the presence
of a resonance, the expected experimental accuracy will be such that
radiative corrections will be indispensable.
Moreover radiative corrections allow to obtain information on otherwise not
accessible quantities such as the mass of the top quark or the Higgs boson.
 
One of the next important classes  of
experiments will be the investigation of the $W$-boson
and its nonabelian couplings at LEP200. We have presented the
relevant formulae necessary for the corresponding higher order
calculations. Together with \cite{Bo88,Be91} these cover the complete analytical
expressions for processes with on-shell $W$-bosons. The one-loop virtual
corrections are settled for the polarized differential and total cross
section. Also hard bremsstrahlung has been calculated by  several
authors. We have given an improved Born approximation including the
leading two-loop contributions for the total and
differential cross section. 
The effects of the finite width of the $W$-bosons have been discussed for the
lowest order cross section and the cross section including radiative
corrections. While the
inclusion of the non-photonic corrections is simple the correct
simultaneous treatment of photonic corrections and finite width
effects  involves nonfactorizable box diagrams. These contributions are
under consideration.
 
We have discussed the total and partial $W$-decay widths including all
one-loop and leading two-loop corrections. Because the $W$-boson decays
only into light fermions the widths can be described by a very simple
expression with an accuracy better than 0.6\%.
 
Furthermore we have given results on the $t$-quark decay width. Also in
this case the electroweak corrections can be incorporated into a simple
approximation valid for a top mass below 250 GeV with an accuracy of
about 1.7\%.
 
Apart from giving these explicit results we have
discussed many techniques needed for the calculation of one-loop
corrections. We have compiled a comprehensive set of formulae which are
relevant for the calculation of one-loop radiative corrections within
and outside the standard model.
We have listed the complete set of Feynman rules for the electroweak
standard model including the counter terms. These were expressed by the
self energies of the physical particles in terms of two-point functions.
We have given explicit results for the scalar N-point functions for
$N = 1,\dots,4$ and the relevant formulae for the reduction of the
higher scalar functions and the tensor functions to those. Furthermore
we have outlined a general strategy for the calculation of one-loop
diagrams. Finally we have given the general expressions for the soft
photonic corrections.
 
If the SM will prove to describe the upcoming experimental results
succesfully, further precision checks will become necessary and a lot
more calculations of radiative corrections will be required. These
calculations will be even more involved than the existing ones because
the structure of the corresponding physical processes will in general
be more complicated. The techniques and formulae presented in this
review are general enough to
serve as a basis for the evaluation of radiative corrections to
reactions which will be studied at future colliders such as gauge boson
scattering processes ($W^{+}W^{-}\to W^{+}W^{-}$), electron photon reactions
($e\gamma \to \nu_{e}W$), reactions with three or more final state
particles and so on.
 
The methods described here have been implemented in
the computer algebra package \FC\/. Many of the quoted formulae are
included in this package. We hope that this compilation together with the
packages \FC\/\ and \FA\/\ can serve as a useful tool, facilitating future
calculations of radiative corrections.
 
\vspace{2.5cm}
 
{\bf Acknowledgements}
 
I would like to thank M.~B\"ohm for his support and permanent
encouragement during the past years. Many of the results presented
here were calculated together with T.~Sack who also provided some of the
figures and tables. The results on
$W$-pair-production were obtained in a fruitful collaboration with
W.~Beenakker,
F.~A.~Berends, H.~Kuijf, M.~B\"ohm and T.~Sack.
I have profited from many stimulating
and clarifying discussions with the people mentioned above and
H.~Spiesberger, W.~Hollik, J.~H.~K\"uhn,
F.~Jegerlehner and W.~L.~van Neerven. I am grateful to R.~Mertig,
J.~K\"ublbeck, R.~Guth, H.~Eck, R.~Scharf, U.~Nierste and S.~Dittmaier for their
assistance. I am indebted to my wife R.~Denner
for typing parts of the manuscript and in particular for her patience.
 
%

\appendix
 
\savebox{\Gr}(48,0)[bl]
{ \put(23,0){\vector(1,0){3}} \multiput(0,0)(4.8,0){11}{\circle*1{1}} }
\savebox{\Gtbr}(32,48)[bl]
{ \multiput(0,24)(4,3){9}{\circle*{1}}
 \multiput(0,24)(4,-3){9}{\circle*{1}}
 \put(16,12){\vector(-4,3){3}} \put(16,36){\vector(4,3){3}} }
 
\savebox{\Vr}(48,0)[bl]
{\multiput(3,0)(12,0){4}{\oval(6,4)[t]}
\multiput(9,0)(12,0){4}{\oval(6,4)[b]} }
\savebox{\Vtr}(32,24)[bl]
{\multiput(4,0)(8,6){4}{\oval(8,6)[tl]}
\multiput(4,6)(8,6){4}{\oval(8,6)[br]} }
\savebox{\Vbr}(32,24)[bl]
{\multiput(4,24)(8,-6){4}{\oval(8,6)[bl]}
\multiput(4,18)(8,-6){4}{\oval(8,6)[tr]}}
\savebox{\Vtbr}(32,48)[bl]
{\put(00,24){\usebox{\Vtr}}
\put(00,00){\usebox{\Vbr}}}
 
\savebox{\Sr}(48,0)[bl]
{ \multiput(0,0)(12.5,0){4}{\line(4,0){10}} }
\savebox{\Str}(32,24)[bl]
{ \multiput(-2,-1.5)(12,9){3}{\line(4,3){10}} }
\savebox{\Sbr}(32,24)[bl]
{\multiput(-2,25.5)(12,-9){3}{\line(4,-3){10}} }
\savebox{\Stbr}(32,48)[bl]
{\put(00,24){\usebox{\Str}}
\put(00,00){\usebox{\Sbr}}}
 
\savebox{\Fr}(48,0)[bl]
{ \put(0,0){\vector(1,0){26}} \put(24,0){\line(1,0){24}} }
\savebox{\Ftr}(32,24)[bl]
{ \put(0,0){\vector(4,3){18}} \put(16,12){\line(4,3){16}} }
\savebox{\Fbr}(32,24)[bl]
{ \put(32,0){\vector(-4,3){19}} \put(16,12){\line(-4,3){16}} }
\savebox{\Ftbr}(32,48)[bl]
{\put(00,24){\usebox{\Ftr}}
\put(00,00){\usebox{\Fbr}}}
 
\chapter{Feynman rules}
 
In this appendix we list the Feynman rules of the SM in the
't~Hooft-Feynman gauge including the counter terms in a
way appropriate for the concept of generic diagrams. I.e.\ we write down
generic Feynman rules and give the possible actual insertions.
We omit any field renormalization constants for the unphysical
fields. For brevity we introduce the shorthand notation
\beq
c = \cw,\qquad s= \sw.
\eeq
In the vertices all momenta are considered as incoming.
\vspace{2mm}
 
Propagators:
\vspace{2mm}
 
for gauge bosons $V = \gamma,\,\PZ,\,\PW$
in the 't Hooft Feynman gauge ($\xi_{i}=1$)
\bma
\barr{l}
\framebox{
\begin{picture}(90,24)
\put(33,16){\makebox(20,10)[bl] {$k$}}
\put(-1,9){\makebox(12,10)[bl] {$\PV_{\mu}$}}
\put(67,9){\makebox(12,10)[bl] {$\PV_{\nu}$}}
\put(15,12){\usebox{\Vr}}
\put(15,12){\circle*{4}}
\put(63,12){\circle*{4}}
\end{picture} }
\earr
\barr{l}
\disp = \frac{-ig_{\mu \nu }}{k^{2}-\MV^{2}},
\earr
\ema
 
for Faddeev-Popov ghosts $G = u^{\gamma},\,u^{\PZ},\,u^{\PW}$
\bma
\barr{l}
\framebox{
\begin{picture}(90,24)
\put(34,16){\makebox(20,10)[bl] {$k$}}
\put(0,9){\makebox(12,10)[bl] {$G$}}
\put(67,9){\makebox(12,10)[bl] {$\bar{G}$}}
\put(15,12){\usebox{\Gr}}
\put(15,12){\circle*{4}}
\put(63,12){\circle*{4}}
\end{picture} }
\earr
\barr{l}
\disp = \frac{i}{k^{2}-M_{G}^{2}},
\earr
\ema
 
for scalar fields $S = H,\,\chi,\,\phi$
\bma
\barr{l}
\framebox{
\begin{picture}(90,24)
\put(34,16){\makebox(20,10)[bl] {$k$}}
\put(0,9){\makebox(12,10)[bl] {$S$}}
\put(67,9){\makebox(12,10)[bl] {$S$}}
\put(15,12){\usebox{\Sr}}
\put(15,12){\circle*{4}}
\put(63,12){\circle*{4}}
\end{picture} }
\earr
\barr{l}
\disp= \frac{i}{k^{2}-M_{S}^{2}},
\earr
\ema
 
and for fermion fields $F = f_{i}$
\bma
\rule{2mm}{0mm}\barr{l}
\framebox{
\begin{picture}(90,24)
\put(34,17){\makebox(20,10)[bl] {$p$}}
\put(0,9){\makebox(12,10)[bl] {$F$}}
\put(67,9){\makebox(12,10)[bl] {$\bar{F}$}}
\put(15,12){\usebox{\Fr}}
\put(15,12){\circle*{4}}
\put(63,12){\circle*{4}}
\end{picture} }
\earr
\barr{l}
\disp= \frac{i(\ps + m_{F})}{p^{2}-m_{F}^{2}}.
\earr
\ema
 
In the 't Hooft-Feynman gauge we have the following relations:
\beq
M_{u^{\gamma}} = 0,\qquad
M_{u^{\PZ}} = M_{\chi} = \MZ,\qquad
M_{u^{\pm}} = M_{\phi} = \MW.
\eeq
 
Tadpole:
\bma
\barr{l}
\framebox {
\begin{picture}(72,24)
\put(50,15){\makebox(10,10)[bl]{$S$}}
\put(7.5,6){\line(3,4){9}}
\put(7.5,18){\line(3,-4){9}}
\put(12,12){\usebox{\Sr}}
\end{picture} }
\earr
\barr{l}
\disp = i\delta t.
\earr
\ema
\newcommand{\dt}{\frac{e}{2s}\frac{\delta t}{\MW}}
\newcommand{\dth}{\frac{e}{2s}\frac{\delta t}{\MW\MH^2}}
 
{\samepage
VV-counterterm:
\bma
\barr{l}
\framebox{
\begin{picture}(120,29)
\put(100,19){\makebox(10,20)[bl]{$V_{2,\nu}$}}
\put(0,19){\makebox(10,20)[bl] {$V_{1,\mu},k$}}
\put(55.5,6){\line(3,4){9}}
\put(55.5,18){\line(3,-4){9}}
\put(12,12){\usebox{\Vr}}
\put(60,12){\usebox{\Vr}}
\end{picture} }
\earr
\barr{l}
= -ig_{\mu\nu}\Bigl[C_{1}k^{2} - C_{2} \Bigr]
\earr
\ema
with the actual values of $V_{1},\:V_{2}$ and $C_{1},\:C_{2}$
\beq
\begin{array}[b]{l@{\quad : \quad}ll}
\PWp\PWm & C_{1} = \DZW, & C_{2} = \MW^{2}\DZW + \DMWS \co \\[1ex]
\PZ\PZ & C_{1} = \DZZ, & C_{2} = \MZ^{2}\DZZ + \DMZS \co \\[1ex]
\PA\PZ & C_{1} = \frac{1}{2}\DZAZ + \frac{1}{2}\DZZA , \qquad &
C_{2} = \MZ^{2}\frac{1}{2}\DZZA \co \\[1ex]
\PA\PA & C_{1} = \DZA, & C_{2} = 0.
\earr
\eeq
}
 
{\samepage
SS-counterterm:
\bma
\barr{l}
\framebox{
\begin{picture}(120,27)
\put(100,18){\makebox(10,20)[bl]{$S_{2}$}}
\put(0,18){\makebox(10,20)[bl] {$S_{1},k$}}
\put(55.5,6){\line(3,4){9}}
\put(55.5,18){\line(3,-4){9}}
\put(12,12){\usebox{\Sr}}
\put(60,12){\usebox{\Sr}}
\end{picture} }
\earr
\barr{l}
= i\Bigl[C_{1}k^{2} - C_{2} \Bigr]
\earr
\ema\samepage
with the actual values of $S_{1},\:S_{2}$ and $C_{1},\:C_{2}$
\beq
\begin{array}[b]{l@{\quad : \quad}ll}
\PH\PH & C_{1} = \delta Z_H, \quad & C_{2} = \MH^{2}\DZH + \DMHS \co\\[1em]
\chi\chi & C_{1} = 0, \quad & C_{2} = -\dt + \DMZS \co\\[1em]
\phi^+\phi^- & C_{1} = 0, \quad & C_{2} = -\dt + \DMWS .
\earr
\eeq
}
 
{\samepage
FF-counterterm:
\bma
\barr{l}
\framebox{
\begin{picture}(120,27)
\put(100,18){\makebox(10,20)[bl]{$\bar{F}_{2}$}}
\put(0,18){\makebox(10,20)[bl] {$F_{1},p$}}
\put(55.5,6){\line(3,4){9}}
\put(55.5,18){\line(3,-4){9}}
\put(12,12){\usebox{\Fr}}
\put(60,12){\usebox{\Fr}}
\end{picture} }
\earr
\barr{l}
= i\Bigl[C_{L}\ps\omega_{-} + C_{R}\ps\omega_{+} - C_{S}^{-}\omega_{-} -
C_{S}^{+}\omega_{+} \Bigr]
\earr
\ema\samepage
with the actual values of $F_{1},\:\bar{F}_{2}$ and
$C_{L},\:C_{R},\:C_{S}^{-},\:C_{S}^{+}$
\beq
\begin{array}[b]{l@{\quad : \quad}l}
f_{j} \bar{f}_{i} & \left\{
\barr{l} C_{L} = \frac{1}{2}\left(\delta Z_{ij}^{f,L}+ \delta
Z_{ij}^{f,L\dagger}\right),
\qquad C_{R} = \frac{1}{2}\left(\delta Z_{ij}^{f,R}+ \delta
Z_{ij}^{f,R\dagger}\right)\co\\[1em]
C_{S}^{-} = m_{f,i}\frac{1}{2}\delta Z_{ij}^{f,L}
+ m_{f,j}\frac{1}{2}\delta Z_{ij}^{f,R\dagger} + \delta_{ij}\delta
m_{f,i}\co\\[1em]
C_{S}^{+} = m_{f,i}\frac{1}{2}\delta Z_{ij}^{f,R}
+ m_{f,j}\frac{1}{2}\delta Z_{ij}^{f,L\dagger} + \delta_{ij}\delta
m_{f,i}.
\earr \right.
\earr
\eeq
}
 
{\samepage
VVVV-coupling:
\bma
\barr{l}
\framebox{
\begin{picture}(96,77)(0,-2)
\put(6,65){\makebox(10,20)[bl]{$V_{1,\mu}$}}
\put(72,65){\makebox(10,20)[bl]{$V_{3,\rho}$}}
\put(6,0){\makebox(10,20)[bl]{$V_{2,\nu}$}}
\put(72,0){\makebox(10,20)[bl]{$V_{4,\sigma}$} }
\put(48,36){\circle*{4}}
\put(16,12){\usebox{\Vtr}}
\put(16,36){\usebox{\Vbr}}
\put(48,12){\usebox{\Vtbr}}
\end{picture} }
\earr
\barr{l}
= ie^{2}C \Bigl[2g_{\mu\nu}g_{\sigma \rho } -
g_{\nu\rho}g_{\mu \sigma } - g_{\rho\mu}g_{\nu \sigma }\Bigr]
\earr
\ema\samepage
with the actual values of $V_{1},\:V_{2},\:V_{3},\:V_{4}$ and $C$
\beq
\begin{array}[b]{l@{\quad : \quad}l}
\PWp \PWp \PWm \PWm & C = \frac{1}{s^{2}}
\Bigl[1+2\DZe -2\frac{\delta s}{s}+ 2\DZW
\Bigr]\co\\[1em]
\PWp \PWm \PZ \PZ & C = -\frac{c^{2}}{s^{2}}
\Bigl[1 + 2\DZe - 2\frac{1}{c^{2}}\frac{\delta s}{s}
+ \DZW + \DZZ \Bigr]
+ \frac{c}{s}\DZAZ\co\\[1em]
\PWp \PWm \PA \PZ & \left\{
\barr{lll} C &=& \frac{c}{s}
\Bigl[1 + 2\DZe -\frac{1}{c^{2}}\frac{\delta s}{s}
+ \DZW + \frac{1}{2}\DZZ + \frac{1}{2}\DZA \Bigr] \\[1ex]
&&\quad -\frac{1}{2}\DZAZ
-\frac{1}{2}\frac{c^{2}}{s^{2}}\DZZA \co\\[1ex]
\earr \right.\\[1em]
\PWp \PWm \PA \PA & C = - \Bigl[1 + 2\DZe + \DZW + \DZA \Bigr]
+ \frac{c}{s}\DZZA.
\earr
\eeq
}
 
{\samepage
VVV-coupling:
\bma
\barr{l}
\framebox{
\begin{picture}(96,77)(0,-2)
\put(60,65){\makebox(10,20)[bl]{$V_{2,\nu},k_{2}$}}
\put(0,43){\makebox(10,20)[bl] {$V_{1,\mu},k_{1}$}}
\put(60,0){\makebox(10,20)[bl] {$V_{3,\rho},k_{3}$}}
\put(48,36){\circle*{4}}
\put(0,36){\usebox{\Vr}}
\put(48,12){\usebox{\Vtbr}}
\end{picture} }
\earr
\barr{l}
= -ieC \Bigl[g_{\mu \nu }(k_{2}-k_{1})_{\rho}
+g_{\nu\rho}(k_{3}-k_{2})_{\mu} +g_{\rho\mu}(k_{1}-k_{3})_{\nu}\Bigr]
\earr
\ema\samepage
with the actual values of $V_{1},\:V_{2},\:V_{3}$ and $C$
\beq
\begin{array}[b]{l@{\quad : \quad}l}
\PA \PWp \PWm & C = 1+\DZe+\DZW+\frac{1}{2}\DZA
-\frac{1}{2}\frac{c}{s}\DZZA\co\\[1ex]
\PZ \PWp \PWm & C = -\frac{c}{s}(1 + \DZe
- \frac{1}{c^{2}}\frac{\delta s}{s}
+ \DZW + \frac{1}{2}\DZZ ) + \frac{1}{2}\DZAZ.
\earr
\eeq
}
 
{\samepage
SSSS-coupling:
\bma
\barr{l}
\framebox {
\begin{picture}(96,77)(0,-2)
\put(9,65){\makebox(10,20)[bl]{$S_{1}$}}
\put(75,65){\makebox(10,20)[bl]{$S_{3}$}}
\put(9,0){\makebox(10,20)[bl]{$S_{2}$}}
\put(75,0){\makebox(10,20)[bl]{$S_{4}$}}
\put(48,36){\circle*{4}}
\put(18,14){\usebox{\Str}}
\put(18,34){\usebox{\Sbr}}
\put(48,12){\usebox{\Stbr}}
\end{picture} }
\earr
\barr{l}
= ie^{2}C
\earr
\ema\samepage
with the actual values of $S_{1},\:S_{2},\:S_{3},\:S_{4}$ and $C$
\beq
\begin{array}[b]{l@{\quad : \quad}l@{\quad}l@{\qquad}l@{\quad}l}
\PH \PH \PH \PH & C = - \frac{3}{4s^{2}} \frac{\MH^{2}}{\MW^{2}}
\Bigl[1+2\DZe -2\frac{\delta s}{s}
+\frac{\DMHS}{\MH^{2}} +\dth -\frac{\delta \MW^{2}}{\MW^{2}}
+ 2\DZH \Bigr] \co\\[1ex]
\barr{l} \PH \PH \chi \chi \\ \PH \PH \phi \phi \earr \left.\rule[-
1.6ex]{0mm}{4ex}\right\}
& C = - \frac{1}{4s^{2}} \frac{\MH^{2}}{\MW^{2}}
\Bigl[1+2\DZe -2\frac{\delta s}{s}
+\frac{\DMHS}{\MH^{2}} +\dth -\frac{\delta \MW^{2}}{\MW^{2}}
+ \DZH \Bigr] \co\\[1ex]
\chi \chi \chi \chi & C = - \frac{3}{4s^{2}} \frac{\MH^{2}}{\MW^{2}}
\Bigl[1+2\DZe -2\frac{\delta s}{s}
+\frac{\DMHS}{\MH^{2}} +\dth -\frac{\delta \MW^{2}}{\MW^{2}}
\Bigr] \co\\[1ex]
\chi \chi \phi \phi & C = - \frac{1}{4s^{2}} \frac{\MH^{2}}{\MW^{2}}
\Bigl[1+2\DZe -2\frac{\delta s}{s}
+\frac{\DMHS}{\MH^{2}} +\dth -\frac{\delta \MW^{2}}{\MW^{2}}
\Bigr] \co\\[1ex]
\phi \phi \phi \phi & C = - \frac{1}{2s^{2}} \frac{\MH^{2}}{\MW^{2}}
\Bigl[1+2\DZe -2\frac{\delta s}{s}
+\frac{\DMHS}{\MH^{2}} +\dth -\frac{\delta \MW^{2}}{\MW^{2}}
\Bigr] .
\earr
\eeq
}
 
{\samepage
SSS-coupling:
\bma
\barr{l}
\framebox {
\begin{picture}(96,77)(0,-2)
\put(75,65){\makebox(10,20)[bl]{$S_{2}$}}
\put(0,42){\makebox(10,20)[bl]{$S_{1}$}}
\put(75,0){\makebox(10,20)[bl]{$S_{3}$}}
\put(48,36){\circle*{4}}
\put(0,36){\usebox{\Sr}}
\put(48,12){\usebox{\Stbr}}
\end{picture} }
\earr
\barr{l}
= ieC
\earr
\ema\samepage
with the actual values of $S_{1},\:S_{2},\:S_{3}$ and $C$
\beq
\begin{array}[b]{l@{\quad : \quad}l@{\quad}l@{\qquad}l@{\quad}l}
\PH \PH \PH & C = - \frac{3}{2s} \frac{\MH^{2}}{\MW}
\Bigl[1+\DZe -\frac{\delta s}{s}
+\frac{\DMHS}{\MH^{2}} +\dth 
-\frac{1}{2}\frac{\delta \MW^{2}}{\MW^{2}}
+ \frac{3}{2}\DZH \Bigr] \co\\[1ex]
\barr{l} \PH\chi\chi \\ \PH\phi\phi \earr \left.
\rule[-1.5ex]{0mm}{4ex}\right\}
& C = - \frac{1}{2s} \frac{\MH^{2}}{\MW}
\Bigl[1+\DZe -\frac{\delta s}{s}
+\frac{\DMHS}{\MH^{2}} +\dth
-\frac{1}{2}\frac{\delta \MW^{2}}{\MW^{2}}
+ \frac{1}{2}\DZH \Bigr] .
\earr
\eeq
}
 
{\samepage
VVSS-coupling:
\bma
\barr{l}
\framebox{
\begin{picture}(96,77)(0,-2)
\put(6,65){\makebox(10,20)[bl]{$V_{1,\mu}$}}
\put(6,0){\makebox(10,20)[bl]{$V_{2,\nu}$}}
\put(75,65){\makebox(10,20)[bl]{$S_{1}$}}
\put(75,0){\makebox(10,20)[bl]{$S_{2}$}}
\put(48,36){\circle*{4}}
\put(16,12){\usebox{\Vtr}}
\put(16,36){\usebox{\Vbr}}
\put(48,12){\usebox{\Stbr}}
\end{picture} }
\earr
\barr{l}
= ie^{2}g_{\mu\nu}C
\earr
\ema\samepage
with the actual values of $V_{1},\:V_{2},\:S_{1},\:S_{2}$ and $C$
\beq
\begin{array}[b]{l@{\quad : \quad}l@{\quad}l@{\qquad}l@{\quad}l}
\PWp \PWm \PH \PH & C =  \frac{1}{2s^{2}}
\Bigl[1+2\DZe -2\frac{\delta s}{s}
+ \DZW + \DZH\Bigr] \co\\[1ex]
\barr{l} \PWp \PWm \chi \chi \\ \PWp \PWm \phi^{+} \phi^{-} \earr
\left.\rule[-1.5ex]{0mm}{4ex}\right\}
& C =  \frac{1}{2s^{2}}
\Bigl[1+2\DZe -2\frac{\delta s}{s}
+ \DZW \Bigr] \co\\[1ex]
\PZ \PZ \phi^{+} \phi^{-} & C =
\frac{(s^{2}-c^{2})^{2}}{2s^{2}c^{2}}
\Bigl[1+2\DZe
+\frac{2}{(s^{2}-c^{2})c^{2}}\frac{\delta s}{s} + \DZZ \Bigr]
+\frac{s^{2}-c^{2}}{sc} \DZAZ \co\\[1ex]
\PZ\PA\phi^{+} \phi^{-} & \left\{
\barr{l} C =\frac{s^{2}-c^{2}}{sc}  \Bigl[1+2\DZe
+\frac{1}{(s^{2}-c^{2})c^{2}}\frac{\delta s}{s}
+ \frac{1}{2}\DZZ + \frac{1}{2}\DZA  \Bigr] \\[1ex]
\qquad+\frac{(s^{2}-c^{2})^{2}}{2s^{2}c^{2}}\frac{1}{2}\DZZA
+ \DZAZ \co \earr \right.\\[1em]
\PA\PA\phi^{+} \phi^{-} & C =
2\Bigl[1+2\DZe + \DZA \Bigr]
+\frac{s^{2}-c^{2}}{sc} \DZZA \co\\[1ex]
\PZ\PZ\PH\PH& C = \frac{1}{2s^{2}c^{2}}
\Bigl[1+2\DZe
+ 2\frac{s^{2}-c^{2}}{c^{2}}\frac{\delta s}{s}
+ \DZZ+ \DZH \Bigr] \co\\[1ex]
\PZ\PZ\chi \chi & C = \frac{1}{2s^{2}c^{2}}
\Bigl[1+2\DZe
+ 2\frac{s^{2}-c^{2}}{c^{2}}\frac{\delta s}{s}
+ \DZZ \Bigr] \co\\[1ex]
\barr{l} \PZ\PA\PH\PH\\ \PZ\PA\chi \chi \earr
\left.\rule[-1.5ex]{0mm}{4ex}\right\}
& C =  \frac{1}{2s^{2}c^{2}} \frac{1}{2}\DZZA \co\\[1ex]
\PW^{\pm} \PZ \phi^{\mp} \PH & C = - \frac{1}{2c}
\Bigl[1+2\DZe -\frac{\delta c}{c}
 + \frac{1}{2}\DZW + \frac{1}{2}\DZH
+ \frac{1}{2}\DZZ \Bigr]
- \frac{1}{2s}\frac{1}{2}\DZAZ \co\\[1ex]
\PW^{\pm} \PA \phi^{\mp} \PH & C = - \frac{1}{2s}
\Bigl[1+2\DZe -\frac{\delta s}{s}
+ \frac{1}{2}\DZW + \frac{1}{2}\DZH
+ \frac{1}{2}\DZA \Bigr]
- \frac{1}{2c}\frac{1}{2}\DZZA \co\\[1ex]
\PW^{\pm} \PZ \phi^{\mp} \chi & C = \mp \frac{i}{2c}
\Bigl[1+2\DZe -\frac{\delta c}{c}
+ \frac{1}{2}\DZW + \frac{1}{2}\DZZ \Bigr]
\mp \frac{i}{2s}\frac{1}{2}\DZAZ \co\\[1ex]
\PW^{\pm} \PA \phi^{\mp} \chi & C =  \mp\frac{i}{2s}
\Bigl[1+2\DZe -\frac{\delta s}{s}
+ \frac{1}{2}\DZW + \frac{1}{2}\DZA \Bigr]
\mp \frac{i}{2c}\frac{1}{2}\DZZA .
\earr
\eeq
}
 
{\samepage
VSS-coupling:
\bma
\barr{l}
\framebox {
\begin{picture}(96,77)(0,-2)
\put(67,65){\makebox(10,20)[bl]{$S_{1},k_{1}$}}
\put(0,42){\makebox(10,20)[bl]{$V_{\mu}$}}
\put(67,0){\makebox(10,20)[bl]{$S_{2},k_{2}$}}
\put(48,36){\circle*{4}}
\put(0,36){\usebox{\Vr}}
\put(48,12){\usebox{\Stbr}}
\end{picture} }
\earr
\barr{l}
= ieC(k_{1}-k_{2})_{\mu}
\earr
\ema\samepage \samepage
with the actual values of $V,\:S_{1},\:S_{2}$ and $C$
\beq
\begin{array}[b]{l@{\quad : \quad}l@{\quad}l@{\qquad}l@{\quad}l}
\PA \chi \PH & C = - \frac{i}{2cs} \frac{1}{2}\DZZA\co\\[1ex]
\PZ \chi \PH & C = - \frac{i}{2cs}
\Bigl[1+\DZe +\frac{s^{2}-c^{2}}{c^{2}}\frac{\delta s}{s}
+ \frac{1}{2}\DZH + \frac{1}{2}\DZZ\Bigr]\co\\[1ex]
\PA\phi^{+}\phi^{-} & C = -\Bigl[1 + \DZe + \frac{1}{2}\DZA
+\frac{s^{2}-c^{2}}{2sc}\frac{1}{2}\DZZA\Bigl] \co\\[1ex]
\PZ \phi^{+} \phi^{-} & C = - \frac{s^{2}-c^{2}}{2sc}
\Bigl[1+\DZe +\frac{1}{(s^{2}-c^{2})c^{2}}\frac{\delta s}{s}
+\frac{1}{2}\DZZ\Bigr]
- \frac{1}{2}\DZAZ \co\\[1ex]
\PW^{\pm} \phi^{\mp} \PH & C = \mp \frac{1}{2s}
\Bigl[1+\DZe -\frac{\delta s}{s}
+ \frac{1}{2}\DZW + \frac{1}{2}\DZH \Bigr] \co\\[1ex]
\PW^{\pm} \phi^{\mp} \chi & C = - \frac{i}{2s}
\Bigl[1+\DZe -\frac{\delta s}{s}
+ \frac{1}{2}\DZW\Bigr] .
\earr
\eeq
}
 
{\samepage
SVV-coupling:
\bma
\barr{l}
\framebox {
\begin{picture}(96,77)(0,-2)
\put(72,65){\makebox(10,20)[bl]{$V_{1,\mu}$}}
\put(72,0){\makebox(10,20)[bl] {$V_{2,\nu}$}}
\put(0,42){\makebox(10,20)[bl]{$S$}}
\put(48,36){\circle*{4}}
\put(0,36){\usebox{\Sr}}
\put(48,12){\usebox{\Vtbr}}
\end{picture} }
\earr
\barr{l}
= ieg_{\mu\nu}C
\earr
\ema\samepage
with the actual values of $S,\:V_{1},\:V_{2}$ and $C$
\beq
\begin{array}[b]{l@{\quad : \quad}l@{\quad}l@{\qquad}l@{\quad}l}
\PH \PWp \PWm & C =  \MW\frac{1}{s}
\Bigl[1+\DZe -\frac{\delta s}{s}
+\frac{1}{2}\frac{\delta \MW^{2}}{\MW^{2}}
+ \frac{1}{2}\DZH+ \DZW \Bigr] \co\\[1ex]
\PH\PZ\PZ& C = \MW\frac{1}{sc^{2}}
\Bigl[1+\DZe +\frac{2s^{2}-c^{2}}{c^{2}}\frac{\delta s}{s}
+\frac{1}{2}\frac{\delta \MW^{2}}{\MW^{2}}
+ \frac{1}{2}\DZH+ \DZZ \Bigr] \co\\[1ex]
\PH \PZ\PA& C = \MW\frac{1}{sc^{2}} \frac{1}{2}\DZZA \co\\[1ex]
\phi^{\pm} \PW^{\mp} \PZ & C = - \MW\frac{s}{c}
\Bigl[1+\DZe +\frac{1}{c^{2}}\frac{\delta s}{s}
+\frac{1}{2}\frac{\delta \MW^{2}}{\MW^{2}}
+ \frac{1}{2}\DZW + \frac{1}{2}\DZZ \Bigr]
- \MW \frac{1}{2}\DZAZ \co\\[1ex]
\phi^{\pm} \PW^{\mp} \PA & C = - \MW
\Bigl[1+\DZe +\frac{1}{2}\frac{\delta \MW^{2}}{\MW^{2}}
+ \frac{1}{2}\DZW +  \frac{1}{2}\DZA \Bigr]
- \MW\frac{s}{c} \frac{1}{2}\DZZA.
\earr
\eeq
}
 
{\samepage
VFF-coupling:
\bma
\barr{l}
\framebox {
\begin{picture}(96,77)(0,-2)
\put(75,65){\makebox(10,20)[bl]{$\bar{F}_{1}$}}
\put(0,42){\makebox(10,20)[bl]{$V_{\mu}$}}
\put(75,0){\makebox(10,20)[bl]{$F_{2}$}}
\put(48,36){\circle*{4}}
\put(0,36){\usebox{\Vr}}
\put(48,12){\usebox{\Ftbr}}
\end{picture} }
\earr
\barr{l}
= ie\gamma_{\mu}(C^{-}\omega_{-} + C^{+}\omega_{+})
\earr
\ema\samepage
with the actual values of $V,\:\bar{F}_{1}\:,F_{2}$ and $C^{+},\:C^{-}$
\beq
\begin{array}[b]{l@{\quad : \quad}l}
\gamma \bar{f}_{i} f_{j} & \left\{
\barr{l}
C^{+} = -Q_{f}\Bigl[\delta_{ij}\Bigl(1 + \DZe +
\frac{1}{2}\DZA\Bigl)
+ \frac{1}{2}(\delta Z^{f,R}_{ij}+\delta Z^{f,R\dagger}_{ij})\Bigr]
+ \delta_{ij}g_{f}^{+}\frac{1}{2}\DZZA\co
\\[1ex]
C^{-} = -Q_{f}\Bigl[\delta_{ij}\Bigl(1 + \DZe +
\frac{1}{2}\DZA\Bigl)
+ \frac{1}{2}(\delta Z^{f,L}_{ij}+\delta Z^{f,L\dagger}_{ij})\Bigr]
+ \delta_{ij}g_{f}^{-}\frac{1}{2}\DZZA\co
\earr \right.\\[1.6em]
\PZ \bar{f}_{i} f_{j} & \left\{\barr{l}
C^{+} = g_{f}^{+}\Bigl[\delta_{ij}\Bigl(1 + \frac{\delta
g_{f}^{+}}{g_{f}^{+}}+\frac{1}{2}\DZZ\Bigr)
+ \frac{1}{2}(\delta Z^{f,R}_{ij}+\delta Z^{f,R\dagger}_{ij})\Bigr]
-\delta_{ij}Q_{f}\frac{1}{2}\DZAZ \co\\[1ex]
C^{-} = g_{f}^{-}\Bigl[\delta_{ij}\Bigl(1 + \frac{\delta
g_{f}^{-}}{g_{f}^{-}}+\frac{1}{2}\DZZ\Bigr)
+ \frac{1}{2}(\delta Z^{f,L}_{ij}+\delta Z^{f,L\dagger}_{ij})\Bigr]
-\delta_{ij}Q_{f}\frac{1}{2}\DZAZ \co
\earr \right.
\\[1.8em]
\PWp \bar{u}_{i} d_{j} & \left\{ \barr{l}
C^{+} = 0,   \qquad
C^{-} = \frac{1}{\sqrt{2}s}\Bigl[V_{ij}\Bigl(1 + \DZe -
\frac{\delta s}{s}+\frac{1}{2}\DZW\Bigr) + \delta V_{ij} \\[1ex]
\hspace{4cm}
+\frac{1}{2}\sum_{k}(\delta Z_{ik}^{u,L\dagger}V_{kj}+V_{ik}\delta
Z_{kj}^{d,L})\Bigr]\co
\earr \right.
\\[1.6em]
\PWm \bar{d}_{j} u_{i} & \left\{ \barr{l}
C^{+} = 0,   \qquad
C^{-} = \frac{1}{\sqrt{2}s}\Bigl[V_{ji}^\dagger\Bigl(1 + \DZe -
\frac{\delta s}{s}+\frac{1}{2}\DZW\Bigr) + \delta V_{ji}^\dagger \\[1ex]
\hspace{4cm}
+\frac{1}{2}\sum_{k}(\delta Z_{jk}^{d,L\dagger}V_{ki}^\dagger
+V_{jk}^\dagger\delta Z_{ki}^{u,L})\Bigr]\co
\earr \right. \\[1.6em]
\PWp \bar{\nu}_{i} l_{j} &
C^{+} = 0,   \qquad
C^{-} = \frac{1}{\sqrt{2}s}\delta_{ij}\Bigl[1 + \DZe -
\frac{\delta s}{s}+\frac{1}{2}\DZW
+\frac{1}{2}(\delta Z_{ii}^{\nu,L\dagger}+\delta Z_{ii}^{l,L})\Bigr]\co
\\[1em]
\PWm \bar{l}_{j} \nu_{i} &
C^{+} = 0,  \qquad
C^{-} = \frac{1}{\sqrt{2}s}\delta_{ij}\Bigl[1 + \DZe -
\frac{\delta s}{s}+\frac{1}{2}\DZW
+\frac{1}{2}(\delta Z_{ii}^{l,L\dagger}+\delta Z_{ii}^{\nu,L})\Bigr]\co
\earr
\eeq
}
where
\beq \label{geZ}
\barr{ll}
g_{f}^{+} = -\frac{s}{c} Q_{f}, &
\delta g_{f}^{+} = -\frac{s}{c} Q_{f}\Bigl[\DZe +
\frac{1}{c^{2}}\frac{\delta s}{s}\Bigr]\co \\[1.2em]
g_{f}^{-} =  \frac{I_{W,f}^{3}-s^{2}Q_{f}}{sc},\qquad &
\delta g_{f}^{-} =  \frac{I_{W,f}^{3}}{sc}\Bigl[\DZe +
\frac{s^{2}-c^{2}}{c^{2}}\frac{\delta s}{s}\Bigr] + \delta g_{f}^{+}.
\earr
\eeq
The vector and axial vector couplings of the $Z$-boson are given by
\beq \label{vfaf}
\barr{l}
v_{f} = \frac{1}{2}(g_{f}^{-} + g_{f}^{+}) = \frac{I_{W,f}^{3}-
2s^{2}Q_{f}}{2sc}, \quad
a_{f} = \frac{1}{2}(g_{f}^{-} - g_{f}^{+}) = \frac{I_{W,f}^{3}}{2sc}.
\earr
\eeq
 
{\samepage
SFF-coupling:
\bma
\barr{l}
\framebox {
\begin{picture}(96,77)(0,-2)
\put(75,65){\makebox(10,20)[bl]{$\bar{F}_{1}$}}
\put(0,42){\makebox(10,20)[bl]{$S$}}
\put(75,0){\makebox(10,20)[bl]{$F_{2}$}}
\put(48,36){\circle*{4}}
\put(0,36){\usebox{\Sr}}
\put(48,12){\usebox{\Ftbr}}
\end{picture} }
\earr
\barr{l}
= ie(C^{-}\omega_{-} + C^{+}\omega_{+})
\earr
\ema\samepage
with the actual values of $S,\:\bar{F}_{1}\:,F_{2}$ and $C^{+},\:C^{-}$
\beq
\begin{array}[b]{l@{\quad : \quad}l}
\PH \bar{f}_{i} f_{j} & \left\{ \barr{l}
C^{+} = - \frac{1}{2s}\frac{1}{\MW}\Bigl[\delta_{ij}m_{f,i}
\Bigl(1 + \DZe -\frac{\delta s}{s} +
\frac{\delta m_{f,i}}{m_{f,i}} - \frac{\delta
\MW}{\MW}+\frac{1}{2}\DZH\Bigr) \\[1ex]
\hspace{3cm}
{}+ \frac{1}{2}(m_{f,i}\delta Z^{f,R}_{ij}+\delta Z^{f,L\dagger}_{ij}m_{f,j})
\Bigr]\co \\[1ex]
C^{-} = - \frac{1}{2s}\frac{1}{\MW}\Bigl[\delta_{ij}m_{f,i}
\Bigl(1 + \DZe -\frac{\delta s}{s} +
\frac{\delta m_{f,i}}{m_{f,i}} - \frac{\delta
\MW}{\MW}+\frac{1}{2}\DZH\Bigr) \\[1ex]
\hspace{3cm}
{}+ \frac{1}{2}(m_{f,i}\delta Z^{f,L}_{ij}+\delta Z^{f,R\dagger}_{ij}m_{f,j})
\Bigr]\co \earr\right.\\[3.7em]
\chi \bar{f}_{i} f_{j} & \left\{ \barr{l}
C^{+} = i\frac{1}{2s}2I_{W,f}^{3}\frac{1}{\MW}\Bigl[\delta_{ij}m_{f,i}
\Bigl(1 + \DZe -\frac{\delta s}{s} +
\frac{\delta m_{f,i}}{m_{f,i}} - \frac{\delta \MW}{\MW}\Bigr)
\\[1ex] \hspace{3cm}
{}+ \frac{1}{2}(m_{f,i}\delta Z^{f,R}_{ij}+\delta Z^{f,L\dagger}_{ij}m_{f,j})
\Bigr]\co
\\[1ex]
C^{-} = -i\frac{1}{2s}2I_{W,f}^{3}\frac{1}{\MW}\Bigl[\delta_{ij}m_{f,i}
\Bigl(1 + \DZe -\frac{\delta s}{s} +
\frac{\delta m_{f,i}}{m_{f,i}} - \frac{\delta \MW}{\MW}\Bigr)
\\[1ex] \hspace{3cm}
{}+ \frac{1}{2}(m_{f,i}\delta Z^{f,L}_{ij}+\delta Z^{f,R\dagger}_{ij}m_{f,j})
\Bigr]\co \earr\right.\\[3.7em]
\phi^{+} \bar{u}_{i} d_{j} &
\left\{ \barr{l}
C^{+} = -\frac{1}{\sqrt{2}s}\frac{1}{\MW}
\Bigl[V_{ij}m_{d,j}\Bigl(1 + \DZe -\frac{\delta s}{s} +
\frac{\delta m_{d,j}}{m_{d,j}} - \frac{\delta \MW}{\MW}\Bigl) +
\delta V_{ij}m_{d,j}\\[1ex]
\hspace{4cm}
+\frac{1}{2}\sum_{k}(\delta Z_{ik}^{u,L\dagger}V_{kj}m_{d,j}+V_{ik}m_{d,k}
\delta Z_{kj}^{d,R})\Bigr]\co \\[1ex]
C^{-} = \frac{1}{\sqrt{2}s}\frac{1}{\MW}
\Bigl[m_{u,i}V_{ij}\Bigl(1 + \DZe -\frac{\delta s}{s} +
\frac{\delta m_{u,i}}{m_{u,i}} - \frac{\delta \MW}{\MW}\Bigl) +
m_{u,i}\delta V_{ij} \\[1ex]
\hspace{4cm}
+\frac{1}{2}\sum_{k}(\delta Z_{ik}^{u,R\dagger}m_{u,k}V_{kj}
+m_{u,i}V_{ik}\delta Z_{kj}^{d,L})\Bigr]\co
\earr\right.\\[3.8em]
\phi^{-} \bar{d}_{j} u_{i} &
\left\{ \barr{l}
C^{+} = \frac{1}{\sqrt{2}s}\frac{1}{\MW}
\Bigl[V_{ji}^\dagger m_{u,i}\Bigl(1 + \DZe -\frac{\delta s}{s} +
\frac{\delta m_{u,i}}{m_{u,i}} - \frac{\delta \MW}{\MW}\Bigl) +
\delta V_{ji}^\dagger m_{u,i}\\[1ex]
\hspace{4cm}
+\frac{1}{2}\sum_{k}(\delta Z_{jk}^{d,L\dagger}V_{ki}^\dagger m_{u,i}
+V_{jk}^\dagger m_{u,k}\delta Z_{ki}^{u,R})\Bigr]\co\\[1ex]
C^{-} = -\frac{1}{\sqrt{2}s}\frac{1}{\MW}
\Bigl[m_{d,j}V_{ji}^\dagger \Bigl(1 + \DZe -\frac{\delta s}{s} +
\frac{\delta m_{d,j}}{m_{d,j}} - \frac{\delta \MW}{\MW}\Bigl) +
m_{d,j}\delta V_{ji}^\dagger \\[1ex]
\hspace{4cm}
+\frac{1}{2}\sum_{k}(\delta Z_{jk}^{d,R\dagger}m_{d,k}V_{ki}^\dagger
+m_{d,j}V_{jk}^\dagger \delta Z_{ki}^{u,L})\Bigr]\co
\earr\right. \\[3.8em]
\phi^{+} \bar{\nu}_{i} l_{j} & \left\{\barr{l}
C^{+} = -\frac{1}{\sqrt{2}s}\frac{m_{l,i}}{\MW}
\delta_{ij}\Bigl[ 1 + \DZe -\frac{\delta s}{s} +
\frac{\delta m_{l,i}}{m_{l,i}} - \frac{\delta \MW}{\MW}
+\frac{1}{2}(\delta Z_{ii}^{\nu,L\dagger}+\delta Z_{ii}^{l,R})\Bigr]\co
\\[1ex]
C^{-} = 0\co
\earr \right.
\\[1.5em]
\phi^{-} \bar{l}_{j} \nu_{i} & \left\{\barr{l}
C^{+} = 0\co\\[1ex]
C^{-} =  -\frac{1}{\sqrt{2}s}\frac{m_{l,i}}{\MW}
\delta_{ij}\Bigl[1 + \DZe -\frac{\delta s}{s} +
\frac{\delta m_{l,i}}{m_{l,i}} - \frac{\delta \MW}{\MW}
+\frac{1}{2}(\delta Z_{ii}^{l,R\dagger}+\delta Z_{ii}^{\nu,L})\Bigr].
\earr \right.
\earr
\eeq
}

{\samepage
VGG-coupling:
\bma
\barr{l}
\framebox {
\begin{picture}(96,77)(0,-2)
\put(67,65){\makebox(10,20)[bl]{$\bar{G_{1}},k_{1}$}}
\put(0,42){\makebox(10,20)[bl]{$V_{\mu}$}}
\put(75,0){\makebox(10,20)[bl]{$G_{2}$}}
\put(48,36){\circle*{4}}
\put(0,36){\usebox{\Vr}}
\put(48,12){\usebox{\Gtbr}}
\end{picture} }
\earr
\barr{l}
= iek_{1,\mu}C
\earr
\ema\samepage
with the actual values of $V,\:\bar{G}_{1}\:,G_{2}$ and $C$
\beq
\begin{array}[b]{l@{\quad : \quad}l@{\quad}l@{\qquad}l@{\quad}l}
\PA \bar{u}^{\pm} u^{\pm} & C = \pm
\Bigl[1 + \DZe + \frac{1}{2}\DZA\Bigl]
\mp \frac{c}{s}\frac{1}{2}\DZZA \co\\[1ex]
\PZ \bar{u}^{\pm} u^{\pm} & C = \mp \frac{c}{s}
\Bigl[1 + \DZe - \frac{1}{c^{2}}\frac{\delta s}{s}
+ \frac{1}{2}\DZZ\Bigl]
\pm \frac{1}{2}\DZAZ \co\\[1ex]
\PW^{\pm} \bar{u}^{\pm} u^{\PZ} & C = \pm \frac{c}{s}
\Bigl[1+\DZe  - \frac{1}{c^{2}}\frac{\delta s}{s} +
\frac{1}{2}\DZW\Bigr] \co\\[1ex]
\PW^{\pm} \bar{u}^{\PZ} u^{\mp} & C = \mp \frac{c}{s}
\Bigl[1+\DZe  - \frac{1}{c^{2}}\frac{\delta s}{s} +
\frac{1}{2}\DZW\Bigr] \co\\[1ex]
\PW^{\pm} \bar{u}^{\pm} u^{\gamma} & C = \mp
\Bigl[1+\DZe  + \frac{1}{2}\DZW\Bigr] \co\\[1ex]
\PW^{\pm} \bar{u}^{\gamma} u^{\mp} & C = \pm
\Bigl[1+\DZe  + \frac{1}{2}\DZW\Bigr] .
\earr
\eeq
}
 
{\samepage
SGG-coupling:
\bma
\barr{l}
\framebox {
\begin{picture}(96,77)(0,-2)
\put(75,65){\makebox(10,20)[bl]{$\bar{G_{1}}$}}
\put(0,42){\makebox(10,20)[bl]{$S$}}
\put(75,0){\makebox(10,20)[bl]{$G_{2}$}}
\put(48,36){\circle*{4}}
\put(0,36){\usebox{\Sr}}
\put(48,12){\usebox{\Gtbr}}
\end{picture} }
\earr
\barr{l}
= ieC
\earr
\ema\samepage
with the actual values of $S,\:\bar{G}_{1}\:,G_{2}$ and $C$
\beq
\begin{array}[b]{l@{\quad : \quad}l@{\quad}l@{\qquad}l@{\quad}l}
\PH \bar{u}^{\PZ} u^{\PZ} & C = -\frac{1}{2sc^{2}}\MW
\Bigl[1 + \DZe + \frac{2s^{2}-c^{2}}{c^{2}}
\frac{\delta s}{s} +\frac{1}{2}\frac{\delta \MW^{2}}{\MW^{2}}
+ \frac{1}{2}\DZH \Bigl] \co\\[1ex]
\PH \bar{u}^{\pm} u^{\pm} & C = -\frac{1}{2s}\MW
\Bigl[1 + \DZe - \frac{\delta s}{s}
+\frac{1}{2}\frac{\delta \MW^{2}}{\MW^{2}}
+ \frac{1}{2}\DZH \Bigl] \co\\[1ex]
\chi \bar{u}^{\pm} u^{\pm} & C = \mp i \frac{1}{2s}\MW
\Bigl[1 + \DZe - \frac{\delta s}{s}
+\frac{1}{2}\frac{\delta \MW^{2}}{\MW^{2}} \Bigl] \co\\[1ex]
\phi^{\pm} \bar{u}^{\PZ} u^{\pm} & C =  \frac{1}{2sc}\MW
\Bigl[1 + \DZe + \frac{s^{2}-c^{2}}{c^{2}}
\frac{\delta s}{s}
+\frac{1}{2}\frac{\delta \MW^{2}}{\MW^{2}} \Bigl] \co\\[1ex]
\phi^{\pm} \bar{u}^{\pm} u^{\PZ} & C =
\frac{s^{2}-c^{2}}{2sc}\MW
\Bigl[1 + \DZe + \frac{1}{(s^{2}-c^{2})c^{2}}
\frac{\delta s}{s}
+\frac{1}{2}\frac{\delta \MW^{2}}{\MW^{2}} \Bigl] \co\\[1ex]
\phi^{\pm} \bar{u}^{\pm} u^{\gamma} & C =  \MW
\Bigl[1 + \DZe
+\frac{1}{2}\frac{\delta \MW^{2}}{\MW^{2}} \Bigl] .
\earr
\eeq
}

\chapter{Self energies}
 
In this appendix we list all self energies of the physical fields.
 
The gauge boson self energies read
\beqar
\nn \lefteqn{\Sigma^{AA}_{T}(k^2)  =   - \frac{\alpha}{4\pi} \Biggl\{
\frac{2}{3} \sum_{f,i} N_{C}^{f}2Q_{f}^{2}
\Bigl[-
(k^{2}+2m_{f,i}^{2})B_{0}(k^{2},m_{f,i},m_{f,i})}\qquad\qquad\qquad\quad\\[1em]
&&\mbox{}\qquad\qquad
+2m_{f,i}^{2}B_{0}(0,m_{f,i},m_{f,i}) +\frac{1}{3}k^{2} \Bigr] \\[1em]
\nn&& \mbox{}+ \biggl\{
\Bigl[3 k^{2} + 4M_{W}^{2}\Bigr] B_{0}(k^{2},M_{W},M_{W})
 - 4M_{W}^{2}B_{0}(0,M_{W},M_{W}) \biggr\}
\Biggr\},
\eeqar
\beqar
\nn \lefteqn{\Sigma^{AZ}_{T}(k^2)  =   - \frac{\alpha}{4\pi} \Biggl\{
\frac{2}{3} \sum_{f,i} N_{C}^{f}(-Q_{f})\Bigl(g_{f}^{+}+g_{f}^{-}\Bigr)
\Bigl[-
(k^{2}+2m_{f,i}^{2})B_{0}(k^{2},m_{f,i},m_{f,i})}\qquad\qquad\qquad\quad\\[1em]
\nn&&\mbox{}\qquad\qquad\qquad
+2m_{f,i}^{2}B_{0}(0,m_{f,i},m_{f,i}) +\frac{1}{3}k^{2} \Bigr] \\[1em]
\nn&& \mbox{}-\frac{1}{3s_{W}c_{W}} \biggl\{
\Bigl[(9c_{W}^{2} + \frac{1}{2}) k^{2} + (12c_{W}^{2} + 4) M_{W}^{2}\Bigr]
B_{0}(k^{2},M_{W},M_{W}) \\[1em]
&& \qquad\qquad\mbox{} -(12c_{W}^{2} - 2) M_{W}^{2}B_{0}(0,M_{W},M_{W})
+ \frac{1}{3}k^{2} \biggr\}
\Biggr\},
\eeqar
\beqar
\nn \lefteqn{\Sigma^{ZZ}_{T}(k^2)  =   - \frac{\alpha}{4\pi} \Biggl\{
\frac{2}{3} \sum_{f,i} N_{C}^{f}
\biggl\{\Bigl((g_{f}^{+})^{2}+(g_{f}^{-})^{2}\Bigr)
\Bigl[-(k^{2}+2m_{f,i}^{2})B_{0}(k^{2},m_{f,i},m_{f,i})
} \\[1em]
\nn&& \qquad\qquad
\mbox{}+2m_{f,i}^{2}B_{0}(0,m_{f,i},m_{f,i}) +\frac{1}{3}k^{2} \Bigr]
+\frac{3}{4s_{W}^{2}c_{W}^{2}} m_{f,i}^{2}
B_{0}(k^{2},m_{f,i},m_{f,i}) \biggr\} \\[1em]
\nn&& \mbox{}+\frac{1}{6s_{W}^{2}c_{W}^{2}} \biggl\{
\Bigl[(18c_{W}^{4} + 2c_{W}^{2} -\frac{1}{2}) k^{2}
+ (24c_{W}^{4} + 16c_{W}^{2} -10) M_{W}^{2}\Bigr]
B_{0}(k^{2},M_{W},M_{W}) \\[1em]
\nn&& \qquad\qquad\mbox{} -(24c_{W}^{4} - 8c_{W}^{2} + 2)
M_{W}^{2}B_{0}(0,M_{W},M_{W})
+ (4c_{W}^{2}-1) \frac{1}{3}k^{2} \biggr\} \\[1em]
\nn&&\mbox{} +\frac{1}{12s_{W}^{2}c_{W}^{2}} \biggl\{
\Bigl(2M_{H}^{2} - 10M_{Z}^{2} -k^{2}\Bigr) B_{0}(k^{2},M_{Z},M_{H}) \\[1em]
&& \qquad\qquad \mbox{}-2M_{Z}^{2}B_{0}(0,M_{Z},M_{Z}) -
2M_{H}^{2}B_{0}(0,M_{H},M_{H}) \\[1em]
\nn&& \qquad\qquad\mbox{}-\frac{(M_{Z}^{2}-M_{H}^{2})^{2}}{k^{2}}
\Bigl( B_{0}(k^{2},M_{Z},M_{H}) - B_{0}(0,M_{Z},M_{H})\Bigr)
-\frac{2}{3}k^{2} \biggr\}
\Biggr\} ,
\eeqar
\beqar
\nn \lefteqn{\Sigma^{W}_{T}(k^2)  =   - \frac{\alpha}{4\pi} \Biggl\{
\frac{2}{3}\frac{1}{2s_{W}^{2}} \sum_{i}
\biggl[-\Bigl(k^{2}-\frac{m_{l,i}^{2}}{2}\Bigr)B_{0}(k^{2},0,m_{l,i})
+\frac{1}{3}k^{2}} \\[1em]
\nn&& \qquad\qquad\qquad \mbox{}+m_{l,i}^{2}B_{0}(0,m_{l,i},m_{l,i})
+\frac{m_{l,i}^{4}}{2k^{2}}
\Bigr(B_{0}(k^{2},0,m_{l,i})-B_{0}(0,0,m_{l,i})\Bigr)\biggr]
\\[1em]
\nn&& \mbox{}+\frac{2}{3}\frac{1}{2s_{W}^{2}} 3 \sum_{i,j} |V_{ij}|^{2}
\biggl[-\Bigl(k^{2}-\frac{m_{u,i}^{2}+m_{d,j}^{2}}{2}\Bigr)
B_{0}(k^{2},m_{u,i},m_{d,j})  +\frac{1}{3}k^{2} \\[1em]
\nn&& \qquad\qquad \mbox{}+m_{u,i}^{2}B_{0}(0,m_{u,i},m_{u,i})+
m_{d,j}^{2}B_{0}(0,m_{d,j},m_{d,j}) \\[1em]
\nn&& \qquad\qquad \mbox{}+\frac{(m_{u,i}^{2}-m_{d,j}^{2})^{2}}{2k^{2}}
\Bigr(B_{0}(k^{2},m_{u,i},m_{d,j})-B_{0}(0,m_{u,i},m_{d,j})\Bigr)\biggr]
\\[1em]
\nn&& \mbox{}+\frac{2}{3} \biggl\{
\Bigl(2M_{W}^{2} + 5k^{2}\Bigr) B_{0}(k^{2},M_{W},\lambda)
 -2M_{W}^{2}B_{0}(0,M_{W},M_{W})  \\[1em]
\nn&& \qquad\qquad\mbox{}-\frac{M_{W}^{4}}{k^{2}}
\Bigl( B_{0}(k^{2},M_{W},\lambda) - B_{0}(0,M_{W},\lambda)\Bigr)
+\frac{1}{3}k^{2} \biggr\} \\[1em]
\nn&& \mbox{}+\frac{1}{12s_{W}^{2}} \biggl\{
\Bigl[(40c_{W}^{2} -1) k^{2} + (16c_{W}^{2} + 54 -10c_{W}^{-2}) M_{W}^{2}
\Bigr] B_{0}(k^{2},M_{W},M_{Z}) \\[1em]
\nn&& \qquad\qquad \mbox{}-(16c_{W}^{2} + 2) \Bigl[M_{W}^{2}B_{0}(0,M_{W},M_{W})
+M_{Z}^{2}B_{0}(0,M_{Z},M_{Z}) \Bigr]
+ (4c_{W}^{2}-1) \frac{2}{3}k^{2}  \\[1em]
\nn&& \qquad\qquad\mbox{}-(8c_{W}^{2}+1)\frac{(M_{W}^{2}-M_{Z}^{2})^{2}}{k^{2}}
\Bigl(B_{0}(k^{2},M_{W},M_{Z}) - B_{0}(0,M_{W},M_{Z})\Bigr)\biggr\} \\[1em]
\nn&& \mbox{}+\frac{1}{12s_{W}^{2}} \biggl\{
\Bigl(2M_{H}^{2} - 10M_{W}^{2} -k^{2}\Bigr) B_{0}(k^{2},M_{W},M_{H}) \\[1em]
&& \qquad\qquad \mbox{}-2M_{W}^{2}B_{0}(0,M_{W},M_{W}) -
2M_{H}^{2}B_{0}(0,M_{H},M_{H}) \\[1em]
\nn&& \qquad\qquad \mbox{}-\frac{(M_{W}^{2}-M_{H}^{2})^{2}}{k^{2}}
\Bigl( B_{0}(k^{2},M_{W},M_{H}) - B_{0}(0,M_{W},M_{H})\Bigr)
-\frac{2}{3}k^{2} \biggr\}
\Biggr\} .
\eeqar
 
For the self energy of the physical Higgs boson we obtain
\beqar
\nn \quad \lefteqn{\Sigma^{H}_{T}(k^2)  =   - \frac{\alpha}{4\pi} \Biggl\{
\sum_{f,i} N_{C}^{f}\frac{m_{f,i}^{2}}{2s_{W}^{2}M_{W}^{2}}
\Bigl[2A_{0}(m_{f,i}) + (4m_{f,i}^{2} -
k^{2})B_{0}(k^{2},m_{f,i},m_{f,i})\Bigr] }\\[1em]
\nn&& \qquad \mbox{}-\frac{1}{2s_{W}^{2}}
\biggl[\Bigl(6M_{W}^{2}-2k^{2}+\frac{M_{H}^{4}}{2M_{W}^{2}}\Bigr)
B_{0}(k^{2},M_{W},M_{W})
+\Bigr(3+\frac{M_{H}^{2}}{2M_{W}^{2}}\Bigl)A_{0}(M_{W})-6M_{W}^{2}\biggr]
 \\[1em]
\nn&& \qquad \mbox{}-\frac{1}{4s_{W}^{2}c_{W}^{2}}
\biggl[\Bigl(6M_{Z}^{2}-2k^{2}+\frac{M_{H}^{4}}{2M_{Z}^{2}}\Bigr)
B_{0}(k^{2},M_{Z},M_{Z})
+\Bigr(3+\frac{M_{H}^{2}}{2M_{Z}^{2}}\Bigl)A_{0}(M_{Z})-6M_{Z}^{2}\biggr]
 \\[1em]
&&\qquad \mbox{}-\frac{3}{8s_{W}^{2}}
\biggl[3\frac{M_{H}^{4}}{M_{W}^{2}}B_{0}(k^{2},M_{H},M_{H})
+\frac{M_{H}^{2}}{M_{W}^{2}}A_{0}(M_{H})\biggr]
\Biggr\}.
\eeqar
 
The fermion self energies are given by
\beqar
\nn \Sigma^{f,L}_{ij}(p^2) & = &  - \frac{\alpha}{4\pi}
 \Biggl\{ \delta_{ij} Q_{f}^{2} \biggl[ 2 B_1(p^2,m_{f,i},\lambda) +1
 \biggl] \\[1em]
\nn &&\quad\mbox{} +\delta_{ij}(g_{f}^-)^2 \biggl[ 2 B_1(p^2,m_{f,i},M_Z)
 +1 \biggl] \\[1em]
\nn &&\quad\mbox{} + \delta_{ij}\frac{1}{2s_{W}^2} \frac{1}{2}
\frac{m_{f,i}^2}{M_W^2}
  \biggl[ B_1(p^2,m_{f,i},M_Z)  + B_1(p^2,m_{f,i},M_H) \biggl]  \\[1em]
&&\quad\mbox{} + \frac{1}{2s_{W}^2}\sum_{k} V_{ik}V_{kj}^\dagger  \biggl[
 \Bigl( 2+ \frac{m_{f',k}^2}{M_W^2} \Bigl) B_1(p^2,m_{f',k},M_W) +1
\biggl] \Biggl\}, \\[1em]
\nn \Sigma^{f,R}_{ij}(p^2) & = &  - \frac{\alpha}{4\pi}
 \Biggl\{ \delta_{ij}Q_{f}^{2} \biggl[ 2 B_1(p^2,m_{f,i},\lambda) +1
 \biggl] \\[1em]
\nn &&\quad\mbox{} +\delta_{ij}(g_{f}^+)^2 \biggl[ 2 B_1(p^2,m_{f,i},M_Z)
 +1 \biggl] \\[1em]
\nn &&\quad\mbox{} + \delta_{ij}\frac{1}{2s_{W}^2} \frac{1}{2}
\frac{m_{f,i}^2}{M_W^2}
  \biggl[ B_1(p^2,m_{f,i},M_Z)  + B_1(p^2,m_{f,i},M_H) \biggl]\\[1em]
&&\quad\mbox{} + \frac{1}{2s_{W}^2} \frac{m_{f,i}m_{f,j}}{M_W^2}
\sum_{k} V_{ik}V_{kj}^\dagger B_1(p^2,m_{f',k},M_W) \Biggl\}, \\[1em]
\nn \Sigma^{f,S}_{ij}(p^2) & = & - \frac{\alpha}{4\pi}
 \Biggl\{ \delta_{ij}Q_{f}^{2} \biggl[ 4 B_0(p^2,m_{f,i},\lambda)
 -2 \biggl] \\[1em]
\nn &&\quad\mbox{} +\delta_{ij}g_{f}^+ g_{f}^-
\biggl[ 4 B_0(p^2,m_{f,i},M_Z) -2 \biggl] \\[1em]
\nn &&\quad\mbox{}+\delta_{ij} \frac{1}{2s_{W}^2} \frac{1}{2}
\frac{m_{f,i}^2}{M_W^2}
  \biggl[ B_0(p^2,m_{f,i},M_Z) - B_0(p^2,m_{f,i},M_H) \biggl] \\[1em]
&&\quad\mbox{} + \frac{1}{2s_{W}^2} \sum_{k} V_{ik}V_{kj}^\dagger
\frac{m_{f',k}^2}{M_W^2}  B_0(p^2,m_{f',k},M_W) \Biggl\} .
\eeqar
$f'$ is the isospin partner of the fermion $f$ and $N_{C}^{f}$ the
colour factor. $i,\:j,\:k$ run over the fermion generations.
For down-type quarks $ V_{ik}V_{kj}^\dagger$ has to be replaced by 
$ V_{ik}^\dagger V_{kj}$.

The two-point function $B_0$ was given in Sect.~\ref{sectensca}.
For $B_{1}$ we find
\begin{equation} \label{B1}
\barr{lll}
B_1(p^{2},m_{0},m_{1}) &=& \disp
\frac{m_1^2-m_0^2}{2 p^{2}} \Bigl(B_{0}(p^{2},m_0,m_1)-
B_{0}(0,m_0,m_1)\Bigr)\\[1em]
&&-\frac{1}{2}B_0(p^{2},m_0,m_1).
\earr
\end{equation}
 
For the field renormalization constants one needs in addition the
derivatives of the self energies with respect to $k^{2}$ or $p^{2}$,
respectively. These are easily obtained from the expressions above.
$\frac{\partial B_{0}}{\partial p^{2}}$ was given in
Sect.~\ref{sectensca}, $\frac{\partial B_{1}}{\partial p^{2}}$ can be
calculated from (\ref{B1}) as
\beq
\barr{lll}
\disp \frac{\partial}{\partial p^{2}} B_{1}(p^{2},m_{0},m_{1}) &=&
\disp -\frac{m_1^2-m_0^2}{2 p^{4}} \Bigl(B_{0}(p^{2},m_0,m_1)-
B_{0}(0,m_0,m_1)\Bigr)\\[1em]
&&\disp +\frac{m_1^2-m_0^2-p^{2}}{2 p^{2}}
\frac{\partial}{\partial p^{2}} B_{0}(p^{2},m_{0},m_{1}) .
\earr
\eeq
 
These derivatives become IR-singular for $m_{0}^{2}=p^{2}$ and
$m_{1}^{2}=0$ or vice versa. This leads to IR-singular contributions in
the field renormalization constants of charged particles arising from
photonic corrections to the corresponding self energies. Because these
reduce to very simple expressions we give the photonic contributions to
the field renormalization constants of the $W$-boson and the charged
fermions explicitly
\beqar
\delta Z_{W}|_{\mbox{photonic}} &=&
-\frac{\alpha }{\pi }\log\frac{\lambda }{M_{W}}
+\frac{\alpha }{6\pi } \left(\frac{1}{3} + 5\Bigl(\Delta +1 -
\log\frac{M_{W}^{2}}{\mu ^{2}}\Bigr) \right),\\[1em]
\nn \delta Z^{f,L}_{ii}|_{\mbox{photonic}} &=&\delta
Z^{f,R}_{ii}|_{\mbox{photonic}}\\[1em]
&=& -\frac{\alpha }{4\pi }Q_{f}^{2}
\left[\Delta - \log\frac{m_{f,i}^{2}}{\mu ^{2}} +4 +4\log\frac{\lambda
}{m_{f,i}}\right] .
\eeqar

\chapter{Vertex formfactors}
 
The vertex formfactors ${\cal V}$, ${\cal W}$, ${\cal X}$, can be
expressed by the scalar one-loop integrals $B_0(m_{0}^{2},M_1,M_2)$,
$C_0(m_1^2,m_{0}^{2},m_2^2,M_0,M_1,M_2)$
and the scalar coefficients of the
vector and tensor integrals $B_1(m_{0}^{2},M_1,M_2)$,
$C_{i(j)}(m_1^2,m_{0}^{2},m_2^2,M_0,M_1,M_2)$
 
\begin{eqnarray}
{\cal V}_a (m_1^2,m_{0}^{2},m_2^2,M_0,M_1,M_2) & = &  B_0(m_{0}^{2},M_1,M_2)
      -2 - ( M_0^2-m_1^2-M_1^2) C_{1} \hspace{10mm} \mbox{ }\\[1ex]
& & \hspace{-25mm} -(M_0^2 -m_2^2-M_2^2 ) C_{2}
-2 (m_{0}^{2}-m_1^2-m_2^2) (C_{1}+C_{2} +C_0),  \nn \\[1em]
{\cal V}^-_b (m_1^2,m_{0}^{2},m_2^2,M_0,M_1,M_2) & = &
3 B_0(m_{0}^{2},M_1,M_2) +4 M_0^2
 C_0 \nn \\[1ex]
& & \hspace{-20mm} + (4m_1^2+2m_2^2-2m_0^2+M_0^2-M_1^2) C_{1} \\[1ex]
& & \hspace{-20mm} + (4m_2^2+2m_1^2-2m_0^2+M_0^2-M_2^2) C_{2},
  \nn \\[1em]
{\cal V}^+_b (m_1^2,m_{0}^{2},m_2^2,M_0,M_1,M_2) & = & 3 m_1^2 C_0,
\\[1em]
{\cal V}_c (m_1^2,m_{0}^{2},m_2^2,M_0,M_1,M_2) & = &
- 2 \frac{m_1^2 m_2^2}{M_W^2}
(C_{1}+C_{2}+2C_0), \\[1em]
{\cal V}_d (m_1^2,m_{0}^{2},m_2^2,M_0,M_1,M_2) & = &
m_1^2  (C_{1}-C_0) ,\\[1em]
{\cal V}_e (m_1^2,m_{0}^{2},m_2^2,M_0,M_1,M_2) & = &
\frac{m_1^2}{M_W^2} C_{00} , \\[1em]
{\cal V}^-_f (m_1^2,m_{0}^{2},m_2^2,M_0,M_1,M_2) & = &
m_1^2 C_0 + m_2^2 C_{2} , \\[1em]
{\cal V}^+_f (m_1^2,m_{0}^{2},m_2^2,M_0,M_1,M_2) & = & m_1^2  C_{1} , \\[1em]
{\cal W}^-_a (m_1^2,m_{0}^{2},m_2^2,M_0,M_1,M_2) & = &
2 ( C_{1} +C_{2}+C_0) , \\[1em]
{\cal W}^+_a (m_1^2,m_{0}^{2},m_2^2,M_0,M_1,M_2) & = & - 2 C_0 , \\[1em]
{\cal W}^-_b (m_1^2,m_{0}^{2},m_2^2,M_0,M_1,M_2) & = & 3 ( C_{1} +C_{2})
, \\[1em]
{\cal W}^+_b (m_1^2,m_{0}^{2},m_2^2,M_0,M_1,M_2) & = & 3 C_0 , \\[1em]
{\cal W}^-_c (m_1^2,m_{0}^{2},m_2^2,M_0,M_1,M_2) & = & \frac{1}{2M_W^2}
\Bigl[ B_0(m_{0}^{2},M_1,M_2) -1 - M_0^2 ( C_{1} + C_{2}) \Bigl] , \\[1em]
{\cal W}^+_c (m_1^2,m_{0}^{2},m_2^2,M_0,M_1,M_2) & = &
\frac{2}{M_W^2} \Bigl[ m_1^2
C_{1} +m_2^2 C_{2} \Bigl] , \\[1em]
{\cal W}_d (m_1^2,m_{0}^{2},m_2^2,M_0,M_1,M_2) & = &  - C_{2} , \\[1em]
{\cal W}_e (m_1^2,m_{0}^{2},m_2^2,M_0,M_1,M_2) & = &
\frac{-1}{M_W^2}  C_{00} , \\[1em]
{\cal W}^-_f (m_1^2,m_{0}^{2},m_2^2,M_0,M_1,M_2) & = & - C_{1}  , \\[1em]
{\cal W}^+_f (m_1^2,m_{0}^{2},m_2^2,M_0,M_1,M_2) & = &
-C_{2} - C_0 , \\[1em]
{\cal X}^-_a (m_1^2,m_{0}^{2},m_2^2,M_0,M_1,M_2) & = &
-4 \Bigl[  C_{11} +C_{12}
+2 C_{1}+C_{2}+C_0 \Bigl],
 \hspace{10mm}\mbox{ } \\[1em]
{\cal X}^+_a (m_1^2,m_{0}^{2},m_2^2,M_0,M_1,M_2) & = &
4 \Bigl[C_{1}+C_{2}+ C_0
\Bigl]   , \\[1em]
{\cal X}^-_b (m_1^2,m_{0}^{2},m_2^2,M_0,M_1,M_2) & = &
2 \Bigl[ 2 C_{11} +2 C_{12}
-C_{2} \Bigl] , \\[1em]
{\cal X}^+_b (m_1^2,m_{0}^{2},m_2^2,M_0,M_1,M_2) & = &
6 \Bigl[ C_{1}+C_{2} \Bigl]
, \\[1em]
{\cal X}^-_c (m_1^2,m_{0}^{2},m_2^2,M_0,M_1,M_2) & = &
- 2 \frac{m_2^2}{M_W^2} \Bigl[
 C_{22} + C_{12} \Bigl] , \\[1em]
{\cal X}^+_c (m_1^2,m_{0}^{2},m_2^2,M_0,M_1,M_2) & = &
-4 \frac{m_2^2}{M_W^2} C_{2}
, \\[1em]
{\cal X}_d (m_1^2,m_{0}^{2},m_2^2,M_0,M_1,M_2) & = & 2 C_{2} , \\[1em]
{\cal X}^-_e (m_1^2,m_{0}^{2},m_2^2,M_0,M_1,M_2) & = &
\frac{m_1^2}{M_W^2} \Bigl[
    C_{11}+C_{12} +C_{1} \Bigl] , \\[1em]
{\cal X}^0_e (m_1^2,m_{0}^{2},m_2^2,M_0,M_1,M_2) & = &  \Bigl[
    C_{22}+C_{12} +C_{2} \Bigl] , \\[1em]
{\cal X}^+_e (m_1^2,m_{0}^{2},m_2^2,M_0,M_1,M_2) & = &
\frac{m_1^2}{M_W^2} \Bigl[
    C_{1}+C_{2} +C_0 \Bigl] , \\[1em]
{\cal X}_f (m_1^2,m_{0}^{2},m_2^2,M_0,M_1,M_2) & = & - 2 C_{1} .
\end{eqnarray}
 
Using the reduction methods decribed in Chap.~\ref{chatenint}
the vector and tensor coefficients can be expressed by scalar integrals.
For illustration we give the explicit reduction formulae.
 
The vertex function is defined as
\beq
\barr{lll}
C_{\cdots}&=&C_{\cdots}(p_{1},p_{2},M_{0},M_{1},M_{2}) =
C_{\cdots}(m_{1}^{2},m_{0}^{2},m_{2}^{2},M_{0},M_{1},M_{2}) \\[1em]
&=&\disp\frac{(2\pi \mu)^{4-D}}{i\pi^{2}}\int d^{D}q \frac{\cdots}
{[q^{2}-M_{0}^{2}][(q+p_{1})^{2}-M_{1}^{2}][(q+p_{2})^{2}-M_{2}^{2}]}
\earr
\eeq
with
\beq
p_{1}^{2}=m_{1}^{2}, \quad p_{2}^{2}=m_{2}^{2}, \quad
p_{1}p_{2}=-\frac{1}{2}(m_{0}^{2}-m_{1}^{2}-m_{2}^{2}).
\eeq
For the three-point vector functions (\ref{tenst}) yields ($P=1$, $M=N-
1=2$)
\beq
C_{k}=T^{3}_{k}=(X_{2}^{-1})_{kk'}R^{3,k'}
\eeq
with  $k,\: k' = 1,\,2$ and
\beq
X_{2}=
\left(\barr{cc}
m_{1}^{2} & \quad \frac{1}{2}(m_{1}^{2}+m_{2}^{2}-m_{0}^{2}) \\
\frac{1}{2}(m_{1}^{2}+m_{2}^{2}-m_{0}^{2})\quad & m_{2}^{2}
\earr \right) .
\eeq
Evaluating $X_{2}^{-1}$ this gives
\begin{eqnarray}
\nn C_{1} & = &\displaystyle -\frac{4}{\kappa^2} \Bigl[
m_2^2 R^{3,1} +\frac{1}{2} (m_{0}^{2}-m_1^2-m_2^2) R^{3,2} \Bigl] ,\\[1em]
C_{2} & = &\displaystyle -\frac{4}{\kappa^2} \Bigl[
\frac{1}{2} (m_{0}^{2}-m_1^2-m_2^2) R^{3,1} + m_1^2 R^{3,2} \Bigl],
\end{eqnarray}
where
\beq
\kappa = \kappa(m_{0}^{2},m_{1}^{2},m_{2}^{2}) ,
\eeq
from (\ref{kappa}). The $R$'s are obtained from (\ref{tensr}) as
\begin{eqnarray} \hspace{-5mm}
\nn R^{3,1} & = &\displaystyle \frac{1}{2}\Bigl[B_0(m_2^2,M_0,M_2)
- (m_1^2-M_1^2+M_0^2) C_0-B_0(m_{0}^{2},M_2,M_1)\Bigr], \hspace{10mm}
\mbox{ } \\[1ex]
R^{3,2} & = & \frac{1}{2}\Bigl[B_0(m_1^2,M_0,M_1)
- (m_2^2-M_2^2+M_0^2) C_0-B_0(m_{0}^{2},M_2,M_1)\Bigr]   .
\end{eqnarray}
The tensor coefficients are evaluated analogously as ($P=2$, $M=2$)
\begin{eqnarray}
\nn C_{00} & = & \frac{1}{D-2} \Bigl[ R^{3,00} - R^{3,1}_{1} -
R^{3,2}_{2} \Bigr] \\[1em]
C_{ki}&=&T^{3}_{ki}=(X_{2}^{-1})_{kk'}[R^{3,k'}_{i} - \delta _{i}^{k'}
C_{00}]
\end{eqnarray}
or more explicitly
\begin{eqnarray}
\nn C_{00} & = & \frac{1}{4} \Bigl[ B_0(m_{0}^{2},M_2,M_1)
+ (M_0^2-M_1^2+m_1^2) C_{1} \\
\nn & & \hspace{3mm} + (M_0^2-M_2^2+m_2^2) C_{2}
+1 + 2 M_0^2 C_0 \Bigl], \\[1em]
\nn C_{11} & = & -\frac{4}{\kappa^2} \Bigl[
m_2^2 (R_{1}^{3,1}-C_{00})
+\frac{1}{2} (m_{0}^{2}-m_1^2-m_2^2) R_{1}^{3,2} \Bigl] ,\\[1em]
\nn C_{21} & = & -\frac{4}{\kappa^2} \Bigl[
\frac{1}{2} (m_{0}^{2}-m_1^2-m_2^2) (R_{1}^{3,1}-C_{00})
 + m_1^2 R_{1}^{3,2} \Bigl] = \\[1em]
\nn C_{12} & = & -\frac{4}{\kappa^2} \Bigl[m_2^2 R_{2}^{3,1}
+\frac{1}{2} (m_{0}^{2}-m_1^2-m_2^2) (R_{2}^{3,2}-C_{00})\Bigl] ,\\[1em]
C_{22} & = & -\frac{4}{\kappa^2} \Bigl[
\frac{1}{2} (m_{0}^{2}-m_1^2-m_2^2) R_{2}^{3,1}
+ m_1^2 (R_{2}^{3,2}-C_{00}) \Bigl]
\end{eqnarray}
with
\begin{eqnarray} \hspace{-5mm}
\nn R^{3,00} & = & M_0^2 C_{0} + B_0(m_{0}^{2},M_2,M_1), \mbox{\hfill}
\\[1ex]
\nn R_{1}^{3,1} & = & \frac{1}{2} \Bigl[
\phantom{B_1(m_2^2,M_0,M_2)} -(m_1^2-M_1^2+M_0^2) C_{1}
- B_1(m_{0}^{2},M_2,M_1)\Bigr], \mbox{\hfill} \\[1ex]
\nn R_{2}^{3,1} & = & \frac{1}{2}\Bigl[B_1(m_2^2,M_0,M_2)
-(m_1^2-M_1^2+M_0^2) C_{2}+ (B_{0}+B_1)(m_{0}^{2},M_2,M_1)\Bigr] ,
\hspace{10mm} \mbox{ }\\[1ex]
\nn R_{1}^{3,2} & = & \frac{1}{2}\Bigl[B_1(m_1^2,M_0,M_1)
-(m_2^2-M_2^2+M_0^2) C_{1}- B_1(m_{0}^{2},M_2,M_1)\Bigr] ,\\[1ex]
 R_{2}^{3,2} & = & \frac{1}{2}\Bigl[
\phantom{B_1(m_2^2,M_0,M_2)}-(m_2^2-M_2^2+M_0^2) C_{2}
+ (B_{0}+B_1)(m_{0}^{2},M_2,M_1) \Bigr] .
\end{eqnarray}
Note that $C_{12}=C_{21}$ can be calculated in two different ways.
In the evaluation of $C_{00}$ we used (\ref{TNdiv})
 
$B_{1}$ was given in (\ref{B1}).
The results for the scalar integrals can again be found in
Sect.~\ref{sectensca}.

\chapter{Bremsstrahlung integrals}
 
For the decay width of a massive particle with momentum $p_{0}$ and
mass $m_{0}$ into two massive particles with momenta $p_{1}$, $p_{2}$
and masses $m_{1}$, $m_{2}$ and a photon with momentum $q$ and mass
$\lambda $ we need the
following phase space integrals
\beq \label{intbs}
I^{j_{1},\ldots,j_{m}}_{i_{1},\ldots,i_{n}}(m_{0},m_{1},m_{2}) =
\frac{1}{\pi^{2}}\int \frac{d^{3}p_{1}}{2p_{10}}\frac{d^{3}p_{2}}{2p_{20}}
\frac{d^{3}q}{2q_{0}}\delta(p_{0}-p_{1}-p_{2}-q)
\frac{(\pm2qp_{j_{1}})\cdots(\pm2qp_{j_{m}})}
{(\pm2qp_{i_{1}})\cdots(\pm2qp_{i_{n}})}.
\eeq
Here $j_{k},\:i_{l}= 0,\,1,\,2$ and the plus signs belong to
$p_{1},p_{2}$, the minus signs to $p_{0}$.
 
Introducing the abbreviations
\beq
\kappa =\kappa (m_{0}^{2},m_{1}^{2},m_{2}^{2}),
\eeq
as defined in (\ref{kappa}) and
\beq
\barr{llllll}
\beta_{0} &=&
\disp\frac{m_{0}^{2}-m_{1}^{2}-m_{2}^{2}+\kappa }{2m_{1}m_{2}}, \\[1em]
\beta_{1} &=& \disp\frac{m_{0}^{2}-m_{1}^{2}+m_{2}^{2}-\kappa
}{2m_{0}m_{2}}, \quad &
\beta_{2} &=&
\disp\frac{m_{0}^{2}+m_{1}^{2}-m_{2}^{2}-\kappa }{2m_{0}m_{1}},
\earr
\eeq
with
\beq
\beta_{0}\beta_{1}\beta_{2}=1,
\eeq
we get compact expressions for the final results.
From (\ref{intbs}) it is evident that the integrals with the indices 1
and 2 interchanged are obtained by interchanging $m_{1}$ and $m_{2}$.
We list only the independent integrals.
The IR-singular ones are given by
\beqar
I_{00} & = & \frac{1}{4 m_{0}^4} \biggl[ \kappa
\log \Bigl( \frac{\kappa^2}{\lambda m_{0}m_1m_2} \Bigl) - \kappa
-(m_1^2-m_2^2) \log \Bigl(\frac{\beta_{1}}{\beta_{2}}\Bigr)
- m_{0}^2  \log (\beta_{0} )\biggl], \\[1em]
I_{11} & = & \frac{1}{4 m_1^2 m_{0}^2} \biggl[ \kappa
\log \Bigl( \frac{\kappa^2}{\lambda m_{0} m_1 m_2}\Bigl) -\kappa
-(m_{0}^2-m_2^2) \log \Bigl(\frac{\beta_0}{\beta_{2}} \Bigl)
-m_1^2 \log (\beta_1) \biggl], \\[1em]
\nn I_{01} & = & \frac{1}{4m_{0}^2} \biggl[ - 2\log \Bigl(
\frac{\lambda m_{0} m_1m_2}{\kappa^2} \Bigl)
 \log ( \beta_2) +2 \log^2 (\beta _{2}) -
\log^{2}(\beta_0)-\log^{2}(\beta_1) \\[1ex]
&& \qquad\qquad+ 2 Sp( 1 - \beta_{2}^{2}) - Sp( 1-\beta_{0}^{2})
-Sp ( 1-\beta_{1}^{2})\biggr],  \\[1em]
\nn I_{12} & = & - I_{01} - I_{02} \\[1ex]
\nn & = & \frac{1}{4m_{0}^2} \biggl[ - 2 \log \Bigl(
\frac{\lambda m_{0} m_1m_2}{\kappa^2} \Bigl)
 \log ( \beta_0) + 2 \log^2 (\beta _{0}) -
\log^{2}(\beta_1)-\log^{2}(\beta_2) \\[1ex]
&& \qquad\qquad+ 2 Sp( 1 - \beta_{0}^{2}) - Sp( 1-\beta_{1}^{2})
-Sp ( 1-\beta_{2}^{2})\biggr].
\eeqar
 
For the IR finite integrals we obtain
\beqar
\nn I & = & \frac{1}{4m_{0}^2} \biggl[
\frac{\kappa}{2} (m_{0}^2+m_1^2+m_2^2)
+2m_{0}^2 m_1^2 \log( \beta_{2})
+ 2m_{0}^2 m_2^2 \log( \beta_{1}) + 2m_{1}^2 m_2^2 \log( \beta_{0})
\biggl] , \\[1em]
I_{0} & = & \frac{1}{4m_{0}^2} \biggl[ -2m_1^2 \log (\beta_{2})
 - 2m_2^2 \log (\beta _{1}) - \kappa \biggl] , \\[1em]
I_{1} & = & \frac{1}{4m_{0}^2} \biggl[  -2m_0^2 \log (\beta_{2})
 - 2m_2^2 \log (\beta _{0}) - \kappa \biggl] , \\[1em]
\nn I_{0}^{1} & = & \frac{1}{4m_{0}^2} \biggl[ m_1^4 \log (\beta_{2})
 - m_2^2 (2m_{0}^2-2m_1^2+m_2^2) \log (\beta _{1}) \\[1ex]
  && \qquad- \frac{\kappa}{4} ( m_{0}^2-3m_1^2+5m_2^2 )  \biggl], \\[1em]
\nn I_{1}^{0} & = & \frac{1}{4m_{0}^2} \biggl[ m_{0}^4 \log (\beta_{2})
 - m_2^2 (2m_1^2-2m_{0}^2+m_{2}^{2}) \log (\beta_0)  \\[1ex]
&& \qquad - \frac{\kappa}{4} (m_1^2-3m_{0}^2+5m_2^2) \biggl], \\[1em]
\nn I^{1}_{2} & = & - I - I^{0}_{2}  \\[1ex]
\nn &=&  \frac{1}{4m_{0}^2} \biggl[ m_1^4 \log (\beta_{0})
 - m_0^2 (2m_{2}^2-2m_1^2+ m_0^2) \log (\beta _{1}) \\[1ex]
  && \qquad- \frac{\kappa}{4} ( m_{2}^2-3m_1^2+5m_0^2 )  \biggl], \\[1em]
I_{00}^{12} & = & -\frac{1}{4m_{0}^2} \biggl[
   m_1^4 \log (\beta _2) + m_2^4 \log (\beta_{1})
+\frac{\kappa^{3}}{6m_{0}^{2}} + \frac{\kappa }{4}(3m_{1}^2+3m_2^2
-m_0^2) \biggl] , \\[1em]
I^{02}_{11} & = & -\frac{1}{4m_{0}^2} \biggl[
 m_0^4 \log (\beta _2) + m_2^4 \log (\beta_{0})
+\frac{\kappa^{3}}{6m_{1}^{2}} + \frac{\kappa }{4}(3m_{0}^2+3m_2^2
-m_1^2) \biggl] , \\[1em]
I_{11}^{00} & = &  - I^{0}_{1} - I^{02}_{11} =
\frac{1}{4m_{0}^2} \biggl[
  2m_{2}^{2}(m_1^2 + m_{2}^{2} - m_0^2) \log (\beta_0)
+\frac{\kappa^{3}}{6m_{1}^{2}} +2\kappa m_{2}^{2} \biggl],\\[1em]
I_{00}^{11} & = &  - I^{1}_{0} - I^{12}_{00} =
\frac{1}{4m_{0}^2} \biggl[
  2m_{2}^{2}(m_0^2 + m_{2}^{2} - m_1^2) \log (\beta_1)
+\frac{\kappa^{3}}{6m_{0}^{2}} +2\kappa m_{2}^{2} \biggl],\\[1em]
I_{11}^{22} & = &  - I^{2}_{1} - I^{02}_{11} =
\frac{1}{4m_{0}^2} \biggl[
  2m_{0}^{2}(m_0^2 + m_{1}^{2} - m_2^2) \log (\beta_2)
+\frac{\kappa^{3}}{6m_{1}^{2}} +2\kappa m_{0}^{2} \biggl].
\eeqar
Note the symmetries in $0 \leftrightarrow 1$ and $0 \leftrightarrow 2$.


 
\end{document}